\pdfoutput=1
\documentclass[3p]{elsarticle}
\usepackage{graphicx}
\usepackage{amsmath}
\usepackage{amssymb}
\usepackage{psfrag}

\usepackage{color}

\usepackage{epic}
\usepackage{pst-all}

\usepackage[utf8]{inputenc}
\usepackage[T1]{fontenc}
\usepackage{url}

\biboptions{sort&compress}

\journal{Physics Reports}

\begin{document}

\begin{frontmatter}



\title{Ultracold bosons with short-range interaction\\ in regular optical lattices}


\author{Konstantin V.~Krutitsky}


\address
{
Fakult\"at f\"ur Physik der Universit\"at Duisburg-Essen,
Campus Duisburg, Lotharstra{\ss}e 1, 47048 Duisburg, Germany
}

\begin{abstract}
During the last decade, many exciting phenomena have been experimentally observed
and theoretically predicted for ultracold atoms in optical lattices.
This paper reviews these rapid developments concentrating mainly on the theory.
Different types of the bosonic systems in homogeneous lattices of different dimensions
as well as in the presence of harmonic traps are considered.
An overview of the theoretical methods used for these investigations as well as of the obtained results is given.
Available experimental techniques are presented and discussed in connection with theoretical considerations.
Eigenstates of the interacting bosons in homogeneous lattices and in the presence of harmonic confinement are analyzed.
Their knowledge is essential for understanding of quantum phase transitions at zero and finite temperature.
\end{abstract}


\end{frontmatter}

\date{\today}


\tableofcontents

\clearpage

\newcounter{nfig}

\section{Introduction}

Ultracold atoms in optical lattices opened a new era in the study of quantum many-body phenomena.
In contrast to other condensed-matter systems, they provide a unique opportunity of control.
Using interference of laser beams propagating in different directions, one can create various types
of periodic potentials with the amplitude proportional to the laser intensity~\cite{Jessen96,WS13}
that are free of defects and dissipative channels.

Optical lattices provide an efficient tool to control the system dimensionality.
Apart from the three-dimensional geometry~\cite{GMEHB02}, it is possible
to reach very strong spatial confinement in certain spatial directions
and reduce system dimensionality creating single or periodically arranged
linear~\cite{PWMMFCSHB04,SMSKE04,HHMDRGDPN10}
and planar~\cite{SPP07,GZHC09,SWECBK10,BPTMSGFPG10,WESCSFBK11,ZHTC12}
structures.
Rapidly moving laser beams allow to arrange ring lattices and two-dimensional periodic structures
with point-like defects~\cite{HRMB09}.
Disorder with known statistical properties and tunable parameters can be also introduced into the system~\cite{FFI08,SL10,Modugno10,Shapiro12}
either by optical means using incommensurate optical lattices~\cite{DZREFMMI10}
and laser speckles~\cite{DZSZL03}
or through the interaction with other atomic species localized at random positions~\cite{GPRVS11}.

Most of experiments in optical lattices are performed with alkali-metal atoms, mainly with Rb
(see, e.g.,~\cite{GMEHB02,GMGWRHB03,SMSKE04,PWMMFCSHB04,SPP07,SPP08,JCLPPS10,TPGSBPST10,NMDM12})
and also with Li~\cite{CMLSSSXK06,ZMMEM11}, Na~\cite{XLAMCMKJT05,HKGLR06}, K~\cite{RSBHLMHBS13}, Cs~\cite{HHMDRGDPN10,HZHTGC2011,ZHTC12}.
In recent years this list was extended and includes also Yb~\cite{FSSTT09}, which belongs to the atoms of the alkaline-earth metals,
and Cr~\cite{PSCMHPSGVL13} which is a transition metal.
The choice of the atoms is mainly determined by the fact that their electronic transitions lie in a convenient spectral range
allowing efficient manipulation by an optical laser.
The atoms can be trapped either in a single state or in a manifold of the electronic ground states
and cooled below the temperature of quantum degeneracy.
If the total number of electrons, protons and neutrons which constitute the atoms is even,
the latters are bosons, otherwise they are fermions.
In this review, we shall consider only bosons.

Two-body interactions of these atoms, except Cr, are of short range and the effective strength can be controlled
by the intensity of the laser creating the optical lattice or by Feshbach resonances~\cite{KGJ2006,CGJT10,Kokkelmans14}.
The latter is accompanied by the formation of molecules that are converted back to atoms.

In experiments, ultracold atomic system can be controlled on a macroscopic as well as microscopic levels.
Macroscopic measurements are based on the time-of-flight imaging~\cite{GMEHB02,GBMHE01}
and Bragg spectroscopy~\cite{CFFFI2009,CFFFI2010,EGKPLPS2010}
which provide information about the energy spectrum and the state of the system in the momentum space.
More recently, new techniques have been developed to perform in situ measurements
on a microscopic level~\cite{HZHTGC2011} with the spatial resolution of the order of one lattice period or even less.
Tremendous progress has been also achieved in the single-site and single-atom addressability~\cite{GZHC09,IVLBMGS2010,BGPFG09,BPTMSGFPG10,SWECBK10}
which is important for applications in quantum technology.

In deep periodic potentials, atoms can move from one potential well (lattice site)
to the next one by quantum tunneling which gives rise to the discrete lattice models.
In the case of one-component spinless bosons, the lattice system is described by the Bose-Hubbard model.
This model was originally introduced in a rather heuristic manner in order to describe the differences
in the ground state and low-energy excitations of interacting bosons in a homogeneous space
under density variations and the associated solid-superfluid transition in $^4$He~\cite{GK63,GT65,FG66}.
Later it was derived in the context of the solid state physics~\cite{Mullin71} and
motivated by experiments on $^4$He absorbed in porous media or Cooper pairs in granular media~\cite{FWGF89}.
The presence of spin degrees of freedom and Feshbach resonances lead to extensions of the standard Bose-Hubbard model.
In the case of cold atoms, the parameters of the corresponding lattice model can be derived from
first principles which allows direct comparison of the theoretical predictions with experimental data.

A remarkable feature of the Bose-Hubbard model is that it reveals a quantum phase transition from the superfluid
to the Mott insulator~\cite{FWGF89,Sachdev} that results from the competition between the kinetic energy and on-site interaction.
It is characterized by a natural order parameter -- the superfluid fraction.
In the case of spinless bosons, it is a second-order transition.
In the superfluid phase, the spectrum of excitations has no gap and
the particle-number statistics is described by a broad Poisson-like distribution.
In two and three dimensions, the one-body density matrix (two-point correlation function of the first order)
shows the off-diagonal long-range order and decays as a power law in one dimension.
In the Mott-insulator phase, there is a finite gap in the excitation spectrum,
particle-number fluctuations are suppressed, and one-body density matrix decays exponentially in all dimensions.
In low dimensions, Mott insulator possesses nonlocal string order~\cite{TBA2006,BTGA2008}.
Superfluid--Mott-insulator transition in optical lattices has been experimentally observed first in three dimensions~\cite{GMEHB02}
and then in one and two dimensions.
The presence of spin degrees of freedom and Feshbach resonances leads to qualitatively new features.

In spite of a big progress in theory, complete description of the interacting quantum systems is still a challenge.
Though at first glance seemingly very simple, even the standard Bose-Hubbard model is not analytically solvable in general.
Exact analytical solutions are known only in very special situations,
as the case of vanishing or infinitely strong interaction.
Approximate analytical results are obtained by systematic expansions
in powers of small parameters. However, they have always their limitations.
For instance, strong-coupling expansion~\cite{FM94,FM96,DZ06,FKKKT2009}
is valid in arbitrary dimensions and for arbitrary filling factors but limited to
small tunneling rates. In addition, it can be easily implemented only
for fillings close to commensurate due to the degeneracy of the superfluid state.
The expansion in powers of the inverse filling factor~\cite{SUXF06,FSU08}
is valid in arbitrary dimensions and for arbitrary tunneling rates
but cannot be applied if the filling factor is of the order of one.

Parallel to the analytical studies, different exact numerical methods have been developed for the analysis of the Bose-Hubbard model.
The most straightforward and easiest to implement is
exact diagonalization~\cite{LHS88,KLA07,EKO94,KPS95,KPPS96,KS96,RB03r,RB03,RB04,HSTR07,KTEG08,WF2008,ZD10}.
However, due to exponential growth of the Hilbert space with the system size,
the method can be used if the number of particles $N$ and the number of lattices sites $L$ are rather small.
The largest system of bosons reported in the literature was $L=N=18$~\cite{KLA07},
although with the restriction that no more than four particles can occupy one lattice site.

More sophisticated deterministic method is the density-matrix renormalization group~\cite{Sch2005,Sch2011}
which is based on the fact that usually the state of the system occupies only a small subspace
of the exponentially large Hilbert space. This approach is quite successful in one dimension
and allows to treat larger systems ($N$ and $L$ of the order of $1000$~\cite{KWM00}) but it fails in higher dimension,
where quantum Monte Carlo methods~\cite{SBG91,Pollet2012} become more efficient.
By stochastic sampling they allow to treat stationary states in realistic experimental situations
of $N=3\times 10^5$ bosons in a three-dimensional lattice~\cite{TPGSBPST10}.

Mean-field theory plays an important role in the studies of the Bose-Hubbard model.
It is based on the Gutzwiller ansatz~\cite{G63} which takes into account local fluctuations
but neglects quantum correlations of different lattice sites.
This approach is exact in infinite dimensions and provides a useful insight into the physics in large finite dimensions
but it fails for low-dimensional systems.
Attempts to correct the mean-field theory incorporating distance-dependent quantum correlations were undertaken by several authors.
These include random phase approximation~\cite{SD05,MT08}, cluster mean field~\cite{Luehmann13},
method of effective potential~\cite{THHE09R,THHE09}, dynamical mean-field theory (DMFT).
The latter is probably the most successful approach but computationally quite demanding.
It was originally developed for fermions~\cite{GKKR96}
and recently for bosons~\cite{HSH09,LBHH11,LBHH12,AGPTW10,AGPTW11}.
It can be derived as an expansion in powers of the inverse coordination number~\cite{AGPTW11}
and reduces to the solution of an impurity problem on a single site or a cluster of sites.
This is a difficult computational problem which requires application of exact methods like
exact diagonalization, DMRG, QMC.

In recent years, excellent reviews on cold atoms in optical lattices were published~\cite{MO06,LSADSS07,BDZ08,Y09,LSA12,BK14,DGHLLMSZ15}.
However, many important aspects were not properly discussed and the field continues to grow.
The plan of this review is the following.
In section~\ref{sec-SAPP}, we discuss the basic mechanism for the creation of external potentials for neutral polarizable atoms by optical laser fields
and consider eigenstates of single atoms in periodic potentials of different types.
This serves as a preliminary step for the derivation of the Bose-Hubbard models of various types.
In section~\ref{sec-SBI}, we derive the Bose-Hubbard Hamiltonian for the simplest case of spinless bosons and discuss its symmetries.
In section~\ref{sec-BD}, we provide definitions of basic physical quantities that are used in the theory of low-temperature phenomena in a lattice.
Section~\ref{sec-MET} provides theoretical background of the main experimental techniques for cold atoms in optical lattices.
In section~\ref{sec-SSC}, we present exact results for bosonic many-body systems in the simplest special cases that allow analytical treatments.
Section~\ref{sec-PTLSI} gives an overview of the perturbative results for the ground state and lowest excited states
in the regime of strong interactions.
Section~\ref{sec-CPBHMSMIT} is devoted to the criticality of the Bose-Hubbard model and the quantum phase transition
from the superfluid to the Mott insulator.
In section~\ref{sec-ENR}, we present exact numerical results for macroscopic and microscopic quantities across the quantum critical point.
In section~\ref{sec-MFT}, we review the mean-field theory based on the Gutzwiller approximation.
In section~\ref{sec-SBNFR}, we consider a system of lattice bosons near a Feshbach resonance.
Section~\ref{sec-S1B} deals with physics of spin-1 bosons.

\section{\label{sec-SAPP}Single atom in a periodic potential}

\begin{figure}[t]

\centering

\stepcounter{nfig}
\includegraphics[page=\value{nfig}]{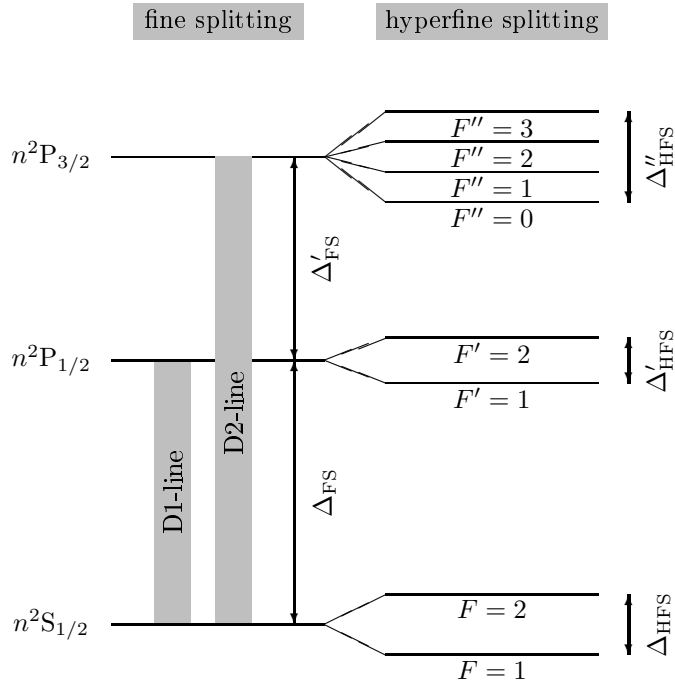}

\caption
{
Scheme of the electronic levels for atoms with nuclear spin $I=3/2$ (not to scale).
}
\label{es}
\end{figure}

Alkali atoms consist of a spherically symmetric atomic residue and one outermost electron (spin $S=1/2$)
in the state with a principal quantum number $n$, that maybe different for different atoms,
and orbital angular momentum $L$.
The full scheme of the relevant electronic levels for atoms
with nuclear spin $I=3/2$ as in the case of $^7$Li, $^{23}$Na, $^{39}$K, $^{41}$K, and $^{87}$Rb,
is shown in Fig.~\ref{es}.
The ground state is an S-state ($L=0$).
Spin-orbit coupling leads to the fine splitting of the first excited level (P-state with $L=1$)
into two states separated by the energy $\Delta_{\rm FS}$.
The states are distinguished by the values of the electronic angular momentum
${\bf J}={\bf L}+{\bf S}$\footnote{For given $L$ and $S$, $J$ takes the values in the range $J=|L-S|,\dots,L+S$.}
and form the D-line doublet $^2 {\rm S}_{1/2} \to {^2}{\rm P}_{1/2}$~(D1),
${^2}{\rm S}_{1/2} \to {^2}{\rm P}_{3/2}$~(D2).\footnote{Here we use a standard spectroscopic notation $n^{2S+1}L_J$ for the states
with the principal quantum number $n$, orbital angular momentum $L={\rm S,P},\dots$, spin $S$, and electronic angular momentum $J$.}
The coupling of the electronic spin to the nuclear spin then leads to the hyperfine splitting
of both ground and excited states with the energies $\hbar\Delta_{\rm HFS}$, $\hbar\Delta_{\rm HFS}'$, and $\hbar\Delta_{\rm HFS}''$.
The additional coupling provides hyperfine levels with the total angular momentum (hyperfine spin)
${\bf F}={\bf J}+{\bf I}$~\footnote{In analogy to the total angular momentum $J$, $F$ takes the values in the range $F=|J-I|,\dots,J+I$.}
which are manifolds of $2F+1$ degenerate states characterized by the magnetic quantum numbers
$m_F=0,\pm 1, \dots, \pm F$.
The energies of the hyperfine splittings are five orders of magnitude smaller than for the fine splitting.

Laser field acting on the atom causes different transitions between electronic levels,
which are determined by the frequency and polarization of the laser wave.
The transitions from ${^2}{\rm S}_{1/2}$ to ${^2}{\rm P}_{1/2}$ and ${^2}{\rm P}_{3/2}$ are electric dipole transitions.
They are allowed if the selection rules $\Delta m_F=0$ for linear polarization
or $\Delta m_F=\pm 1$ for circular polarization are fulfilled~\cite{MS99}.
If the detuning of the laser frequency $\omega_L$ is much larger than the spontaneous emission rate,
one can adiabatically eliminate all the excited states in the spectrum of atoms
denoted by $F'$ and $F''$ in Fig.~\ref{es}.
This leads to the effective potential acting only on the ground state sublevels labeled by $F$
(see, e.g.,~\cite{GWO00,DDCN04} and references therein)
\begin{equation}
V^{\rm las}_{\alpha\beta}({\bf x})
=
\sum_{\gamma}
\frac
{
\left(
    {\bf E}({\bf x})
    \cdot
    {\bf d}_{\gamma\alpha}
\right)^*
\left(
    {\bf E}({\bf x})
    \cdot
    {\bf d}_{\gamma\beta}
\right)
}
{
\hbar
\left(
    \omega_L-\omega_\gamma
\right)
}
\;,
\end{equation}
where ${\bf E}({\bf x})$ is the electric field strength of the laser field,
${\bf d}_{\gamma\alpha}$ is the dipole matrix element between the ground state sublevel
$\alpha$ and the excited state sublevel $\gamma$ of energy $\hbar\omega_\gamma$.
This allows to create controlled potentials for neutral polarizable atoms which can be of
completely different types ranging from random and quasi-random to perfectly periodic.
In this review, we will be dealing with the potentials of the latter type known under the name optical lattices.
The overview of the geometries of the optical lattices was given in
Refs.~\cite{PCG94,Jessen96,GR01,WS13}.
Here we consider mainly hypercubic lattices but focus more on the effects coming from the interference
of the excited electronic levels which have various manifestations depending on the laser frequency as well as polarization.

\subsection{\label{1Dlattice}One-dimensional lattice in the case of large detuning}

We consider a pair of counterpropagating laser beams with the wavevectors ${\bf k}_{\rm L}$ and $-{\bf k}_{\rm L}$
along the $x_1$-direction. If the detuning is much larger than the hyperfine splitting of the electronic levels,
this laser configuration does not lead to any coupling of the internal ground states.
It creates a one-dimensional periodic potential which is the same for all ground-state sublevels and has the form
\begin{equation}
\label{pp1d}
V_{\rm L}(x)
=
V_0
\cos^2
\left(
    \pi \frac{x}{a}
\right)
\;,
\end{equation}
where $a=\pi/k_{\rm L}=\lambda_{\rm L}/2$ is the period (lattice constant).
If the two laser beams intersect at an angle $\varphi<\pi$, one can create a one-dimensional lattice with a larger period given by
$a=\lambda_{\rm L}/\left(2\sin(\varphi/2)\right)$.
Using this technique, optical lattices with $a$ up to $80\ \mu$m were demonstrated in experiments with $^{87}$Rb in the field of Ti:Sa laser
emitting at the wavelength $\lambda_{\rm L}=820$~nm~\cite{FFLI05}.

\subsubsection{\label{sBB}Bloch bands}

We suppose that the system consists of $L$ potential wells and impose periodic
boundary conditions on the wavefunction of the atom $\psi(x+La)=\psi(x)$
which satisfies the Schr\"odinger equation
\begin{equation}
\label{Sch-eq}
\left[
    -
    \frac{\hbar^2}{2M} \frac{d^2}{dx^2}
    +
    V_{\rm L}(x)
\right]
\psi(x)
=
E
\psi(x)
\;,\quad
x \in
\left[
    -\frac{La}{2},\frac{La}{2}
\right]
\;.
\end{equation}
According to the Bloch theorem, the solution has the following form
\begin{equation}
\label{psib}
\psi(x)
\equiv
\psi_b(x;k)
=
u_b(x;k)
e^{ikx}
\;,\quad
E
\equiv
E_b(k)
\;,
\end{equation}
where $u_b(x;k)$ is a periodic function of $x$ with the period $a$ and $b$ is the band index.
The wavenumber $k\equiv k_q=2\pi q/(La)$ takes in general discrete values determined
by the integer $q$ which is defined up to modulo $L$.
If we do not want to care about the differences between even and odd $L$, we can assume that $q=0,\dots,L-1$.
In this case, $k=0,\dots,2\pi(L-1)/(La)$.
However, usually the first Brillouin zone (1BZ) is defined as $k\in[-\pi/a,\pi/a]$ and we will also follow this convention.
In the limit of infinite lattice ($L\to\infty$), $k$ becomes a continuous variable.

The solution of the eigenvalue problem can be expressed in terms of Mathieu functions.
Despite they are rather well studied in the mathematical literature~\cite{McLachlan,MS54,MSW80},
exact results can be obtained only numerically.
One can use {\it Mathematica} (see, e.g., Ref.~\cite{DLT13} for notes on that)
but in order to have full flexibility it is better to write an own program, for instance, in C/C++ or Fortran.
With this purpose in mind we use the Fourier series expansion
\begin{equation}
\label{ub}
u_b(x;k)
=
\frac{1}{\sqrt{a}}
\sum_{n=-\infty}^\infty
c_{b n}(k)
\exp
\left(
    i 2 \pi n \frac{x}{a}
\right)
\;,
\end{equation}
where the coefficients $c_{b n}$ are the solutions of the eigenvalue problem
\begin{eqnarray}
\label{evp-bands}
&&
\sum_{n'=-\infty}^\infty
{\cal H}_{nn'}(k)
c_{b n'}(k)
=
E_b(k)
c_{b n}(k)
\;,\quad
n=-\infty,\dots,\infty
\;,
\\
&&
{\cal H}_{nn'}(k)
=
\left[
    E_{\rm R}
    \left(
        \frac{ka}{\pi} + 2 n
    \right)^2
    +
    \frac{V_0}{2}
\right]
\delta_{nn'}
+
\frac{V_0}{4}
\left(
    \delta_{n,n'-1}
    +
    \delta_{n,n'+1}
\right)
\;,
\nonumber
\end{eqnarray}
where $E_{\rm R}=\hbar^2 k_{\rm L}^2/(2M)$ is the recoil energy.
They satisfy the orthonormality condition
\begin{equation}
\sum_{n=-\infty}^\infty
c_{b_1 n}^*(k)
c_{b_2 n}(k)
=
\delta_{b_1 b_2}
\;.
\end{equation}
The solutions of Eq.~(\ref{evp-bands}) are periodic functions of $k$ with the period $2\pi/a$.
This is because the shift of the wavenumber $k\to k+2\pi/a$ can be compensated by the corresponding shift of the index $n\to n-1$.
Using this property one can show that $\psi_b(x;k_q)$ and $\psi_b(x;k_q')$ are orthogonal, unless $(q'-q)/L$ is an integer.

Since matrix ${\cal H}(k)$ is real and symmetric, the coefficients $c_{b n}(k)$ can be chosen to be real which provides unique solutions.
In addition, it is tridiagonal and the eigenvalue problem can be solved numerically using efficient algorithms~\cite{NumRec}.
In addition, the off-diagonal terms in ${\cal H}(k)$ coincide with the lowest-order approximation of the second derivative via
a finite difference and the diagonal terms are the same as a discrete harmonic potential.
Therefore, one can expect that the coefficients $c_{bn}$ decrease exponentially with $|n|$ and the infinite-dimensional matrix
${\cal H}(k)$ can be safely truncated to a moderate finite dimension.
The fact that the system is finite leads only to the discretization of $k$ but does not change the values of $c_{b n}(k)$ and $E_b(k)$.

\begin{figure}[t]

\centering

\stepcounter{nfig}
\includegraphics[page=\value{nfig}]{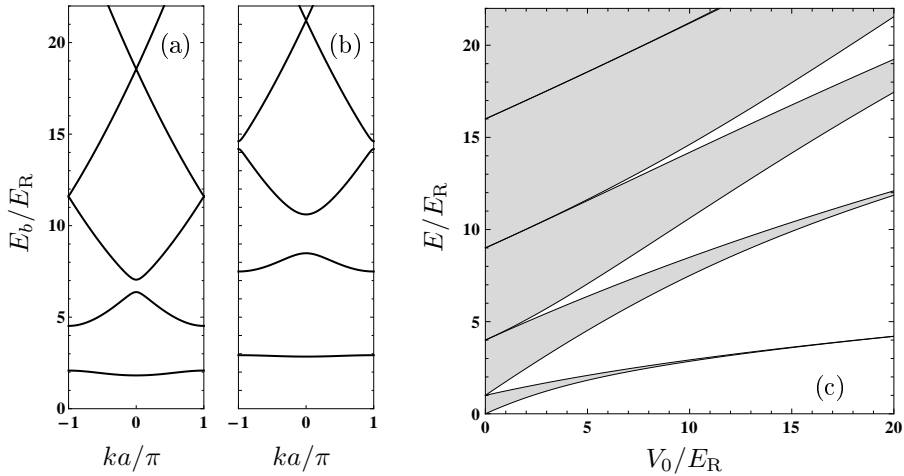}

\caption
{
(a)
Lowest Bloch bands for $V_0/E_{\rm R}=5$~(a), $10$~(b).
(c)
Regions of allowed (gray) and forbidden (white) energies.
}
\label{bands}
\end{figure}

Energy spectrum $E_b(k)$ within the first Brillouin zone, is shown in Fig.~\ref{bands}.
At each value of $k$ the spectrum is discrete and all the eigenvalues
are distinct~\cite{AS72}.
The functions $E_b(k)$ take their extremal values at $k=0,\pm\pi/a$.
$k=0$ is a minimum for even $b$ and maximum, if $b$ is odd.
With the increase of the amplitude of the periodic potential $V_0$,
the energy bands $E_b(k)$ become more flat and the gaps between the bands grow.
In the limit $V_0\gg E_{\rm R}$, the width of the bands is given by
the asymptotic expression~\cite{MS54}
\begin{eqnarray}
\label{w_large_V0}
&&
\frac
{
\left|
    E_b(\pi/a)
    -
    E_b(0)
\right|
}
{E_{\rm R}}
=
\frac{2^{3b+4}}{b!\sqrt{\pi}}
\left(
    \frac{V_0}{E_{\rm R}}
\right)^{\frac{b}{2}+\frac{3}{4}}
\exp
\left(
    -2
    \sqrt{\frac{V_0}{E_{\rm R}}}
\right)
\nonumber\\
&&
\times
\left[
    1 -
    \frac{6 b^2 + 14 b + 7}{16}
    \sqrt{\frac{E_{\rm R}}{V_0}}
    +
    O
    \left(
        \frac{E_{\rm R}}{V_0}
    \right)
\right]
\;,\quad
b=0,1,\dots
\end{eqnarray}

\begin{figure}[t]

\centering

\stepcounter{nfig}
\includegraphics[page=\value{nfig}]{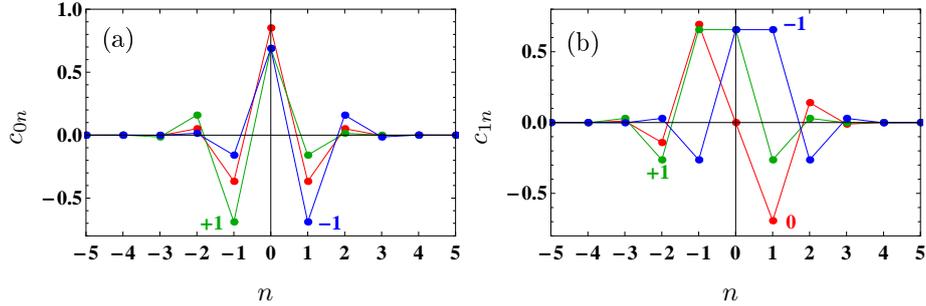}

\caption
{
(color online)
Eigenstates of Eq.~(\ref{evp-bands}) for the lowest two bands with $b=0$~(a) and $b=1$~(b).
In both panels, $V_0=10\,E_{\rm R}$ and $ka/\pi=0$~(red), $+1$~(green), $-1$~(blue).
The lines are guide to the eye.
}
\label{cbn}
\end{figure}

The coefficients $c_{bn}(k)$ which represent the eigenstates in Eq.~(\ref{evp-bands}) are shown in Fig.~\ref{cbn}
for the lowest two bands ($b=0,1$). For $b=0$ and $k=0$, $c_{bn}$ is an even function of $n$.
However, if we move towards the edges of the Brillouin zone, this symmetry is destroyed.
Due to the analogy of Eq.~(\ref{evp-bands}) to the Schr\"odinger equation for the harmonic oscillator mentioned above,
one would expect that $c_{0n}$ should be positive. The fact that $c_{0n}$ take negative values for odd $n$'s
is simply because we have chosen $V_0>0$. In the opposite case ($V_0<0$), $c_{0n}$ are indeed always positive.
In the next energy band ($b=1$), $c_{bn}(0)$ is an odd function of $n$.
$c_{1n}(k)$ becomes symmetric with respect to $n=\pm 0.5$ for $ka/\pi=\mp 1$.
Similar features can be observed in the higher energy bands.

\subsubsection{\label{sWF}Wannier functions}

Bloch functions $\psi_b(x;k)$ are extended over the whole lattice for any $b$ and $k$.
An alternative basis suitable for the description of single particles at individual lattice sites
is provided by Wannier functions defined via the Fourier transform~\cite{Wannier}
\begin{equation}
\label{Wannier-def}
W_{b\ell}(x)
\equiv
W_{b}(x-x_\ell)
=
\frac{1}{L}
\sum_{k\in 1{\rm BZ}}
\psi_b(x;k)
e^{-i k x_\ell}
\;,
\end{equation}
where $x_\ell=x_0+a\ell$, with $\ell$ being an integer, are the minima of the periodic potential.
$x_0=a/2$ if $V_0$ in Eq.~(\ref{pp1d}) is positive but $x_0=0$ for negative $V_0$.
The summation in Eq.~(\ref{Wannier-def}) is over the values of $k$ within the first Brillouin zone.
The functions~(\ref{Wannier-def}) satisfy the orthonormality condition~\cite{Koster}
\begin{equation}
\label{W_ortho}
\int_{-\frac{La}{2}}^\frac{La}{2}
W_{b_1\ell_1}^*(x)
W_{b_2\ell_2}(x)
\,dx
=
\delta_{b_1 b_2}
\delta_{\ell_1\ell_2}
\end{equation}
and form a complete set~\cite{KS54}.
They possess the symmetry
$W_{b}(-x)=(-1)^b W_{b}(x)$.
In finite lattices, $W_b(x)$ are periodic functions: $W_b(x+La)=W_b(x)$.
In the limit of infinite lattice, the sum in Eq.~(\ref{Wannier-def}) can be replaced by the integral:
\begin{equation}
\frac{1}{L}
\sum_{k\in 1{\rm BZ}}
\to
\frac{a}{2\pi}
\int_{-\pi/a}^{\pi/a}
dk
\;.
\nonumber
\end{equation}
As it was proven in Ref.~\cite{Kohn59} for a general case of separated energy bands in one dimension,
the Wannier functions are uniquely defined by their symmetry properties and asymptotic behavior at large distances
(see the discussion below) that guaranties their minimal width.
In the following we shall consider the Wannier functions for the lowest Bloch band $W_0(x)$.

\begin{figure}[t]

\centering

\stepcounter{nfig}
\includegraphics[page=\value{nfig}]{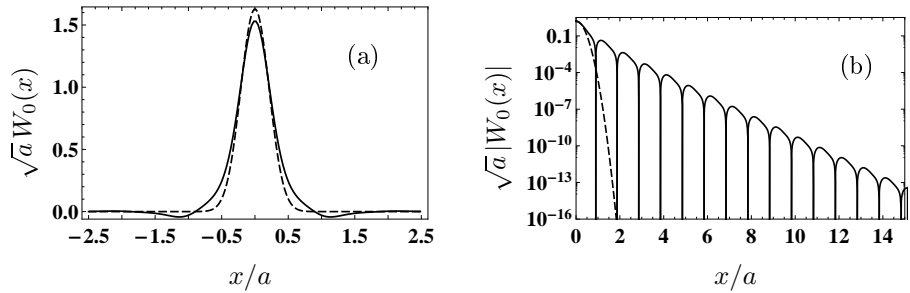}

\caption
{
Wannier function in the first Bloch band for $V_0=5\;E_{\rm R}$.
Solid line is exact result and dashed line is a Gaussian approximation~(\ref{WGauss}).
(b) is the same as (a) but with a logarithmic scale.
}
\label{Wannier_V0_5}
\end{figure}

In the limit of vanishing potential ($V_0\to0$), the eigenvalue problem~(\ref{evp-bands})
has a very simple analytical solution which leads to the following result for an infinite lattice~\cite{Parzen53}:
\begin{equation}
\label{W0_V=0}
W_0(x)
=
\frac{1}{\sqrt{a}}
\frac
{
\sin
\left(
    \pi x/a
\right)
}
{\pi x/a}
\;.
\end{equation}
This function oscillates with the amplitude decreasing with the distance $x$,
and this type of behavior is typical for the Wannier functions (see Fig.~\ref{Wannier_V0_5}).
We would like to stress that $W_0(x)$ is not a ground-state eigenfunction of any Hamiltonian
and the nodes appear to be necessary in order to satisfy the orthogonality condition~(\ref{W_ortho}).

At finite $V_0$, $1/x$ decay of the envelope of the function $W_0(x)$ is preserved only for $|x|\ll x_{\rm c}$.
For $|x|\gg x_{\rm c}$, the asymptotics of the envelope acquires a different form~\cite{Kohn59,HV01}:
\begin{equation}
\label{W0largex}
W_0(x)
\sim
|x|^{-3/4}
\exp
\left(
    -h_0 |x|
\right)
\;.
\end{equation}
The crossover distance $x_{\rm c}$, which is infinite for $V_0=0$, becomes finite for nonvanishing $V_0$ and decreases with $V_0$.
$h_0$ is a constant which vanishes in the limit $V_0\to 0$ and grows with the lattice depth $V_0$.
For shallow and deep lattices it can be calculated analytically~\cite{Kohn59} and the result reads~\cite{BGHH07}
\begin{eqnarray}
\frac{h_0 a}{\pi}
=
\left\{
    \begin{tabular}{cl}
    ${V_0}/{E_{\rm R}}$
    &
    for $V_0\ll E_{\rm R}$\;,
    \\
    $
        \sqrt{{V_0}/{E_{\rm R}}}
        -
        {1}/{4}
    $
    &
    for $V_0\gg E_{\rm R}$\;.
    \end{tabular}
\right.
\end{eqnarray}

\begin{figure}[t]

\centering

\stepcounter{nfig}
\includegraphics[page=\value{nfig}]{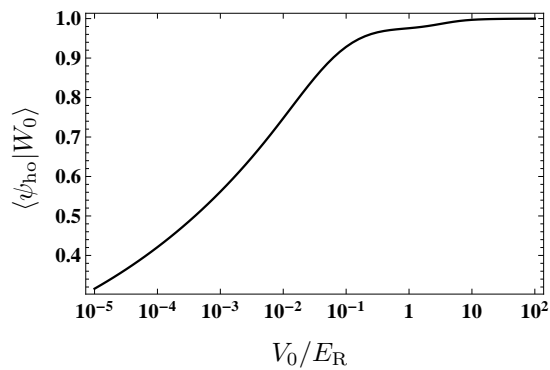}

\caption
{
Overlap of the Wannier function for the lowest Bloch band with the Gaussian approximation~(\ref{WGauss}).
$\langle \psi_{\rm ho}|W_0\rangle$ vanishes for $V_0=0$ and grows rapidly with the increase of $V_0$.
}
\label{overlap}
\end{figure}

In the case of a deep optical lattice, each lattice site can be described by a harmonic potential with the frequency~\cite{Slater52}
\begin{equation}
\label{omega_ho}
\omega_{\rm ho}
=
2\frac{E_{\rm R}}{\hbar}\sqrt{\frac{V_0}{E_{\rm R}}}
\;.
\end{equation}
Then the solution~(\ref{psib}) of the Schr\"odinger equation~(\ref{Sch-eq}) can be approximated
by the eigenfunctions of the harmonic oscillator. This leads to the Gaussian
approximation for the lowest-band Wannier function
\begin{equation}
\label{WGauss}
W_0(x)
\approx
\psi_{\rm ho}(x)
=
\left(
    \frac{1}{\pi a_{\rm ho}^2}
\right)^{1/4}
\exp
\left(
    - \frac{x^2}{2 a_{\rm ho}^2}
\right)
\end{equation}
of the width
\begin{equation}
a_{\rm ho}
=
\sqrt{\frac{\hbar}{M \omega_{\rm ho}}}
=
\frac{a}{\pi}
\left(
    \frac{E_{\rm R}}{V_0}
\right)^{1/4}
\;.
\end{equation}
We compare it with exact $W_0(x)$ in Fig.~\ref{Wannier_V0_5}.
One can see that the Gaussian approximation overestimates the maximum height and fails to reproduce
the detailed structure of the Wannier functions. In addition, it violates the orthogonality condition~(\ref{W_ortho}) at any finite $V_0$.
The Gaussian approximation becomes exact only in the limit of infinitely large $V_0$, when (\ref{WGauss})
takes the form of the $\delta$-function.
Nevertheless, the overlap with the exact Wannier function
\begin{equation}
\langle \psi_{\rm ho}|W_0\rangle
=
\int_{-\infty}^\infty
\psi_{\rm ho}^*(x)
W_0(x)
\,dx
\end{equation}
is close to one even if the amplitude $V_0$ of the periodic potential is of the order of few recoil energies
(see Fig.~\ref{overlap} and Ref.~\cite{BGHH07}).
Due to its simplicity, the Gaussian approximation is often used in order to obtain analytical estimations of the parameters of
the Bose-Hubbard model~\cite{JBCGZ98,OSS01,Zwerger03,Zwerger04,BBZ03,MS04,MGI06,BGHH07,BS10,LZJ12}.

\subsubsection{\label{SecTunMat}Tunneling matrix}

In the basis of the Wannier functions the single-particle Hamiltonian can be represented in the form of the tunneling matrix ${\cal J}$
with the matrix elements defined as
\begin{equation}
\label{Jmat}
{\cal J}_{\ell_1 \ell_2}^{b_1 b_2}
=
\int_{-La/2}^{La/2}
W_{b_1 \ell_1}^*(x)
\left[
    -
    \frac{\hbar^2}{2M}
    \frac{\partial^2}{\partial x^2}
    +
    V_{\rm L}(x)
\right]
W_{b_2 \ell_2}(x)
\,dx
\;.
\end{equation}
Using the definition of the Wannier functions and the Bloch theorem,
one can show that the matrix elements depend on the distance $s=\left|\ell_1-\ell_2\right|$
and do not vanish only for $b_2=b_1\equiv b$.
Eq.~(\ref{Jmat}) can also be rewritten in the form
\begin{equation}
\label{Jsb}
{\cal J}_{s}^{b}
=
\frac{1}{L}
\sum_{k\in 1{\rm BZ}}
E_b(k)
\exp
\left(
    i ka s
\right)
\;,
\end{equation}
where we observe that ${\cal J}_{s}^{b}$ is a Fourier transform of the band structure.
${\cal J}_{0}^{b}$ is the average energy of the band.

Numerical calculations show that for $s>0$ the sign of ${\cal J}_{s}^{b}$ alternates with the distance $s$ and with the band index $b$.
The latter can be easily seen for $s=1$ in the limit of infinite lattice. Replacing the sum in Eq.~(\ref{Jsb}) by the integral
and integrating by parts, we obtain
\begin{equation}
{\cal J}_1^b
=
-\frac{1}{\pi}
\int_0^{\pi/a}
\frac{dE_b(k)}{dk}
\sin(ka)
dk
\;.
\end{equation}
Since $\sin(ka)\ge0$ in the integration interval, the sign of ${\cal J}_1^b$ is determined by the derivative $d E_b(k)/dk$
which is positive for even $b$ and negative for odd $b$. In the lowest Bloch band, ${\cal J}_1^{b=0}$ is negative.
Since this quantity plays an important role, we give it a special notation $J\equiv-{\cal J}_1^{0}$.

\begin{figure}[t]

\centering

\stepcounter{nfig}
\includegraphics[page=\value{nfig}]{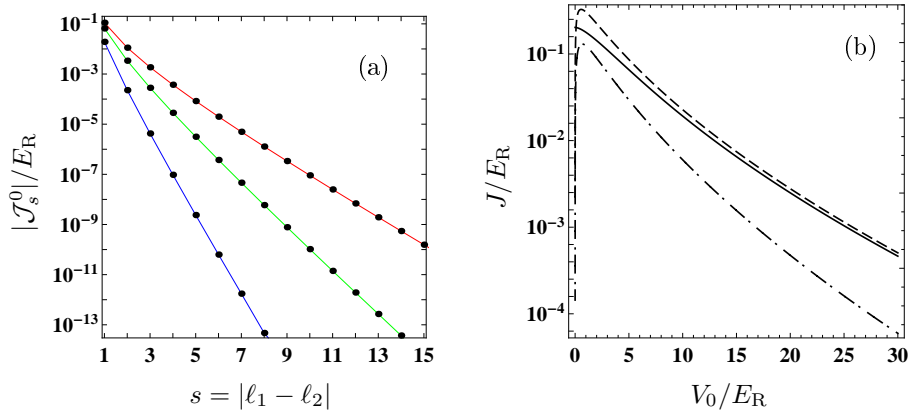}

\caption
{
(color online)
Tunneling matrix element in the lowest Bloch band.
(a)
The distance dependence for (from top to the bottom)
$V_0/E_{\rm R}=3$ (red), $5$ (green), $10$ (blue).
(b)
Tunneling matrix element for the nearest lattice sites.
Solid line is exact numerical result.
Calculations within the Gaussian approximation [Eqs.~(\ref{JGauss})] are shown by dashed-dotted line.
Dashed line is the asymptotic formula~(\ref{Jasymp}).
}
\label{tunmat}
\end{figure}

Typical behavior of ${\cal J}_{s}^{b=0}$ is shown in Fig.~\ref{tunmat}.
It is a decreasing function of $V_0$ and $s$. Asymptotics of ${\cal J}_{s}^{0}$ at large distances $s$
is similar to that of the Wannier functions and given by~\cite{HV01}
\begin{equation}
{\cal J}_{s}^{0}
\sim
|as|^{-3/2}
\exp
\left(
    -h_0 |as|
\right)
\;,
\end{equation}
where the constant $h_0$ is the same as in Eq.~(\ref{W0largex}).

The dependence of $J$ on the lattice amplitude $V_0$ is shown in Fig.~\ref{tunmat}(b).
In the limit $V_0\to0$, $J$ can be calculated analytically using Eq.~(\ref{W0_V=0}) and the result is
\begin{equation}
\label{J_V0_0}
\lim_{V_0\to0}
\frac{J}{E_{\rm R}}
=
\frac{2}{\pi^2}
\;.
\end{equation}
In the tight-binding limit $V_0\gg E_{\rm R}$ which leads to Eq.~(\ref{HBH}),
the width of the lowest Bloch band $E_0(\pi/a)-E_0(0)=4J$.
Then from Eq.~(\ref{w_large_V0}) we obtain asymptotic expression
\begin{equation}
\label{Jasymp}
\frac{J}{E_{\rm R}}
\approx
\frac{4}{\sqrt{\pi}}
\left(
    \frac{V_0}{E_{\rm R}}
\right)^{3/4}
\exp
\left(
    - 2 \sqrt{\frac{V_0}{E_{\rm R}}}
\right)
\;.
\end{equation}
As one can see in Fig.~\ref{tunmat}(b), Eq.~(\ref{Jasymp}) gives quite accurate results for $V_0/E_{\rm R}\gtrsim 10$.

In the Gaussian approximation~(\ref{WGauss}) which is supposed to be valid for large $V_0$, the tunneling matrix element is given by
\begin{equation}
\label{JGauss}
\frac{J}{E_{\rm R}}
\approx
\left(
    \frac{\pi^2}{4} - 1
\right)
\frac{V_0}{E_{\rm R}}
\exp
\left(
    -\frac{\pi^2}{4}
    \sqrt{\frac{V_0}{E_{\rm R}}}
\right)
\;.
\end{equation}
Although this equation describes correctly qualitative behavior,
the magnitude of $J$ appears to be underestimated compared to the exact numerical results.

Exact numerical results for $J(V_0)$ were also fitted by the function with three parameters
\begin{equation}
\label{Jfit}
\frac{J}{E_{\rm R}}
=
p_1
\left(
\frac{V_0}{E_{\rm R}}
\right)^{p_2}
\exp
\left(
    -p_3
    \sqrt{\frac{V_0}{E_{\rm R}}}
\right)
\end{equation}
which is motivated by Eqs.~(\ref{Jasymp}) and (\ref{JGauss}).
For instance, in Ref.~\cite{Rey04} we find $p_1=1.39666$, $p_2=1.051$, $p_3=2.12104$,
and Ref.~\cite{GWFMGB05a} suggests $p_1=1.43$, $p_2=0.98$, $p_3=2.07$.
Although both sets of parameters give indeed quite a high accuracy for $V_0/E_{\rm R}\gtrsim 3$,
the function~(\ref{Jfit}) fails to reproduce the correct behavior for small $V_0$.
Here we suggest a fit which has more parameters but works very well for small and for large $V_0$:
\begin{equation}
\frac{J}{E_{\rm R}}
=
p_1
\left(
\frac{V_0}{E_{\rm R}}
\right)^{p_2}
\exp
\left[
    -p_3
    \left(
        \frac{V_0}{E_{\rm R}}
    \right)^{p_4}
\right]
+
\frac{2}{\pi^2}
\exp
\left[
    -p_5
    \left(
        \frac{V_0}{E_{\rm R}}
    \right)^{p_6}
\right]
\end{equation}
with $p_1=0.116828$, $p_2=1.16938$, $p_3=1.11717$, $p_4=0.63$, $p_5=0.369658$, $p_6=1.01448$.

\subsection{Multi-dimensional lattices}

The generalization of the theory of one-dimensional lattices presented in Section~\ref{1Dlattice}
to arbitrary dimension $d$ can be done employing a standard approach from the solid state physics~\cite{AM76}.
Let ${\bf x}_{\bf l}$ denote the global minima of the periodic potential $V_{\rm L}({\bf x})$.
We assume that the vectors ${\bf x}_{\bf l}$ form a Bravais lattice and, therefore, have the form
${\bf x}_{\bf l}=\sum_{\nu=1}^d \ell_\nu{\bf a}_\nu$,
where ${\bf a}_\nu$ are primitive vectors that are in general not orthogonal to each other.
The potential $V_{\rm L}({\bf x})$ can be represented in the form of a Fourier series
\begin{equation}
\label{VL-Fourier}
V_{\rm L}({\bf x})
=
\sum_{\bf j}
\tilde V_{\bf j}
\exp
\left(
    i {\bf g}_{\bf j} \cdot {\bf x}
\right)
\;,
\end{equation}
where the coefficients $\tilde V_{\bf j}$ are given by
\begin{equation}
\label{Vj}
\tilde V_{\bf j}
=
\frac{1}{v}
\int_v
d{\bf x}
\exp
\left(
    - i {\bf g}_{\bf j} \cdot {\bf x}
\right)
V_{\rm L}({\bf x})
\;.
\end{equation}
${\bf g}_{\bf j}$'s in Eqs.~(\ref{VL-Fourier}),~(\ref{Vj}) are vectors of the reciprocal lattice determined by the conditions
$\exp\left(i {\bf g}_{\bf j} \cdot {\bf x}_{\bf l}\right)=1$.
In terms of the primitive vectors of the reciprocal lattice ${\bf b}_\nu$,
defined by the identities ${\bf a}_{\nu_1}\cdot{\bf b}_{\nu_2}=2\pi\delta_{\nu_1\nu_2}$,
${\bf g}_{\bf j}$ have the following representation:
${\bf g}_{\bf j}=\sum_{\nu=1}^d j_\nu{\bf b}_\nu$.
The integration in Eq.~(\ref{Vj}) is over one primitive cell of the volume $v$.

According to the Bloch theorem, the wavefunction of the stationary Schr\"odinger equation has the form analogous to (\ref{psib}),~(\ref{ub}):
\begin{eqnarray}
\psi_b({\bf x};{\bf k})
=
u_b({\bf x};{\bf k})
e^{i{\bf k}\cdot{\bf x}}
\;,\quad
u_b({\bf x};{\bf k})
=
\frac{1}{\sqrt{v}}
\sum_{\bf j}
c_{b{\bf j}}({\bf k})
\exp
\left(
    i{\bf g}_{\bf j}\cdot{\bf x}
\right)
\;.
\end{eqnarray}
Imposing Born-von~Karman boundary conditions $\psi_b({\bf x}+L_\nu{\bf a}_\nu;{\bf k})=\psi_b({\bf x};{\bf k})$,
we deduce that the wavevector ${\bf k}$ takes the values
\begin{equation}
{\bf k}_{\bf q}
=
\sum_{\nu=1}^d
\frac{q_\nu}{L_\nu}
{\bf b}_\nu
\;,\quad
q_\nu \in \mathbb{Z}
\;.
\end{equation}
The coefficients $c_{b{\bf j}}({\bf k})$ and the energy eigenvalues $E_b({\bf k})$ are obtained from the solution of the eigenvalue problem
\begin{equation}
\frac{\hbar^2}{2M}
\left(
    {\bf g}_{\bf j}
    +
    {\bf k}
\right)^2
c_{b{\bf j}}({\bf k})
+
\sum_{{\bf j}'}
\tilde V_{{\bf j}'}
c_{b,{\bf j}-{\bf j}'}({\bf k})
=
E_b({\bf k})
c_{b{\bf j}}({\bf k})
\;.
\end{equation}
The Wannier functions can be determined in analogy to Eq.~(\ref{Wannier-def}) as
\begin{equation}
W_b({\bf x}-{\bf x}_{\bf l})
=
\left(
    \prod_{\nu=1}^d
    L_\nu
\right)^{-1}
\sum_{{\bf k}\in{1{\rm BZ}}}
\psi_b({\bf x};{\bf k})
\,
e^{-i{\bf k}\cdot{\bf x}_{\bf l}}
\end{equation}
with the orthonormality condition similar to Eq.~(\ref{W_ortho}).
They allow to define the tunneling matrix ${\cal J}_{{\bf l}_1{\bf l}_2}^b$.

In what follows, we restrict ourselves to hypercubic lattices that are created by $d$ pairs of laser beams propagating along the $x_\nu$ axes.
In order to create square or cubic lattices one has to avoid the interference of laser beams propagating in the
orthogonal directions which is achieved if their frequencies are sufficiently different.
This setup allows to create multi-dimensional lattices described by the potential
\begin{equation}
V_{\rm L}({\bf x})
=
\sum_{\nu=1}^d
V_{0\nu}
\cos^2
\left(
    \pi \frac{x_\nu}{a}
\right)
\;,
\end{equation}
where $d=1,2,3$ is the lattice dimension.
The variables $x_\nu$ in the Schr\"odinger equation can be separated and the solutions
are obtained from the one-dimensional ones according to the rules
\begin{equation}
\label{Ebd}
E
\equiv
E_{\bf b}({\bf k})
=
\sum_{\nu=1}^d
E_{b_\nu}(k_\nu)
\;,\quad
\psi_{\bf b}({\bf x};{\bf k})
=
\prod_{\nu=1}^d
\psi_{b_\nu}
\left(
    x_\nu;k_\nu
\right)
\;.
\end{equation}
Then the Wannier functions will be also given by the products
\begin{equation}
\label{Wannier-d}
W_{{\bf b} {\bf l}}({\bf x})
=
\prod_{\nu=1}^d
W_{b_\nu \ell_\nu}
\left(
    x_\nu
\right)
\;.
\end{equation}
Multidimensional analogues of Eqs.~(\ref{Jmat}),~(\ref{Jsb}) together with Eqs.~(\ref{Ebd}),~(\ref{Wannier-d})
allow to express the tunneling matrix in terms of the one-dimensional quantities as
\begin{equation}
\label{Jijb}
{\cal J}_{{\bf l}_1 {\bf l}_2}^{\bf b}
=
\sum_{\nu=1}^d
{\cal J}_{{l}_{1\nu} {l}_{2\nu}}^{b_\nu}
\prod_{\nu'\ne\nu}
\delta_{{l}_{1\nu'},{l}_{2\nu'}}
\;,
\end{equation}
which formally shows that ${\cal J}_{{\bf l}_1{\bf l}_2}^{\bf b}$ do not vanish only along the lattice axes.

\subsection{State-dependent potentials}

\begin{figure}[t]
\centering

\stepcounter{nfig}
\includegraphics[page=\value{nfig}]{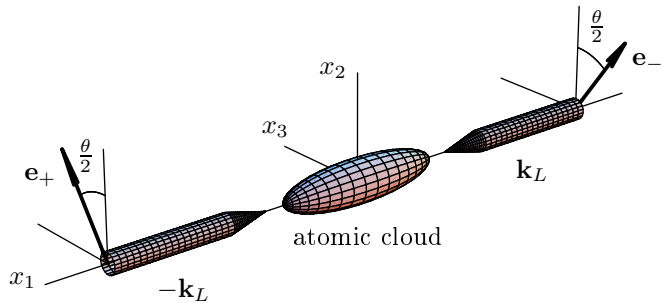}

\caption
{
(color online)
Lin-$\theta$-lin laser configuration.
}
\label{ltl}
\end{figure}

We consider two counterpropagating linearly polarized laser waves of equal amplitudes and frequencies with the wave number $k_{\rm L}$,
and the angle $\theta$ (with $0\le\theta\le\pi/2$) between the polarization vectors
\begin{eqnarray}
{\bf e}_\pm
=
\cos\frac{\theta}{2}\;{\bf e}_2
\pm
\sin\frac{\theta}{2}\;{\bf e}_3
\;.
\end{eqnarray}
The running laser waves form left- and right- polarized standing waves.
This setup shown in Fig.~\ref{ltl} is called lin-$\theta$-lin laser configuration~\cite{GR01}.
Here we consider a one-dimensional lattice but generalizations to higher dimensions are also possible~\cite{GR01,KG04}.

If the laser is tuned between the ${\rm P}_{1/2}$ and ${\rm P}_{3/2}$ electronic levels, the effective potentials acting
on the ground-state sublevels of ${\rm S}_{1/2}$ are different~\cite{BCJD99,JBCGZ99,GMGWRHB03,GMGWRHB03n,JZ05}.
For instance,
\begin{eqnarray}
V_{|F=2,m_F=\pm2\rangle}(x)
&=&
V_{\pm}(x)
\;,
\\
V_{|F=1,m_F=\pm1\rangle}(x)
&=&
\left[
    3 V_{\pm}(x) + V_{\mp}(x)
\right]/4
\;,
\nonumber
\end{eqnarray}
where $V_\pm(x)=V_0\cos^2(k_{\rm L} x \pm \theta/2)$.
These potentials have the same period $a=\pi/k_{\rm L}$.
For $\theta=0$, they coincide but for other values of $\theta$ their amplitudes are different and the positions of minima are shifted.
This can be easily seen, if we rewrite $V_{|1,\pm1\rangle}(x)$ in the form
\begin{eqnarray}
&&
V_{|F=1,m_F=\pm1\rangle}(x)
=
V_0^{\rm eff}
\cos^2(k_{\rm L} x \pm \phi)
+A_0
\;,
\\
&&
V_0^{\rm eff}
=
\frac{V_0}{2}
\sqrt{1+3\cos^2\theta}
\;,\quad
\phi
=
\arctan
\left(
    \frac{1}{2}
    \tan\theta
\right)
\;,
\nonumber\\
&&
A_0
=
\frac{V_0}{4}
\left(
    2 - \sqrt{1+3\cos^2\theta}
\right)
\;.
\nonumber
\end{eqnarray}
The band structure and all other single-particle states remain the same as in the case of large detuning
discussed in section~\ref{1Dlattice}.

\subsection{\label{lin-t-lin}Atoms with coupled ground states}

\begin{figure}[t]
\centering

\stepcounter{nfig}
\includegraphics[page=\value{nfig}]{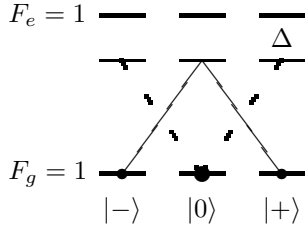}

\caption
{
Scheme of the electronic transitions for lin-$\theta$-lin laser configuration.
}
\label{ltl-scheme}
\end{figure}

We consider again lin-$\theta$-lin laser configuration and turn to the case
when the laser detuning $\Delta$ is comparable to the hyperfine splitting of the electronic levels
but still larger than the spontaneous emission rate.
Assuming that only the ground states with $F=1$ are populated and the laser frequency is close
to the $F=1\to F'=1$ transition frequency of the D1-line,
the left- and right- polarized standing laser waves
will couple internal ground and excited states with magnetic quantum numbers $m_F=0,\pm1$
by $V$ and $\Lambda$ transitions, see Fig.~\ref{ltl-scheme}.

The $V$ and $\Lambda$ laser-induced transitions lead to two sets
of orthogonal Bloch eigenmodes which we denote by the indices $0$ and $\Lambda$, respectively.
The solutions of the Schr\"odinger equation are the three-component spinors of the form
\begin{equation}
\label{3c-spinors}
{\bf \Psi}^{(0)}=(0,\psi_{0},0)^T
,
\quad
{\bf \Psi}^{(\Lambda)}=(\psi_{+},0,\psi_{-})^T
\;,
\end{equation}
which allow also to determine the Wannier spinors as well as the tunneling matrices for both types of modes
according to Eqs.~(\ref{Wannier-def}),~(\ref{Jmat}),~(\ref{Jsb}).

\subsubsection{$0$-modes}

The effective potential acting on the atoms in the internal state $\alpha=0$ is given by
\begin{equation}
\label{VB}
V_{\rm B}(x)
=
\frac{V_0}{2}
\left[
    1
    +
    \cos\theta
    \cos
    \left(
        2\pi\frac{x}{a}
    \right)
\right]
\;,\quad
a=\pi/k_{\rm L}
\;.
\end{equation}
Therefore, these modes are exactly the same as those discussed
in Sections~\ref{sBB},~\ref{sWF}, provided that the amplitude $V_0$ is replaced by $V_0\cos\theta$.

\subsubsection{$\Lambda$-modes}

The non-vanishing components of the spinor ${\bf \Psi}^{(\Lambda)}$
are solutions of the Schr\"odinger equation with the effective potential
which has the form of a $2\times 2$ matrix:
\begin{equation}
\label{s-pot}
\hat V_{\rm L}(x)
=
\frac{V_0}{2}
\left(
    \begin{array}{cc}
       \Omega_+^2        &  \Omega_+ \Omega_-\\
       \Omega_+ \Omega_- &  \Omega_-^2
    \end{array}
\right)
\;,\quad
\Omega_\pm(x)
=
\cos
\left(
    \pi\frac{x}{a} \pm \frac{\theta}{2}
\right)
\;.
\end{equation}
Due to the periodicity of the potential~(\ref{s-pot}),
they can be written down in the form of Eqs.~(\ref{psib}),~(\ref{ub})
with the scalars $u_b$ and $c_{bn}$ replaced by two-component vectors
${\bf u}_b^{(\Lambda)}$ and ${\bf c}_{bn}^{(\Lambda)}$,
where the coefficients ${\bf c}_{b n}$ are solutions of the eigenvalue problem
\begin{eqnarray}
\label{evp-bands-Lambda}
&&
\sum_{n'=-\infty}^\infty
{\bf H}_{nn'}^{(\Lambda)}
{\bf c}_{b n'}^{(\Lambda)}
=
E_b^{(\Lambda)}
{\bf c}_{b n}^{(\Lambda)}
\;,
\\
&&
{\bf H}_{nn'}^{(\Lambda)}
=
\left[
    E_{\rm R}
    \left(
        \frac{k}{k_{\rm L}} + 2 n
    \right)^2
    +
    \frac{V_0}{4}
    \left(
        \begin{array}{cc}
           1          & \cos\theta\\
           \cos\theta & 1
        \end{array}
    \right)
\right]
\delta_{nn'}
\nonumber\\
&&
+
\frac{V_0}{8}
\left[
    \left(
        \begin{array}{cc}
           e^{i\theta} & 1\\
           1            & e^{-i\theta}
        \end{array}
    \right)
    \delta_{n',n-1}
    +
    \left(
        \begin{array}{cc}
           e^{-i\theta} & 1\\
           1            & e^{i\theta}
        \end{array}
    \right)
    \delta_{n',n+1}
\right]
\;.
\nonumber
\end{eqnarray}

\begin{figure}[t]

\centering

\stepcounter{nfig}
\includegraphics[page=\value{nfig}]{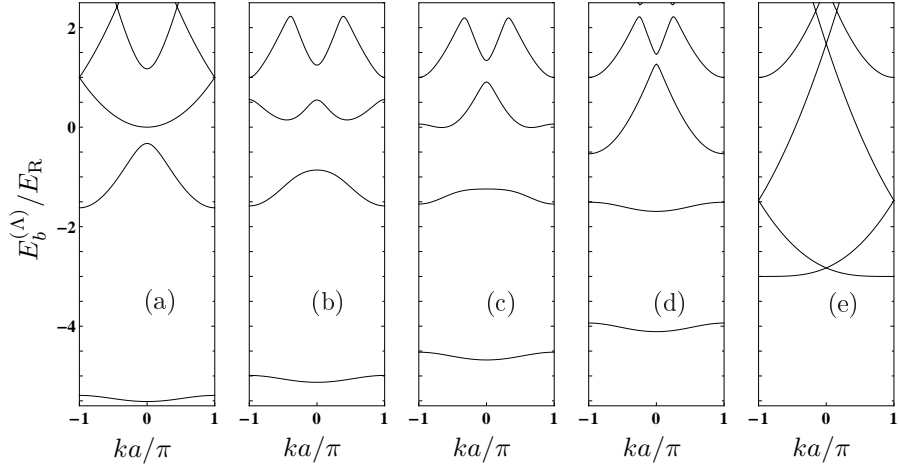}

\caption
{
Band structure of the $\Lambda$-modes.
$V_0=-8\,E_{\rm R}$,
$\theta=0^\circ$~(a), $30^\circ$~(b), $45^\circ$~(c), $60^\circ$~(d), $90^\circ$~(e).
}
\label{L-bands-neg}
\end{figure}

\begin{figure}[t]

\centering

\stepcounter{nfig}
\includegraphics[page=\value{nfig}]{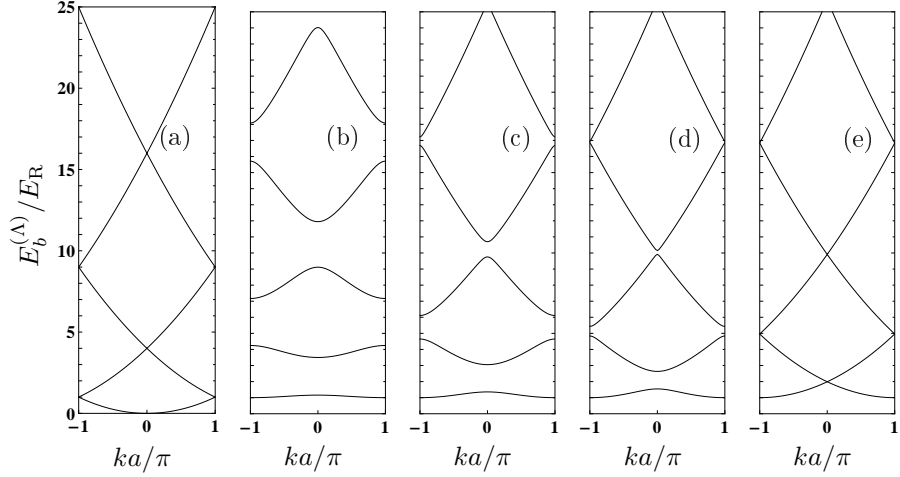}

\caption
{
Band structure of the $\Lambda$-modes.
$V_0=800\,E_{\rm R}$,
$\theta=0^\circ$~(a), $30^\circ$~(b), $45^\circ$~(c), $60^\circ$~(d), $90^\circ$~(e).
}
\label{L-bands-pos}
\end{figure}

The lowest Bloch bands obtained by the numerical solution of Eq.~(\ref{evp-bands-Lambda})
are shown in Figs.~\ref{L-bands-neg},~\ref{L-bands-pos} (see also Refs.~\cite{MDTZSP94,DO,DTR99}).
In contrast to the spinless case considered in section~\ref{sBB} the bands can overlap.
There is a strong dependence on the angle $\theta$ and the results are drastically different for positive and negative $V_0$.
In the case of positive $V_0$ the band gaps remain of the order of $E_{\rm R}$ or vanish even for very large values of $V_0$
in contrast to the case of negative $V_0$.

\begin{figure}[t]
\centering

\stepcounter{nfig}
\includegraphics[page=\value{nfig}]{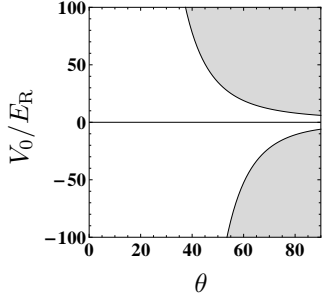}

\caption
{
Boundary between the regions of the normal and anomalous dispersion in the lowest Bloch band
for the $\Lambda$-modes obtained from the condition $E_0^{(\Lambda)}(0)=E_0^{(\Lambda)}(\pi/a)$.
In the shaded region, $E_0^{(\Lambda)}(0)>E_0^{(\Lambda)}(\pi/a)$.
}
\label{nadispersion}
\end{figure}

The principal difference between the cases of positive and negative $V_0$
can be understood if we apply the unitary transformation
\begin{equation}
\label{U}
\hat {\cal U}
=
\frac{1}{\Omega}
\left(
    \begin{array}{rr}
        \Omega_+ & \Omega_- \\
       -\Omega_- & \Omega_+
    \end{array}
\right)
\,,\quad
\Omega
=
\sqrt{\Omega_+^2 + \Omega_-^2}
\;,
\end{equation}
to the spinor $(\psi_+,\psi_-)^T$.
After the transformation we end up with the bright and dark states~\cite{DO},
which are not degenerate in contrast to the original ones.
The important point is that only the bright state is directly coupled to
the electromagnetic field and influenced by the potential~(\ref{VB}).

We consider first the case $\theta=0$, when the transformation $\hat {\cal U}$ does not depend on the position $z$.
In this case the Hamiltonian matrix ${\bf H}^{(\Lambda)}$ is diagonal in the basis of bright and dark states
and the dark state does not ``feel" any periodic potential.
Since for $V_0>0$ the dark state has lower energy than the bright state the ground state is the same as for free atoms.
If $V_0<0$, the situation is reversed:
The energy of the bright state is lower than that of the dark one and only
the bright state is populated by the atoms. Therefore, increasing $|V_0|$
one can strongly influence the lowest energy bands as in the case of spinless atoms.

In the case $\theta \neq 0$, $\hat {\cal U}$ is a position-dependent transformation.
The atomic center-of-mass motion leads to the gauge potential
\begin{equation}
\label{Vg}
V_{\rm g}
=
E_{\rm R}
\left[
    \frac
    {\sin\theta}
    {
     1+
     \cos\theta
     \cos
     \left(
         2\pi{x}/{a}
     \right)
    }
\right]^2
\,,
\end{equation}
acting on the bright and dark atomic states and to the motional coupling
of the states~\cite{DO}. The transformation $\hat {\cal U}$ does not allow to diagonalize the Hamiltonian
in the case $\theta \neq 0$. Nevertheless, it helps to understand what is going on, assuming that
$\left|V_0\right| \gg E_{\rm R}$.
In this approximation $V_{\rm g} \ll \left|V_{\rm B}\right|$,
and one can neglect the gauge potential for the bright state as well as
the motional coupling between the bright and dark
states~\cite{DO}. Then the only potentials acting on the bright and dark states are
given by Eqs.~(\ref{VB}) and ({\ref{Vg}}), respectively.
On the basis of the same argument as in the case $\theta=0$ we see that
the low-energy eigenstates in the cases $V_0<0$ and $V_0>0$ are determined by the potentials $V_{\rm B}$ and $V_{\rm g}$, respectively.
Accordingly in the case $V_0<0$ the quantity $V_0\cos\theta$ defines the strength of the periodic potential,
while in the opposite case the potential does not depend on $V_0$.

This simplified description provides a correct physical insight but does not describe
such an important feature of the $\Lambda$-modes as the change of the type of the dispersion relation
$E_0^{(\Lambda)}(k)$ in the lowest Bloch band under variation of the angle $\theta$.
In the case of spinless atoms, one always has a normal dispersion in the lowest Bloch band, i.e., $d E_0/dk \ge 0$ for $0 \le k \le \pi/a$.
In the case we are dealing with, one can get anomalous dispersion, i.e., $d E_0^{(\Lambda)}/dk < 0$ for $0 \le k \le \pi/a$, as well,
see Figs.~\ref{L-bands-neg}(e),~\ref{L-bands-pos}(b-e).
The change of the dispersion type happens at the points $(V_0,\theta)$ indicated in Fig.~\ref{nadispersion} by the solid lines
separating white and shaded areas.

\begin{figure}[t]
\centering

\stepcounter{nfig}
\includegraphics[page=\value{nfig}]{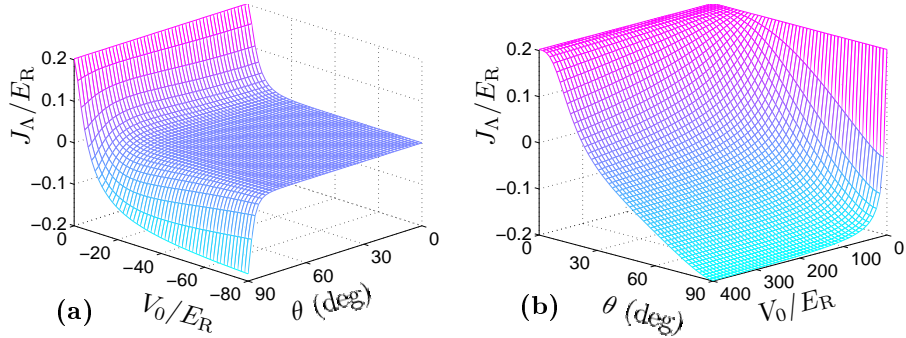}

\caption
{
(color online)
Tunneling matrix element for the $\Lambda$-modes in the lowest Bloch band
for the nearest neighbors defined by Eq.~(\ref{JLambda}) in the case of negative (a) and positive (b) $V_0$.
}
\label{JL_t_V0}
\end{figure}

The change of the dispersion types leads to the fact that the tunneling matrix elements
for the nearest-neighboring sites in the lowest Bloch band
\begin{equation}
\label{JLambda}
J_\Lambda
=
-\int
{\bf W}_{\ell+1}^{(\Lambda)\dagger}(x)\cdot
\left(
-
\frac{\hbar^2}{2M}
\frac{\partial^2}{\partial x^2}
+
\hat V_{\rm L}
\right)
\cdot
{\bf W}_{\ell}^{(\Lambda)}(x)
\,dx
\;,
\end{equation}
can take positive and negative values.
This can be seen from Eq.~(\ref{Jsb}), which remains valid in the present situation as well,
and it is demonstrated in Fig.~\ref{JL_t_V0} by exact numerical calculations.
For negative $V_0$ and small $\theta$, $J_\Lambda \approx J_0$ but in general they are completely different.

\section{\label{sec-SBI}Spinless bosons with interactions}

We consider spin-polarized interacting bosons of the mass $M$ in a periodic potential $V_{\rm L}({\bf x})$.
In all experiments with cold atoms in optical lattices there is also a (harmonic) trapping potential
which can be of two different types. One of them is produced by the magneto-optical trap in three dimensions
and has trapping frequencies of the order of $10\dots100$~Hz. This kind of trapping leads to the inhomogeneity
of density profile of the atomic cloud which extends over several tens of the lattice periods.
Another type of trapping can be produced by an additional optical lattice of large amplitude $V_{0\perp}$
in one or two dimensions which suppresses tunnelings in the corresponding directions and leads to effective
trapping frequencies $\omega_\perp$ up to $\sim 1$~MHz as described by Eq.~({\ref{omega_ho}}).
The latter provides an opportunity to create
(quasi) two-dimensional~\cite{GZHC09,SWECBK10,BPTMSGFPG10,WESCSFBK11,ZHTC12}
and (quasi) one-dimensional~\cite{PWMMFCSHB04,SMSKE04,HHMDRGDPN10}
lattice systems and to reach experimentally the Tonks-Girardeau regime~\cite{PWMMFCSHB04,KWW04}.

In order to take into account all these possibilities, we will denote by $V_{\rm T}({\bf x})$
the trapping potential of the first type and assume that the system's dimension $d$ can be less than three.
Therefore, ${\bf x}$ is a vector in a $d$-dimensional space with $d=1,2,3$.
In the second quantization, the Hamiltonian has the form
\begin{eqnarray}
\label{Hs0}
\hat H
&=&
\int
\hat\Psi^\dagger({\bf x})
\left[
    -
    \frac{\hbar^2}{2M} \nabla^2
    +
    V_{\rm L}({\bf x})
    +
    V_{\rm T}({\bf x})
\right]
\hat\Psi({\bf x})
\,d{\bf x}
\\
&+&
\frac{1}{2}
\int
\int
\hat\Psi^\dagger({\bf x})
\hat\Psi^\dagger({\bf x}')
    V_{\rm at}({\bf x}-{\bf x}')
\hat\Psi({\bf x}')
\hat\Psi({\bf x})
\,d{\bf x}
\,d{\bf x}'
\;,
\nonumber
\end{eqnarray}
where $\hat\Psi({\bf x})$ is a field operator which annihilates one atom at the point ${\bf x}$.
If the atomic interactions are short-range, they can be described by two-body contact potential
\begin{equation}
\label{Vat}
V_{\rm at}({\bf x}-{\bf x}')
=
g_d
\delta({\bf x}-{\bf x}')
\;,\quad
g_d
=
\frac
{g_3}
{
 \left(
     a_\perp \sqrt{2\pi}
 \right)^{3-d}
}
\;,\quad
g_3
=
\frac{4\pi\hbar^2 a_{\rm s}}{M}
\;,
\end{equation}
with $a_{\rm s}$ being s-wave scattering length in three dimensions and
$a_\perp$ the harmonic oscillator length corresponding to the frequency $\omega_\perp$.
Eq.~(\ref{Vat}) implies that the transverse confinement is not too tight, otherwise,
confinement-induced resonances~\cite{HGMDHPN09,HMHDRMSN10}
may lead to modifications of the interaction parameter $g_d$~\cite{Olshanii98,DLO01,HHZETC13}.

\subsection{Derivation of the Bose-Hubbard model}

The matter-field operator can be written down in terms of the Wannier functions as
\begin{equation}
\label{FO}
\hat\Psi({\bf x})
=
\sum_{{\bf b},{\bf l}}
W_{\bf b l}({\bf x})
\hat a_{\bf b l}
\;,
\end{equation}
where the annihilation and creation operators for the band ${\bf b}$ at site ${\bf l}$,
$\hat a_{\bf b l}$ and $\hat a_{\bf b l}^\dagger$,
obey the bosonic commutation relations
\begin{equation}
\left[
    \hat a_{{\bf b}_1 {\bf l}_1}
    ,
    \hat a_{{\bf b}_2 {\bf l}_2}^\dagger
\right]
=
\delta_{{\bf b}_1 {\bf b}_2}
\delta_{{\bf l}_1 {\bf l}_2}
\;,\quad
\left[
    \hat a_{{\bf b}_1 {\bf l}_1}
    ,
    \hat a_{{\bf b}_2 {\bf l}_2}
\right]
=
0
\;.
\end{equation}
Substituting (\ref{FO}) into the second-quantized Hamiltonian of interacting bosons
in continuum~(\ref{Hs0}), we obtain its discrete representation
\begin{eqnarray}
\label{Hs0d}
\hat H
=
\sum_{\bf b}
\sum_{{\bf l}_1 {\bf l}_2}
{\cal J}_{{\bf l}_1 {\bf l}_2}^{\bf b}
\hat a_{{\bf b} {\bf l}_1}^\dagger
\hat a_{{\bf b} {\bf l}_2}^{\phantom{\dagger}}
+
\sum_{{\bf b}_1 {\bf b}_2}
\sum_{{\bf l}_1 {\bf l}_2}
v_{{\bf l}_1 {\bf l}_2}^{{\bf b}_1 {\bf b}_2}
\hat a_{{\bf b}_1 {\bf l}_1}^\dagger
\hat a_{{\bf b}_2 {\bf l}_2}^{\phantom{\dagger}}
+
\frac{1}{2}
\sum_{{\bf b}_1 {\bf b}_2 {\bf b}_3 {\bf b}_4}
\sum_{{\bf l}_1 {\bf l}_2 {\bf l}_3 {\bf l}_4}
U_{{\bf l}_1 {\bf l}_2 {\bf l}_3 {\bf l}_4}^{{\bf b}_1 {\bf b}_2 {\bf b}_3 {\bf b}_4}
\hat a_{{\bf b}_1 {\bf l}_1}^\dagger
\hat a_{{\bf b}_2 {\bf l}_2}^\dagger
\hat a_{{\bf b}_3 {\bf l}_3}^{\phantom{\dagger}}
\hat a_{{\bf b}_4 {\bf l}_4}^{\phantom{\dagger}}
\;,
\end{eqnarray}
\begin{equation}
\label{epsb}
v_{{\bf l}_1 {\bf l}_2}^{{\bf b}_1 {\bf b}_2}
=
\int
W_{{\bf b}_1 {\bf l}_1}^*({\bf x})
V_{\rm T}({\bf x})
W_{{\bf b}_2 {\bf l}_2}({\bf x}')
\,d{\bf x}
\;,
\end{equation}
\begin{equation}
\label{Ulb}
U_{{\bf l}_1 {\bf l}_2 {\bf l}_3 {\bf l}_4}^{{\bf b}_1 {\bf b}_2 {\bf b}_3 {\bf b}_4}
=
\int
\!\!\!
\int
W_{{\bf b}_1 {\bf l}_1}^*({\bf x})
W_{{\bf b}_2 {\bf l}_2}^*({\bf x}')
    V_{\rm at}({\bf x}-{\bf x}')
W_{{\bf b}_3 {\bf l}_3}({\bf x}')
W_{{\bf b}_4 {\bf l}_4}({\bf x})
\,d{\bf x}
\,d{\bf x}'
\;.
\end{equation}
By doing different approximations in the Hamiltonian~(\ref{Hs0d}) one can derive lattice models of different types,
such as multiband~\cite{LCM09,LDZ13} or extended~\cite{LJS12} Bose-Hubbard models.
These non-standard quantum lattice models were recently reviewed in Ref.~\cite{DGHLLMSZ15}.

The Hamiltonian~(\ref{Hs0d}) can be simplified in the tight-binding limit $V_0\gg E_{\rm R}$
when the width of the Bloch bands becomes small and the gaps grow.
In this regime we can keep only the terms corresponding to the lowest Bloch band ${\bf b}=0$.
This is valid, if all the energy scales are less than the energy gap separating the first two Bloch bands.
Since we will be dealing further only with the lowest Bloch band, we will drop the band index ${\bf b}$.

In this regime, the tunneling matrix elements ${\cal J}_s^0$ with $s\ge 2$ become much smaller than ${\cal J}_1^0$ (see Fig.~\ref{tunmat})
and, therefore, can be neglected. If the interactions are short-range, we can keep only local terms in the second part of Eq.~(\ref{Ulb}).
Since the spatial scale of the potential $V_{\rm T}({\bf x})$ is much larger than the lattice period $a$
and the Wannier functions are strongly localized, one can take $V_{\rm T}({\bf x})$ in Eq.~(\ref{epsb})
out of the integral and use the orthonormality condition~(\ref{W_ortho}).
In the isotropic cubic lattice, these approximations lead to the standard Bose-Hubbard Hamiltonian~\cite{FWGF89}
\begin{eqnarray}
\label{HBH}
\hat H_{\rm BH}
=
-J
\sum_{\nu=1}^d
\sum_{\bf l}
\left(
    \hat a_{\bf l}^\dagger
    \hat a_{{\bf l}+{\bf e}_\nu}^{\phantom{\dagger}}
    +
    {\rm h.c.}
\right)
+
\frac{U_d}{2}
\sum_{\bf l}
\hat a^\dagger_{\bf l}
\hat a^\dagger_{\bf l}
\hat a_{\bf l}^{\phantom{\dagger}}
\hat a_{\bf l}^{\phantom{\dagger}}
+
\sum_{\bf l}
V_{\rm T}({\bf x}_{\bf l})
\hat a^\dagger_{\bf l}
\hat a_{\bf l}^{\phantom{\dagger}}
\;,
\end{eqnarray}
where ${\bf e}_\nu$ is a unit vector on the lattice in the direction $\nu$,
$
J
\equiv
-{\cal J}_{s=1}^0
$
is the tunneling matrix element for the nearest neighbors,
$
U_d
\equiv
U_{\bf l\,l\,l\,l}^{0000}
$
is the on-site atom-atom interaction energy.
For contact interaction~(\ref{Vat}) it takes the form
\begin{equation}
\label{U_d}
U_d
=
g_d
\int
\left|
    W_{\bf l}({\bf x})
\right|^4
\,d{\bf x}
\;.
\end{equation}

\begin{figure}[t]

\centering

\stepcounter{nfig}
\includegraphics[page=\value{nfig}]{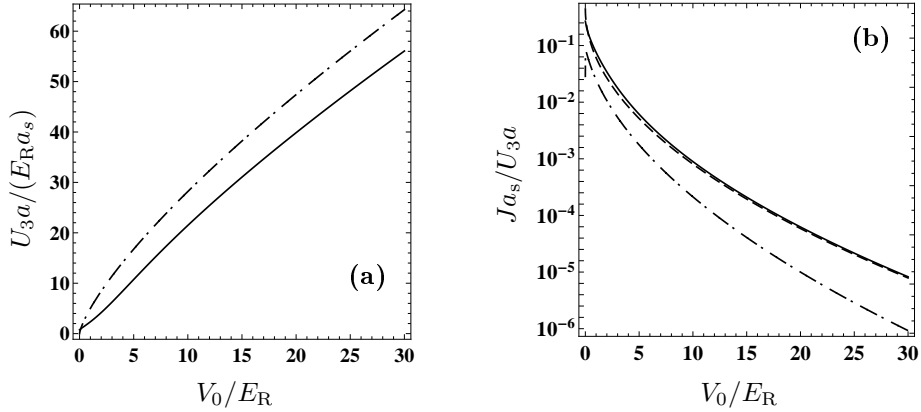}

\caption
{
(a)
On-site interaction constant in a three-dimensional lattice.
Solid line is exact numerical result obtained according to Eq.~(\ref{Ulb}).
Calculations within the Gaussian approximation [Eq.~(\ref{UGauss})] are shown by dashed-dotted line.
(b)
The ratio of the tunneling matrix element for the nearest neighbors in the lowest Bloch band to the on-site interaction constant.
Solid line -- exact numerical result,
dashed line -- estimations from Eqs.~(\ref{Jasymp}),~(\ref{UGauss}),
dashed-dotted line -- Gaussian approximation~(\ref{JGauss}),~(\ref{UGauss}).
}
\label{JandU}
\end{figure}

The tunneling parameter $J$ was already discussed in section~\ref{SecTunMat}.
The dependence of $U$ on the lattice amplitude $V_0$ is shown in Fig.~\ref{JandU}.
While $J$ rapidly decreases, $U$ grows.
In the case $V_0\to0$, Eq.~(\ref{W0_V=0}) yields analytical result for $U_d$:
\begin{equation}
\lim_{V_0\to0}
\label{Ud_V0_0}
\frac{U_d}{E_{\rm R}}
=
g_d
\left(
    \frac{2}{3a}
\right)^d
=
\left(
    \frac{4}{3}
\right)^d
\pi^{\frac{1-d}{2}}
\left(
    \frac{\hbar\omega_\perp}{E_{\rm R}}
\right)^{\frac{3-d}{2}}
\frac{a_{\rm s}}{a}
\;.
\end{equation}
Eqs.~(\ref{J_V0_0}) and~(\ref{Ud_V0_0}) show that the maximal value of the ratio $J/U_d$ can be of the order of $10$ in realistic experiments.
Since $J$ rapidly decreases with $V_0$ and $U_d$ grows, one can easily reach very small ratios of $J/U_d$
which allows to access different regimes of the Bose-Hubbard model.

In the Gaussian approximation~(\ref{WGauss}) the interaction parameter takes the form
\begin{equation}
\label{UGauss}
\frac{U_d}{E_{\rm R}}
\approx
2^{d/2}
\sqrt{\pi}
\left(
    \frac{\hbar\omega_\perp}{E_{\rm R}}
\right)^{\frac{3-d}{2}}
\left(
    \frac{V_0}{E_{\rm R}}
\right)^{d/4}
\frac{a_{\rm s}}{a}
\;.
\end{equation}
The comparison with exact numerical results in Fig.~\ref{JandU} shows that this expression overestimates the value of $U_d$,
although it predicts correct qualitative behavior.

Exact numerical results for the interaction parameter in three dimensions were fitted by
\begin{equation}
\frac{U_3}{E_{\rm R}}
=
p_1
\left(
    \frac{V_0}{E_{\rm R}}
\right)^{p_2}
\frac{a_{\rm s}}{a}
\end{equation}
with $p_1=2.985$, $p_2=0.88$~\cite{GWFMGB05a}.
However, this equation as well as Eq.~(\ref{UGauss}) would imply that $U_d$ vanishes in the limit $V_0\to0$ which contradicts to Eq.~(\ref{Ud_V0_0}).
In order to find a fit which works well also for small $V_0$, we observe from Eqs.~(\ref{U_d}) and (\ref{Ud_V0_0})
that $U_d(V_0)$ can be represented in the form
\begin{equation}
\frac{U_d}{E_{\rm R}}
=
\frac{8}{\pi}
\left(
    \frac{\pi}{4}
    \frac{\hbar\omega_\perp}{E_{\rm R}}
\right)^{\frac{3-d}{2}}
\left[
    u
    \left(
        \frac{V_0}{E_{\rm R}}
    \right)
\right]^{d/3}
\frac{a_{\rm s}}{a}
\;.
\end{equation}
The numerical data are very accurately reproduced by a polynomial function $u(x)=\sum_{i=0}^6 p_i x^i$
with the coefficients $p_0=8/27$, $p_1=0.554092$, $p_2=8.01432\times10^{-2}$, $p_3= -8.94513\times10^{-3}$,
$p_4= 4.55577\times10^{-4}$, $p_5= -1.12896\times10^{-5}$, $p_6= 1.09512\times10^{-7}$.

\subsection{Particle-number conservation}

Since the Bose-Hubbard Hamiltonian commutes with the operator of the total number of particles
\begin{equation}
\label{N-operator}
\hat N
=
\sum_{\ell}
\hat a_{\ell}^\dagger
\hat a_{\ell}^{\phantom{\dagger}}
\;,
\end{equation}
the latter is a good quantum number. Therefore, the eigenstates of the Hamiltonian can be represented
as superpositions of the eigenstates of the operator (\ref{N-operator})
which are given by the product of Fock states
\begin{equation}
\label{Fock}
|{\bf n}_\Gamma\rangle
=
\bigotimes_{\ell}
\frac
{
 \left(
 \hat a_{\ell}^\dagger
 \right)^{n_{\Gamma{\ell}}}
}
{
 \sqrt{n_{\Gamma{\ell}}!}
}
|0\rangle
\;,
\quad
\Gamma=1,\dots,{\cal D}
\;,\quad
{\cal D}=\frac{(N+L-1)!}{N!(L-1)!}
\;,
\end{equation}
where $\Gamma$ labels the configuration of the bosons and the occupation numbers 
of individual lattice sites $n_{\Gamma{\ell}}$ satisfy the condition
\begin{equation}
\sum_{\ell}
n_{\Gamma{\ell}}
=
N
\end{equation}
for any $\Gamma$.

\subsection{\label{T-ivariance}Translational invariance}

In the case of homogeneous lattices with periodic boundary conditions,
the Hamiltonian commutes with the translation operator $\hat{\cal T}$ which shifts
the indices of the bosonic operators by one:
$\hat{\cal T} \hat a_{\ell_1}\hat a_{\ell_2}\cdots \to \hat a_{\ell_1+1}\hat a_{\ell_2+1}\cdots$,
and it holds
$\hat a_{\ell}=\hat{\cal T}^{\ell}\hat a_{0}\hat{\cal T}^{-\ell}$.
In order to simplify the notations, we consider here one-dimensional lattices but
the formalism can be easily generalized to higher dimensions.
Obviously, after $L$ translations we always return to the original site, i.e., $\hat{\cal T}^L=\mathbb{I}$.
Therefore, the eigenvalues of $\hat{\cal T}$ are given by
$\tau_K=\exp(-iKa)$, where $\hbar K$ are the eigenvalues of the total momentum which takes discrete values
$K\equiv K_q=\frac{2\pi}{La}q$ determined by an integer $q$.

The operator $\hat{\cal T}$ commutes with the total-number operator $\hat N$.
The common eigenstates of $\hat{\cal T}$ and $\hat N$ can be obtained
acting by the projection operator~\cite{GM1997,WF2008}
\begin{equation}
\frac{1}{L}
\sum_{j=0}^{L-1}
\left(
    \frac
    {\hat {\cal T}}
    {\tau_K}
\right)^j
\end{equation}
on the states~(\ref{Fock}) which creates linear combinations of the form
\begin{eqnarray}
\label{basis}
|{\bf n}_{K\Gamma}\rangle
&=&
\sum_{m=-\infty}^{\infty}
\delta_{q,m\frac{L}{\nu_\Gamma}}
\frac{1}{\sqrt{\nu_\Gamma}}
\sum_{j=0}^{\nu_\Gamma-1}
\left(
    \frac
    {\hat {\cal T}}
    {\tau_{K}}
\right)^j
|{\bf n}_\Gamma\rangle
\;,
\end{eqnarray}
where $\nu_\Gamma$ is a minimal number of translations required to map the state $|{\bf n}_\Gamma\rangle$
into itself which has to be a divider of $L$.
Eq.~(\ref{basis}) contains only those states $|{\bf n}_\Gamma\rangle$ that cannot be obtained
from the others by cyclic permutations:
$|{\bf n}_\Gamma\rangle\ne\hat{\cal T}^{j}|{\bf n}_{\Gamma'}\rangle$ for $j=1,\dots,L-1$.
The states (\ref{basis}) satisfy the orthonormality condition
$
\langle
{\bf n}_{K\Gamma} | {\bf n}_{K'\Gamma'}
\rangle
=
\delta_{\Gamma\Gamma'}
\delta_{KK'}
$.
The sum over $m$ in Eq.~(\ref{basis}) implies that the states $|{\bf n}_{K\Gamma}\rangle$ do not necessarily exist for all values of $K_q$.
For instance, the states with equal occupation numbers of all sites have $\nu_\Gamma=1$ and, therefore, exist only for $q=mL$, $m=0,\pm1,\dots$
On the other hand, the states with distinct occupation numbers have $\nu_\Gamma=L$ and, therefore, exist for any $q$.
Eq.~(\ref{basis}) can be also rewritten in the form
\begin{eqnarray}
\label{basis_L}
|{\bf n}_{K\Gamma}\rangle
&=&
\sum_{m=-\infty}^{\infty}
\delta_{q,m\frac{L}{\nu_\Gamma}}
\frac{\sqrt{\nu_\Gamma}}{L}
\sum_{j=0}^{L-1}
\left(
    \frac
    {\hat {\cal T}}
    {\tau_{K}}
\right)^j
|{\bf n}_\Gamma\rangle
\;.
\end{eqnarray}

Thus, the eigenstates of the homogeneous system under periodic boundary conditions
can be labeled by two indices: $K$ and $\Omega$, where the second index $\Omega$ distinguishes between the states with the same value of $K$.
The states~(\ref{basis}) are used as a basis in the exact diagonalization because they allow to reduce
the dimension of the Hamiltonian matrix by a factor of the order of the number of lattice sites.

\subsection{Momentum operators}

The translation operator $\hat{\cal T}$ is a unitary operator and, therefore, can be represented in the form
\begin{equation}
\hat{\cal T}
=
\exp
\left(
    - \frac{i}{\hbar} a
    \sum_{\nu=1}^d
    \hat\Pi_\nu
\right)
\;,
\end{equation}
where $\hat\Pi_\nu$ is a component of the momentum operator which is a generator of translations on a discrete lattice
in the direction $\nu$.
Operators $\hat\Pi_\nu$ should commute with $\hat H$ as well as $\hat N$, and have the eigenvalues $\hbar K$
discussed in the previous section which are restricted modulo $2\pi\hbar/a$.

We consider first the usual momentum operator in a continuous space
\begin{equation}
\hat{\bf P}
=
\int
\hat\Psi^\dagger({\bf x})
\left(
    -i\hbar\nabla
\right)
\hat\Psi({\bf x})
\,d{\bf x}
\;.
\end{equation}
A lattice version of $\hat{\bf P}$ can be obtained using the basis of Wannier functions.
In the tight-binding approximation, we obtain
\begin{equation}
\label{momentum-tba}
\hat{\bf P}
=
i P_0
\sum_{\nu=1}^d
{\bf e}_\nu
\sum_{\bf l}
\left(
    \hat a_{{\bf l}+{\bf e}_\nu}^\dagger
    \hat a_{{\bf l}}^{\phantom{\dagger}} - {\rm h.c.}
\right)
\;,
\end{equation}
where
\begin{equation}
P_0
=
\hbar
\int_{-\frac{La}{2}}^{\frac{La}{2}}
W(x)
\frac{\partial}{\partial x}
W(x-a)
\,dx
\;.
\nonumber
\end{equation}
In the Fourier space introduced via the transformation
\begin{equation}
\label{ak}
\hat{\tilde a}_{\bf k}
=
\sum_{\bf l}
\varphi_{\bf k l}^*
\hat{a}_{\bf l}^{\phantom{\dagger}}
\;,\quad
\varphi_{\bf k l}
=
\frac{1}{\sqrt{L^d}}
\exp
\left(
    i {\bf k} \cdot {\bf x_l}
\right)
\;,
\quad
k_\nu = \frac{2\pi}{La}q_\nu
\;,
\end{equation}
the momentum operator $\hat{\bf P}$ takes the following form
\begin{equation}
\hat{\bf P}
=
2 P_0
\sum_{\nu=1}^d
{\bf e}_\nu
\sum_{\bf k}
\hat{\tilde a}_{\bf k}^\dagger
\hat{\tilde a}_{\bf k}^{\phantom{\dagger}}
\sin
\left(
    k_\nu a
\right)
\;.
\end{equation}
Note that the operators in the momentum space $\hat{\tilde a}_{\bf k}$
satisfy standard bosonic commutation relations and have the following property
\begin{equation}
\sum_{{\bf k}\in 1{\rm BZ}}
\hat{\tilde a}_{\bf k}^\dagger
\hat{\tilde a}_{\bf k}^{\phantom{\dagger}}
=
\sum_{\bf l}
\hat a_{\bf l}^\dagger
\hat a_{\bf l}^{\phantom{\dagger}}
=
\hat N
\;.
\end{equation}

Operator $\hat{\bf P}$ cannot be identified with $\hat{\bf\Pi}$,
because it does not commute with the Hamiltonian~(\ref{HBH}) due to the interaction term.
Instead, the quasi-momentum operator
\begin{displaymath}
\hat{\bf Q}
=
\sum_{{\bf k}\in 1{\rm BZ}}
\hbar
{\bf k}
\hat{\tilde a}_{\bf k}^\dagger
\hat{\tilde a}_{\bf k}^{\phantom{\dagger}}
\;,
\end{displaymath}
is introduced which commutes with the Hamiltonian as well as with the translation operator.
However, it is not restricted modulo $2\pi\hbar/a$.
A proper momentum operator which satisfies all the conditions is given by~\cite{GM1997,EFGKK05}
\begin{eqnarray}
\hat\Pi_\nu	
=
\frac{2\pi\hbar}{La}
\sum_{\ell=1}^{L-1}
\left[
    \frac{1}{2}
    +
    \frac
    {
     \hat{\cal T}_\nu^\ell
    }
    {
     \exp
     \left(
         i \frac{2\pi}{L} \ell
     \right)
     - 1
    }
\right]
\;.
\end{eqnarray}

Although being of fundamental importance, the explicit form of the momentum operator does not play
a role for the interpretation of the experiments with ultracold atoms in optical lattices.
What is more relevant is the quasi-momentum distribution~\cite{KPS} determined as
\begin{equation}
\label{md-def}
P({\bf k})
=
\frac{1}{N}
\langle
     \hat{\tilde\Psi}^\dagger({\bf k})
     \hat{\tilde\Psi}({\bf k})
\rangle
\;,\quad
\hat{\tilde\Psi}({\bf k})
=
\frac{1}{(La)^{d/2}}
\int
\hat\Psi({\bf x})
\exp
\left(
    - i {\bf k}\cdot{\bf x}
\right)
\,d{\bf x}
\;,
\end{equation}
where $N$ should be understood in general as an expectation value of the operator $\hat N$.
Taking into account Eqs.~(\ref{FO}) and (\ref{ak}) it can be written down in the form
\begin{equation}
\label{md-real}
P({\bf k})
=
\left|
    \tilde W({\bf k})
\right|^2
\frac
{
\langle
     \hat{\tilde a}_{\bf k}^\dagger
     \hat{\tilde a}_{\bf k}^{\phantom{\dagger}}
\rangle
}
{N}
\;,\quad
\sum_{{\bf k}}
P({\bf k})
=
1
\;,
\end{equation}
where
\begin{equation}
\tilde W({\bf k})
=
\frac{1}{a^{d/2}}
\int
W({\bf x})
\exp
\left(
    - i {\bf k}\cdot {\bf x}
\right)
\,d{\bf x}
\;,
\end{equation}
and the values of ${\bf k}$ are not restricted to the first Brillouin zone.
The normalization constants in Eqs.~(\ref{ak}),~(\ref{md-def}) are chosen such that
\begin{eqnarray}
\sum_{\bf k}
\hat{\tilde\Psi}^\dagger({\bf k})
\hat{\tilde\Psi}({\bf k})
=
\int
\hat\Psi^\dagger({\bf x})
\hat\Psi({\bf x})
\,
d{\bf x}
=
\hat N
\;.
\nonumber
\end{eqnarray}
In the Gaussian approximation~(\ref{WGauss}), $\tilde W({\bf k})$ has the form
\begin{equation}
\tilde W({\bf k})
=
\left(
    4\pi
\right)^{d/4}
\left(
    \frac
    {a_{\rm ho}}
    {a}
\right)^{d/2}
\exp
\left(
    -\frac{1}{2}
    k^2
    a_{\rm ho}^2
\right)
\;.
\end{equation}

Usually in the theoretical analysis of the lattice problems based on the Bose-Hubbard model~(\ref{HBH}),
the Wannier functions are omitted. Then the expression for the quasi-momentum distribution is considered to be
\begin{equation}
\label{md-lat}
\tilde P({\bf k})
=
\frac
{
\langle
     \hat{\tilde a}_{\bf k}^\dagger
     \hat{\tilde a}_{\bf k}^{\phantom{\dagger}}
\rangle
}
{N}
\;,\quad
\sum_{{\bf k}\in 1{\rm BZ}}
\tilde P({\bf k})
=
1
\;.
\end{equation}
In an infinite system, ${\bf k}$ is a continuous variable and it makes sense to redefine the quasi-momentum distribution as
$\tilde P_\infty({\bf k})=(La/2\pi)^d \tilde P({\bf k})$
in order to have a proper normalization:
$\int_{{\bf k}\in 1{\rm BZ}} \tilde P_\infty({\bf k})d{\bf k}=1$.

Eq.~(\ref{ak}) allows to express the momentum distribution in terms of the operators in real space in the form of
a double sum over the lattice sites. In a translationally invariant system, the double sum can be converted into a single sum
and Eq.~(\ref{md-lat}) takes the form of the discrete Fourier transform:
\begin{equation}
\tilde P({\bf k}_{\bf q})
=
\frac{1}{N}
\sum_{\ell_1=0}^{L-1}
\cdots
\sum_{\ell_d=0}^{L-1}
    \exp
    \left[
        i {\bf k}_{\bf q}
        \cdot
        \left(
            {\bf x}_{\bf l} - {\bf x}_{\bf 0}
        \right)
    \right]
    \langle
        \hat a_{\bf 0}^\dagger
        \hat a_{\bf l}^{\phantom{\dagger}}
    \rangle
\;.
\end{equation}
One can show that $\tilde P({\bf k}_{\bf q})$ in the last equation is real-valued.

In general, the quasi-momentum distribution $\tilde P_\infty({\bf k})$ defined by Eq.~(\ref{md-lat})
is an even and periodic function of $k_\nu$ with the period $2\pi/a$.
For $J>0$, it takes maximal (minimal) values at $ka=\pi m$, $m=0,\pm 2,\pm 4,\dots$ ($m=\pm 1,\pm 3,\dots$).
In the case of negative $J$, the positions of the minima and maxima are reversed.
However, the presence of $|\tilde W({\bf k})|^2$ in Eq.~(\ref{md-real}) for the true quasi-momentum distribution
destroys this periodic structure resulting in smaller heights of the peaks with larger values of ${\bf k}$.
With the decrease of the hopping parameter $J$ the spatial correlations of bosons
become weaker which leads to the broadening of the momentum distribution.

\section{\label{sec-BD}Basic definitions}

\subsection{\label{sec-Tq}Thermodynamic quantities}

We remind the definitions of the basic thermodynamic quantities which are often used in the studies
of the many-body lattice problems. Starting with the grand-canonical partition function
\begin{equation}
{\cal Z}(\mu,T)
=
{\rm Tr}
\exp
\left(
    - \frac{\hat H - \mu \hat N}{k_{\rm B}T}
\right)
\end{equation}
the free energy is defined as
\begin{equation}
{\cal F}(\mu,T)
=
-k_{\rm B}T
\ln
{\cal Z}(\mu,T)
\;.
\end{equation}
The derivative of the free energy gives the mean number of particles per lattice site (filling factor)
\begin{equation}
\langle\hat n_{\bf l}\rangle
=
-
L^{-d}
\frac{\partial{\cal F}}{\partial\mu}
\;,
\quad
\langle\hat n_{\bf l}\rangle
=
\frac{N}{L^d}
\;,
\end{equation}
and the derivative of the latter gives the (isothermal) compressibility
\footnote
{
This is slightly different from the standard definition~\cite{Schwabl-SM}
\begin{equation}
\kappa_{T}
=
-
\frac{1}{V}
\left(
    \frac{\partial V}{\partial P}
\right)_{T,N}
=
\frac{V}{N^2}
\left(
    \frac{\partial N}{\partial \mu}
\right)_{T,V}
=
\frac{a^d}{\langle\hat n_{\bf l}\rangle^2}
\left(
    \frac{\partial \langle\hat n_{\bf l}\rangle}{\partial \mu}
\right)_{T,V}
=
\frac{a^d}{\langle\hat n_{\bf l}\rangle^2}
\kappa
\;.
\nonumber
\end{equation}
}
\begin{equation}
\label{kappa}
\kappa
=
\left(
    \frac{\partial\langle\hat n_{\bf l}\rangle}{\partial\mu}
\right)_T
\;.
\end{equation}
The compressibility is related to the fluctuations of the total number of particles:
\begin{equation}
\label{kappa-fluct}
\kappa
=
\frac{1}{L^d}
\frac
{
 \langle
     \hat N^2
 \rangle
 -
 N^2
}
{k_{\rm B}T}
\;.
\end{equation}

At zero temperature, the free energy takes the form
\begin{equation}
\label{F_T0}
{\cal F}(\mu,T=0)
=
\min_N
\left(
    E_N - \mu N
\right)
\;,
\end{equation}
where $E_N$ is the ground-state energy of $N$ particles with $N$ being a non-negative integer.
If $N$ can be considered as a continuous variable, the minimization in Eq.~(\ref{F_T0}) gives
\begin{equation}
\label{mu}
\mu(N)
=
\frac{\partial E_N}{\partial N}
\;.
\end{equation}
However, strictly speaking $N$ in Eq.~(\ref{F_T0}) is discrete and the minimization leads to the fact that
the chemical potential $\mu$ is enclosed in the interval $[\mu_-,\mu_+]$, where the boundaries are given by
\begin{equation}
\label{mu-pm}
\mu_\pm(N)
=
\pm
\left(
    E_{N\pm1}
    -
    E_N
\right)
\;.
\end{equation}

Discrete analogue of Eq.~(\ref{kappa}) has the form
\begin{equation}
\kappa^{-1}
=
L^d
\left[
    \mu_+(N)
    -
    \mu_-(N)
\right]
=
L^d
\Delta_{\rm c}
\end{equation}
and defines the one-particle ('charge') gap
\begin{equation}
\label{Delta_c}
\Delta_{\rm c}
=
E_{N+1}+E_{N-1}-2E_N
\;.
\end{equation}
This is different from the `neutral' gap $\Delta_{\rm n}$ which is defined as the energy difference
of the lowest excited state and the ground state with the same number of particles $N$.

In inhomogeneous lattices, it is useful to consider local quantities. For instance, local (on-site) compressibility~\cite{BRSRMDT02,WATB04}
\begin{equation}
\label{loc-comp}
\kappa_{\bf l}
=
\frac
{\partial \langle\hat n_{\bf l}\rangle}
{\partial\mu_{\bf l}}
\end{equation}
quantifies the response of local density to the local variations of the chemical potential.

\subsection{Superfluidity}

Superfluidity is one of the most fascinating phenomena which can occur in many-body quantum systems at low temperature~\cite{Leggett99,Leggett06}.
It was discovered first in the experiments with liquid helium~\cite{Kapitza38,AM38}
and also observed with ultracold atomic gases in traps~\cite{ORVACK00,RACNHP07,RWMZHLHPC11,DCYLBWD12}
and in shallow optical lattices~\cite{BCFMICT01,FSLMSFI04,SFLMSFI05}.
Superfluidity is usually related to the flow properties of a quantum system which is assumed to be composed
of a normal and superfluid components distinguished through their behavior in the presence of moving boundaries.
If the system is enclosed in a narrow region between two moving walls, the normal component is dragged by the walls,
whereas the superfluid remains at rest. In other words, the superfluid is at rest in the lab frame, while the normal component
is at rest in the frame of the moving walls.

The superfluidity is quantified imposing twisted boundary conditions on the many-body wavefunction~\cite{FBJ73,SS90,Krauth91}
\begin{equation}
\Psi(\dots,{\bf x}_j+L'{\bf e}_\nu,\dots)
=
e^{i\theta}
\Psi(\dots,{\bf x}_j,\dots)
\;,
\end{equation}
where $L'$ is the linear size of the system\footnote{In the case of a periodic potential, $L'=La$.}
and the twist angle $\theta\in(0,\pi)$.
This requirement leads to the increase of the free energy ${\cal F}$ which is attributed to the kinetic energy of the superfluid.
Since the corresponding velocity is fixed by the value of theta, the number of particles in the superfluid component $N_{\rm s}$ is determined as
\begin{equation}
\label{Ns-def}
{\cal F}({\bf k}_{\rm s})
-
{\cal F}(0)
=
\frac{\hbar^2 k_{\rm s}^2}{2M}
N_{\rm s}
\;,\quad
{\bf k}_{\rm s}
=
{\bf e}_\nu
\frac{\theta}{L'}
\;,
\end{equation}
which readily gives the superfluid fraction determined as $\nu_s=N_{\rm s}/N$.
According to this definition, $\nu_{\rm s}\in[0,1]$ and the upper bound corresponds to noninteracting particles in free space at zero temperature.
In the limit $k_{\rm s}\to0$, Eq.~(\ref{Ns-def}) yields
\begin{equation}
\label{nus}
\nu_{\rm s}
=
\frac{M}{\hbar^2 N}
\left.
\nabla_{{\bf k}_{\rm s}}^2
{\cal F}
\right|_{k_{\rm s}=0}
\;.
\end{equation}
Then in the case of noninteracting particles in a periodic potential at $T=0$, we get $\nu_{\rm s}=M/M_*$~\cite{Eggington77,AHNS80},
where $M_*$ is the effective mass, which follows from the dispersion relation of non-interacting particles.
In the tight-binding regime, $M_*$ is given by
\begin{equation}
\label{meff}
M_*=\hbar^2/(2Ja^2)
\;,
\end{equation}
and $\nu_{\rm s}=\pi^2 J/E_{\rm R}$ is exponentially small.
This makes the measurement of the superfluid fraction in deep optical lattices rather difficult.
Nevertheless, this quantity plays a fundamental role in the theoretical studies due to the possibility of the superfluid-insulator transition.

In a lattice model, the dispersion relation of noninteracting particles
$\epsilon_{\bf k}$ has a quadratic form only for small ${\bf k}$.
For the description of the superfluid properties of the system it is convenient to introduce the superfluid stiffness
in analogy to Eq.~(\ref{Ns-def}):
\begin{equation}
\label{fs-def}
f^{\rm s}_N
=
\frac
{
{\cal F}({\bf k}_{\rm s})
-
{\cal F}(0)
}
{
N
\left(
    \epsilon_{{\bf k}_{\rm s}}
    -
    \epsilon_0
\right)
}
\;,\quad
{\bf k}_{\rm s}
=
{\bf e}_\nu
\frac{\theta}{La}
\;.
\end{equation}
For noninteracting particles in a translationally invariant lattice, $f^{\rm s}_N\equiv1$ for arbitrary $k_{\rm s}$.
For small $k_{\rm s}$, Eq.~(\ref{fs-def}) can be rewritten in the form~(\ref{nus}) with $M$ replaced by the effective mass $M_*$.
In the hydrodynamic approach the superfluid stiffness is related to isothermal compressibility $\kappa$ as
\begin{equation}
\label{fs_kappa_cs}
f_N^{\rm s}
=
M_*
\kappa
\,
c_{\rm s}^2
/
\langle\hat n_{\bf l}\rangle
\;,
\end{equation}
where $c_{\rm s}$ is the sound velocity.

From the practical point of view, twisted boundary conditions can be introduced in slightly different ways.
One of the often used possibilities is to impose antiperiodic boundary conditions~\cite{RSZ99}
which corresponds to $\theta=\pi$.
This allows to avoid complex arithmetics in the numerical calculations.
Another option is to introduce Peierls phase factors $\exp(i\theta/L)$ into the hopping part of the Hamiltonian~(\ref{HBH})
by means of transformation
$
\hat a_{\bf l}
\to
\hat a_{\bf l}
\exp
\left(
    i
    {\bf k}_{\rm s}
    \cdot
    {\bf x}_{\bf l}
\right)
$.
It is interesting to note that the Peierls phases of arbitrary magnitude can be created by periodic shaking of the optical lattice
along a chosen direction~\cite{EHSBSL10} as it was demonstrated experimentally in Ref.~\cite{SOWHSELSW12}.

One can also consider the limit $\theta\to0$ which leads to the concept of winding number~\cite{PC87} used in QMC calculations.
In addition, in this case one can express the superfluid stiffness at zero temperature as~\cite{RB03r,RB03,SM10}
\begin{equation}
\label{fs-exc}
f^{\rm s}_N
=
-\frac{1}{2NJ}
\langle
\Psi_0
\left|
    \hat H_\nu^{\rm kin}
\right|
\Psi_0
\rangle
-
\frac{1}{NJ}
\sum_{\lambda\ne0}
\frac
{
 \left|
     \langle
     \Psi_\lambda
     \left|
         \hat J_\nu
     \right|
     \Psi_0
     \rangle
 \right|^2
}
{E_\lambda-E_0}
\;,
\end{equation}
where
\begin{equation}
\hat J_\nu
=
iJ
\sum_{\bf l}
\left(
    \hat a_{\bf l}^\dagger
    \hat a_{{\bf l}+{\bf e}_\nu}^{\phantom\dagger}
    -
    \hat a_{{\bf l}+{\bf e}_\nu}^\dagger
    \hat a_{\bf l}^{\phantom\dagger}
\right)
\end{equation}
is the current operator for the direction $\nu$ which up to a constant prefactor coincides with the operator~(\ref{momentum-tba}).
The operator $\hat J_\nu$ can be formally obtained taking the derivative with respect to $\theta$ in the Hamiltonian
with the Peierls phase factors and considering the limit $\theta\to0$ (see, e.g., Ref.~\cite{NRDS14}).
Eq.~(\ref{fs-exc}) shows that the superfluidity is not just a property of the ground state but also contains
information about all excited states described by the second term which always gives a negative contribution.

\subsection{Bose-Einstein condensation}

The Bose-Einstein condensation is in general defined as a macroscopic population of an eigenstate of
the one-body density matrix~\cite{PS2003,PO56}
$
\rho_1({\bf x},{\bf x}')
=
\langle
\hat\Psi^\dagger({\bf x})
\hat\Psi({\bf x}')
\rangle
$.
Formally, this is determined from the solution of the eigenvalue problem
\begin{equation}
\label{OBDM-evp-c}
\int
\rho_1({\bf x},{\bf x}')
\phi_i({\bf x}')
d{\bf x}'
=
N_i
\phi_i({\bf x})
\;,
\end{equation}
where index $i$ labels the eigenstates.

In the tight-binding regime, the eigenfunctions $\phi_i({\bf x})$ as well as the field operator
$\hat\Psi({\bf x})$ are decomposed in the basis of the Wannier functions for the lowest Bloch band
which leads to the lattice version of Eq.~(\ref{OBDM-evp-c}):
\begin{equation}
\label{OBDM-evp}
\sum_{{\bf l}'}
\langle
\hat a_{\bf l}^\dagger
\hat a_{{\bf l}'}^{\phantom{\dagger}}
\rangle
\phi_{i{\bf l}'}
=
N_i
\phi_{i{\bf l}}
\;,
\end{equation}
where the eigenvalues $N_i$ remain the same.
The discrete eigenfunctions satisfy the orthonormality condition
\begin{equation}
\sum_{\bf l}
\phi_{i{\bf l}}^*
\phi_{j{\bf l}}^{\phantom{*}}
=
\delta_{ij}
\;,
\end{equation}
and the sum of all $N_i$ gives the total number of particles $N$.
If the largest eigenvalue is labeled by $i=0$, the condensate fraction is determined as $f_N^{\rm c}=N_0/N$.
In a homogeneous lattice with periodic boundary conditions,
the discrete one-body density matrix with the entries
$F_a({\bf l}_1,{\bf l}_2)\equiv\langle \hat a_{{\bf l}_1}^\dagger \hat a_{{\bf l}_2}^{\phantom{\dagger}} \rangle$
depends only on ${\bf l}_2-{\bf l}_1$.
The eigenfunction corresponding to the largest eigenvalue is constant,
$\phi_{0{\bf l}}=L^{-d/2}$, and
\begin{equation}
N_0
=
\sum_{\ell_1=0}^{L-1}
\dots
\sum_{\ell_d=0}^{L-1}
\langle
   \hat a_{\bf 0}^\dagger
   \hat a_{\bf l}^{\phantom{\dagger}}
\rangle
\;.
\end{equation}
Therefore, $f_N^{\rm c}=\tilde P({\bf 0})$.

A sufficient condition for the existence of a Bose Einstein condensate is the off-diagonal long-range order
of the one-body density matrix~\cite{Yang62,BDZ08,CBY01}.
If the asymptotic value
\begin{equation}
N_0^{{\bf k}=0}
=
\lim_{|{\bf x}-{\bf x}'|\to\infty}
\rho_1({\bf x},{\bf x}')
\equiv
N\,P({\bf k}=0)
\label{ODLRO}
\end{equation}
does not vanish, there is a finite fraction of particles with zero momentum.
In the tight-binding regime and in a translationally invariant lattice,
\begin{equation}
\label{k0fc}
N_0^{{\bf k}=0}/N
=
\left|\tilde W({\bf 0})\right|^2 f_N^{\rm c}
\;.
\end{equation}
Since $\left|\tilde W({\bf 0})\right|^2$ is smaller than one, the fraction of particles with zero momentum is smaller than the condensate fraction
in agreement with a general statement that $N_0^{{\bf k}=0}$ is a lower estimate of $N_0$~\cite{AK11}.
It was also demonstrated in two dimensional lattices without using tight-binding approximation
that the difference between $N_0^{{\bf k}=0}/N$ and $f_N^{\rm c}$ can be large~\cite{AK11}.
In the experiments with ultracold atoms, $P({\bf k})$ is usually measured which gives a direct access to $N_0^{{\bf k}=0}$
(see, e.g.,~\cite{XLMCSK06,SPP08,JCLPPS10,NMDM12}). However, due to inhomogeneities caused, for instance, by harmonic confinement,
there is no simple relation like~(\ref{k0fc}) between $N_0^{{\bf k}=0}$ and $f_N^{\rm c}$ and more careful analysis is necessary in order to
extract $f_N^{\rm c}$ from the experimental data.

\section{\label{sec-MET}Main experimental techniques}

Possibility of experimental control is an attractive feature of ultracold atoms in optical lattices.
Below we briefly discuss the main experimental methods which provide information about the state of the system
and give motivation for theoretical studies.

\subsection{Time-of-flight imaging}

Spatial coherence of ultracold atoms is usually probed by the time-of-flight imaging of expanding atomic cloud
after sudden switching-off the lattice and confining potentials~\cite{GBMHE01,GMEHB02,FCFFI2011}.
The images obtained by the resonant absorption of photons are directly related to the density profiles.
When the phase coherence length is at least of the order of several lattice spacings, the density
distribution of an expanding cloud shows an interference pattern which has the symmetry of the reciprocal lattice.
This is usually interpreted as a signature of superfluidity.
If the phase coherence length is of the order of one lattice spacing, the density distribution becomes broad
and does not show any peaks~\cite{GBMHE01,LSA12}.

In the experiments, the expansion times are sufficiently long such that the interaction effects during the expansion
are negligible~\cite{GTFSTWBPTCPS08}. For
\begin{equation}
\label{tint}
\omega_{\rm R} t
\gg
\frac{R_0}{a}
\sqrt
{
 \frac{E_{\rm R}}{\hbar\omega_{\rm ho}}
}
\;,
\end{equation}
where $R_0$ is the characteristic size of the atomic cloud before expansion, and $\omega_{\rm ho}$ is the effective oscillation
frequency at the bottom of a lattice well given by Eq.~(\ref{omega_ho}),
the matter-field operator takes the form~\cite{NCK06,GTFSTWBPTCPS08}
\begin{equation}
\label{psiToF}
\hat\psi({\bf x},t)
\approx
\left(
    \frac{M}{\hbar t}
\right)^{d/2}
\tilde W({\bf k})
\sum_{{\bf l}}
\exp
\left(
    - i {\bf k} \cdot {\bf x}_{\bf l}
    + i
    \frac{M}{2\hbar t} {\bf x}_{\bf l}^2
\right)
\hat a_{\bf l}^{\phantom\dagger}
\;,\quad
{\bf k}
=
\frac{M{\bf x}}{\hbar t}
\;,
\end{equation}
where we omitted unimportant phase factors. The expectation value of the density operator
$\langle\hat\psi^\dagger({\bf x},t)\hat\psi({\bf x},t)\rangle\equiv \rho_{\rm ToF}({\bf x},t)$
is given by
\begin{equation}
\label{nToF}
\rho_{\rm ToF}({\bf x},t)
=
\left(
    \frac{M}{\hbar t}
\right)^d
\left|
    \tilde W({\bf k})
\right|^2
{\cal S}({\bf k})
\;,
\end{equation}
\begin{equation}
\label{Sk}
{\cal S}({\bf k})
=
\sum_{{\bf l}_1,{\bf l}_2}
\exp
\left[
    i
    {\bf k}
    \cdot
    \left(
        {\bf x}_{{\bf l}_1}
        -
        {\bf x}_{{\bf l}_2}
    \right)
    - i
    \frac{M}{2\hbar t}
    \left(
        {\bf x}_{{\bf l}_1}^2
        -
        {\bf x}_{{\bf l}_2}^2
    \right)
\right]
\langle
    \hat a_{{\bf l}_1}^\dagger
    \hat a_{{\bf l}_2}^{\phantom\dagger}
\rangle
\;.
\end{equation}
The second term of the exponential function in Eq.~(\ref{Sk}) can be neglected, if
\begin{equation}
\label{FF}
\omega_{\rm R} t
\gg
\frac{\ell_{\rm c}R_0}{a^2}
\frac{\pi^2}{2}
\;,
\end{equation}
where $\ell_{\rm c}$ is a characteristic coherence length of the system which is of the order of few $a$
in the Mott-insulator (MI) phase and of the order of $R_0$ in the superfluid (SF) phase.
Under condition~(\ref{FF}),
${\cal S}({\bf k})=L^d \langle\hat {\tilde a}_{\bf k}^\dagger \hat {\tilde a}_{\bf k}^{\phantom{\dagger}}\rangle$
and, therefore, describes the quasi-momentum distribution~(\ref{md-lat}).
However, the expansion times in the experiments are usually shorter than those given by
the ``far-field" condition~(\ref{FF}), although Eq.~(\ref{tint}) is fulfilled.
Therefore, for the interpretation of the experimental data it is important to keep the second term in the exponential function in Eq.~(\ref{Sk})
as it was demonstrated in Ref.~\cite{GTFSTWBPTCPS08}.
It is also necessary to keep in mind that experimentally one observes a two-dimensional column density $\rho_\perp({\bf r}_\perp,t)$
obtained from $\rho_{\rm ToF}({\bf r},t)$ by the integration along the probe line of sight.

The structure of the density distribution of the time-of-flight images is quantitatively described by visibility~\cite{GWFMGB05,GTFSTWBPTCPS08}
\begin{equation}
\label{visibility}
{\cal V}
=
\frac
{
 \rho_\perp({\bf k}_{\rm max})
 -
 \rho_\perp({\bf k}_{\rm min})
}
{
 \rho_\perp({\bf k}_{\rm max})
 +
 \rho_\perp({\bf k}_{\rm min})
}
\;.
\end{equation}
In a three-dimensional setup, a special choice of the two-dimensional vectors ${\bf k}_{\rm max}$ and ${\bf k}_{\rm min}$
allows to cancel the contribution of the function $\tilde W_0({\bf k})$ and replace $\rho_\perp({\bf k})$ in Eq.~(\ref{visibility})
by ${\cal S}_\perp(\bf k)$. This is achieved, provided that ${\bf k}_{\rm max}$ is in the center of the second Brillouin zone, i.e.,
${\bf k}_{\rm max}a=(2\pi,0)$, and ${\bf k}_{\rm min}$ is along a diagonal and has the same length as ${\bf k}_{\rm max}$, i.e.,
${\bf k}_{\rm min}a=\sqrt{2}(\pi,\pi)$.
Eqs.~(\ref{nToF}),~(\ref{Sk}),~(\ref{visibility}) show that the one-body density matrix plays an important role in the experiments
with ultracold atoms in optical lattices.

An interesting aspect of the time-of-flight images is that in each experimental measurement a single realization of the density
distribution is observed rather than the expectation value~\cite{ADL04}. This allows to extract the density-density correlation function
\begin{eqnarray}
\label{noisecorr}
{\cal G}({\bf x}_1,{\bf x}_2)
=
\langle\hat\psi^\dagger({\bf x}_1,t)\hat\psi({\bf x}_1,t)\hat\psi^\dagger({\bf x}_2,t)\hat\psi({\bf x}_2,t)\rangle
-
\langle\hat\psi^\dagger({\bf x}_1,t)\hat\psi({\bf x}_1,t)\rangle
\langle\hat\psi^\dagger({\bf x}_2,t)\hat\psi({\bf x}_2,t)\rangle
\end{eqnarray}
from the experimental data and constitutes the basic idea of noise-correlation interferometry~\cite{FGWMGB05,GFFFSI08,F14}.
It provides information on spatial order in the lattice that is absent in the average density.

In order to simplify equations, we consider the regime of large expansion times such that the condition~(\ref{FF}) is fulfilled.
Using equal-time commutation relations for bosonic field operators and Eq.~(\ref{psiToF}), we can rewrite Eq.~(\ref{noisecorr})
in the form~\cite{RSC06,F14}
\begin{eqnarray}
{\cal G}({\bf x}_1,{\bf x}_2)
&=&
\left(
    \frac{M}{\hbar t}
\right)^{2d}
\left|
    \tilde W({\bf k}_1)
\right|^2
\left\{
    \left|
        \tilde W({\bf k}_2)
    \right|^2
    \sum_{{\bf l}_1 {\bf l}_2 {\bf l}_3 {\bf l}_4}
    e^{
       i{\bf k}_1
       \cdot
       \left(
           {\bf x}_{{\bf l}_1}
           -
           {\bf x}_{{\bf l}_3}
       \right)
      }
    e^{
       i{\bf k}_2
       \cdot
       \left(
           {\bf x}_{{\bf l}_2}
           -
           {\bf x}_{{\bf l}_4}
       \right)
      }
    \langle
    \hat a_{{\bf l}_1}^\dagger
    \hat a_{{\bf l}_2}^\dagger
    \hat a_{{\bf l}_3}^{\phantom\dagger}
    \hat a_{{\bf l}_4}^{\phantom\dagger}
    \rangle
\right.
\nonumber\\
    &+&
\left.
    \delta
    \left(
        {\bf k}_1 - {\bf k}_2
    \right)
    \sum_{{\bf l}_1 {\bf l}_2}
    e^{
       i{\bf k}_1
       \cdot
       \left(
           {\bf x}_{{\bf l}_1}
           -
           {\bf x}_{{\bf l}_2}
       \right)
      }
    \langle
    \hat a_{{\bf l}_1}^\dagger
    \hat a_{{\bf l}_2}^{\phantom\dagger}
    \rangle
    -
    \left|
        \tilde W({\bf k}_2)
    \right|^2
    {\cal S}\left( {\bf k}_1 \right)
    {\cal S}\left( {\bf k}_2 \right)
\right\}
\;,
\end{eqnarray}
where the first term is the second-order correlator, while the second term which gives delta-peak for vanishing
relative momentum corresponds to the correlations of an atom with itself (autocorrelation).
In the case of fermions, this term would enter with the minus sign.

In terms of the operators~(\ref{ak}) in the quasi-momentum space, we get
\begin{eqnarray}
&&
{\cal G}({\bf x}_1,{\bf x}_2)
=
\left(
    \frac{M}{\hbar t}
\right)^{2d}
L^{2d}
\left|
    \tilde W({\bf k}_1)
\right|^2
\\
&&
\times
\left\{
    \left|
        \tilde W({\bf k}_2)
    \right|^2
    \left[
        \langle
            \hat n_{{\bf k}_1}
            \hat n_{{\bf k}_2}
        \rangle
        -
        \langle
            \hat n_{{\bf k}_1}
        \rangle
        \langle
            \hat n_{{\bf k}_2}
        \rangle
        -
        \langle
            \hat n_{{\bf k}_1}
        \rangle
        \delta_{{\bf k}_1-{\bf k}_2,{\bf q}\frac{2\pi}{La}}
    \right]
    +
    \langle
        \hat n_{{\bf k}_1}
    \rangle
    \delta_{{\bf k}_1{\bf k}_2}
\right\}
\;,
\nonumber
\end{eqnarray}
where ${\bf q}$ is a $d$-dimensional vector of arbitrary integers.
This leads to the following lattice version of the noise-correlation function used in theoretical studies~\cite{RSC06,HR2011}:
\begin{equation}
{\cal G}_{\rm L}({\bf k}_1,{\bf k}_2)
=
\langle
    \hat n_{{\bf k}_1}
    \hat n_{{\bf k}_2}
\rangle
-
\langle
    \hat n_{{\bf k}_1}
\rangle
\langle
    \hat n_{{\bf k}_2}
\rangle
-
\langle
    \hat n_{{\bf k}_1}
\rangle
\delta_{{\bf k}_1-{\bf k}_2,{\bf q}\frac{2\pi}{La}}
\;,\quad
{\bf q}\ne{\bf 0}
\;.
\end{equation}

If the lattice potential is switched off slowly enough such that no transitions between the Bloch bands take place,
the quasi-momentum of atoms is projected to the usual momentum.
As a consequence, the population of the $n$th Bloch band is mapped into the momentum interval corresponding to the $n$th Brillouin zone.
This is used in the bandmapping technique which allows momentum-resolved measurements of the populations of the lowest Bloch bands~\cite{GBMHE01,MWD09}.

\subsection{Optical Bragg spectroscopy}

Information about the excitation spectrum of the many-body system can be obtained with
the aid of optical Bragg spectroscopy~\cite{PS02book,PS2003,SMSKE04,PRB06,CFFFI2009,FCFFMSI2009,CFFFI2010,EGKPLPS2010,DWYRHLW2010,FCFFI2011}.
In ultracold atomic gases it is performed using two laser beams with the wavevectors ${\bf k}_1$ and ${\bf k}_2$
and frequencies $\omega_1$ and $\omega_2$. This setup allows to transfer the momentum
$\hbar{\bf k}=\hbar({\bf k}_1-{\bf k}_2)$ and the energy $\hbar\omega=\hbar(\omega_1-\omega_2)$ to the atomic sample
which can be tuned independently~\cite{EGKPLPS2010}.
The measurement is performed again as a time-of-flight imaging of the expanding atomic cloud.

This kind of perturbation is described by the Hamiltonian
\begin{eqnarray}
\hat H_{\rm Bragg}
&=&
V_{\rm Bragg}
\int
d{\bf x}
\cos
\left(
    {\bf k}\cdot{\bf x}
    -
    \omega t
\right)
\hat\psi^\dagger({\bf x},t)
\hat\psi({\bf x},t)
\\
&=&
\frac{V_{\rm Bragg}}{2}
\left[
    \hat{\tilde\rho}^\dagger({\bf k},t)
    \exp
    \left(
        - i \omega t
    \right)
    +
    \hat{\tilde\rho}({\bf k},t)
    \exp
    \left(
        i \omega t
    \right)
\right]
\;,
\nonumber
\end{eqnarray}
where
\begin{equation}
\label{tilde-rho}
\hat{\tilde\rho}({\bf k},t)
=
\int
d{\bf x}
\exp
\left(
    -i{\bf k}\cdot{\bf x}
\right)
\hat\psi^\dagger({\bf x},t)
\hat\psi({\bf x},t)
\;,\quad
\hat{\tilde\rho}^\dagger({\bf k},t)
=
\hat{\tilde\rho}(-{\bf k},t)
\;.
\end{equation}

In the linear-response regime, the fluctuation of the density induced by the perturbation is given by the susceptibility $\chi({\bf k},\omega)$
which depends only on the properties of the system in the absence of perturbation.
It is determined by the relation~\cite{PS2003}
\begin{equation}
\label{lin-resp}
\langle
\hat{\tilde\rho}({\bf k},t)
\rangle_{V_{\rm Bragg}}
-
\langle
\hat{\tilde\rho}({\bf k},t)
\rangle_{0}
=
\left[
    \chi({\bf k},\omega)
    e^{-i\omega t}
    +
    \chi({\bf k},-\omega)
    e^{i\omega t}
\right]
V_{\rm Bragg}
\;.
\end{equation}
Due to the causality of the response to the perturbation, the susceptibility $\chi({\bf k},\omega)$
is an analytic function of $\omega$ in the upper half of the complex plane and, therefore, satisfies
the Kramers-Kronig relation
\begin{equation}
\label{KKrelation}
\chi({\bf k},\omega)
=
\frac{1}{i\pi}
\int_{-\infty}^\infty
\chi({\bf k},\omega')
\frac{\cal P}{\omega'-\omega}
d\omega'
\;,
\end{equation}
where ${\cal P}$ denotes the principal value.
Eq.~(\ref{KKrelation}) establishes the relation between the real and imaginary parts of $\chi$.

The probability to transfer the momentum $\hbar{\bf k}$ and energy $\hbar\omega$
into the many-body system is proportional to the dynamic structure factor
\begin{equation}
\label{dsf}
S({\bf k},\omega)
=
\frac{1}{2\pi}
\int_{-\infty}^\infty
dt
\,
\langle
    \Delta\hat{\tilde\rho}({\bf k},0)
    \Delta\hat{\tilde\rho}(-{\bf k},t)
\rangle
\exp(-i \omega t)
\;,
\end{equation}
where $\Delta\hat{\tilde\rho}({\bf k},t)$ is the spatial Fourier transform of the density-fluctuation operator
\begin{equation}
\Delta\hat\rho({\bf x},t)
=
\hat\psi^\dagger({\bf x},t)
\hat\psi({\bf x},t)
-
\langle
\hat\psi^\dagger({\bf x},t)
\hat\psi({\bf x},t)
\rangle
\;.
\end{equation}
In terms of $S({\bf k},\omega)$, the susceptibility function is given by
\begin{eqnarray}
\operatorname{Re}[\chi({\bf k},\omega)]
&=&
\int_{-\infty}^\infty
\left[
    S({\bf k},\omega')
    \frac{\cal P}{\omega'-\omega}
    +
    S(-{\bf k},\omega')
    \frac{\cal P}{\omega'+\omega}
\right]
d\omega'
\;,
\nonumber\\
\operatorname{Im}[\chi({\bf k},\omega)]
&=&
\pi
\left[
    S({\bf k},\omega)
    -
    S(-{\bf k},-\omega)
\right]
\;.
\end{eqnarray}
If we denote the eigenstates of the unperturbed Hamiltonian by $|\Psi_\lambda\rangle$
with $\lambda=0$ corresponding to the ground state, the expression~(\ref{dsf}) at zero temperature can be rewritten in the form
\begin{equation}
\label{dsf-lambda}
S({\bf k},\omega)
=
\frac{1}{\hbar}
\sum_{\lambda}
\left|
    \langle\Psi_\lambda|
        \Delta\hat{\tilde\rho}(-{\bf k})
    |\Psi_0\rangle
\right|^2
\delta
\left(
    \omega-\omega_{\lambda 0}
\right)
\;.
\end{equation}
A finite-temperature generalization of Eq.~(\ref{dsf-lambda}) is given by
\begin{equation}
\label{dsf-lambda-T}
S({\bf k},\omega)
=
\frac{1}{\hbar{\cal Z}}
\sum_{\lambda_1\lambda_2}
\exp
\left(
    - \frac{E_{\lambda_2}}{k_{\rm B}T}
\right)
\left|
    \langle\Psi_{\lambda_1}|
        \Delta\hat{\tilde\rho}(-{\bf k})
    |\Psi_{\lambda_2}\rangle
\right|^2
\delta
\left(
    \omega-\omega_{\lambda_1 \lambda_2}
\right)
\;,
\end{equation}
where ${\cal Z}$ is the partition function.

The dynamic structure factor obeys certain sum-rules~\cite{PS2003,Schwabl-QMII}.
For instance, the energy-weighted moment is given by
\begin{equation}
\label{f-sum-rule}
\hbar^2
\int_{-\infty}^{\infty}
S({\bf k},\omega)
\omega
d\omega
=
\frac{1}{2}
\langle
  \left[
      \Delta\hat{\tilde\rho}^\dagger({\bf k}) ,
      \left[
          \hat H, \Delta\hat{\tilde\rho}({\bf k})
      \right]
  \right]
\rangle
=
N
\frac{\hbar^2 k^2}{2M}
\;,
\end{equation}
which is known as f-sum rule.
It holds for a wide class of many-body systems independent on the external potential and temperature.
Eq.~(\ref{f-sum-rule}) is directly related to the equation of continuity and the conservation of the particles number.
The inverse energy-weighted moment in the long-wavelength limit has the form
\begin{equation}
\label{comp-sum-rule}
\lim_{{\bf k}\to0}
\int_{-\infty}^{\infty}
\frac{S({\bf k},\omega)}{\omega}
d\omega
=
N
\frac{\kappa_{\rm T}}{2}
\;,
\end{equation}
where $\kappa_{\rm T}$ is the isothermal compressibility (see footnote in Sec.~\ref{sec-Tq}).
Eq.~(\ref{comp-sum-rule}) is referred to as the compressibility sum rule.
The zeroth moment of $S({\bf k},\omega)$ yields the static structure factor:
\begin{equation}
\label{ssf-def}
S_0({\bf k})
=
\frac{\hbar}{N}
\int_{-\infty}^\infty S({\bf k},\omega)
d\,\omega
=
\frac{1}{N}
\langle
    \Delta\hat{\tilde\rho}({\bf k})
    \Delta\hat{\tilde\rho}(-{\bf k})
\rangle
\end{equation}
which is the Fourier transform of the density-density correlation function in real space.
Together with Eq.~(\ref{tilde-rho}), this implies
\begin{equation}
\label{S0k0}
S_0(0)
=
\left(
    \langle
       \hat N^2
    \rangle
    -
    N^2
\right)/N
\;.
\end{equation}
If the fluctuations of the total number of particles vanish, which is always the case in the canonical ensemble, $S_0(0)$ vanishes as well.

If we restrict ourselves to the lowest Bloch band, the static structure factor takes the form
\begin{equation}
\label{S0-lowestBB}
S_0({\bf k})
=
1
+
\frac{1}{N}
\sum_{{\bf l}_1{\bf l}_2{\bf l}_3{\bf l}_4}
\!\!
G_{{\bf l}_1{\bf l}_2}({\bf k})
G_{{\bf l}_4{\bf l}_3}^*({\bf k})
\left[
    \langle
        \hat a_{{\bf l}_1}^\dagger
        \hat a_{{\bf l}_3}^\dagger
        \hat a_{{\bf l}_2}^{\phantom{\dagger}}
        \hat a_{{\bf l}_4}^{\phantom{\dagger}}
    \rangle
    -
    \langle
        \hat a_{{\bf l}_1}^\dagger
        \hat a_{{\bf l}_2}^{\phantom{\dagger}}
    \rangle
    \langle
        \hat a_{{\bf l}_3}^\dagger
        \hat a_{{\bf l}_4}^{\phantom{\dagger}}
    \rangle
\right]
\;,
\end{equation}
where
\begin{equation}
G_{{\bf l}_1{\bf l}_2}({\bf k})
=
\int
d{\bf x}
\,
W_{{\bf l}_1}^*({\bf x})
W_{{\bf l}_2}^{\phantom{*}}({\bf x})
\exp
\left(
    - i{\bf k}\cdot{\bf x}
\right)
\end{equation}
with $W_{\bf l}({\bf x})\equiv W({\bf x}-{\bf x}_{\bf l})$ being the Wannier function for the lowest Bloch band.
Thus, $S_0({\bf k})$ contains correlations of four lattice points.
It is easy to see that $G_{{\bf l}_1{\bf l}_2}({\bf k}=0)=\delta_{{\bf l}_1{\bf l}_2}$ and Eq.~(\ref{S0k0}) is fulfilled.

The absolute values of $G_{{\bf l}_1{\bf l}_2}({\bf k})$ decrease with increasing distance between the lattice points ${\bf l}_1$ and ${\bf l}_2$.
Taking into account only the dominant terms with ${\bf l}_2={\bf l}_1$ and ${\bf l}_4={\bf l}_3$ in Eq.~(\ref{S0-lowestBB}),
we come to the expression
\begin{eqnarray}
\label{S0lBB}
S_0({\bf k})
=
1
+
G_0^2({\bf k})
\left[
    \tilde S_0({\bf k})
    -1
\right]
\;,
\end{eqnarray}
where
\begin{eqnarray}
\label{S0discrete}
\tilde S_0({\bf k})
=
\frac{1}{N}
\sum_{{\bf l}_1,{\bf l}_2}
F_n({\bf l}_1,{\bf l}_2)
\exp
\left[
    i{\bf k}
    \cdot
    \left(
        {\bf x}_{{\bf l}_2}
        -
        {\bf x}_{{\bf l}_1}
    \right)
\right]
\end{eqnarray}
with the particle-number correlation function
\begin{equation}
\label{nncorr}
F_n({\bf l}_1,{\bf l}_2)
=
\langle
    \hat n_{{\bf l}_1}
    \hat n_{{\bf l}_2}
\rangle
-
\langle
    \hat n_{{\bf l}_1}
\rangle
\langle
    \hat n_{{\bf l}_2}
\rangle
\end{equation}
is the discrete analogue of $S_0({\bf k})$.
It corresponds to the fluctuations of the particle number described by the operator
$\Delta\hat{n}_{\bf l}=\hat{n}_{\bf l}-\langle\hat{n}_{\bf l}\rangle$
and contains only particle-number correlations of two sites.

$G_0({\bf k})$ in Eq.~(\ref{S0lBB}) is defined as
\begin{equation}
\label{G0}
G_0({\bf k})
\equiv
\left|
G_{\bf ll}({\bf k})
\right|
=
\int
d{\bf x}
\left|
    W({\bf x})
\right|^2
\exp
\left(
    -i {\bf k}\cdot{\bf x}
\right)
\;.
\end{equation}
In the Gaussian approximation~(\ref{WGauss}) it takes the form
\begin{equation}
G_0({\bf k})
\approx
\exp
\left[
    -
    \left(
        \frac{k a}{2\pi}
    \right)^2
    \sqrt{\frac{E_{\rm R}}{V_0}}
\right]
\;.
\end{equation}

The ``discrete" dynamic structure factor $\tilde S({\bf k},\omega)$ corresponding to the static structure factor $\tilde S_0({\bf k})$
is determined by Eqs.~(\ref{dsf}),~(\ref{dsf-lambda}),~(\ref{dsf-lambda-T}) with the operator $\Delta\hat{\tilde\rho}({\bf k},t)$
replaced by
\begin{equation}
\label{dnk}
\Delta\hat{\tilde n}({\bf k})
=
\sum_{\bf l}
\left(
    \hat n_{\bf l}
    -
    \langle
        \hat n_{\bf l}
    \rangle
\right)
\exp
\left(
    -i {\bf k}\cdot{{\bf x}_{\bf l}}
\right)
\;.
\end{equation}
This corresponds to the perturbation described by the Hamiltonian
\begin{eqnarray}
\hat{\tilde H}_{\rm Bragg}
=
V_{\rm Bragg}
\sum_{\bf l}
\cos
\left(
    {\bf k}\cdot{\bf x}_{\bf l}
    -
    \omega t
\right)
\hat n_{\bf l}
=
\frac{V_{\rm Bragg}}{2}
\left[
    \hat{\tilde n}^\dagger({\bf k})
    e^{-i\omega t}
    +
    \hat{\tilde n}({\bf k})
    e^{i\omega t}
\right]
\;.
\end{eqnarray}

The sum rule~(\ref{ssf-def}) is also fulfilled for $\tilde S({\bf k},\omega)$ and $\tilde S_0({\bf k})$:
\begin{eqnarray}
\label{S0-lat}
\tilde S_0({\bf k})
=
\frac{\hbar}{N}
\int_{-\infty}^\infty \tilde S({\bf k},\omega)
d\,\omega
=
\frac
{1}
{N}
\langle
    \Delta\hat{\tilde n}({\bf k})
    \Delta\hat{\tilde n}(-{\bf k})
\rangle
\;,
\end{eqnarray}
and Eq.~(\ref{S0k0}) holds for $\tilde S_0({\bf k}=0)$ too.
However, the f-sum rule~(\ref{f-sum-rule}) and the compressibility sum rule~(\ref{comp-sum-rule}) take
for $\tilde S({\bf k},\omega)$ slightly different forms.
The f-sum rule becomes~\cite{RB04}
\begin{eqnarray}
\label{fsum}
\hbar^2
\int_{-\infty}^\infty
\tilde S({\bf k},\omega)
\omega
d\omega
&=&
\frac{1}{2}
\langle
\left[
    \left[
        \Delta\hat{\tilde n}({\bf k}) , \hat H_{\rm BH}
    \right],
    \Delta\hat{\tilde n}^\dagger({\bf k})
\right]
\rangle
\nonumber\\
&=&
\sum_{\nu=1}^d
\left[
    \cos
    \left(
        q_\nu a
    \right)
    -1
\right]
\langle
   \hat H_\nu^{\rm kin}
\rangle
\;,
\end{eqnarray}
where
\begin{equation}
\hat H_\nu^{\rm kin}
=
-J
\sum_{\bf l}
\left(
    \hat a_{\bf l}^\dagger
    \hat a_{{\bf l}+{\bf e}_\nu}^{\phantom\dagger}
    +
    \hat a_{{\bf l}+{\bf e}_\nu}^\dagger
    \hat a_{\bf l}^{\phantom\dagger}
\right)
\end{equation}
is the kinetic-energy part of the Hamiltonian for the direction $\nu$.
In a homogeneous isotropic lattice, Eq.~(\ref{fsum}) can be rewritten in the form
\begin{equation}
\label{fsum-hom}
\hbar^2
\int_{-\infty}^\infty
\tilde S({\bf k},\omega)
\omega
d\omega
=
\langle
    \hat a_{\bf l}^\dagger
    \hat a_{{\bf l}+{\bf e}_\nu}^{\phantom{\dagger}}
\rangle
L^d
\epsilon_{\bf k}
\;,
\end{equation}
where
\begin{equation}
\label{e1p-lat}
\epsilon_{\bf k}
=
4J
\sum_{\nu=1}^d
\sin^2
\left(
    \frac{k_\nu a}{2}
\right)
\end{equation}
is the energy of free particles.
Comparing with Eq.~(\ref{f-sum-rule}) we see that the single-particle dispersion relation in continuous space
$\hbar^2 k^2/(2M)$
is replaced by the corresponding tight-binding dispersion relation $\epsilon_{\bf k}$ and instead of the total particle number $N$
we have
$
\langle
    \hat a_{\bf l}^\dagger
    \hat a_{{\bf l}+{\bf e}_\nu}^{\phantom{\dagger}}
\rangle
L^d
$
which is less than $N$.
The compressibility sum rule in a lattice has the form
\begin{equation}
\label{comp-sum-rule-lat}
\lim_{{\bf k}\to0}
\int_{-\infty}^{\infty}
\frac{\tilde S({\bf k},\omega)}{\omega}
d\omega
=
L^d
\frac{\kappa}{2}
\;,
\end{equation}
where $\kappa$ is defined by Eq.~(\ref{kappa}).

In some rather general cases, the dynamic structure factor is determined by a single excitation mode with the energy $\hbar\omega_{\bf k}$.
Then at $T=0$ it can be approximated as
\begin{equation}
\tilde S({\bf k},\omega)
=
Z_{\bf k}
\delta(\omega-\omega_{\bf k})
\;.
\end{equation}
Using the sum rules~(\ref{S0-lat}),~(\ref{fsum-hom}) we obtain a lattice analogue of the Feynman relation
\begin{equation}
\label{Feynman}
\hbar\omega_{\bf k}
=
\frac
{
 \langle
    \hat a_{\bf l}^\dagger
    \hat a_{{\bf l}+{\bf e}_\nu}^{\phantom{\dagger}}
 \rangle
}
{
 \langle
    \hat n_{\bf l}
 \rangle
}
\frac
{\epsilon_{\bf k}}
{\tilde S_0({\bf k})}
\;.
\end{equation}
Employing in addition the compressibility sum rule~(\ref{comp-sum-rule-lat}) we obtain
\begin{equation}
\label{csound}
\lim_{{\bf k}\to0}
\omega_{\bf k}
=
c_{\rm s}
\left|
    {\bf k}
\right|
\;,\quad
c_{\rm s}
=
\frac{a}{\hbar}
\sqrt
{
 \frac{2J}{\kappa}
 \langle
    \hat a_{\bf l}^\dagger
    \hat a_{{\bf l}+{\bf e}_\nu}^{\phantom{\dagger}}
 \rangle
}
\;,
\end{equation}
where $c_{\rm s}$ is the sound velocity.
From Eqs.~(\ref{Feynman}),~(\ref{csound}) we get
\begin{equation}
\label{S0small-k}
\lim_{{\bf k}\to0}
\tilde S_0({\bf k})
=
\frac
{\hbar\kappa c_{\rm s}}
{
 2
 \langle
    \hat n_{\bf l}
 \rangle
}
\left|
    {\bf k}
\right|
\;.
\end{equation}
For small $\left|{\bf k}\right|$, $G_0({\bf k})$ can be decomposed in powers of $\left|{\bf k}\right|^2$.
Therefore, the behavior of $S_0({\bf k})$ in the limit ${\bf k}\to0$ is also described by Eq.~(\ref{S0small-k}).

\subsection{In-situ imaging}

If the absorption images are taken without time-of-flight, they reveal {\it in situ}
atomic density distribution~\cite{GZHC09,IVLBMGS2010}. This technique allows to achieve the spatial resolution
of the order of few microns which makes possible to resolve individual sites only in some special cases~\cite{FFLI05,AWF10}.
Below we give a brief overview of other methods that reach higher resolution and in addition allow efficient manipulation on a single-atom level.
These methods allow to access the particle-number statistics of the individual sites~\cite{GZHC09,IVLBMGS2010,BPTMSGFPG10,SWECBK10}
as well as nonlocal correlation functions~\cite{HZHTGC2011,ECFWSGMBPBK11,CBPESFGBKK12}.

\subsubsection{Microwave spectroscopy}

Microwave spectroscopy is based on the resonant transfer of atoms from one internal state to the other
with the aid of two-photon pulses composed of microwave and radiofrequency photons.
In the experiment of Ref.~\cite{CMBMLMPK06}, these were the states
$|1\rangle\equiv|F=1,m_F=-1\rangle$ and $|2\rangle\equiv|F=2,m_F=1\rangle$ of $^{87}$Rb.
In the presence of interactions, the transition frequency is shifted by
$\Delta\omega\sim \rho({\bf x})(a_{21}-a_{11})$, where $\rho({\bf x})$ is the density,
$a_{21}$ is the scattering length of two atoms in the states $|1\rangle$ and $|2\rangle$,
$a_{11}$ is the scattering length of two atoms in the state $|1\rangle$.
After the pulse, the atoms in the state $|2\rangle$ are detected by light absorption imaging
showing the spatial distribution of the atoms with a given density.
Changing the photon frequency and repeating the measurements one can reconstruct the complete density profile.

\subsubsection{Scanning electron microscopy}

The fundamental physical process of the scanning electron microscopy is ionization of atoms by a focused electron beam.
The ionized atoms are extracted from the atomic sample with an electrostatic field and subsequently registered
by an ion detector. The diameter of the electron beam determines the spatial resolution which was $100-150$~nm
in the experiments of Refs.~\cite{GWRLO08,WLGKO09}.
The ion detection provides single atom sensitivity and allows to count atoms on individual sites.
This method can be used not only for measurements of the atomic distribution but also to produce arbitrary
patterns of occupied lattice sites~\cite{WLGKO09}.

\subsubsection{Fluorescence imaging}

Fluorescence imaging is a technique that allows to measure occupation numbers of the lattice sites modulo two.
The atomic sample is illuminated with a near-resonant light which provides simultaneously
sub-Doppler cooling~\cite{BGPFG09,BPTMSGFPG10,SWECBK10,WESCSFBK11,ECFWSGMBPBK11}.
During this process, pairs of atoms undergo light assisted collisions and quickly leave the trap before they can be detected.
After that only single atoms remain on the lattice sites with odd initial populations
and the sites with initially even populations are empty.
The remaining atoms scatter several thousand photons during the exposure time and can be detected with
high fidelity~\cite{WESCSFBK11}.
Spatial resolution of the obtained images is about $600-700$~nm~\cite{BGPFG09,BPTMSGFPG10,SWECBK10}.

The physical observable measured by the fluorescence imaging is described by the parity operator
\begin{equation}
\label{s-op}
\hat s_{{\bf l}}
=
\exp
\left(
    i \pi
    \hat n_{\bf l}
\right)
=
(-1)^{\hat n_{\bf l}}
\end{equation}
which yields $-1$ and $+1$ for odd and even occupation numbers, respectively.
Measuring the parities at different lattice sites and averaging over many experimental realizations
allows to determine the parity correlation function~\cite{KM10}
\begin{equation}
\label{parcorr-def}
F_{(-1)^n}
\left(
    {\bf l}_1, {\bf l}_2
\right)
=
\langle
    \hat s_{{\bf l}_1}
    \hat s_{{\bf l}_2}
\rangle
-
\langle
    \hat s_{{\bf l}_1}
\rangle
\langle
    \hat s_{{\bf l}_2}
\rangle
\;.
\end{equation}
It was measured in one- and two-dimensional lattices~\cite{ECFWSGMBPBK11,CBPESFGBKK12}
and provides an important information about correlated particle-hole pairs.

The limitations imposed by the parity projection were circumvented in bilayer Bose-Hubbard systems.
Using the interaction blockade in double wells~\cite{CTFSMFB08}, the occupation-sensitive transport between the two lattice planes
was engineered that made possible to resolve the occupation numbers of lattice sites in one plane $n_{\bf l}=0,1,2,3$
and to observe the spin ordering in a mixture of two hyperfine states~\cite{PMTSG15}.

\section{\label{sec-SSC}Simple special cases}

In this section, we discuss limiting cases which allow exact analytical solutions
or do not require much computational effort.

\subsection{Ideal Bose gas}

We start the discussion of the many-body phenomena
considering first an ideal gas of $N$ bosons in a finite lattice of $L^d$ sites under periodic boundary conditions
(${\hat a}_{{\bf l}+{\bf e}_\nu L} \equiv {\hat a}_{\bf l}$).
In this non-interacting limit ($U=0$), the system is governed by the first term of the Hamiltonian~(\ref{HBH}).
In the homogeneous lattice, it is convenient to work in the momentum representation~(\ref{ak}),
where the Hamiltonian takes the diagonal form
\begin{equation}
\hat H
=
\sum_{\bf k}
\epsilon_{\bf k}
\hat{\tilde a}_{\bf k}^\dagger
\hat{\tilde a}_{\bf k}^{\phantom{\dagger}}
\;,
\end{equation}
which readily gives the energy eigenvalues of a single particle
\begin{equation}
\label{e1p}
\epsilon_{\bf k}
=
-2J
\sum_{\nu=1}^d
\cos
\left(
    k_\nu a
\right)
\;,
\end{equation}
and the corresponding eigenstates
\begin{equation}
\label{1pp}
|{\bf k}\rangle
=
\hat{\tilde a}_{\bf k}^\dagger
|0\rangle
\;.
\end{equation}

\subsubsection{Energy spectrum}

From the solution of the single-particle eigenvalue problem, one can construct
the eigenstates of the many-body system as products of the Fock states
\begin{equation}
|\tilde {\bf n}\rangle
=
\bigotimes_{\bf k}
|\tilde n_{\bf k}\rangle
\;,\quad
\sum_{\bf k}
\tilde n_{\bf k}
=
N
\;,
\end{equation}
with the occupation numbers $\tilde n_{\bf k}$ of the ${\bf k}$-modes.
They have the energies
\begin{equation}
E_{\tilde{\bf n}}
=
\sum_{\bf k}
\tilde n_{\bf k}
\epsilon_{\bf k}
\;.
\end{equation}
In the ground state, all the bosons occupy the single-particle mode with ${\bf k}=0$.
In the lowest excited state, $N-1$ bosons occupy the mode with ${\bf k}=0$ and one boson
is in the mode with $k_\nu a=\pm 2\pi/L$ in any direction $\nu$.
Thus, the lowest excited state is $2d$-fold degenerate and has the total momentum
$\pm 2\pi\hbar/(La)$.
If $L$ is large, all the eigenenergies are infinitesimally close to each other and, therefore,
there is no gap in the excitation spectrum in the thermodynamic limit.
Note that the energy of particle-hole excitations vanishes for any lattice size.
In general, the energy spectrum has a band structure and its explicit form depends on the number of particles.
In order to demonstrate this, we consider two examples.
For the sake of simplicity, we shall restrict ourselves to one dimension but the generalization to higher dimensions is straightforward.

In the case of two atoms, the energy spectrum is obtained by addition of contributions from two free particles with the momenta
$P_\pm=\left(\frac{P}{2}\pm p\right)$,
where
$P=P_+ + P_-=\hbar K$
is the center-of-mass momentum and
$p=(P_+ - P_-)/2=\hbar k$
is the relative momentum.
The energies are given by
\begin{equation}
\label{EKk}
E^{K\Omega}_{2}
\equiv
E_{K,k}
=
\epsilon_{\frac{K}{2}-k}
+
\epsilon_{\frac{K}{2}+k}
=
-q_K
\cos
\left(
    k a
\right)
\;,\quad
q_K
=
4J
\cos
\left(
    \frac{Ka}{2}
\right)
\;,
\end{equation}
and form a quasi-continuous band with the boundaries $E_{K,\pm{\pi}/{a}} = \pm q_K$.

In the case of a commensurate filling,
$N=nL$, with $n$ being an integer, the lower boundary of the band is given by
\begin{equation}
\label{Emin}
E^{K,{\rm min}}_{nL}
=
-2JN
\left[
    1 -
    \sin^2
    \left(
        \frac{\pi}{L}
    \right)
    \frac{\left|K a\right|}{\pi n}
\right]
\;\quad
\pi \le K a \le \pi
\;,
\end{equation}
and the upper one
\begin{equation}
E^{K{\rm max}}_{nL}
=
\left\{
    \begin{tabular}{ll}
      $-E^{K{\rm min}}_{nL}$ &, if $L$ is even\;,\\
      $2JN\cos\left(\frac{\pi}{L}\right)$  &, if $L$ is odd\;.
    \end{tabular}
\right.
\end{equation}
In the thermodynamic limit, all the eigenstates of the $N$-body system
are enclosed in the interval $[-2JN,2JN]$ for any value of $K$.

\subsubsection{Ground-state properties}

The ground state of $N$ ideal bosons
\begin{equation}
\label{GSibg}
|\tilde{\bf n}_{\bf 0}\rangle
=
\bigotimes_{\bf k}
|N \delta_{{\bf k},{\bf 0}}\rangle
=
\frac
{
 \left(
     \hat{\tilde a}_{\bf 0}^\dagger
 \right)^N
}
{
 \sqrt{N!}
}
|0\rangle
\end{equation}
is unique and has the energy $E_{\tilde{\bf n}_{\bf 0}}=-2dJN$.
It is a perfect superfluid state with $f^{\rm s}_N=1$.
Taking into account Eq.~(\ref{ak}),
the expression~(\ref{GSibg}) for the ground state can be rewritten in terms of the occupation numbers of
the individual lattice sites $n_{\bf l}$ as
\begin{equation}
|\tilde{\bf n}_{\bf 0}\rangle
=
\sum_{\bf n}
\sqrt{\frac{N!}{L^{dN}}}
\bigotimes_{{\bf l}}
\frac
{|n_{\bf l}\rangle}
{\sqrt{n_{\bf l}!}}
\;.
\end{equation}
Therefore, the number of bosons $n_{\bf l}$ at any lattice site ${\bf l}$ follows the binomial probability distribution
\begin{equation}
\label{binomial}
p(n_{\bf l}=n)
=
\frac{N!}{n!(N-n)!}
\left(
    \frac{1}{L^d}
\right)^n
\left(
    1-\frac{1}{L^d}
\right)^{N-n}
\;,\quad
n=0,1,\dots,N
\;,
\end{equation}
with the mean value $\langle \hat n_{\bf l}\rangle=N/L^d$ and standard deviation
\begin{equation}
\label{sigma-ibg}
\sigma_{n_{\bf l}}
=
\sqrt{\langle \hat n_{\bf l}^2\rangle-\langle \hat n_{\bf l}\rangle^2}
=
\sqrt
{
 \langle \hat n_{\bf l}\rangle
 \left(
     1 - \frac{1}{L^d}
 \right)
}
\;.
\end{equation}
Eq.~(\ref{binomial}) reflects the fact that in the absence of interactions every boson
can be placed on a particular site independent of the others with the probability $1/L^d$.
In the thermodynamic limit, the probability distribution~(\ref{binomial}) takes the form
\begin{equation}
\label{Poisson}
p(n_{\bf l}=n)
=
e^{-\langle \hat n_{\bf l}\rangle}
\frac{\langle \hat n_{\bf l}\rangle^n}{n!}
\;.
\end{equation}
This result is known as Poisson limit theorem.

The one-body density matrix has the entries
\begin{equation}
\label{F_a-ibg}
\langle
  \hat a_{{\bf l}_1}^{\dagger}
  \hat a_{{\bf l}_2}^{\phantom{\dagger}}
\rangle
=
\langle \hat n_{\bf l}\rangle
\end{equation}
for any ${\bf l}_1$ and ${\bf l}_2$ and does not show any dependence on the system size and dimensionality.
Therefore, the condensate fraction is exactly one.

The particle-number correlation function has the following form:
\begin{equation}
\label{F_n-ibg}
F_n
\left(
    {\bf l}_1, {\bf l}_2
\right)
=
\langle \hat n_{\bf l}\rangle
\left(
    \delta_{{\bf l}_1, {\bf l}_2}
    -
    \frac{1}{L^d}
\right)
\;.
\end{equation}
For ${\bf l}_1 = {\bf l}_2$, Eq.~(\ref{F_n-ibg}) is consistent with~(\ref{sigma-ibg}).
The term $1/L^d$ makes no contribution in the thermodynamic limit and $F_n({\bf l}_1, {\bf l}_2\ne{\bf l}_1)$ vanishes.
However, it gives singular contributions to the static structure factor.
From Eqs.~(\ref{S0lBB}) and (\ref{S0discrete}) we obtain
\begin{equation}
\tilde S_0({\bf k})
=
1-\delta_{{\bf k},\frac{2\pi}{a}{\bf q}}
\;,\quad
S_0({\bf k})
=
1-G_0^2({\bf k})\delta_{{\bf k},\frac{2\pi}{a}{\bf q}}
\;.
\end{equation}
$\tilde S_0({\bf k})$ and $S_0({\bf k})$ are equal to one for all ${\bf k}$, except those that coincide with the vectors of reciprocal lattice
$\frac{2\pi}{a}{\bf q}$. At these points $\tilde S_0$ always drops to zero, while $S_0$ drops to zero only for ${\bf k}=0$
and remains different from zero for finite ${\bf k}$.

The parity correlation function for a finite system is given by
\begin{equation}
\label{parcorr-ibg}
F_{(-1)^n}
\left(
    {\bf l}_1, {\bf l}_2
\right)
=
\left(
    1 - \frac{4\langle \hat n_{\bf l}\rangle}{N}
\right)^N
-
\left(
    1 - \frac{2\langle \hat n_{\bf l}\rangle}{N}
\right)^{2N}
\;,
\end{equation}
provided that ${\bf l}_1 \ne {\bf l}_2$. If ${\bf l}_1 = {\bf l}_2$, the first term on the right-hand side of Eq.~(\ref{parcorr-ibg})
is replaced by one. Approaching the thermodynamic limit,
$
F_{(-1)^n}
\left(
    {\bf l}_1, {\bf l}_2
\right)
$
vanishes as
\begin{equation}
\label{parcorr-ibg-tdl}
F_{(-1)^n}
\left(
    {\bf l}_1 \ne {\bf l}_2
\right)
\approx
-
\frac{4}{L^d}
\langle \hat n_{\bf l}\rangle
\exp
\left(
    - 4 \langle \hat n_{\bf l}\rangle
\right)
\;.
\end{equation}
Eqs.~(\ref{sigma-ibg}),~(\ref{F_n-ibg}),~(\ref{parcorr-ibg-tdl}) show that the leading finite-size corrections
scale as $\sim\langle \hat n_{\bf l}\rangle/L^d$.

\subsubsection{\label{sec:TcIBG}Critical temperature for the condensation}

For the ideal Bose gas, the total number of particles in the grand-canonical ensemble is given by
\begin{equation}
\label{Ngc}
N
=
\sum_{{\bf k}\in 1{\rm BZ}}
\frac
{1}
{
 \exp
 \left[
     \left(
         \epsilon_{\bf k} - \mu
     \right)/k_{\rm B} T
 \right]
 -1
}
\;.
\end{equation}
The critical temperature $T_{\rm c}$ is defined by the condition $\mu=\epsilon_{\bf 0}$~\cite{PS02book,PS2003}.
In the limit of infinite lattice, the summation in Eq.~(\ref{Ngc}) can be replaced by an integral
which leads to the following equation for the critical temperature:
\begin{equation}
\label{Tc-eq}
\langle
   \hat n_{\bf l}
\rangle
=
\frac
{N}{L^d}
=
\sum_{j=1}^\infty
\left[
    \exp
    \left(
        - \frac{2J}{k_{\rm B} T_{\rm c}} j
    \right)
    I_0
    \left(
        \frac{2J}{k_{\rm B} T_{\rm c}} j
    \right)
\right]^d
\;,
\end{equation}
where
\begin{equation}
I_0(x)
=
\frac{1}{2\pi}
\int_{-\pi}^\pi
\exp
\left(
    x\cos\varphi
\right)
d\varphi
\end{equation}
is the modified Bessel function of the first kind.
At large values of the argument, the asymptotics of $I_0(x)$ is given by
\begin{equation}
\label{I0-large}
I_0(x)
\approx
\exp(x)/\sqrt{2\pi x}
\;,
\end{equation}
which implies that the series in Eq.~(\ref{Tc-eq}) converges only for $d\ge 3$, provided that $T_{\rm c}$ is finite.
Therefore, in an infinite homogeneous lattice the finite-temperature Bose-Einstein condensate exists only in three dimensions similarly
to the case without any external potentials.

\begin{figure}[tb]
\centering

\stepcounter{nfig}
\includegraphics[page=\value{nfig}]{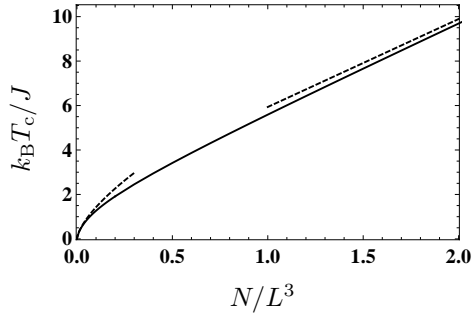}

\caption
{
Critical temperature of the ideal Bose gas in a three-dimensional lattice.
Solid line is an exact numerical solution of Eq.~(\ref{Tc-eq}).
Dashed lines are approximate analytical results for small and large fillings
described by Eqs.~(\ref{Tc-small}) and (\ref{Tc-large}), respectively.
}
\label{Tc-ibg}
\end{figure}

Numerical solution of Eq.~(\ref{Tc-eq}) presented in Fig.~\ref{Tc-ibg} shows that $T_{\rm c}$ grows with the filling factor
$\langle\hat n_{\bf l}\rangle=N/L^3$.
For $\langle\hat n_{\bf l}\rangle \ll 1$, $T_{\rm c}$ is small and $I_0(2Jj/k_{\rm B}T_{\rm c})$ in Eq.~(\ref{Tc-eq}) can be well approximated
by the asymptotic expression~(\ref{I0-large}).
This gives the result
\begin{equation}
\label{Tc-small}
\frac{k_{\rm B}T_{\rm c}}{J}
=
4\pi
\left[
    \frac
    {\langle\hat n_{\bf l}\rangle}
    {\zeta(3/2)}
\right]^{2/3}
\;,\quad
\zeta(3/2)=2.612\dots
\;,
\end{equation}
which can be also obtained from the well-known expression for the ideal Bose gas in a homogeneous continuous space~\cite{PS02book}
replacing the mass $M$ by the effective mass~(\ref{meff})
which follows from the dispersion relation of non-interacting particles in a lattice~(\ref{e1p-lat}).

In the opposite limit, $\langle\hat n_{\bf l}\rangle \gg 1$, $T_{\rm c}$ is large and the sum in Eq.~(\ref{Tc-eq}) can be replaced by an integral.
In this manner, we obtain
\begin{equation}
\label{Tc-large}
\frac{k_{\rm B}T_{\rm c}}{J}
=
\frac
{2 \langle\hat n_{\bf l}\rangle +1}
{I_3}
\;,\quad
I_3
=
\int_0^\infty
\left[
    \exp(-x)
    I_0(x)
\right]^3
dx
\;.
\end{equation}
The integral $I_3$ was calculated in fact by G.~N.~Watson in 1939~\cite{W39} and the result can be expressed in terms
of the complete elliptic integral of the first kind
\begin{equation}
K(x)
=
\int_{0}^{\pi/2}
\left(
    1 - x^2 \sin^2 t
\right)^{-1/2}
dt
\end{equation}
as~\cite{W39,PBM-book}
\begin{eqnarray}
I_3
=
\frac{4}{\pi^2}
\left(
    18 + 12\sqrt{2} - 10\sqrt{3} - 7\sqrt{6}
\right)
K^2
\left[
    \left(
        2-\sqrt{3}
    \right)
    \left(
        \sqrt{3}-\sqrt{2}
    \right)
\right]
\;,
\end{eqnarray}
which results in the numerical value $I_3=0.505\dots$.
In the case of unit filling, $\langle\hat n_{\bf l}\rangle=1$, Eq.~(\ref{Tc-eq}) gives $k_{\rm B} T_{\rm c}/J=5.591$~\cite{CPS2007},
and for $\langle\hat n_{\bf l}\rangle=2$ we get $k_{\rm B} T_{\rm c}/J=9.69$.

\subsubsection{Harmonic trap}

In the presence of confining potential the translational invariance is broken.
Since in the case of harmonic trap the problem is separable,
we restrict ourselves to one dimension. In addition, we shall consider open boundary conditions.

\begin{figure}[tb]
\centering

\stepcounter{nfig}
\includegraphics[page=\value{nfig}]{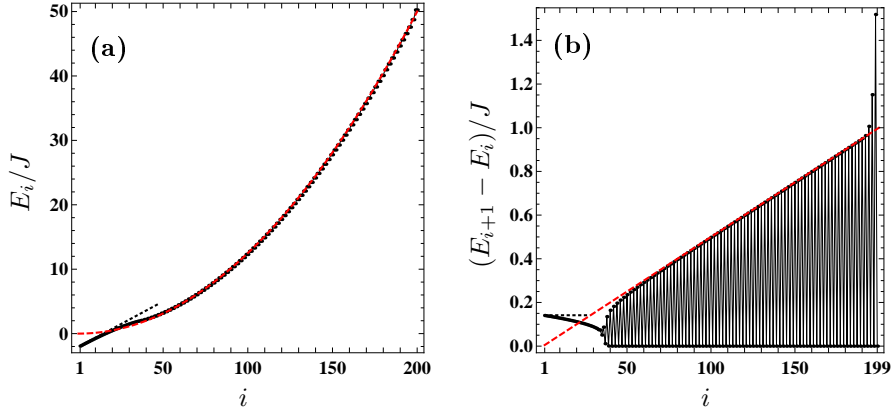}

\caption
{
(color online)
Energy eigenvalues (a) and the level spacing (b) for a single particle in a lattice of $L=200$ sites
in the presence of harmonic potential with $V_{\rm T}/J=0.005$.
Black dotted lines show the results given by Eq.~(\ref{Eiho})
and red dashed lines are determined by $E_i=V_{\rm T}(i/2)^2$ which corresponds to the local energies
of the harmonic confinement.
(For interpretation of the references to colour in this figure legend, the reader is referred to the web version of this article.)
}
\label{Fig:e_trap_1p}
\end{figure}

\begin{figure}[tb]
\centering

\stepcounter{nfig}
\includegraphics[page=\value{nfig}]{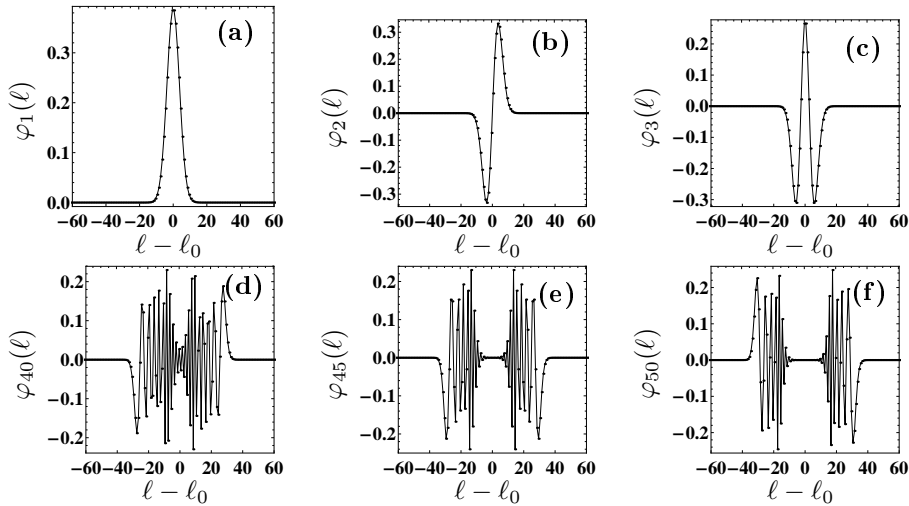}

\caption
{
Eigenmodes of a single particle in a lattice of $L=200$ sites
in the presence of harmonic potential with $V_{\rm T}/J=0.005$.
}
\label{Fig:wf_trap_1p}
\end{figure}

In the special case $V_{\rm T}=0$, the solution of the eigenvalue problem is given by
\begin{eqnarray}
E_i
&=&
-2J
\cos
\left(
    \frac{\pi i}{L+1}
\right)
\;,\quad
i=1,\dots,L
\;,
\\
\varphi_i(\ell)
&=&
\sqrt{\frac{2}{L+1}}
\sin
\left(
    \frac{\pi i}{L+1} \ell
\right)
\;,\quad
\ell=1,\dots,L
\;.
\nonumber
\end{eqnarray}
In the case of nonvanishing $V_{\rm T}$ the single-particle eigenstates become qualitatively different~\cite{HQ04,RM04a,RPCW05,VP08epl}.
They can be separated into two parts and some transient regime between those.
The lowest part of the energy spectrum is well described by the textbook result
for the harmonic oscillator in continuum (see Fig.~\ref{Fig:e_trap_1p}). In terms of the parameters of the Bose-Hubbard model it reads
\begin{equation}
\label{Eiho}
E_i
=
-2J
+
2
\sqrt{JV_{\rm T}}
\left(
    i - \frac{1}{2}
\right)
\;,\quad
i=1,2,\dots,
\end{equation}
and the agreement becomes better for smaller values of $V_{\rm T}/J$.
The corresponding wavefunctions are localized in the middle of the trap, see Fig.~\ref{Fig:wf_trap_1p}(a,b,c).

If the energies grow, the level spacing is not constant anymore.
For $E_i>2J$, the states become doubly degenerate which results in the rapid oscillations
of the level spacing~\cite{RM04a},~Fig.~\ref{Fig:e_trap_1p}(b).
This part of the spectrum is determined mainly by the local energies of the trapping potential because the contribution
of the hopping term is suppressed.
The corresponding wavefunctions vanish at the trap center and oscillate in the regions between the classical turning points
$\ell_c=\ell_0\pm\sqrt{(E_i+2J)/V_{\rm T}}$ and the turning points
$\ell_B=\ell_0\pm\sqrt{(E_i-2J)/V_{\rm T}}$ associated with Bragg reflection~\cite{HQ04,RM04a}.
With the increase of energy these regions move towards the trap edges and shrink
which can be seen in Fig.~\ref{Fig:wf_trap_1p}(d,e,f).
The Bragg reflection modifies significantly the density of states of a single atom
even in the limit $V_{\rm T}\to0$ compared to the homogeneous case $V_{\rm T}=0$ with periodic boundary conditions:
the square-root singularity in one dimension is replaced by a logarithmic one,
and the logarithmic van~Hove singularity in two dimensions disappears altogether~\cite{HQ04}.

\subsection{\label{Section_J=0}The limit of vanishing tunneling}

\subsubsection{Eigenstates}

In the absence of tunneling ($J=0$), the eigenstates of the Hamiltonian are given by Eq.~(\ref{basis})
and the corresponding eigenvalues
\begin{equation}
E^{{\bf K}\Gamma}_N
=
\frac{U}{2}
\sum_{\bf l}
n_{\Gamma{\bf l}}
\left(
    n_{\Gamma{\bf l}}-1
\right)
\end{equation}
do not depend on ${\bf K}$.
The properties of the eigenstates appear to be different for commensurate and incommensurate fillings.

We consider first the case of the commensurate filling, $N=nL^d$, where $n$ is a positive integer.
The ground state has equal occupation numbers at each lattice site, that is $n_{\Gamma{\bf l}}\equiv n$
and, therefore, the particle-number distribution is a Kronecker delta.
Since any translation maps the state into itself, we have $\nu_\Gamma=1$ in Eq.~(\ref{basis}) which means that
such a state exists only at $K=0$. It has the energy 
\begin{equation}
\label{E00}
E^{00}_{nL^d}
=
L^d
\frac{U}{2}n(n-1)
\,.
\end{equation}
This is a perfect insulator ($f^{\rm s}_{nL^d}\equiv 0$) and all spatial correlation functions vanish for this state.

The excited states are degenerate and form flat energy bands.
The lowest band contains, at each value of ${\bf K}$, $L^d-1$ degenerate eigenstates with the energies
$E^{{\bf K}\Gamma}_{nL^d}=E^{00}_{nL^d}+U$, $\Gamma=1,\dots,L^d-1$.
These states correspond to bosonic configurations with the same 
occupation numbers $n$ at any site except two, one of which contains 
$n-1$ bosons (hole excitation) and the other one $n+1$ (particle excitation).
They have $\nu_\Gamma=L$ for each spatial direction and, therefore, exist at any value of ${\bf K}$.
The highest band contains $L^d$ degenerate states (one state for each value of ${\bf K}$)
with all atoms sitting at the same lattice site.
These states have the energy $E^{{\bf K}\Gamma}_N=UN(N-1)/2$.
As we will see later, finite hopping rate $J$ lifts the degeneracy,
the bands acquire finite widths and can even overlap if the tunneling parameter is large enough.
If the filling is incommensurate, not only excited states but also the ground state are degenerate.

In the case of large interaction but incommensurate filling $N=n L^d + N'$, $N'<L^d$,
the ground state is degenerate.
These are $\frac{(N'+L^d-1)!}{N'!(L^d-1)!}$ states which are obtained from the state
$|\psi_{n L^d}\rangle$ creating one additional particle on $N'$ sites.
Their energy is
\begin{equation}
E_N
=
(L^d-N')
\frac{U}{2}
n
\left(
    n-1
\right)
+
N'
\frac{U}{2}
\left(
    n + 1
\right)
n
\;.
\end{equation}
The degeneracy is lifted at least partially, if we switch on infinitesimally small hopping $J$.
The energy of the first excited state will be infinitesimally close
to that of the ground state, i.e., there will be no energy gap in the excitation spectrum.

\subsubsection{Finite temperature}

The grand-canonical partition function in the limit $J=0$ has the form
${\cal Z}(\mu)={\cal Z}_0^{L^d}(\mu)$~\cite{RPP06}, where
\begin{equation}
\label{Z0}
{\cal Z}_0(\mu)
=
\sum_{n=0}^\infty
\exp
\left(
    - \frac{E_n-\mu n}{k_{\rm B}T}
\right)
\;,\quad
E_n
=
\frac{U}{2}
n(n-1)
\end{equation}
is the partition function of a single lattice site.
From this we can deduce the on-site particle-number distribution
\begin{equation}
\label{pnTJ0}
p(n_{\bf l}=n)
=
\frac{1}{{\cal Z}_0(\mu)}
\exp
\left(
    - \frac{E_n-\mu n}{k_{\rm B}T}
\right)
\;.
\end{equation}
At finite temperature, there is always a one-to-one correspondence between the chemical potential $\mu$ and
the filling $\langle\hat n_{\bf l}\rangle$ determined by the equation
\begin{equation}
\label{mean-pn}
\langle\hat n_{\bf l}\rangle
=
\sum_{n=0}^\infty
n \, p(n_{\bf l}=n)
\;.
\end{equation}
In general, this equation has to be solved numerically but in some special cases it allows also analytical solutions.

If the temperature is low, $k_{\rm B}T \ll U$, the particle-number distribution is strongly peaked near $n=n_0$,
where $n_0$ is an integer which is close to $\langle\hat n_{\bf l}\rangle$, and the probabilities of
the occupation numbers different from $n_0,n_0\pm1$ are negligible. In this case, we obtain the following expression
for the chemical potential
\begin{eqnarray}
\exp
\left(
    \frac{\mu}{k_{\rm B}T}
\right)
=
\frac
{
 \exp
 \left(
     \frac{Un_0}{k_{\rm B}T}
 \right)
}
{
2
\left(
     1 + n_0 - \langle\hat n_{\bf l}\rangle
\right)
}
\left[
    \langle\hat n_{\bf l}\rangle - n_0
    +
    \sqrt
    {
     \left(
        \langle\hat n_{\bf l}\rangle - n_0
     \right)^2
     +
     4
     \left[
         1
         -
         \left(
            \langle\hat n_{\bf l}\rangle - n_0
         \right)^2
     \right]
     e^{-\frac{U}{k_{\rm B}T}}
    }
\right]
\end{eqnarray}
which generalizes the solution obtained in Ref.~\cite{SRU06} for fillings $\langle\hat n_{\bf l}\rangle$ close to one.
The nonvanishing probabilities are given by
\begin{eqnarray}
\label{pn-lowT}
p(n_{\bf l}=n_0)
&=&
\frac
{
 \left(
      1 + n_0 - \langle\hat n_{\bf l}\rangle
 \right)
 \exp
 \left(
     \frac{\mu}{k_{\rm B}T}
 \right)
}
{
 \exp
 \left(
     \frac{\mu}{k_{\rm B}T}
 \right)
 +
 2
 \exp
 \left[
     \frac{U(n_0-1)}{k_{\rm B}T}
 \right]
}
\\
p(n_{\bf l}=n_0-1)
&=&
\frac
{
 \left(
      1 + n_0 - \langle\hat n_{\bf l}\rangle
 \right)
 \exp
 \left[
     \frac{U(n_0-1)}{k_{\rm B}T}
 \right]
}
{
 \exp
 \left(
     \frac{\mu}{k_{\rm B}T}
 \right)
 +
 2
 \exp
 \left[
     \frac{U(n_0-1)}{k_{\rm B}T}
 \right]
}
\nonumber\\
p(n_{\bf l}=n_0+1)
&=&
1 - p(n_{\bf l}=n_0) - p(n_{\bf l}=n_0-1)
\;.
\nonumber
\end{eqnarray}
If the filling is integer, $\langle\hat n_{\bf l}\rangle=n_0$, $\mu=U(n_0-1/2)$ and these expressions simplify as
\begin{equation}
\label{pn-lowT-int}
p(n_{\bf l}=n_0)
=
\frac
{1}
{
 1
 +
 2
 \exp
 \left(
     - \frac{U}{2k_{\rm B}T}
 \right)
}
\;,\quad
p(n_{\bf l}=n_0\pm1)
=
\frac
{
 \exp
 \left(
     - \frac{U}{2k_{\rm B}T}
 \right)
}
{
 1
 +
 2
 \exp
 \left(
     - \frac{U}{2k_{\rm B}T}
 \right)
}
\;.
\end{equation}

\begin{figure}[tb]
\centering

\stepcounter{nfig}
\includegraphics[page=\value{nfig}]{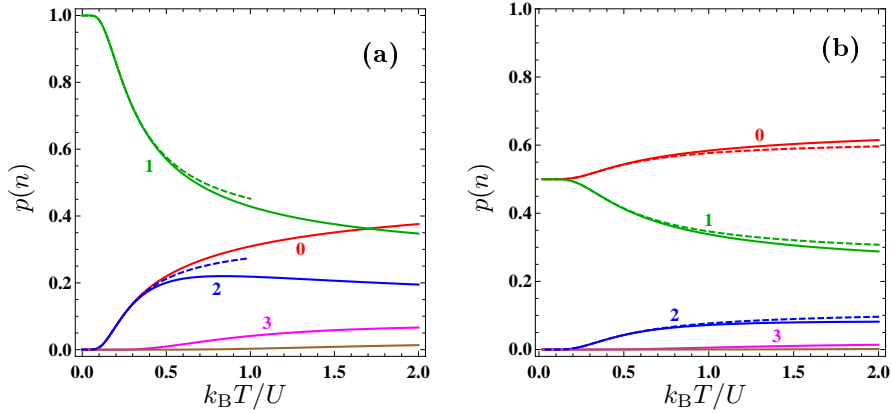}

\caption
{
(color online)
Probabilities $p(n_{\bf l}=n)$ of the occupation numbers $n=0$~(red), $1$~(green), $2$~(blue), $3$~(magenta), $4$~(brown)
for $J=0$ and filling $\langle\hat n_{\bf l}\rangle=1$~{\bf(a)}, $\langle\hat n_{\bf l}\rangle=0.5$~{\bf(b)}
obtained according to Eq.~(\ref{pnTJ0}) with the aid of numerical solution of Eq.~(\ref{mean-pn})
for the chemical potential $\mu$.
Dashed lines are the low-temperature limit~(\ref{pn-lowT}) with $n_0=1$.
}
\label{Fig:pn-T-J_0}
\end{figure}

In the high-temperature limit, $k_{\rm B}T \gg U$, and in the case of arbitrary filling,
the chemical potential is negative and has the form~\cite{RPP06}
\begin{equation}
\mu
=
-k_{\rm B}T
\ln
\left(
    \frac
    {1 + \langle\hat n_{\bf l}\rangle}
    {\langle\hat n_{\bf l}\rangle}
\right)
\;.
\end{equation}
In this regime, the particle-number statistics is described by the geometric distribution
\begin{equation}
p(n_{\bf l}=n)
=
\left(
    \frac
    {\langle\hat n_{\bf l}\rangle}
    {1 + \langle\hat n_{\bf l}\rangle}
\right)^n
\frac
{1}
{1 + \langle\hat n_{\bf l}\rangle}
\;,\quad
n=0,1,\dots
\end{equation}
The complete temperature dependence of $p(n_{\bf l}=n)$ in the case of unit and half filling
is shown in Fig.~\ref{Fig:pn-T-J_0}, see also Ref.~\cite{RPP06}.

\subsection{\label{S_BS2atoms}Scattering and bound states of two interacting atoms}

We consider two atoms in a one-dimensional lattice under periodic boundary conditions.
We assume that $L$ is odd
and look for the eigenstates of the Hamiltonian~(\ref{HBH})
in the form of the superposition
\begin{equation}
\label{apsi_k}
|K\Omega\rangle
=
\sum_{\Gamma=0}^{(L-1)/2}
c_{K\Omega\Gamma}
|{\bf n}_{K\Gamma}\rangle
\;,
\end{equation}
where the basis states~(\ref{basis}) are explicitly given by
\begin{eqnarray}
\label{nk2}
|{\bf n}_{K0}\rangle
&=&
\frac{1}{\sqrt{L}}
\sum_{j=0}^{L-1}
\left(
    \frac{\hat {\cal T}}{\tau_K}
\right)^j
|2 \underbrace{0 \dots 0}_{L-1} \rangle
\;,\quad
\\
|{\bf n}_{K\Gamma}\rangle
&=&
\frac{1}{\sqrt{L}}
\sum_{j=0}^{L-1}
\left(
    \frac{\hat {\cal T}}{\tau_K}
\right)^j
|1 \underbrace{0 \dots 0}_{\Gamma-1} 1 \underbrace{0 \dots 0}_{L-\Gamma-1}\rangle
\;,
\quad
\Gamma=1,\dots,\frac{L-1}{2}
\;,
\nonumber
\end{eqnarray}
where the index $\Gamma$ in this particular system
has a meaning of the interatomic distance and we have introduced the notation
$|n_1 \dots n_L\rangle \equiv \bigotimes_{\ell=1}^L |n_\ell\rangle$.
The eigenvalue problem for the Hamiltonian~(\ref{HBH}) can be written down in the form
\begin{equation}
\label{aevp}
\sum_{\Gamma'=0}^{(L-1)/2}
H_K^{\Gamma,\Gamma'}
c_{K\Omega\Gamma'}
=
E^{K\Omega}_2
c_{K\Omega\Gamma}
\;.
\end{equation}
The nonvanishing entries of the tridiagonal $(L+1)/2 \times (L+1)/2$ matrix $H_K$
are given by~\cite{Scott99}
\begin{eqnarray}
\label{HK-2a}
&&
H_K^{00}=U
\;,
\\
&&
H_K^{(L-1)/2,(L-1)/2}
=
- J
\left[
    \tau_K^{(L+1)/2}
    +
    \tau_K^{(L-1)/2}
\right]
\;,
\nonumber\\
&&
H_K^{01}
=
\left(
    H_K^{10}
\right)^*
=
- J \sqrt{2}
\left(
    1 + \tau_K
\right)
\;,
\nonumber\\
&&
H_K^{\Gamma,\Gamma+1}
=
\left(
    H_K^{\Gamma+1,\Gamma}
\right)^*
=
- J
\left(
    1 + \tau_K
\right)
\;,\quad
\Gamma=1,\dots,\frac{L-3}{2}
\;.
\nonumber
\end{eqnarray}
The eigenvectors in Eq.~(\ref{aevp}) satisfy the normalization condition
\begin{eqnarray}
\label{anorma}
\sum_{\Gamma=0}^{(L-1)/2}
\left|
    c_{K\Omega\Gamma}
\right|^2
&=&
1
\;.
\end{eqnarray}
In the case of even $L$, the summation in Eq.~(\ref{apsi_k}) is over $\Gamma=0,\dots,L/2$.
The basis state with $\Gamma=L/2$ can be rewritten in the form
\begin{equation}
|{\bf n}_{K,\Gamma=L/2}\rangle
=
\sum_{m=-\infty}^\infty
\delta_{q,2m}
\sqrt{\frac{2}{L}}
\sum_{j=0}^{\frac{L}{2}-1}
\left(
    \frac{\hat {\cal T}}{\tau_K}
\right)^j
|1 \underbrace{0 \dots 0}_{\frac{L}{2}-1} 1 \underbrace{0 \dots 0}_{\frac{L}{2}-1}\rangle
\;,
\end{equation}
which shows that it exists only for even values of $K$.
The form of matrix $H_K$ remains almost the same as in Eq.~(\ref{HK-2a}).
The only difference is that now all diagonal elements $H_K^{\Gamma,\Gamma}$ with $\Gamma > 0$ vanish
and the last off-diagonal matrix element
\begin{equation}
H_K^{L/2-1,L/2}
=
\left(
    H_K^{L/2,L/2-1}
\right)^*
=
- J \sqrt{2}
\left(
    1 + \tau_K
\right)
\sum_{m=-\infty}^\infty
\delta_{q,2m}
\end{equation}
for even $q$ is the same as $H_K^{01}$.

\begin{figure}[tb]
\centering

\stepcounter{nfig}
\includegraphics[page=\value{nfig}]{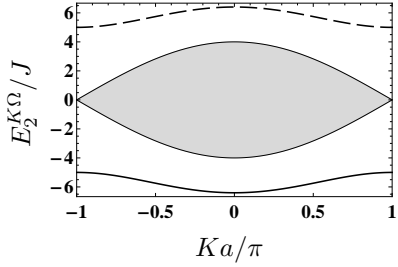}

\caption
{
Energy spectrum of two interacting atoms.
Shaded region is the scattering continuum described by Eq.~(\ref{EKk}).
Solid line below the scattering continuum and the dashed line above it
are the energies of the bound state~(\ref{E_k})
with $U/J=-5$ and $U/J=+5$, respectively.
}
\label{as}
\end{figure}

The eigenvalue problem~(\ref{aevp}) was solved in Refs.~\cite{SEG94,Scott99,E03}
analytically and numerically for negative $U$
but the result can be easily generalized to arbitrary $U$.
Later it was discussed in the context of ultracold
atoms~\cite{NPF07,PM07,PSAF07,VP08,JCS09,PV10,DSSAC12}.
The energy spectrum consists of the scattering states of a pair of asymptotically free particles and the bound state.

The energies of the scattering states are given by Eq.~(\ref{EKk}). In the limit $L\to\infty$,
they form a continuous band shown in Fig.~\ref{as} with the boundaries
$E_{K,0} = -q_K$ and $E_{K,{\pi}/{a}} = q_K$.
The corresponding eigenstates~(\ref{apsi_k}) are described by the coefficients
\begin{equation}
c_{K,k,\Gamma}
=
c_{K,k,0}
\sqrt{2}
\left[
    \cos
    \left(
        k a \Gamma
    \right)
    +
    \frac
    {
     U
     \sin
     \left(
         ka\Gamma
     \right)
    }
    {
     q_K
     \sin
     \left(
         ka
     \right)
    }
\right]
\exp
\left(
    i \frac{Ka}{2} \Gamma
\right)
\;,
\end{equation}
and the normalization has been discussed in Ref.~\cite{DSSAC12}.
In the limit $L\to\infty$, the bound state is given by
\begin{eqnarray}
\label{cKG}
c_{K0}
&=&
\sqrt
{
 \frac
 {1-b_K^2}
 {1+b_K^2}
}
\;,\quad
b_K
=
\frac{U-{\cal E}_K}{q_K}
\;,
\\
c_{K\Gamma}
&=&
\sqrt{2}
c_{K0} b_K^\Gamma
\exp
\left(
    i
    \frac{Ka}{2}
    \Gamma
 \right)
\;,\quad
\Gamma=1,2,\dots,\infty
\;,
\nonumber
\end{eqnarray}
and has the energy
\begin{eqnarray}
\label{E_k}
E^{K\Omega}_2
\equiv
{\cal E}_K
=
U
\sqrt
{
1
+
\left(
    \frac{q_K}{U}
\right)^2
}
\;,
\end{eqnarray}
which is also shown in Fig.~\ref{as}.

In the absence of interactions ($U=0$), $b_K=\pm 1$ and the normalization condition
(\ref{anorma}) cannot be fulfilled which means that the bound state does not exist in this case.
As soon as the interactions are present ($U\ne 0$), $|b_K|<1$ which guarantees the normalization.

In the case of attractive interactions ($U<0$), $b_K>0$ and all the coefficients
$c_{K\Gamma}$ are positive. The wavefunction does not have nodes since we are dealing with
the ground state of the system. In the case of repulsive interactions ($U>0$),
$b_K<0$ and the sign of the coefficients $c_{K\Gamma}$ alternates.
The wavefunction has infinitely many nodes reflecting the fact that this is a highly excited state.

\begin{figure}[tb]
\centering

\stepcounter{nfig}
\includegraphics[page=\value{nfig}]{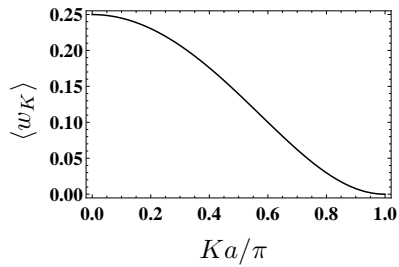}

\caption
{
Mean distance between two atoms in the bound state~(\ref{w_K}) with $U/J=\pm 5$.
}
\label{wb-2}
\end{figure}

The distance between the atoms $w$ is a random variable which takes the values
$w=\Gamma$, with the probabilities
$
\left|
     c_{K\Gamma}
\right|^2
$.
The mean interatomic distance in the bound state~(\ref{cKG}) is given by
\begin{equation}
\label{w_K}
\langle
   w_K
\rangle
=
\frac{2 b_K^2}{1 - b_K^4}
\;,
\end{equation}
which is shown in Fig.~\ref{wb-2}.
$
\langle
   w_K
\rangle
$
takes its maximal value at $K=0$ but vanishes at $Ka=\pi$.

\begin{figure}[tb]
\centering

\stepcounter{nfig}
\includegraphics[page=\value{nfig}]{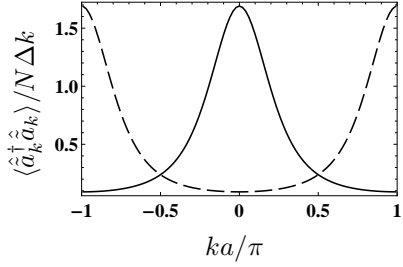}

\caption
{
Quasi-momentum distribution of two interacting atoms in the bound state
with $K=0$ and $U/J=-5$~(solid), $+5$~(dashed).
}
\label{mdb}
\end{figure}

Momentum distribution appears to be drastically different for different types of the eigenstates.
For the scattering states, it has two sharp peaks corresponding to the momenta of the atoms.
The bound states are characterized by broad momentum distributions shown in Fig.~\ref{mdb}.
In the case of attractive interactions, the momentum distribution takes its maximal value
at $k=0$, while for repulsive interactions, the maxima appear at $k=\pm\pi/a$.
With the increase of the interaction strength $|U|$, the momentum distribution
of the bound states becomes broader~\cite{WTLGDDKBZ06}.

The characteristic feature of the bound state is that
$
  \left|
      c_{K0}
  \right|^2
  >
  \left|
      c_{K\Gamma}
  \right|^2
$,
$\Gamma=1,\dots,(L-1)/2$, i.e., the probability of finding two atoms on the same lattice site
is higher than all the other ones.
In the case of attractive interactions ($U<0$),
this sort of localization corresponds to the soliton solution of the discrete nonlinear
Schr\"odinger equation and, therefore, the discrete level can be called a ``soliton band"~\cite{Scott99}.

The existence of the bound state in the case of repulsive interaction ($U>0$)
is quite unusual from the point of view of classical physics.
This is a purely quantum phenomenon which occurs in a structured environment produced
by the periodic potential in the absence of dissipation.
Repulsively bound pairs of $^{87}$Rb atoms were created in the experiment~\cite{WTLGDDKBZ06}
using a magnetic-field sweep across a Feshbach resonance and demonstrated by measurements
of the momentum distribution and the binding energy.

It is interesting to compare the solutions obtained for identical bosons with those for distinguishable particles.
In this case the energy spectrum remains the same, although the states are formally different.
They are linear combinations
\begin{eqnarray}
|K\Omega\rangle
=
\sum_{\Gamma=0}^{L-1}
c_{K\Omega\Gamma}
|{\bf n}_{K\Gamma}\rangle
\end{eqnarray}
of the basis states
\begin{eqnarray}
|{\bf n}_{K0}\rangle
&=&
\frac{1}{\sqrt{L}}
\sum_{j=0}^{L-1}
\left(
    \frac{\hat {\cal T}}{\tau_K}
\right)^j
|1_1+1_2, \underbrace{0 \dots 0}_{L-1} \rangle
\;,\quad
\nonumber\\
|{\bf n}_{K\Gamma}\rangle
&=&
\frac{1}{\sqrt{L}}
\sum_{j=0}^{L-1}
\left(
    \frac{\hat {\cal T}}{\tau_K}
\right)^j
|1_1 \underbrace{0 \dots 0}_{\Gamma-1} 1_2 \underbrace{0 \dots 0}_{L-\Gamma-1}\rangle
\;,
\quad
\Gamma=1,\dots,L-1
\;,
\nonumber
\end{eqnarray}
where the indices $1$ and $2$ label the two atoms.
The bound state in the limit of large $L$ is described by the coefficients
\begin{eqnarray}
c_{K\Gamma}
=
c_{K0}
b_K^\Gamma
\exp
\left(
    i
    \frac{Ka}{2}
    \Gamma
\right)
\;,\quad
c_{K,L-\Gamma}=c_{K\Gamma}^*
\;,\quad
\Gamma=0,1,\dots,L/2
\;,
\end{eqnarray}
where $c_{K0}$ is given by Eq.~(\ref{cKG}).
In spite of this difference, the distribution of the interparticle distances,
which is given now by $\left|c_{K0}\right|^2$ for $\Gamma=0$ and
$\left|c_{K\Gamma}\right|^2+\left|c_{K,L-\Gamma}\right|^2$ for $\Gamma=1,\dots,L/2$,
remains the same as in the case of the indistinguishable bosons.

\subsection{Hard-core bosons}

In the limit of infinite repulsion ($U=\infty$), the occupation  numbers of the individual lattice sites are
restricted by $0$ and $1$. Formally, this leads to the constrains for the bosonic operators
$\hat a_{\bf l}^{\dagger 2} = \hat a_{\bf l}^{2} = 0$ and
$
 \left\{
     \hat a^{\phantom \dagger}_{\bf l} , \hat a_{\bf l}^\dagger
 \right\}
 =1
$
and also implies that the total number of atoms $N$ cannot be larger than the total number of the lattice sites $L^d$.
The state with $N=L^d$ is trivially an insulator as no hopping on individual sites can take place.
Non-trivial physics is possible only for $N<L^d$.

In this limit, bosonic operators can be mapped into spin-$1/2$ operators by means of
the Holstein-Primakoff transformation~\cite{HP40}
\begin{equation}
\label{HPt}
\hat\sigma_{\bf l}^+
=
\hat a_{\bf l}^\dagger
\sqrt{1 - \hat a_{\bf l}^\dagger \hat a_{\bf l}^{\phantom{\dagger}}}
\;,\quad
\hat\sigma_{\bf l}^-
=
\sqrt{1 - \hat a_{\bf l}^\dagger \hat a_{\bf l}^{\phantom{\dagger}}}
\,\hat a_{\bf l}^{\phantom{\dagger}}
\;,\quad
\hat\sigma_{\bf l}^z
=
2 \hat a_{\bf l}^\dagger \hat a_{\bf l}^{\phantom{\dagger}} - 1
\;,
\end{equation}
where
$\hat\sigma_{\bf l}^\pm = \hat\sigma_{\bf l}^x \pm i \hat\sigma_{\bf l}^y$
are raising and lowering operators and $\hat\sigma_{\bf l}^a$, $a=x,y,z$, are Pauli matrices.
In what follows we assume that the products of the bosonic operators $\hat a_{\bf l}^\dagger$, $\hat a_{\bf l}^{\phantom{\dagger}}$
are arranged in the normal order, that is, creation operators are placed to the left of the annihilation operators.
Then the square roots in Eq.~(\ref{HPt}) become identities and the bosonic operators can be directly replaced by
$\hat\sigma_{\bf l}^\pm$~\cite{RSC06,HR2011}.
The transformation~(\ref{HPt}) maps the Bose-Hubbard Hamiltonian into the XY spin-$1/2$ Hamiltonian:
\begin{equation}
\hat H_{\rm XY}
=
-2 J
\sum_{\bf l}
\sum_{\nu=1}^d
\left(
    \hat\sigma_{\bf l}^x
    \hat\sigma_{{\bf l}+{\bf e}_\nu}^x
    +
    \hat\sigma_{\bf l}^y
    \hat\sigma_{{\bf l}+{\bf e}_\nu}^y
\right)
+
\sum_{\bf l}
\epsilon_{\bf l}
\frac
{\hat\sigma_{\bf l}^z + 1}
{2}
\;.
\end{equation}

\subsubsection{Bose-Fermi mapping in one-dimensional lattices}

The limit of infinite repulsion of bosons in a one-dimensional lattice is called Tonks-Girardeau regime.
It is exactly solvable via the Jordan-Wigner transformation~\cite{JW,LSM}
\begin{equation}
\label{JWT}
\hat a_\ell
=
\exp
\left(
    i \pi
    \sum_{\ell'<\ell}
    \hat c_{\ell'}^\dagger
    \hat c_{\ell'}^{\phantom \dagger}
\right)
\hat c_\ell
=
\prod_{\ell'<\ell}
\left(
    1 -
    2
    \hat c_{\ell'}^\dagger
    \hat c_{\ell'}^{\phantom \dagger}
\right)
\hat c_\ell
\;,
\end{equation}
where $\hat c_\ell$ and $\hat c^\dagger_\ell$
are the fermionic annihilation and creation operators.
Under this transformation, the hopping term of the Bose-Hubbard Hamiltonian as well as the local
particle-number operators
$\hat a_\ell^\dagger \hat a_\ell^{\phantom \dagger}$
remain invariant and the Hamiltonian~(\ref{HBH}) in the presence of an arbitrary external potential $\epsilon_\ell$
is mapped onto the one of noninteracting fermions:
\begin{equation}
\label{HHC}
\hat H_{\rm BH}
\equiv
\hat H_{\rm F}
=
-J
\sum_{\ell=1}^L
\left(
    \hat c_\ell^\dagger
    \hat c_{\ell+1}^{\phantom \dagger}
    +
    \hat c_{\ell+1}^\dagger
    \hat c_\ell^{\phantom \dagger}
\right)
+
\sum_{\ell=1}^L
\epsilon_\ell
\hat c^\dagger_\ell
\hat c_\ell^{\phantom \dagger}
\;.
\end{equation}
Periodic boundary conditions for bosons are equivalent to the requirement
\begin{equation}
\hat c_{L+1}
=
\exp
\left(
    - i \pi
    \sum_{\ell=1}^L
    \hat c_\ell^\dagger
    \hat c_\ell^{\phantom \dagger}
\right)
\hat c_1
\;,
\end{equation}
which implies periodic boundary conditions for fermions if the number of particles $N$ is odd,
otherwise one should use antiperiodic boundary conditions in the Hamiltonian~(\ref{HHC}).
This feature requires some care in the studies of the hard-core bosons with periodic boundary conditions
in the grand-canonical ensemble because the states with even and odd particle numbers should be treated in general separately~\cite{CC03,R05}.
In the case of open boundary conditions for bosons, the boundary conditions for the equivalent fermionic system remain the same.
Comparison with exact numerical solutions for the one-dimensional Bose-Hubbard model for arbitrary interaction $U$ shows that
the fermionization approach gives quantitatively correct results for $U/J\gtrsim200$~\cite{PRD04}.

The $N$-particle eigenstates $|\psi_{\rm F}\rangle$ of the Hamiltonian~(\ref{HHC}) can be constructed as products of the single-particle eigenstates
\begin{equation}
|\alpha\rangle
\equiv
\hat{\tilde c}_\alpha^\dagger
\,
|0\rangle
=
\sum_{\ell=1}^L
\varphi_{\alpha\ell}
\,
\hat c_\ell^\dagger
\,
|0\rangle
\;,
\end{equation}
with the eigenenergies $\varepsilon_\alpha$, $\alpha=1,\dots,L$.
If the energies are labeled in the ascending order, i.e.,
$\varepsilon_1 \le \varepsilon_2 \le \dots \le \varepsilon_L$,
the ground state is given by
$|\psi_F^G\rangle=\prod_{\alpha=1}^N |\alpha\rangle$.

The superfluid stiffness is defined by Eq.~(\ref{fs-def}).
In the limit $\theta\to0$ and at $T=0$ it is given by Eq.~(\ref{fs-exc}).
For hard-core bosons in 1D, the latter takes the following form~\cite{KTEG08}:
\begin{eqnarray}
\label{fshc}
f^{\rm s}_N
=
\frac{1}{2N}
\sum_{\ell=1}^L
\sum_{\alpha=1}^N
\left(
    \varphi_{\alpha\ell}^*
    \varphi_{\alpha,\ell+1}^{\phantom*}
    +
    {\rm c.c.}
\right)
-
\frac{J}{N}
\sum_{\alpha=N+1}^L
\sum_{\beta=1}^N
\frac
{1}
{\varepsilon_\alpha-\varepsilon_\beta}
 \left|
     \sum_{\ell=1}^L
     \left(
         \varphi_{\alpha\ell}^*
         \varphi_{\beta,\ell+1}^{\phantom*}
         -
         \varphi_{\alpha,\ell+1}^*
         \varphi_{\beta\ell}^{\phantom*}
     \right)
 \right|^2
\;,
\end{eqnarray}
where $\varphi_{\alpha,L+1}=(-1)^{N+1}\varphi_{\alpha 1}$.

From Eq.~(\ref{JWT}), it follows that the bosonic $L\times L$ one-body density matrix with the entries
$\langle\hat a_\ell^\dagger \hat a_{\ell'}^{\phantom\dagger}\rangle$
cannot be simply identified with the fermionic one
$\langle\hat c_\ell^\dagger \hat c_{\ell'}^{\phantom\dagger}\rangle$.
Although for $\ell'=\ell,\ell\pm 1$,
$\langle\hat a_\ell^\dagger \hat a_{\ell'}^{\phantom\dagger}\rangle$
coincide with
$\langle\hat c_\ell^\dagger \hat c_{\ell'}^{\phantom\dagger}\rangle$,
in general they are not the same.
However, they are related to each other and for $\ell'>\ell$ the quantities
$\langle\hat a_\ell^\dagger \hat a_{\ell'}^{\phantom\dagger}\rangle$
can be worked out as determinants of the
$(\ell'-\ell)\times(\ell'-\ell)$ matrices $G^{(\ell,\ell')}$~\cite{LSM,EL75,CC03,PWMMFCSHB04,MTEG05}:
\begin{equation}
\label{ada-det}
\langle\hat a_\ell^\dagger \hat a_{\ell'}^{\phantom\dagger}\rangle
=
2^{\ell'-\ell-1}
\det G^{(\ell,\ell')}
\;,
\end{equation}
where the entries of $G^{(\ell,\ell')}$ are given by
\begin{equation}
G^{(\ell,\ell')}_{i,j}
=
\langle
    \hat c_{\ell'-i}^\dagger
    \hat c_{\ell'+1-j}^{\phantom\dagger}
\rangle
-
\frac{1}{2}
\delta_{j,i+1}
\;,\quad
i,j=1,\dots,(\ell'-\ell)
\;.
\end{equation}
At zero temperature, fermionic one-body density matrix
$\langle \hat c_\ell^\dagger \hat c_{\ell'}^{\phantom\dagger} \rangle$
can be calculated using the solution of the single-particle eigenvalue problem as
\begin{equation}
\label{cdc_T0}
\langle \hat c_\ell^\dagger \hat c_{\ell'}^{\phantom\dagger} \rangle
=
\sum_{\alpha=1}^N
\varphi^*_{\alpha\ell}
\,
\varphi^{\phantom*}_{\alpha\ell'}
\;.
\end{equation}
If the temperature is finite and for open boundary conditions, Eq.~(\ref{cdc_T0}) is generalized by 
\begin{equation}
\label{obdm-fermi}
\langle \hat c_\ell^\dagger \hat c_{\ell'}^{\phantom\dagger} \rangle
=
\sum_{\alpha=1}^L
\varphi^*_{\alpha\ell}
\,
\varphi^{\phantom*}_{\alpha\ell'}
f_{\rm FD}(\varepsilon_\alpha)
\;,
\end{equation}
where $f_{\rm FD}$ is the Fermi-Dirac distribution function:
\begin{equation}
f_{\rm FD}(\varepsilon)
=
\frac
{1}
{
 \exp
 \left[
     \left(
         \varepsilon - \mu
     \right)/k_{\rm B} T
 \right]
 +
 1
}
\;.
\end{equation}
The chemical potential $\mu$ is fixed by the requirement to have the desired number of particles
$N=\sum_\ell \langle \hat a_\ell^\dagger \hat a_\ell^{\phantom\dagger} \rangle$.

Alternatively, the bosonic one-body density matrix at $T=0$ can be represented in the form~\cite{RM04,RM05}
\begin{equation}
\label{OBDMhcP}
\langle\hat a_\ell^\dagger \hat a_{\ell'}^{\phantom\dagger}\rangle
=
\det
\left[
    P^\dagger(\ell')
    P(\ell)
\right]
\;,
\end{equation}
where the $L\times (N+1)$ matrix $P(\ell)$ is determined as
\begin{eqnarray}
\label{P_ial}
P_{i\alpha}(\ell)
=
\left\{
    \begin{tabular}{rlll}
    $-\varphi_{\alpha i}$ & for & $i=1,\dots,\ell-1$, & $\alpha=1,\dots,N$;\\
     $\varphi_{\alpha i}$ & for & $i=\ell,\dots,L$,   & $\alpha=1,\dots,N$;\\
     $\delta_{i\ell}$     & for & $i=1,\dots,L$,      & $\alpha=N+1$.
    \end{tabular}
\right.
\end{eqnarray}
Eq.~(\ref{OBDMhcP}) can be also rewritten in the form~\cite{NI11}
\begin{equation}
\label{OBDMhcA}
\langle\hat a_\ell^\dagger \hat a_{\ell'}^{\phantom\dagger}\rangle
=
\det
\left[
    A(\ell,\ell')
\right]
\sum_{\alpha,\beta=1}^N
\varphi^{\phantom*}_{\alpha\ell'}
A^{-1}_{\alpha\beta}(\ell,\ell')
\varphi^*_{\beta\ell}
\;,
\end{equation}
where the entries of the $N\times N$ matrix $A(\ell,\ell')$ are given by
\begin{equation}
\label{matrixA}
A_{\alpha\beta}(\ell,\ell')
=
\delta_{\alpha\beta}
-2
\sum_{i=\ell}^{\ell'-1}
\varphi^*_{\alpha i}
\varphi^{\phantom*}_{\beta i}
\;,\quad
\ell < \ell'
\;.
\end{equation}
The advantage of Eqs.~(\ref{OBDMhcA}),~(\ref{matrixA}) is that the numerical calculation of
$\langle\hat a_\ell^\dagger \hat a_{\ell'}^{\phantom\dagger}\rangle$
can be done efficiently thanks to the recurrence relation
\begin{equation}
A_{\alpha\beta}(\ell,\ell'+1)
=
A_{\alpha\beta}(\ell,\ell')
-2
\varphi^*_{\alpha\ell'}
\varphi^{\phantom*}_{\beta\ell'}
\;.
\end{equation}
It is interesting to note that Eqs.~(\ref{OBDMhcA}),~(\ref{matrixA}) are the discrete version of the ones for
the Tonks-Girardeau gas in the continuum~\cite{PB07}.

For finite temperature and in the grand-canonical ensemble with open boundary conditions,
the bosonic one-body density matrix for $\ell\ne\ell'$ can be also obtained as~\cite{R05}
\begin{eqnarray}
\label{OBDMhcT}
\langle\hat a_\ell^\dagger \hat a_{\ell'}^{\phantom\dagger}\rangle
=
\frac{1}{\cal Z}
\left\{
    \det
    \left[
        I + (I+A) O(\ell') U R \, U^\dagger O(\ell)
    \right]
    -
    \det
    \left[
        I + O(\ell') U R \, U^\dagger O(\ell)
    \right]
\right\}
\;,
\end{eqnarray}
where
\begin{eqnarray}
{\cal Z}
=
\det(I+R)
=
\prod_{\alpha=1}^L
\left[
    1
    +
    \exp
    \left(
        - \frac{\varepsilon_\alpha-\mu}{k_{\rm B}T}
    \right)
\right]
\end{eqnarray}
is the partition function, $I$ is the identity matrix,
$U$ is the unitary $L\times L$ matrix containing the eigenvectors of $\hat H_{\rm F}$ for $N=1$ in its columns
($U_{\ell\alpha}=\varphi_{\alpha\ell}$),
$O(\ell)$ is a diagonal matrix with the first $(l-1)$ elements equal to $-1$ and the others equal to $+1$.
In addition, we defined
$
R=
\exp
\left[
    - (\varepsilon-\mu I)/k_{\rm B}T
\right]
$
with $\varepsilon$ being a diagonal matrix of the eigenenergies $\varepsilon_\alpha$.
The diagonal elements of the one-body density matrix are the same as for noninteracting fermions and can be easily calculated
according to Eq.~(\ref{obdm-fermi}) with $\ell'=\ell$, which can be also rewritten as~\cite{R05}
\begin{equation}
\langle\hat a_\ell^\dagger \hat a_\ell^{\phantom\dagger}\rangle
=
\langle\hat c_\ell^\dagger \hat c_\ell^{\phantom\dagger}\rangle
=
\left[
    U (I+R)^{-1} U^\dagger
\right]_{\ell\ell}
\;.
\end{equation}
After simple transformations, Eq.~(\ref{OBDMhcT}) can be rewritten in the form analogous to~(\ref{OBDMhcA})~\cite{NI11}
\begin{equation}
\label{OBDMhcB}
\langle\hat a_\ell^\dagger \hat a_{\ell'}^{\phantom\dagger}\rangle
=
\sum_{\alpha,\beta=1}^L
\varphi^*_{\alpha\ell}
B_{\alpha\beta}(\ell,\ell')
\varphi^{\phantom*}_{\beta\ell'}
\;,
\end{equation}
where the $L\times L$ matrix $B(\ell,\ell')$ is defined by
\begin{equation}
\label{matrixB}
B(\ell,\ell')
=
(-1)^{\ell'-\ell}
\frac
{\det(A+R)}
{\det(I+R)}
\left(
    A^T + R
\right)^{-1}
\;,
\end{equation}
and $A$ is the same as in Eq.~(\ref{matrixB}) but extended to $\alpha,\beta=1,\dots,L$.

The particle-number correlation function of hard-core bosons is formally the same as for non-interacting fermions:
\begin{equation}
\label{Fn-HCB}
F_n(\ell,\ell')
=
\langle\hat n_\ell\rangle
\delta_{\ell\ell'}
-
\left|
    \langle\hat c_\ell^\dagger \hat c_{\ell'}^{\phantom\dagger}\rangle
\right|^2
\;.
\end{equation}
At finite $T$ and under open boundary conditions, it has the form
\begin{equation}
F_n(\ell,\ell')
=
\sum_{\alpha=1}^L
\varphi^*_{\alpha\ell}
\,
\varphi^{\phantom*}_{\alpha\ell'}
f_{\rm FD}(\varepsilon_\alpha)
\sum_{\beta=1}^L
\varphi^*_{\beta\ell'}
\,
\varphi^{\phantom*}_{\beta\ell}
\left[
    1 -
    f_{\rm FD}(\varepsilon_\beta)
\right]
\end{equation}
which allows to express the dynamic structure factor in terms of the single-particle eigenmodes as~\cite{VM01}
\begin{eqnarray}
\label{dsfhc}
\tilde S(k,\omega)
=
\sum_{\alpha,\beta}
\left|
    \sum_\ell
    \varphi_{\alpha\ell}^*
    \varphi_{\beta\ell}
    e^{ika\ell}
\right|^2
f_{\rm FD}(\varepsilon_\alpha)
\left[
    1 - f_{\rm FD}(\varepsilon_\beta)
\right]
\delta
\left(
    \omega-\frac{\varepsilon_\beta-\varepsilon_\alpha}{\hbar}
\right)
\;.
\end{eqnarray}
The parity operator~(\ref{s-op}) has a form $\hat s_{\ell}=1-2\hat c_{\ell}^\dagger\hat c_{\ell}^{\phantom{\dagger}}$
and, therefore, the parity correlation is $F_{(-1)^n}(\ell_1,\ell_2)=4F_n(\ell_1,\ell_2)$.

Calculation of the noise correlations reduces to the calculation of the one-body density matrix and of the four-point correlation function
$\langle \hat a_{\ell_1}^\dagger \hat a_{\ell_2}^\dagger \hat a_{\ell_3}^{\phantom{\dagger}} \hat a_{\ell_4}^{\phantom{\dagger}}\rangle$.
The former has been already discussed above and the latter at $T=0$ can be computed as~\cite{HR2011}
\begin{equation}
\langle \hat a_{\ell_1}^\dagger \hat a_{\ell_2}^\dagger \hat a_{\ell_3}^{\phantom{\dagger}} \hat a_{\ell_4}^{\phantom{\dagger}}\rangle
=
\det
\left[
    P^\dagger(\ell_3,\ell_4)
    P(\ell_1,\ell_2)
\right]
\;,
\end{equation}
where the $L\times(N+2)$ matrix $P(\ell_1,\ell_2)$ is given by
\begin{equation}
P_{i\alpha}(\ell_1,\ell_2)
=
\left\{
    \begin{tabular}{rlll}
    $-P_{i\alpha}(\ell_2)$ & for & $i=1,\dots,\ell_1-1$, & $\alpha=1,\dots,N+1$;\\
    $P_{i\alpha}(\ell_2)$  & for & $i=\ell_1,\dots,L$,   & $\alpha=1,\dots,N+1$;\\
    $\delta_{i\ell_1}$     & for & $i=1,\dots,L$,        & $\alpha=N+2$;
    \end{tabular}
\right.
\end{equation}
and $P_{i\alpha}(\ell)$ are determined by Eq.~(\ref{P_ial}).
A different approach to the computation of noise correlations based on Wick theorem was developed in Ref.~\cite{RSC06}.
However, it appears to be less efficient.

Finally, we would like to note that the time evolution of an arbitrary initial state
\begin{equation}
|\psi_F(0)\rangle
=
\prod_{\alpha=1}^N
\sum_{\ell=1}^L
\phi_{\alpha\ell}(0)
\,
\hat c_\ell^\dagger
\,
|0\rangle
\end{equation}
can be easily calculated~\cite{RM04L}
\begin{equation}
|\psi_F(t)\rangle
=
\exp
\left(
    - i \frac{\hat H_{\rm F}}{\hbar} t
\right)
|\psi_F(0)\rangle
=
\prod_{\alpha=1}^N
\sum_{\ell=1}^L
\varphi_{\alpha\ell}(t)
\,
\hat c_\ell^\dagger
\,
|0\rangle
\;,
\end{equation}
which remains to be a product of single-particle states similarly to the ground state $|\psi_F^G\rangle$.
The only difference is that the coefficients $\varphi_{\alpha\ell}$ become time-dependent:
\begin{equation}
\varphi_{\alpha\ell}(t)
=
\sum_{\ell'=1}^L
\phi_{\alpha\ell'}(0)
\sum_{\beta=1}^L
\varphi^*_{\beta\ell'}
\,
\varphi^{\phantom*}_{\beta\ell}
\,
\exp
\left(
    - i \frac{\varepsilon_\beta}{\hbar}t
\right)
\;.
\end{equation}
Therefore, the time evolution of the observables is described by the same equations as for the expectation values
in the ground state (at $T=0$) but with $\varphi_{\alpha\ell}$ replaced by $\varphi_{\alpha\ell}(t)$.

\subsubsection{\label{HC1Dhom}Homogeneous lattice}

Now we apply the general formalism described in the previous section to the homogeneous lattices ($\epsilon_\ell\equiv 0$).
In this case, the solution of the single-particle problem
has the form~(\ref{e1p}),~(\ref{1pp}) with $k_\alpha a = 2q_\alpha \pi/L$, $q_\alpha=0,\dots,L-1$, for periodic boundary conditions,
and $k_\alpha a = (2 q_\alpha + 1) \pi/L$ for antiperiodic.

The ground-state energy has the same form for even and odd $N$:
\begin{equation}
E^{00}_N
=
-2J
\,
\frac
{
 \sin
 \left(
     \pi
     \langle
         \hat n_\ell
     \rangle
 \right)
}
{
 \sin
 \left(
     \pi/L
 \right)
}
\;,\quad
\langle
    \hat n_{\ell}
\rangle
=
\frac{N}{L}
\;.
\end{equation}
In the thermodynamic limit, this leads to the following expression for the chemical potential:
\begin{equation}
\label{mu-hc1d-hom}
\mu
=
-2J
\cos
\left(
    \pi
    \langle
        \hat n_\ell
    \rangle
\right)
\end{equation}
and, therefore, the compressibility is given by
\begin{equation}
\label{kappahc1D}
\kappa
=
\left[
    2 \pi J
    \sin
    \left(
        \pi
        \langle
            \hat n_\ell
        \rangle
    \right)
\right]^{-1}
\;.
\end{equation}
Note that $\mu$ decreases with $J$ for $0\le\langle\hat n_\ell\rangle<1/2$ but increases for $1/2<\langle\hat n_\ell\rangle\le 1$.

The energy spectrum of the $N$-particle system in the thermodynamic limit is continuous and has no gaps
with the minimal and maximal energies equal to
$
\pm
2JL
\sin
\left(
    \pi
    \langle
        \hat n_\ell
    \rangle
\right)
/\pi
$.
The lowest excitation mode at small momentum has a linear dispersion relation characterized by the sound velocity~\cite{C04}
\begin{equation}
\label{cshc1D}
c_{\rm s}
=
2a
\frac{J}{\hbar}
\sin
\left(
    \pi
    \langle
        \hat n_{\ell}
    \rangle
\right)
\end{equation}
which coincides with the Fermi velocity
\begin{equation}
v_{\rm F}
=
\frac{1}{\hbar}
\left.
    \frac{\partial\epsilon_k}{\partial k}
\right|_{k=k_{\rm F}}
\;,
\end{equation}
where the Fermi momentum $k_{\rm F}$ is defined as
\begin{equation}
\label{kF}
k_{\rm F}\equiv k_{q=N/2}=\frac{\pi}{a}\langle\hat n_{\ell}\rangle
\;.
\end{equation}

The superfluid stiffness is described by Eq.~(\ref{fshc}).
In the case of a homogeneous lattice, the second term vanishes and we obtain
\begin{equation}
\label{fshchom}
f^{\rm s}_N
=
\frac
{
 \sin
 \left(
     \pi
     \langle
         \hat n_{\ell}
     \rangle
 \right)
}
{
 N
 \sin
 \left(
     \pi/L
 \right)
}
\;.
\end{equation}
In the thermodynamic limit, this gives the well-known result (see, e.g., Ref.~\cite{RRSS05})
\begin{equation}
\label{fshc1D}
f^{\rm s}_\infty
=
\frac
{
 \sin
 \left(
     \pi
     \langle
         \hat n_{\ell}
     \rangle
 \right)
}
{
 \pi
 \langle
     \hat n_{\ell}
 \rangle
}
\;.
\end{equation}
From Eqs.~(\ref{kappahc1D}),~(\ref{cshc1D}) and (\ref{fshc1D}), one can easily see that the superfluid stiffness
is related to the sound velocity as~\cite{RARHBS06}
\begin{equation}
\label{rhos_cs}
f_\infty^{\rm s}
=
\frac
{M_* c_{\rm s} a}
{\pi\hbar\langle\hat n_\ell\rangle}
\;,
\end{equation}
and the hydrodynamic relation~(\ref{fs_kappa_cs}) is also fulfilled.

In the ground state, the onsite particle-number statistics is described by a binary distribution
\begin{eqnarray}
p(n_\ell = n)
=
\left(
    1
    -
    \langle
        \hat n_{\ell}
    \rangle
\right)
\delta_{n,0}
+
\langle
    \hat n_{\ell}
\rangle
\delta_{n,1}
\;.
\end{eqnarray}
This leads to the result for the particle-number fluctuations
\begin{eqnarray}
\sigma_{n_\ell}
=
\sqrt
{
 \langle
     \hat n_{\ell}
 \rangle
 \left(
     1
     -
     \langle
         \hat n_{\ell}
     \rangle
 \right)
}
\end{eqnarray}
which follows also from Eq.~(\ref{Fn-HCB}).

The fermionic one-body density matrix defined by Eq.~(\ref{cdc_T0}) is given by the explicit expression
\begin{equation}
\label{OBDM-Fermi}
\langle \hat c_\ell^\dagger \hat c_{\ell'}^{\phantom\dagger} \rangle
=
\frac
{
 \sin
 \left[
     \pi
     \langle
         \hat n_{\ell}
     \rangle
     (\ell-\ell')
 \right]
}
{
 L
 \sin
 \left[
     \frac{\pi}{L} (\ell-\ell')
 \right]
}
\;,
\end{equation}
which appears to be the same for even and odd $N$.
Therefore, matrices $G^{(\ell,\ell')}$ needed for the calculation of the bosonic one-body density matrix in Eq.~(\ref{ada-det})
acquire a Toeplitz form
\begin{equation}
G^{(\ell,\ell')}
\equiv
G^{(\ell'-\ell)}
=
\begin{pmatrix}
g_1            & g_0    & g_1    & \dots  & g_{\ell'-\ell-2}\\
g_2            & g_1    & g_0    & \ddots & \vdots          \\
g_3            & g_2    & \ddots & \ddots & g_1\\
\vdots         & \ddots & \ddots & g_1    & g_0\\
g_{\ell'-\ell} & \dots  & g_3    & g_2    & g_1
\end{pmatrix}
\;,
\end{equation}
where
\begin{equation}
g_n
=
\frac
{
 \sin
 \left(
     \pi
     \langle
         \hat n_{\ell}
     \rangle
     n
 \right)
}
{
 L
 \sin
 \left(
     \frac{\pi}{L} n
 \right)
}
-
\frac{1}{2}
\delta_{n,0}
\;.
\end{equation}

In the case of half filling, $ \langle\hat n_{\ell}\rangle$=1/2, the coefficients $g_n$ with even $n$ vanish
and the bosonic one-body density matrix is given by~\cite{Ovchinnikov02,Ovchinnikov04}
\begin{eqnarray}
\label{ada-0.5}
\langle \hat a_\ell^\dagger \hat a_{\ell'}^{\phantom\dagger} \rangle
=
\left\{
    \begin{tabular}{ll}
     $\frac{1}{2}R^2_{\frac{\ell'-\ell}{2}}$ & for even $\ell'-\ell$;\\
     $\frac{1}{2}R_{\frac{\ell'-\ell-1}{2}}R_{\frac{\ell'-\ell+1}{2}}$ & for odd $\ell'-\ell$;
    \end{tabular}
\right.
\end{eqnarray}
where
\begin{equation}
R_q
=
\left(
    \frac{2}{\pi}
\right)^q
\prod_{k=1}^{q-1}
\left\{
    \frac
    {
     \sin^2
     \left(
         2k\pi/L
     \right)
    }
    {
     \sin
     \left[
         \left(
             2k+1
         \right)
         \pi/L
     \right]
     \sin
     \left[
         \left(
             2k-1
         \right)
         \pi/L
     \right]
    }
\right\}^{q-k}
\;.
\end{equation}
In the thermodynamic limit, $R_q$ simplifies as
\begin{equation}
\label{Rq-termodyn}
R_q
=
\left(
    \frac{2}{\pi}
\right)^q
\prod_{k=1}^{q-1}
\left[
    \frac
    {
     \left(
         2k
     \right)^2
    }
    {
     \left(
         2k+1
     \right)
     \left(
         2k-1
     \right)
    }
\right]^{q-k}
\;.
\end{equation}
At large distances $|\ell'-\ell|$, the bosonic one-body density matrix has the following asymptotics~\cite{Ovchinnikov02,Ovchinnikov04}:
\begin{equation}
\label{ada-asympt}
\langle \hat a_\ell^\dagger \hat a_{\ell'}^{\phantom\dagger} \rangle
\approx
\frac{C}{\sqrt{|\ell'-\ell|}}
\left[
    1 -
    \frac
    {
     (-1)^{|\ell'-\ell|}
    }
    {
     8
     \left|
         \ell'-\ell
     \right|^2
    }
\right]
\;,\quad
C=0.294176\dots
\end{equation}

\begin{figure}[tb]
\centering

\stepcounter{nfig}
\includegraphics[page=\value{nfig}]{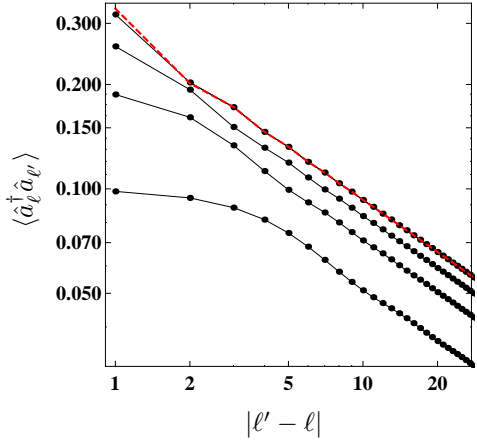}

\caption
{
One-body density matrix of hard core bosons in a one-dimensional homogeneous lattice with periodic boundary conditions
in the thermodynamic limit for $N/L=0.5,0.3,0.2,0.1$ (from top to the bottom).
Red dashed line shows the asymptotics at large distances in the case of $N/L=0.5$ according to Eq.~(\ref{ada-asympt}).
(For interpretation of the references to colour in this figure legend, the reader is referred to the web version of this article.)
}
\label{aacorr-hc1d}
\end{figure}

The dependence of $\langle\hat a_\ell^\dagger \hat a_{\ell'}^{\phantom{\dagger}}\rangle$
on the distance $|\ell'-\ell|$ in the case of half filling described by Eqs.~(\ref{ada-0.5}),~(\ref{Rq-termodyn})
is shown in Fig.~\ref{aacorr-hc1d}.
Surprisingly, the asymptotic formula (\ref{ada-asympt}) is in perfect agreement with the exact result already at small $|\ell'-\ell|$.
For other fillings, the correlation function
$\langle\hat a_\ell^\dagger \hat a_{\ell'}^{\phantom{\dagger}}\rangle$
is smaller and has a different behavior at small distances.
However, at large distances it shows the same asymptotics
$\langle\hat a_\ell^\dagger \hat a_{\ell'}^{\phantom{\dagger}}\rangle \sim |\ell'-\ell|^{-1/2}$
which is a manifestation of quasi-long-range order.
This implies that the quasi-momentum distribution has a $|k|^{-1/2}$ singularity at $k\to 0$~\cite{Lenard64,VT79,Gangardt04}.
The largest eigenvalue of the one-body density matrix which corresponds to the zero quasi-momentum state scales as $\sqrt{N}$~\cite{RM05}.

Any finite temperature has a strong influence on the properties of the one-body density matrix~\cite{R05}.
The long-distance asymptotics becomes exponential,
$
\langle\hat a_{\ell_1}^\dagger \hat a_{\ell_2}^{\phantom{\dagger}}\rangle
\sim
\exp
\left(
    -|\ell_1-\ell_2|/\xi
\right)
$,
and the quasi-long-range order present in the ground state is destroyed.
The correlation length $\xi$ decreases with $T$ and for small temperatures decays as $\xi\sim1/T$.
It has also strong dependence on the filling.

\begin{figure}
\centering

\stepcounter{nfig}
\includegraphics[page=\value{nfig}]{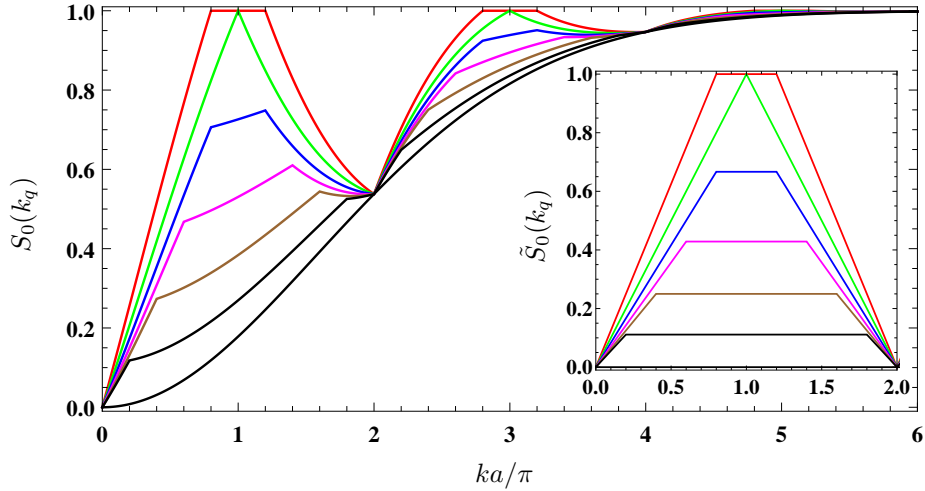}

\caption
{
(color online)
Static structure factor of hard core bosons in a one-dimensional homogeneous lattice of $L=100$ sites with periodic boundary conditions.
$N=40,50,60,70,80,90,100$ from top to the bottom.
The data obtained at discrete points $k_q$ are connected by straight lines.
$\tilde S_0(k_q)$ is given by Eqs.~(\ref{S0hcb1}),~(\ref{S0hcb2}).
$S_0(k_q)$ is obtained from $\tilde S_0(k_q)$ according to Eq.~(\ref{S0lBB}) using $G_0(k)$ for $V_0=10\,E_{\rm R}$.
}
\label{ssf1Dhcb}
\end{figure}

The static structure factor $\tilde S_0(k)$ is a periodic function of $k$ with the period equal to
the vector of the reciprocal lattice $k_{q=L}=\frac{2\pi}{a}$.
It can be easily obtained at $T=0$ from Eqs.~(\ref{S0discrete}),~(\ref{cdc_T0}),~(\ref{Fn-HCB}).
Within one period and for $N=1,\dots,L/2$ it is given by
\begin{equation}
\label{S0hcb1}
\tilde S_0(k_q)
=
\left\{
    \begin{tabular}{ll}
    $\frac{k_q}{2k_{\rm F}}$ & , $k_q=0,\dots,2k_{\rm F}$\;,\\
    $1$ & , $q=2k_{\rm F},\dots,k_L-2k_{\rm F}$\;,\\
    $\frac{k_L-k_q}{2k_{\rm F}}$ & , $k_q=k_L-2k_{\rm F},\dots,k_L$\;,
    \end{tabular}
\right.
\end{equation}
where $k_q=\frac{2\pi}{La}q$ and $k_{\rm F}$ is the Fermi momentum.
In this case, $k_L\ge4k_{\rm F}$.
For $N=\frac{L}{2}+1,\dots,L$ we have instead
\begin{equation}
\label{S0hcb2}
\tilde S_0(k_q)
=
\left\{
    \begin{tabular}{ll}
    $\frac{k_q}{2k_{\rm F}}$ & , $k_q=0,\dots,k_L-2k_{\rm F}$\;,\\
    $\frac{k_L-2k_{\rm F}}{2k_{\rm F}}$ & , $k_q=k_L-2k_{\rm F},\dots,2k_{\rm F}$\;,\\
    $\frac{k_L-k_q}{2k_{\rm F}}$ & , $k_q=2k_{\rm F},\dots,k_L$\;,
    \end{tabular}
\right.
\end{equation}
with $2k_{\rm F}\le k_L < 4k_{\rm F}$.
For $N=1,\dots,L-1$, $\tilde S_0(k_q)$ is a linear function of $k_q$ in finite intervals in the vicinity of $k=0$ and $k=2\pi/a$
with the slope $1/(2k_{\rm F})$ in complete agreement with Eqs.~(\ref{S0small-k}),~(\ref{kappahc1D}),~(\ref{cshc1D}).
In the remaining interval near $k=\pi/a$, $\tilde S_0(k_q)$ takes a constant value which is one for $N=1,\dots,L/2$
and reduces from $(L-2)/(L+2)$ to zero for $N=\frac{L}{2}+1,\dots,L$.
As it follows from Eq.~(\ref{S0hcb2}), $\tilde S_0(k_q)$ vanishes for $N=L$ and the static structure factor in continuum
takes the form $S_0(k)=1-G_0^2(k)$, where $G_0(k)$ is defined by Eq.~(\ref{G0}).
The comparison of $S_0(k)$ and $\tilde S_0(k)$ for different values of $N$ is given in Fig.~\ref{ssf1Dhcb}.

Noise correlations of hard-core bosons in homogeneous lattices possess the following characteristic features~\cite{RSC06,HR2011}.
(i) Very large peaks appear at $k_1=k_2=0$ as well as at integer multiples of the reciprocal lattice vector
as a result of the quasi-condensation and the underlying order induced by a periodic lattice.
The heights of the peaks depend on the filling.
(ii) A line of maxima exists for $k_1=k_2$ due to the bunching typical for bosonic systems.
(iii) There are dips immersed in a negative background along the lines $k_1=0$ and $k_2=0$ which are related to quantum depletion.
(iv) The correlation function for $k_1=k_2=0$ scales linearly with the system size.

\subsubsection{\label{HCB1Dht}Harmonic trap}

\begin{figure}[t]

\centering

\stepcounter{nfig}
\includegraphics[page=\value{nfig}]{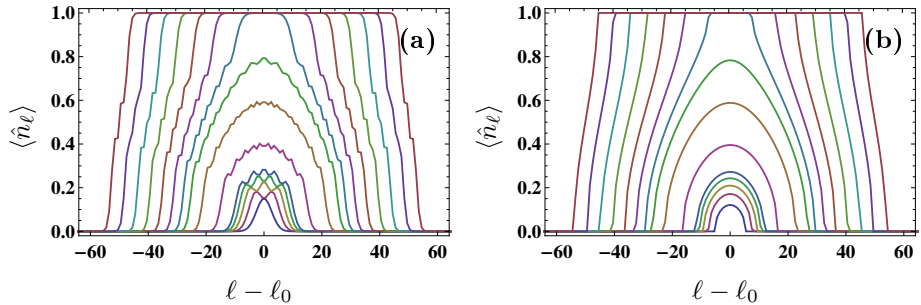}

\caption
{
(color online)
Spatial distribution of the mean occupation numbers of hard-core bosons in a one-dimensional lattice
and in the presence of harmonic confinement with the parameter $J/V_{\rm T}=200$ at $T=0$.
The total number of particles $N=1,2,3,4,5,10,20,30,40,50,60,70,80,90,100$ from bottom to the top.
{(a)} Exact calculations in a lattice of $L=200$ sites (the size of the lattice does not affect the results).
{(b)} Local-density approximation~(\ref{LDAhc1D}).
}
\label{hc1d_trap}
\end{figure}

We consider the effects of the harmonic trapping potential described by the local terms
\begin{equation}
\epsilon_\ell
=
V_{\rm T}
\left(
    \ell - \ell_0
\right)^2
\;,\quad
V_{\rm T}
=
\frac{M\omega_{\rm T}^2 a^2}{2}
\;,
\end{equation}
in the Hamiltonian~(\ref{HHC}), where $\ell_0$ denotes the center of the trap (which is not necessarily an integer).
Spatial distributions of the mean occupation numbers $\langle\hat n_\ell\rangle$
for different total particle numbers $N$ at $T=0$ determined by Eq.~(\ref{cdc_T0})
with $\ell'=\ell$ are shown in Fig.~\ref{hc1d_trap}.
With the increase of $N$, $\langle\hat n_\ell\rangle$'s become larger and the size of the atomic sample grows.
If the particle number exceeds the value $N\approx 2.68\sqrt{J/V_{\rm T}}$, a plateau with $\langle\hat n_\ell\rangle=1$
appears at the trap center~\cite{RM04a,CCGOR11,AMZ05}.
In the example shown in Fig.~\ref{hc1d_trap} this happens for $N\ge38$.
Since the local chemical potential $\mu_\ell=\mu-\epsilon_\ell$ varies from site to site but the occupation
numbers $\langle\hat n_\ell\rangle$ within the plateau do not, the local compressibility~(\ref{loc-comp}) vanishes.

The density profiles can be also obtained within the local density approximation~\cite{PWMMFCSHB04,CV10,MFZ08}
replacing the chemical potential $\mu$ in Eq.~(\ref{mu-hc1d-hom}) by $\mu_\ell=\mu-\epsilon_\ell$.
This leads to the following expression
\begin{eqnarray}
\label{LDAhc1D}
\langle\hat n_\ell\rangle
=
\left\{
    \begin{tabular}{ll}
    $0$ & , $\mu_\ell<-2J$,\\
    $\frac{1}{\pi}\arccos\left(-\frac{\mu_\ell}{2J}\right)$ & , $|\mu_\ell|<2J$,\\
    $1$ & , $\mu_\ell>2J$,
    \end{tabular}
\right.
\end{eqnarray}
where the global chemical potential $\mu$ is determined by the total number of particles as
$\sum_\ell \langle\hat n_\ell\rangle = N$.
Fig.~\ref{hc1d_trap} shows that the local-density approximation works very well for $N\ge10$,
although it does not describe fine details of the exact density profiles.

One-body density matrix as well as natural orbitals in the presence of a trapping potential were studied in Refs.~\cite{RM04,MTEG05}.
It was shown that the power-law decay
$\langle\hat a_\ell^\dagger\hat a_{\ell'}^{\phantom\dagger}\rangle\sim|\ell-\ell'|^{-1/2}$
is preserved for intermediate distances but at larger distances the correlations drop much faster~\cite{RM04}
which removes the singularity in the quasi-momentum distribution for $k\to0$~\cite{MTEG05}.
The harmonic trap sets a momentum scale $p_{\rm T}=\sqrt{M\hbar\omega_{\rm T}}$ below which the momentum distribution is flattened due to
suppression of the long-range correlations~\cite{PRD04,RSC06}.
The presence of the plateau in the density profile is accompanied by the splitting of the natural orbital
of the one-body density matrix with the largest eigenvalue into two parts localized in the side regions with nonvanishing compressibility~\cite{RM04}.

In the experiment of Ref.~\cite{PWMMFCSHB04}, a two-dimensional array of independent one-dimensional chains
with the filling factor smaller than one has been created. Due to the harmonic confinement, the number of
atoms in a chain labeled by a pair of indices $(i,j)$ is given by
\begin{displaymath}
N_{ij}
=
N_{00}
\left[
    1 -
    \frac{5N}{2\pi N_{00}}
    \left(
        i^2 + j^2
    \right)
\right]^{3/2}
\;,
\end{displaymath}
where $N$ is the total number of atoms and $N_{00}$ is the number of atoms in the central chain.
The probability of having a chain with $N'$ atoms reads
\begin{displaymath}
P(N')
=
\frac{2}
{3N_{00}^{2/3}(N')^{1/3}}
\;,\quad
N'\le N_{00}
\;.
\end{displaymath}
The only parameter which determines this distribution was estimated to be $N_{00}=18$.
The quasi-momentum distribution calculated for this setup using the Bose-Fermi mapping at finite temperature
is in excellent agreement with the experimental data~\cite{PWMMFCSHB04,PDH10}.

\subsubsection{Extended fermionization}

As it was mentioned above, standard Bose-Fermi mapping is not valid for $N>L$.
Nevertheless, this formalism can be used within the framework of the extended fermionization model~\cite{PRB06,PRWC06}.
The idea is to divide the whole system into two subsystems:
(i) $n_{\rm B}$ bosons siting at each lattice site, and (ii) $N'=N-n_{\rm B}L$ excess bosons,
where $n_{\rm B}$ is such that $N'<L$. Then the excess particles can be treated as ordinary hard-core bosons
with the effective tunneling parameter $J'=J(n_{\rm B}+1)$.
This approach was employed to study the dynamic structure factor of hard-core bosons
in the homogeneous one-dimensional lattices as well as in the presence of a harmonic trap~\cite{PRB06}
as well as damping of dipole oscillations~\cite{PRWC06}.
In the homogeneous system, it leads to straightforward modifications of the results of section~\ref{HC1Dhom}.
For instance, the expression for the superfluid stiffness becomes
\begin{equation}
\label{fsghchom}
f^{\rm s}_N
=
\frac{n_{\rm B}+1}{N}
\,
\frac
{
 \sin
 \left(
     \pi N'/L
 \right)
}
{
 \sin
 \left(
     \pi/L
 \right)
}
\;.
\end{equation}

\section{\label{sec-PTLSI}Perturbation theory in the limit of strong interaction}

If the tunneling parameter $J$ is small compared to the interaction energy $U$,
perturbative solution of the Bose-Hubbard model can be obtained in powers of $J/U$,
which is called strong-coupling expansion~\cite{FM94,FM96}.
This method works very well for $J/U$ below the critical point $(J/U)_{\rm c}$,
which is less than one in all dimensions,
where all physical quantities are analytical functions.
It was used in the studies of spinless bosons with
local~\cite{FM94,FM96,DZ06,FKKKT2009,EFGMKAL12,EM99,EM99p,F08,BV05,HIR10,BPV04,BPV05,VM11,NFM99}
and nearest-neighbor interactions~\cite{IF09-1,IF09-2}, two-species bosons~\cite{I10},
and spin-1 bosons~\cite{K13,K14}
in different types of regular lattices like hypercubic isotropic~\cite{FM94,FM96,FKKKT2009,IF09-1,IF09-2,I10,K13,HIR10,SG12}
and anisotropic~\cite{F08},
superlattices~\cite{BV05,HIR10}, two-dimensional triangular and kagome lattices~\cite{BPV04,BPV05,VM11},
and in the presence of artificial gauge fields~\cite{NFM99,SG12},
as well as in disordered lattices~\cite{FM96,GWSL05,GWSL05apb,MF08,KTEG08}.
The strong-coupling expansion allows to avoid finite-size effects and shows excellent agreement with exact numerical data.

However, this method can be applied only for commensurate fillings $N=nL^d$, where $n$ is an integer,
or close to it for $N=nL^d \pm 1$ due to the degeneracies
of the zeroth-order eigenstates which may become intractable in the case of arbitrary filling.
Analytical calculations for arbitrary $d$ and $n$ can be done in practice only for few lowest orders.
Higher-order studies require symbolic calculations on a computer.
This has been done for $d=1$ and $n=1$ up to the 14th order in Ref.~\cite{DZ06}
for the ground-state energy, variance of the occupation numbers and two-point correlation functions.
Recently, symbolic calculations were done for $d=1$ and $n=1,2,3$ up to the 16th order and the results were reported
for the energy, two-point correlation functions and the first five moments of the on-site particle-number distribution~\cite{DZ15}.
Numerical calculations along these lines were also performed for $d=1$, $n=1,2$,
and for $d=2$, $n=1$ up to the 13th order~\cite{EM99,EM99p}.
In Refs.~\cite{E09,THHE09R,THHE09}, the method was developed such that it became possible
to do calculations in principle for arbitrary $d$ and $n$.
With the aid of the scaling theory it is possible to extrapolate to the infinite order of $J/U$~\cite{FM96,EM99,FKKKT2009}.
Here we present analytical results for one-dimensional and isotropic hypercubic lattices of arbitrary dimensions.

\subsection{Ground state in the case of commensurate filling}

The ground state in the case of commensurate filling ($N=n L^d$) is denoted as $|{\bf \psi}_{n L^d}^{00}\rangle$.
As it was discussed in Sec.~\ref{Section_J=0}, in the zeroth-order of the perturbation theory,
this state is not degenerate and has the energy~(\ref{E00}).
Therefore, in order to study the ground-state properties of the Mott-insulator,
one has to employ non-degenerate perturbation theory.
For the ground-state energy per lattice site at arbitrary $d$ and $n$ up to the 4th order in $J/U$, this gives~\cite{KQ}
\begin{eqnarray}
\label{energy4order}
\frac{E_{n L^d}^{00}}{UL^d}
=
\frac{n(n-1)}{2}
-
\left(
    \frac{J}{U}
\right)^2
Zn(n+1)
-
\left(
    \frac{J}{U}
\right)^4
Z
\frac{n(n+1)}{12}
\left[
    16 Z - 34 + ( 76 Z - 157 ) n (n+1)
\right]
\;,
\end{eqnarray}
where $Z=2d$ is the coordination number.
Symbolic perturbative expansion on a computer for $d=1$, $n=1$ allowed to obtain the result up to the 14th order~\cite{DZ06}:
\begin{eqnarray}
\label{energy14order}
\frac{E_{L}^{00}}{4UL}
&=&
-
\left(
    \frac{J}{U}
\right)^2
+
\left(
    \frac{J}{U}
\right)^4
+
\frac{68}{9}
\left(
    \frac{J}{U}
\right)^6
-
\frac{1267}{81}
\left(
    \frac{J}{U}
\right)^8
+
\frac{44171}{1458}
\left(
    \frac{J}{U}
\right)^{10}
\nonumber\\
&-&
\frac{4902596}{6561}
\left(
    \frac{J}{U}
\right)^{12}
-
\frac{8020902135607}{2645395200}
\left(
    \frac{J}{U}
\right)^{14}
\;.
\end{eqnarray}
For $d=1$, $n=1$, Eq.~(\ref{energy4order}) reproduces the first two terms in Eq.~(\ref{energy14order}).

The probabilities to have $n_{\bf l}$ particles on a lattice site are given by~\cite{KQ}
\begin{eqnarray}
\label{pnm1-sce}
&&
p(n_{\bf l}=n-1)
=
\left(
    \frac{J}{U}
\right)^2
n(n+1)Z
\\
&&
+
\left(
    \frac{J}{U}
\right)^4
\frac{n(n+1)Z}{18}
\left[
84 Z - 156
+
n (334 Z - 703)
+
n^2 (326 Z - 695)
\right]
\;,
\nonumber\\
\label{pnp1-sce}
&&
p(n_{\bf l}=n+1)
=
\left(
    \frac{J}{U}
\right)^2
n(n+1)Z
\\
&&
+
\left(
    \frac{J}{U}
\right)^4
\frac{n(n+1)Z}{18}
\left[
    76 Z -148
    +
    n(318 Z - 687)
    +
    n^2(326 Z - 695)
\right]
\;,
\nonumber\\
\label{pnm2-sce}
&&
p(n_{\bf l}=n-2)
=
\left(
    \frac{J}{U}
\right)^4
n(n^2-1)
Z
\left[
    \frac{n+2}{16}
    +
    \frac{2}{9}
    (n+1)
    (Z-1)
\right]
\;,
\\
\label{pnp2-sce}
&&
p(n_{\bf l}=n+2)
=
\left(
    \frac{J}{U}
\right)^4
n(n+1)(n+2)
Z
\left[
    \frac{n-1}{16}
    +
    \frac{2}{9}
    n
    (Z-1)
\right]
\;,
\end{eqnarray}
and the probability $p(n_{\bf l}=n)$ can be obtained from the normalization condition
\begin{eqnarray}
\sum_{n_{\bf l}=n-2}^{n+2}
p(n_{\bf l})
=1
\;.
\end{eqnarray}
Probabilities of other occupation numbers vanish in this order of the perturbation theory.
The probabilities $p(n_{\bf l})$ given by Eqs.~(\ref{pnm1-sce})-(\ref{pnp2-sce})
satisfy the relation
\begin{eqnarray}
p(n_{\bf l}=n+1)
-
p(n_{\bf l}=n-1)
+
2
\left[
    p(n_{\bf l}=n+2)
    -
    p(n_{\bf l}=n-2)
\right]
=0
\end{eqnarray}
that follows from the obvious condition $\langle\hat n_{\bf l}\rangle=n$.
Second-order terms in Eqs.~(\ref{energy4order}),~(\ref{pnm1-sce}),~(\ref{pnp1-sce}) were obtained in Refs.~\cite{FM94,FM96,CKPS2007}.

Elements of the one-body density matrix
$
F_a({\bf l}_1,{\bf l}_2)
=
\langle
   \hat a_{{\bf l}_1}^\dagger
   \hat a_{{\bf l}_2}^{\phantom{\dagger}}
\rangle
$
up to the third order in $J/U$ have the form~\cite{FKKKT2009}
\begin{eqnarray}
\label{Fa1}
&&
F_a(s=1)
=
\frac{J}{U}
2
n(n+1)
+
\left(
    \frac{J}{U}
\right)^3
\frac{n(n+1)}{3}
\left[
    16 Z - 34 + ( 76 Z - 157 ) n (n+1)
\right]
\;,
\\
\label{Fa2}
&&
F_a(s=2)
=
\left(
    \frac{J}{U}
\right)^2
3n(n+1)(2n+1)
\;,
\\
&&
F_a(s=\sqrt{2})
=
2 F_a(s=2)
\;,
\nonumber\\
\label{Fa3}
&&
F_a(s=3)
=
\left(
    \frac{J}{U}
\right)^3
4n(n+1)(5n^2+5n+1)
\;,
\\
&&
F_a(s=\sqrt{5})
=
3 F_a(s=3)
\;,\quad
F_a(s=\sqrt{3})
=
6 F_a(s=3)
\;.
\nonumber
\end{eqnarray}
Note that integer distances, $s=1,2,\dots$, are possible in lattices of any dimension $d$,
while irrational distances $s=\sqrt{2},\sqrt{5}$ and $s=\sqrt{3}$ exist only for $d\ge 2$ and $d\ge 3$, respectively.
Eqs.~(\ref{energy4order})-(\ref{Fa1}) can be tested for self-consistency using the identities
\begin{eqnarray}
\label{EgFa1}
\frac{E^{00}_N}{UL^d}
&=&
-
\frac{J}{U}
Z
F_a(s=1)
+
\frac
{
    \langle
         \hat n_{\bf l}^2
    \rangle
    -
    \langle
         \hat n_{\bf l}
    \rangle
}
{2}
\;,\quad
\\
\frac{\partial}{\partial U}
\frac{E^{00}_N}{L^d}
&=&
\frac
{
    \langle
         \hat n_{\bf l}^2
    \rangle
    -
    \langle
         \hat n_{\bf l}
    \rangle
}
{2}
\;,
\nonumber
\end{eqnarray}
that follow from the Hamiltonian.

With the aid of Eqs.~(\ref{Fa1})-(\ref{Fa3}), one can deduce perturbative results for the quasi-momentum distribution in the ground state:
\begin{eqnarray}
\label{md-sce}
\tilde P({\bf k})
=
\frac{1}{N}
\left[
    n
    +
    \sum_{\bf s}
    b_{\bf s}
    \prod_{\nu=1}^d
    \cos(s_\nu k_\nu a)
\right]
\;,
\end{eqnarray}
where $b_{\bf s}\equiv b_{s_1 \dots s_d}$ are invariant under any permutation of indices.
In one dimension, the nonvanishing coefficients in the third order of the perturbation theory are given by
\begin{equation}
\label{md-sce-1D}
b_s = 2 F_a(s)
\;,\quad
s=1,2,3.
\end{equation}
In two dimensions, we have
\begin{eqnarray}
b_{s0}
&=&
2 F_a(s)
\;,\quad
s=1,2,3,
\\
b_{11}
&=&
8 F_a(2)
\;,\quad
b_{12} = 12 F_a(3)
\;,
\nonumber
\end{eqnarray}
and in three dimensions:
\begin{eqnarray}
b_{s00}
&=&
2 F_a(s)
\;,\quad
s=1,2,3,
\\
b_{110}
&=&
8 F_a(2)
\;,\quad
b_{120} = 12 F_a(3)
\;,\quad
b_{111} = 48 F_a(3)
\;,
\nonumber
\end{eqnarray}
with all possible permutations of the indices.

For the particle-number correlation function~(\ref{nncorr}) one finds~\cite{KQ}
\begin{eqnarray}
\label{dd1-sce}
F_{n}(1)
&=&
-
\left(
    \frac{J}{U}
\right)^2
2n(n+1)
\left(
    \frac{J}{U}
\right)^4
\frac{n(n+1)}{18}
\left[
    64Z-190
    +
    n(n+1)(448Z-1069)
\right]
\;,
\\
\label{dd2-sce}
F_{n}(2)
&=&
-
\left(
    \frac{J}{U}
\right)^4
\frac{2n}{9}(11+43n+64n^2+32n^3)
\;,
\\
\label{ddsqrt2-sce}
F_{n}(\sqrt{2})
&=&
-
\left(
    \frac{J}{U}
\right)^4
\frac{4n}{9}(20+79n+118n^2+59n^3)
\,.
\end{eqnarray}
The consistency of Eqs.~(\ref{dd1-sce})-(\ref{ddsqrt2-sce}) with (\ref{energy4order}), (\ref{Fa1})
can be also tested using the fact that the fluctuations of the total number of particles vanish:
$
\langle
   \hat N^2
\rangle
-
N^2
=0
$.
In a translationally invariant system this leads to the condition~\cite{DZ15}
\begin{equation}
\sum_{l_1=0}^{L-1}
\cdots
\sum_{l_d=0}^{L-1}
\left(
    \langle
        \hat n_{\bf 0}
        \hat n_{\bf l}
    \rangle
    -
    \langle
        \hat n_0
    \rangle
    \langle
        \hat n_{\bf l}
    \rangle
\right)
=0
\;.
\end{equation}
Considering only the terms up to the 4th order of the perturbation theory, we get
\begin{eqnarray}
\label{EgFaFn}
    \frac{E_{nL^d}^{00}}{UL^d}
    -
    \frac{n(n-1)}{2}
    +
    \frac{J}{U}
    Z F_a(1)
+
\frac{Z}{2}
\left[
    F_n(1) + F_n(2) + \frac{Z-2}{2} F_n(\sqrt{2})
\right]
=0
\;,
\end{eqnarray}
where we have taken into account Eq.~(\ref{EgFa1}).
Eqs.~(\ref{energy4order}),~(\ref{Fa1}),~(\ref{dd1-sce})-(\ref{ddsqrt2-sce}) satisfy indeed Eq.~(\ref{EgFaFn}).

The parity correlation is given by~\cite{KQ}
\begin{eqnarray}
\label{par1-sce}
F_{(-1)^n}(1)
&=&
\left(
    \frac{J}{U}
\right)^2
8n(n+1)
+
\left(
    \frac{J}{U}
\right)^4
\frac{4n(n+1)}{3}
\left[
    16Z -34
    +
    n(n+1)(40Z-157)
\right]
\;,
\\
\label{par2-sce}
F_{(-1)^n}(2)
&=&
\left(
    \frac{J}{U}
\right)^4
\frac{8n}{9}(7+29n+44n^2+22n^3)
\;,
\\
\label{parsqrt2-sce}
F_{(-1)^n}(\sqrt{2})
&=&
\left(
    \frac{J}{U}
\right)^4
\frac{16n}{9}(16+83n+134n^2+67n^3)
\;.
\end{eqnarray}
The results for the parity correlation function up to the second order in $J/U$ at finite temperature can be found in Ref.~\cite{KM10}.
A multi-point generalization of the two-point parity correlation function $F_{(-1)^n}({\bf l}_1,{\bf l}_2)$,
a so-called string correlator defined by Eq.~(\ref{string}),
was calculated at zero temperature in the second order of the strong-coupling expansion for lattices of arbitrary dimensions
and in the fourth order in one dimension~\cite{RSEZ13}.

\subsection{Lowest excited states}

At each value of the discrete total momentum $\hbar{\bf K}$, the first excitation band consists of $L^d-1$ states.
In the limit $J=0$, the states are degenerate and
correspond to bosonic configurations with the same occupation numbers $n$ at any site except two, one of which contains 
$n-1$ bosons and the other one $n+1$.
In one dimension, they are explicitly given by
\begin{eqnarray}
|\psi^{K\Omega}_{nL}\rangle^{(0)}
\equiv
|{\bf n}_{K\Omega}\rangle
=
\frac{1}{\sqrt{L}}
\sum_{j=0}^{L-1}
\left(
    \frac
    {\hat {\cal T}}
    {\tau_{K}}
\right)^j
|n+1 , \underbrace{n ,\dots ,n}_{\Omega-1} , n-1 , \underbrace{n, \dots, n}_{L-\Omega-1}\rangle
\;,
\end{eqnarray}
where $\Omega=1,\dots,L-1$.
In the first order of the perturbation theory, the degeneracy of these states at a given value of $K$ is completely lifted.
The energies of the excited states for one-dimensional lattices were calculated in the first order of $J/U$ for unit filling $n=1$~\cite{EFGMKAL12}
and arbitrary filling~\cite{BPCK12}.
By doing calculations for arbitrary $n$ up to the second order in $J/U$, we obtain
\begin{eqnarray}
\label{eband-sce}
&&
\frac{E^{K\Omega}_{nL}}{U}
=
\frac{E^{00}_{nL}}{U}
+
1
-
2
\frac{J}{U}
\cos
\left(
    \pi\frac{\Omega}{L}
\right)
\sqrt
{
 1
 +
 4n(n+1)
 \cos^2\left(\frac{Ka}{2}\right)
}
\\
&&
+
\left(
    \frac{J}{U}
\right)^2
\left\{
    5n^2 + 6n + 2
    -
    4n(n+1)
    \cos
    \left(
        2\pi\frac{\Omega}{L}
    \right)
    \frac
    {
     2n(n+1)
     +
     \left(
         2 n^2 + 2n + 1
     \right)
     \cos Ka
    }
    {
     1
     +
     4n(n+1)
     \cos^2(Ka/2)
    }
    \cos Ka
\right\}
\;,
\nonumber
\end{eqnarray}
where $\Omega=1,\dots,L-1$. Eq.~(\ref{eband-sce}) has been derived for sufficiently large
one-dimensional lattices ($Z=2$) and can be easily generalized to arbitrary dimensions.
The first-order term in Eq.~(\ref{eband-sce}) predicts symmetric form of the excitation band with respect to $E=E^{00}+U$.
The second-order term describes the asymmetry of the band.

The lowest excited state is labeled by $K=0$, $\Omega=1$.
From Eq.~(\ref{eband-sce}) it follows that in the thermodynamic limit the gap separating it from the ground state (neutral gap)
is given by
\begin{eqnarray}
\label{Delta_n_2}
\frac{\Delta_{\rm n}}{U}
=
\frac
{
E^{01}_{nL}
-
E^{00}_{nL}
}
{U}
=
1
-
2
\frac{J}{U}
(2n+1)
+
\left(
    \frac{J}{U}
\right)^2
\left(
    n^2+2n+2
\right)
\;.
\end{eqnarray}

\subsection{\label{sec-ph-excitations}Particle-hole excitations}

Excitations considered in the previous section constitute a part of the energy spectrum of a bosonic system with commensurate filling $N=nL$.
They should not be confused with particle and hole excitations which arise, if we add or remove one particle from the system.
The states corresponding to these particle and hole excitations, which are the ground states of the system with the total
number of particles $N=nL\pm 1$, in the lowest order of the perturbation theory are given by
\begin{eqnarray}
|\psi^{K0}_{nL\pm 1}\rangle^{(0)}
\equiv
|{\bf n}_{{K}\Gamma}\rangle
=
\frac{1}{\sqrt{L}}
\sum_{j=0}^{L-1}
\left(
    \frac
    {\hat {\cal T}}
    {\tau_{K}}
\right)^j
|n \pm 1 , \underbrace{n ,\dots ,n}_{L-1}\rangle
\;.
\end{eqnarray}
Up to the third order in $J/U$ the energies of the states $|\psi^{K0}_{nL^d\pm 1}\rangle$ at $K=0$
in a $d$-dimensional lattice are~\cite{FM94,FM96}
\begin{eqnarray}
\label{mupd}
&&
\frac
{
E^{00}_{n L^d + 1}
-
E^{00}_{n L^d}
}
{U}
=
n
-
\frac{ZJ}{U}
(n+1)
-
\left(
    \frac{ZJ}{U}
\right)^2
n
\left[
    n+1
    -
    \frac{5n+4}{2Z}
\right]
\nonumber\\
&&
-
\left(
    \frac{ZJ}{U}
\right)^3
n(n+1)
\left[
    2n+1
    -
    \frac{25n+14}{4Z}
    +
    2
    \frac{2n+1}{Z^2}
\right]
+
O(J^4)
\;,
\\
\label{muhd}
&&
\frac
{
E^{00}_{n L^d}
-
E^{00}_{n L^d - 1}
}
{U}
=
n-1
+
\frac{ZJ}{U}
n
+
\left(
    \frac{ZJ}{U}
\right)^2
(n+1)
\left[
    n
    -
    \frac{5n+1}{2Z}
\right]
\nonumber\\
&&
+
\left(
    \frac{ZJ}{U}
\right)^3
n(n+1)
\left[
    2n+1
    -
    \frac{25n+11}{4Z}
    +
    2
    \frac{2n+1}{Z^2}
\right]
+
O(J^4)
\;,
\end{eqnarray}
where $E^{00}_{n L^d}$ is determined by Eq.~(\ref{energy4order}).

In a one-dimensional lattice with unit filling the dependences of the energies
of the particle and hole excitations on $K$ up to the sixth order in $J/U$ are given by~\cite{EFGMKAL12}
\begin{eqnarray}
\label{epk6}
&&
\frac
{
E^{K0}_{L + 1}
-
E^{00}_{L}
}
{U}
=
1
+
5
\left(
    \frac{J}{U}
\right)^2
-
\frac{513}{20}
\left(
    \frac{J}{U}
\right)^4
-
\frac{80139}{200}
\left(
    \frac{J}{U}
\right)^6
\nonumber\\
&&
+
\left[
    -4
    \frac{J}{U}
    +18
    \left(
        \frac{J}{U}
    \right)^3
    -\frac{137}{150}
    \left(
        \frac{J}{U}
    \right)^5
\right]
\cos(Ka)
+
\left[
    -4
    \left(
        \frac{J}{U}
    \right)^2
    +64
    \left(
        \frac{J}{U}
    \right)^4
    -\frac{426161}{1500}
    \left(
        \frac{J}{U}
    \right)^6
\right]
\cos(2Ka)
\nonumber\\
&&
+
\left[
    -12
    \left(
        \frac{J}{U}
    \right)^3
    +276
    \left(
        \frac{J}{U}
    \right)^5
\right]
\cos(3Ka)
+
\left[
    -44
    \left(
        \frac{J}{U}
    \right)^4
    +1296
    \left(
        \frac{J}{U}
    \right)^6
\right]
\cos(4Ka)
\nonumber\\
&&
-180
\left(
    \frac{J}{U}
\right)^5
\cos(5Ka)
-792
\left(
    \frac{J}{U}
\right)^6
\cos(6Ka)
\;,
\end{eqnarray}
\begin{eqnarray}
\label{ehk6}
&&
\frac
{
E^{K0}_{L - 1}
-
E^{00}_{L}
}
{U}
=
8
\left(
    \frac{J}{U}
\right)^2
-
\frac{512}{3}
\left(
    \frac{J}{U}
\right)^6
+
\left[
    -2
    \frac{J}{U}
    +12
    \left(
        \frac{J}{U}
    \right)^3
    -\frac{224}{3}
    \left(
        \frac{J}{U}
    \right)^5
\right]
\cos(Ka)
\;,
\nonumber\\
&&
+
\left[
    -4
    \left(
        \frac{J}{U}
    \right)^2
    +64
    \left(
        \frac{J}{U}
    \right)^4
    -\frac{1436}{3}
    \left(
        \frac{J}{U}
    \right)^6
\right]
\cos(2Ka)
+
\left[
    -12
    \left(
        \frac{J}{U}
    \right)^3
    +276
    \left(
        \frac{J}{U}
    \right)^5
\right]
\cos(3Ka)
\nonumber\\
&&
+
\left[
    -44
    \left(
        \frac{J}{U}
    \right)^4
    +1296
    \left(
        \frac{J}{U}
    \right)^6
\right]
\cos(4Ka)
-180
\left(
    \frac{J}{U}
\right)^5
\cos(5Ka)
-792
\left(
    \frac{J}{U}
\right)^6
\cos(6Ka)
\;.
\end{eqnarray}
Eqs.~(\ref{epk6}),~(\ref{ehk6}) are consistent with Eqs.~(\ref{mupd}),~(\ref{muhd}) in the same orders of $J/U$.

Eqs.~(\ref{mupd}),~(\ref{muhd}),~(\ref{epk6}),~(\ref{ehk6}) allow to obtain perturbative expansions for the charge gap~(\ref{Delta_c}).
In order to make a comparison with the neutral gap in Eq.~(\ref{Delta_n_2}), we set $d=1$ in Eqs.~(\ref{mupd}),~(\ref{muhd}) and
take into account only the terms up to the second order in $J/U$. This gives
\begin{equation}
\frac{\Delta_{\rm c}}{U}
=
1-
2\frac{J}{U}
\left(
    2n + 1
\right)
+
\left(
    \frac{J}{U}
\right)^2
\left(
    2n^2 + 2n + 1
\right)
\;.
\end{equation}
The first-order term is the same as in Eq.~(\ref{Delta_n_2}).
The second-order term is different for $n\ge2$ and only for $n=1$ it becomes the same.  

\subsection{Phase diagram}

From the energies of particle and hole excitations at $K=0$ one can obtain the boundaries of the regions
in the $(\mu,J)$ plane corresponding to the commensurate fillings. They are determined by Eq.~(\ref{mu-pm})
and, therefore, are readily obtained from Eqs.~(\ref{mupd}),~(\ref{muhd}) up to the third order of $J/U$ for arbitrary $n$ and $d$
or from Eqs.~(\ref{epk6}),~(\ref{ehk6}) up to the 6th order of $J/U$ for $n=1$ and $d=1$.
The region $\mu<\mu_+(0)=-2dJ$ corresponds to $N=0$.

Eqs.~(\ref{mupd}),~(\ref{muhd}) as well as all other perturbative results are valid only in some interval of $J\in[0,J_{\rm c}]$,
where $J_{\rm c}$ is the smallest value of $J$ at which $\mu_-(n)$ and $\mu_+(n)$ become equal.
In this interval the difference $\mu_+(n)-\mu_-(n)$ remains finite and positive which means that the compressibility vanishes.
$J_{\rm c}$ depends on the order up to which the calculations are performed and provides an estimate of the transition point
from the Mott-insulator to the superfluid.

\begin{figure}[t]

\centering

\stepcounter{nfig}
\includegraphics[page=\value{nfig}]{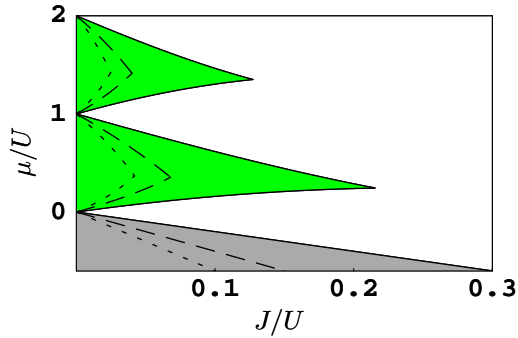}

\caption
{
(color online)
Boundaries of the Mott-insulator phase via strong-coupling expansion up to the third order in $J/U$
for $d=1$~(solid lines), $2$~(dashed lines), $3$~(dotted lines).
The gray region corresponds to $N=0$, and the two green regions correspond to the Mott insulator with $n=1,2$
in one dimension.
}
\label{pd-sc3}
\end{figure}

Perturbative results for the superfluid stiffness $f^{\rm s}_N$ can be obtained from the calculations of the ground-state energies
in the presence of the Peierls factors.
In this manner we get that for $N=nL^d$ the superfluid stiffness $f^{\rm s}_{n L^d}\equiv 0$ in all orders of $J/U$
but the case of an incommensurate filling is different.
For instance, in a one-dimensional lattice with $N=nL\pm 1$, in the lowest orders of $J/U$ we obtain
\begin{eqnarray}
f^{\rm s}_{n L+1}
&=&
\frac{1}{n L+1}
\left[
    n+1
    +
    4
    \frac{J}{U}
    n(n+1)
    +
    \dots
\right]
\;,
\\
f^{\rm s}_{n L-1}
&=&
\frac{1}{n L-1}
\left[
    n
    +
    4
    \frac{J}{U}
    n(n+1)
    +
    \dots
\right]
\;.
\label{fshole}
\end{eqnarray}
In the hard-core limit, $J/U\to0$ and $n=1$, Eq.~(\ref{fshole}) gives the same result as Eq.~(\ref{fshchom}) for $N=L-1$.
Although $f^{\rm s}_{n L\pm 1}$ tend to zero in the thermodynamic limit, they do not vanish in lattices of finite size.
This is an indication that the MI lobes are surrounded by the superfluid.

The phase diagram obtained in the third order of the strong-coupling expansion according to Eqs.~(\ref{mupd}),~(\ref{muhd})
is shown in Fig.~\ref{pd-sc3} for different dimensions $d=1,2,3$ (see also Ref.~\cite{FM94,FM96}).
The size of the insulating regions decreases with the number of dimensions $d$ and with the filling factor $n$.

\section{\label{sec-CPBHMSMIT}Critical properties of the Bose-Hubbard model and the superfluid--Mott-insulator transition}

From the analytical results discussed in the previous sections, it becomes clear that the ground-state properties
in the case of a commensurate filling undergo qualitative changes under variation of the ratio $J/U$.
The superfluid stiffness is equal to one for the ideal Bose gas but vanishes identically in the limit of strong interactions.
The compressibility also vanishes for strong interactions but becomes divergent in the limit of vanishing interactions.
Drastic changes are also observed in the excitation spectrum which is gapless in the case of the ideal gas but becomes gapped
in the strongly interacting regime. In the case of an incommensurate filling, the superfluid stiffness is always finite,
the compressibility does not vanish and there is no gap in the excitation spectrum.

All these characteristic features are manifestations of the SF-MI transition.
Since the transition takes place at zero temperature, this is an example of a quantum phase transition
which is driven by quantum fluctuations in contrast to classical phase transitions driven by thermal fluctuations.
According to the classification adopted in the theory of critical phenomena, the transition at fixed commensurate filling
which is controlled only by $J/U$ and the transition controlled by the variation of the filling factor from the incommensurate
to commensurate belong to different universality classes.

The basic idea is that near the critical point, the spatial correlation length $\xi_x$ as well as
the correlation time $\xi_t\sim\xi_x^z$, where $z$ is the dynamical critical exponent, diverge.
The dependence of $\xi_x$ on the distance $\delta$ from the critical point,
which is defined as $\delta=|J-J_c|/U$ or $\delta=|\mu-\mu_c|/U$ for the two different types of transition,
as well as the value of $z$ do not depend on the microscopic details of the Hamiltonian and are entirely determined
by the symmetries and system dimensionality $d$. The divergence of the correlation time leads to the vanishing energy gap
\begin{equation}
\label{gap-xi}
\Delta
\sim
\xi_t^{-1}
\sim
\xi_x^{-z}
\;.
\end{equation}
Dimensional analysis shows also that near the transition point the superfluid stiffness and the compressibility are determined by
$\xi_t$ and $\xi_x$ as~\cite{FWGF89}
\begin{equation}
\label{fs-xi}
f_{\rm s}
\sim
\xi_t^{-1}
\xi_x^{2-d}
\sim
\xi_x^{2-d-z}
\;,
\end{equation}
\begin{equation}
\label{kappa-xi}
\kappa
\sim
\xi_t
\xi_x^{-d}
\sim
\xi_x^{z-d}
\;.
\end{equation}

In a finite systems of $L^d$ sites, the phase transition becomes a crossover.
The correlation length $\xi_x$ is limited by $L$ and Eqs.~(\ref{gap-xi}),~(\ref{fs-xi}),~(\ref{kappa-xi})
have to be generalized. For instance, for the superfluid stiffness, we have
\begin{equation}
\label{fs-scaling}
f_s
=
L^{2-d-z}
\Phi_s(L/\xi_x)
\;,
\end{equation}
where $\Phi_s$ is a universal scaling function for this particular quantity. The power of $L$ in front of it is the same as
the power of $\xi_x$ in the second part of Eq.~(\ref{fs-xi}). The relation~(\ref{fs-scaling}) implies that the dependences of $f_s$
calculated for different system sizes $L$ and multiplied by $L^{d+z-2}$ should intersect at one point which gives
an estimation of the transition point. The expressions of the form~(\ref{fs-scaling}) with other powers of $L$
also exist for other quantities and constitute the basis of the finite-size scaling analysis which is often used in practice
for the calculations of the phase diagrams.

\subsection{\label{SecTCF}Transition at commensurate fillings}

Now we turn to the transition at commensurate fillings which belongs to the universality class of the $(d+1)$-dimensional
XY-model and characterized by $z=1$ in all dimensions. It has lower critical dimension $d=1$ and upper critical dimension $d=3$.

For $d=1$, the transition is of the Berezinskii-Kosterlitz-Thouless type~\cite{B71,B72,KT73,K74}.
This means that the correlation length has an exponential divergence near the transition point
\begin{equation}
\xi_x
\sim
\exp
\left(
    \frac
    {\rm const}
    {
     \sqrt{J_{\rm c}-J}
    }
\right)
\;,\quad
J<J_{\rm c}
\;,
\end{equation}
and according to Eq.~(\ref{gap-xi}) this leads to the exponentially small energy gap.
This behavior makes the study of the transition by computing the energy gap rather difficult and requires large system sizes.
On the other hand, the superfluid stiffness as well as the compressibility do not depend on $\xi_x$ [see, Eqs.~(\ref{fs-xi}),~(\ref{kappa-xi})]
and, therefore, they do not need to be rescaled in the finite-size scaling analysis.

The low-energy physics of the one-dimensional Bose-Hubbard model in the SF phase can be also described by the effective harmonic-fluid approach~\cite{FWGF89,H81,G92}
and all the properties are determined by the Tomonaga-Luttinger parameter~\cite{FWGF89,Giamarchi04,CCGOR11,Cazalilla04}
\footnote
{
 Very often in the literature the Tomonaga-Luttinger parameter is defined as the inverse of that (${\cal K}_{\rm TL}\to1/{\cal K}_{\rm TL}$).
}
\begin{equation}
\label{KTL}
{\cal K}_{\rm TL}
=
\left(
    2 J \pi^2 f_N^{\rm s}
    \langle\hat n_\ell\rangle
    \kappa
\right)^{-1/2}
\;,
\quad
0 \le {\cal K}_{\rm TL} \le 1
\;.
\end{equation}
This parameter vanishes for the ideal Bose gas and equals to one for hard-core bosons.
The parameter ${\cal K}_{\rm TL}$ determines asymptotic behavior of the correlation functions in the superfluid phase
(see sections~\ref{sec-obdm},~\ref{sec-ddcorr})
and at the transition point SF-MI for the commensurate filling it takes the value
${\cal K}_{\rm TL}={\cal K}_{\rm TL}^{\rm c}=1/2$~\cite{FWGF89,Giamarchi04,CCGOR11,Giamarchi06}.

MI phase in one dimension possesses a nonlocal string order described by the correlator~\cite{KM10,ECFWSGMBPBK11}
\begin{equation}
\label{string}
{\cal O}_{\rm P}^2(\ell)
=
\langle
   \prod_{\ell'=1}^\ell
   \hat s_{\ell'}
\rangle
\;,
\end{equation}
where $\hat s_\ell$ is the parity operator~(\ref{s-op}).
In the case $J=0$, the ground state is a product of Fock states and ${\cal O}_{\rm P}^2(\ell)\equiv 1$ for any $\ell$.
If $J/U$ increases, the ground state contains contributions from the particle-hole pairs created at different distances.
As long as the positions of all particles and holes are within the range covered by the string correlator~(\ref{string}),
${\cal O}_{\rm P}^2(\ell)$ contains only positive contributions.
However, it may happen, for instance, that the particle form one pair will be within the range of the correlator~(\ref{string})
but the corresponding hole not. This will give negative contribution and ${\cal O}_{\rm P}^2(\ell)$ will decrease.
At the transition to the SF and above the critical point $(J/U)_{\rm c}$, the pairs are completely deconfined
resulting in random positive and negative contributions in Eq.~(\ref{string}) and ${\cal O}_{\rm P}^2(\ell)$ vanishes.

In the dimensions larger than one, the correlation length has a power-law dependence near the critical point:
\begin{equation}
\label{xi_delta}
\xi_x
\sim
\delta^{-\nu}
\;,
\end{equation}
which leads to the power-law dependences of the energy gap, superfluid stiffness and compressibility.
For $d=2$, the critical exponent $\nu\approx 2/3$ (see, e.g.,~\cite{CHPRV01} and references therein).
For $d\ge 3$, it takes the mean-field value $\nu=1/2$~\cite{FWGF89}.

Quantum phase transitions can be viewed as fundamental changes of the ground state $|\Psi(g)\rangle$
of a many-body Hamiltonian $\hat H(g)=\hat H_0 + g\hat H_1$ under small variations $\delta_g$ of the control parameter $g$
near the critical point $g=g_{\rm c}$~\cite{Sachdev}. Quantitatively, this can be described by the ground-state fidelity
defined as a scalar product of the two ground states~\cite{ZP06}
\begin{equation}
\Phi(g,\delta_g)
=
\langle
   \Psi(g-\delta_g/2) | \Psi(g+\delta_g/2)
\rangle
\end{equation}
which is expected to exhibit a sharp drop at $g_{\rm c}$. Since $|\delta_g|$ is small, the fidelity can be expanded as
\begin{equation}
\Phi(g,\delta_g)
=
1-
\chi(g)
\frac{\delta_g^2}{2}
+
O(\delta_g^4)
\;,
\end{equation}
where $\chi(g)$ is a fidelity susceptibility~\cite{YLG07}.
The aforementioned drop of $\Phi(g,\delta_g)$ should be accompanied by a divergence of $\chi(g)$ which can be employed
to detect the critical point~\cite{AASC10,Gu10}.
This concept borrowed from the quantum information theory is appealing because it does not require any {\it a priori}
identification of the order parameter and does not rely on the symmetries of the Hamiltonian.
The scaling of fidelity $\Phi$ and the susceptibility $\chi$ near and far from the critical point is determined by
the system's dimensionality $d$ and the critical exponent of the correlation length $\nu$~\cite{ZZL08,AASC10,Gu10,GP10,RaDa11,RamDam11}.
Studies of fidelity for the one-dimensional Bose-Hubbard model of finite size are presented in Refs.~\cite{BV07,CMR13,LDZ14}.
It was observed that the fidelity is very sensitive to the boundary conditions and its minimum is shifted from the critical point
to the Mott-insulator side, where the changes of the ground state under variation of $g\equiv J/U$ are more rapid than
in the superfluid phase.

The quantum phase transitions manifest themselves also through the entanglement properties of the system~\cite{OAFF02,AFOV08}.
This is often quantified by the von~Neumann entanglement entropy
\begin{equation}
S_{A}
=
-{\rm Tr}_{A}
\left(
    \rho_{A}
    \ln \rho_{A}
\right)
\;,
\end{equation}
where $\rho_{A}$ is the reduced density matrix of a subsystem $A$ of the whole system.
If the state of the whole system is a tensor product of the states of the subsystem $A$ and of the remainder, $S_A$ vanishes.
In the case of one-dimensional Bose-Hubbard model, the SF phase is described by the conformal (i.e., relativistic and massless)
field theory with the conformal anomaly number (central charge) ${c}=1$.
In this regime and under periodic boundary conditions, the von~Neumann entropy of a contiguous block of $L_A$ lattice sites ($L_A<L$)
is given by~\cite{CC04}
\begin{equation}
\label{SApbc}
S_A =
\frac{c}{3}
\ln
\left(
    \frac{L}{\pi}
    \sin
    \frac{\pi L_A}{L}
\right)
+
c_1'
\;,
\end{equation}
where $c_1'$ is a non-universal constant. In the case of open boundary conditions, Eq.~(\ref{SApbc}) is replaced by
\begin{equation}
\label{SAobc}
S_A =
\frac{c}{6}
\ln
\left(
    \frac{2L}{\pi}
    \sin
    \frac{\pi L_A}{L}
\right)
+
c_1'
+
S_\Omega
\;,
\end{equation}
where $S_\Omega$ is the boundary entropy~\cite{AL91}.
In the MI phase, the von~Neumann block entropy scales as
\begin{equation}
\label{SAmi}
S_A \sim
\ln
\left(
    \xi_x/a
\right)
\;,
\end{equation}
where the correlation length $\xi_x$ is assumed to obey inequality $1\ll\xi_x/a\ll L_A$, and $a$ is the lattice spacing.
Eqs.~(\ref{SApbc})-(\ref{SAmi}) were used in Refs.~\cite{LK08,PPSJC12,EFGMKAL12} in order to extract the critical value of the one-dimensional
Bose-Hubbard model from the numerical data.
In Ref.~\cite{RLSH12} the critical point in one dimension was obtained using the concept of
bipartite fluctuations~\cite{SLRH11,GK06,SRH10,SRFKLH12}
through the analysis of the particle number fluctuations in the subsystem $A$:
\begin{equation}
{\cal F}_A
=
\langle
   \hat N_A^2
\rangle
-
\langle
   \hat N_A
\rangle^2
\;,\quad
\hat N_A
=
\sum_{{\bf l}\in A}
\hat n_{\bf l}
\;.
\end{equation}
This quantity has similar scaling properties as the von~Neumann entropy and allows quite efficient calculations
of the Tomonaga-Luttinger parameter ${\cal K}_{\rm TL}$.

\begin{table}
\begin{tabular}{lccccc}
\hline
$(J/U)_{\rm c}$    & method & quantity                & $L$        & year & Ref.\\
\hline
$0.215\pm 0.01$    & QMC    & energy gap              & up to 32   & 1990 & \cite{BSZ90}\\
$1/(2\sqrt{3})\approx 0.2887$ & BA & superfluid stiffness & $\infty$ & 1991 & \cite{Krauth91}\\
$0.215$            & RSRG   & fixed point             & $\infty$   & 1992 & \cite{SR92}\\
$0.215$            & SCE    & energy gap              & $\infty$   & 1994 & \cite{FM94}\\
$0.275\pm 0.005$   & ED     & energy gap              & up to 9    & 1994 & \cite{EKO94}\\
$0.22\pm 0.02$     & ED     & energy gap              & up to 11   & 1995 & \cite{Openov95}\\
$0.298\pm 0.002$   & DMRG   & energy gap              & up to 70   & 1996 & \cite{PPKR96}\\
$0.304\pm 0.002$   & ED+RG  & superfluid stiffness    & up to 12   & 1996 & \cite{KS96}\\
$0.300\pm 0.005$   & QMC    & energy gap              & up to 50   & 1996 & \cite{KKS96}\\
$0.265$            & SCE+RG & energy gap              & $\infty$   & 1996 & \cite{FM96}\\
$0.25$             & TDVP   & energy gap              & $\infty$   & 1998 & \cite{AP98}\\
$0.277\pm 0.01$    & DMRG   & $F_a$                   & up to 76   & 1998 & \cite{KM98}\\
$0.26\pm 0.01$     & SCE+PA & energy gap              & $\infty$   & 1999 & \cite{EM99}\\
$0.260\pm 0.005$   & DMRG   & superfluid stiffness    & up to 50   & 1999 & \cite{RSZ99}\\
$0.297\pm 0.01$    & DMRG   & $F_a$                   & up to 1024 & 2000 & \cite{KWM00}\\
$0.283\pm 0.005$   & ED     & superfluid stiffness    & up to 14   & 2004 & \cite{PPC04}\\
$0.305\pm 0.004$   & QMC    & $F_a$                   & 128        & 2005 & \cite{Pollet05}\\
$0.257\pm 0.001$   & ED     & ground-state fidelity   & up to 12   & 2007 & \cite{BV07}\\
$0.238\pm 0.011$   & GFMC   & structure factor        & up to 150  & 2007 & \cite{CBFS07}\\
$0.204\pm 0.004$   & VMC    & structure factor        & up to 150  & 2007 & \cite{CBFS07}\\
$0.303\pm 0.009$   & DMRG   & energy gap              & up to 80   & 2008 & \cite{RBMKSG08}\\
$0.2975\pm 0.0005$ & TEBD   & $F_a$                   & $\infty$   & 2008 & \cite{ZD08} \\
$0.29\dots0.30$    & DMRG   & von Neumann entropy     & up to 1024 & 2008 & \cite{LK08} \\
$0.305\pm 0.001$   & DMRG   & $F_n$                   & up to 1024 & 2011 & \cite{EFG11} \\
$0.319\pm 0.001$   & TEBD   & energy splitting        & up to 48   & 2011 & \cite{DP11} \\
$0.295\dots0.320$  & DMRG   & string correlator       & 216        & 2011 & \cite{ECFWSGMBPBK11} \\
$0.2989\pm 0.0002$ & DMRG   & bipartite fluctuations  & up to 256  & 2012 & \cite{RLSH12} \\
$0.2885\pm 0.0001$ & NBA    & superfluid stiffness    & up to 1400 & 2012 & \cite{GCCL12}\\
$0.30\pm 0.01$     & TEBD   & von Neumann entropy     & $\infty$   & 2012 & \cite{PPSJC12} \\
$0.305\pm 0.003$   & DMRG   & von Neumann entropy     & up to 64   & 2012 & \cite{EFGMKAL12} \\
$0.3050\pm 0.0001$ & DMRG   & energy gap              & up to 700  & 2013 & \cite{CMR13} \\
$0.270\pm 0.008$   & TEBD   & ground-state fidelity   & up to 64   & 2014 & \cite{LDZ14} \\
$0.289\pm 0.008$   & TEBD   & fidelity susceptibility & up to 64   & 2014 & \cite{LDZ14} \\
$0.286\pm 0.005$   & ED     & energy gap              & up to 12   & 2015 & \cite{Sowinski15} \\
\hline
\end{tabular}
\caption
{
The critical value of $J/U$ for the MI-SF phase transition in one dimension at unit filling
obtained by the analysis of different physical quantities in the lattices of size $L$
employing different method such as quantum Monte Carlo~(QMC), Bethe Ansatz~(BA),
numerical solution of Bethe equations~(NBA),
real-space renormalization group~(RSRG),
strong-coupling expansion~(SCE),
exact diagonalization~(ED),
density-matrix renormalization group~(DMRG),
Pade analysis~(PA),
time-evolving block decimation~(TEBD),
mean-field theory based on the time-dependent variational principle~(TDVP),
Green's function Monte Carlo~(GFMC),
variational Monte Carlo~(VMC).
}
\label{Jc1D}
\end{table}

Exact calculation of the critical values of $J/U$ in different dimensions and for different fillings
is a challenging problem and various methods have been applied in order to achieve this goal.
A brief overview of these extensive studies in one dimension at unit filling
was given in Ref.~\cite{PPSJC12}. These results are summarized in Table~\ref{Jc1D}
which includes also some other missing as well as more recent references.
Although the numerical data are quite different, all methods give much larger values than
the prediction of the mean-field theory $(J/U)_{\rm c}\approx 0.086$ [see Eq.~(\ref{Jcmax})].
Since the results obtained by exact diagonalization suffer only from the finite-size effects,
they can be considered as a lower estimate of $(J/U)_{\rm c}$.
Taking into consideration also the most recent studies for larger systems, one can accept that $(J/U)_{\rm c}\approx 0.3$.
Another observation is that analytical methods have a tendency to underestimate the critical value $(J/U)_{\rm c}$.

The critical values of $J/U$ in one dimension and in the case of unit filling were measured first in experiments
with Cs atoms using the lattice modulation spectroscopy~\cite{HHMDRGDPN10}.
For the lattice depths $V_0/E_{\rm R} = 6 \dots 10$ corresponding to the tight-binding regime,
the experimental values of $(J/U)_{\rm c}$ appeared to be less than $0.26$ calculated in Ref.~\cite{RSZ99}.
This systematic underestimation was attributed to the presence of the harmonic trap that leads to the spatial inhomogeneity and finite-size effects.
Similar experiment was performed very recently with $^{39}$K~\cite{BGHKLTIGEGMS15}.
The transition points determined from the measurements of the critical momentum for the occurrence
of a dynamical instability are in good agreement with $(J/U)_{\rm c}\approx 0.297$.

For larger integer fillings, exact diagonalization cannot be applied anymore due to very limited systems size but other methods still can be used
and some of the references listed in Table~\ref{Jc1D} reported also estimations of $(J/U)_{\rm c}$ in one dimension
for $\langle\hat n_\ell\rangle=2,3$.
Analytical studies within the SCE up to the third order in combination with the scaling theory found
$(J/U)_{\rm c}=0.155$ for $\langle\hat n_\ell\rangle=2$ and $(J/U)_{\rm c}=0.111$ for $\langle\hat n_\ell\rangle=3$~\cite{FM96}.
Slightly lower value of $(J/U)_{\rm c}=0.008$ for $\langle\hat n_\ell\rangle=3$ was obtained
in the mean-field theory based on the TDVP~\cite{AP98,AP98prb}.
Numerical study of the correlation function $F_a(s)$ with the aid of the TEBD method designed for infinite systems gave
$(J/U)_{\rm c}=0.175\pm 0.002$ for $\langle\hat n_\ell\rangle=2$~\cite{ZD08} which
is very close to $(J/U)_{\rm c}=0.180\pm 0.001$ obtained from the DMRG calculations
of the correlation function $F_n(s)$ in the systems up to $L=128$~\cite{EFG11} as well as to
$(J/U)_{\rm c}=0.179\pm 0.007$ resulting from the DMRG calculations of the von Neumann entropy in the systems up to $L=64$~\cite{EFGMKAL12}.
Recent DMRG calculations of the energy gap for the systems up to $L=250$ lead to
$(J/U)_{\rm c}=0.1790\pm 0.0003$ for $\langle\hat n_\ell\rangle=2$ and
$(J/U)_{\rm c}=0.12697\pm 0.00003$ for $\langle\hat n_\ell\rangle=3$~\cite{CMR13}.

In Ref.~\cite{DP11}, the critical points in one dimension were calculated for the fillings up to $\langle\hat n_\ell\rangle=1000$
from the energy splitting caused by the tunneling between two states with macroscopically distinct currents
using the TEBD method. The numerical data were well fitted by the function
\begin{equation}
\left(
    \frac{U}{J}
\right)_{\rm c}
=
d
\langle\hat n_\ell\rangle
\left(
    a + b \langle\hat n_\ell\rangle^{-c}
\right)
\;,
\label{Jc1Dfit}
\end{equation}
with the coefficients $a=2.16$, $b=0.97$, $c=2.13$.

\begin{table}
\begin{tabular}{lccccc}
\hline
$(J/U)_{\rm c}$      & method  & quantity           & $L$      & year & Ref.\\
\hline
$0.061\pm 0.003$     & PIMC    & superfluid stiffness & up to 8  & 1991 & \cite{KT91}\\
$0.0564$             & RSRG    & fixed point          & $\infty$ & 1992 & \cite{SR92}\\
$0.0585$             & SCE+RG  & energy gap           & $\infty$ & 1996 & \cite{FM96}\\
$0.0625$             & TDVP    & energy gap           & $\infty$ & 1998 & \cite{AP98}\\
$0.05974\pm 0.00004$ & SCE+PA  & energy gap           & $\infty$ & 1999 & \cite{EM99}\\
$0.05963\pm 0.00001$ & SSE     & superfluid stiffness & up to 20 & 2005 & \cite{SS05}\\
$0.0485\pm 0.0005$   & VMC     & structure factor     & up to 30 & 2008 & \cite{CBFS08}\\
$0.0588\pm 0.0007$   & GFMC    & structure factor     & $16$     & 2008 & \cite{CBFS08}\\
$0.05974\pm 0.00003$ & WA      & energy gap           & up to 80 & 2008 & \cite{CSPS2008}\\
$0.05909$            & MEP+SCE & susceptibility       & $\infty$ & 2009 & \cite{THHE09R}\\
$0.067$              & VCA     & energy gap           & $\infty$ & 2010 & \cite{KAL10}\\
$0.060$              & NPRG    & superfluid stiffness & $\infty$ & 2011 & \cite{RD11}\\
$0.055$              & POA     & superfluid stiffness & $\infty$ & 2011 & \cite{TS11}\\
\hline
\end{tabular}
\caption
{
The critical value of $J/U$ for the MI-SF phase transition in two dimensions at unit filling
obtained by the analysis of different physical quantities in the lattices of linear size $L$
employing different method such as
path integral Monte Carlo~(PIMC),
stochastic series expansion~(SSE),
worm algorithm~(WA),
variational Monte Carlo~(VMC),
Green's function Monte Carlo~(GFMC),
real-space renormalization group~(RSRG),
strong-coupling expansion~(SCE),
Pade analysis~(PA),
mean-field theory based on the time-dependent variational principle~(TDVP),
nonperturbative renormalization group~(NPRG),
method of effective potential~(MEP),
projection-operator approach~(POA).
}
\label{Jc2D}
\end{table}

In higher dimensions, the arsenal of the methods is restricted because exact diagonalization and the DMRG cannot be applied.
Nevertheless, QMC works very well and approximate analytical methods can be also used.
The results for two-dimensional square lattices at unit filling are summarized in Table~\ref{Jc2D}.
It is interesting to note that one (semi)analytical and two different numerical methods used
in Refs.~\cite{EM99,SS05,CSPS2008} give essentially the same value of $(J/U)_{\rm c}\approx 0.0597$.
The method of the effective potential, developed in Refs.~\cite{THHE09R,THHE09}
as a combination of the mean-field theory and high-order SCE,
as well as the nonperturbative renormalization group approach~\cite{RD11}
lead to almost the same results which are considerably larger than $(J/U)_{\rm c}\approx 0.0429$ predicted
by the standard mean-field theory [see Eq.~(\ref{Jcmax})].

For larger fillings in two dimensions, to the best of our knowledge, no exact numerical results were published
but some approximate analytical results are available.
The third-order SCE after extrapolation to the infinite order gives
$(J/U)_{\rm c}=0.0345$ for $\langle\hat n_{\bf l}\rangle=2$ and
$(J/U)_{\rm c}=0.0245$ for $\langle\hat n_{\bf l}\rangle=3$~\cite{FM96},
while the VCA results in $(J/U)_{\rm c}=0.038$ for $\langle\hat n_{\bf l}\rangle=2$~\cite{KAL10}.

In three-dimensional cubic lattices, the critical value of $J/U$ was calculated by the extrapolation of
the third-order SCE to the infinite order which gives
$(J/U)_{\rm c}=0.0337\pm 0.0027$, $0.0200\pm 0.0013$, and $0.0140\pm 0.0007$
for $\langle\hat n_{\bf l}\rangle=1$, $2$, and $3$, respectively~\cite{FM96}.
The numerical value for $\langle\hat n_{\bf l}\rangle=1$ is in excellent agreement with QMC calculations
of the energy gap in the lattices with the linear sizes up to $L=20$, which give
$(J/U)_{\rm c}=0.03408\pm 0.00002$~\cite{CPS2007}.
The NPRG approach yields almost the same result $(J/U)_{\rm c}=0.0339$~\cite{RD11}
and the POA developed in Refs.~\cite{TS11,DTS12} allows to reproduce the results of the QMC calculations with the accuracy $\sim0.05\%$.
Although the mean-field theory is expected to work better in higher dimensions,
Eq.~(\ref{Jcmax}) provides an estimate $(J/U)_{\rm c}\approx 0.0286$ which is again noticeably lower than the exact numerical values.
VMC calculations of the static structure factor for systems with linear sizes up to $L=12$
gave $(J/U)_{\rm c}\approx0.0278$~\cite{CBFS07}.
Recently developed self-consistent standard basis operator approach, which goes significantly beyond the mean-field theory,
yields $(J/U)_{\rm c}\approx 0.03356$ for $\langle\hat n_{\bf l}\rangle=1$ and
$(J/U)_{\rm c}\approx 0.0185$ for $\langle\hat n_{\bf l}\rangle=2$~\cite{SPMR15}.

Theoretical predictions for the critical value $(J/U)_{\rm c}$ in three dimensions were tested in experiments with Cs atoms
analyzing the quasi-momentum distributions in the time-of-flight images~\cite{MHLDDN11}.
It was observed that the width of the central peak as a function of the amplitude of the periodic potential $V_0$
at fixed  scattering length $a_{\rm s}$ has a kink at some value $V_{\rm c}$ that is interpreted as the transition point.
The experimental data obtained in deep lattices ($V_0=8\dots18\;E_{\rm R}$) in the case of unit filling
for different values of $a_{\rm s}$ controlled with the aid of Feshbach resonances are in good agreement with the critical values
obtained by QMC~\cite{CPS2007} as well as in the framework of the mean-field theory within the experimental uncertainty.

In Refs.~\cite{THHE09R,THHE09}, the method of the effective potential combined with the high-order SCE
was used for the calculations of $(J/U)_{\rm c}$ for fillings up to $\langle\hat n_{\bf l}\rangle=10000$
in two and three dimensions. The computed critical values were fitted by
\begin{equation}
\left(
    \frac{J}{U}
\right)_{\rm c}
=
\left(
    \frac{J}{U}
\right)_{\rm c}^{\rm MF}
+
\frac
{0.13}
{
 \sqrt
 {
  \langle\hat n_{\bf l}\rangle
  \left(
      \langle\hat n_{\bf l}\rangle
      +
      1
  \right)
 }
 \,d^{2.5}
}
\;,
\end{equation}
where $(J/U)_{\rm c}^{\rm MF}$ is given by Eq.~(\ref{Jcmax}), with the accuracy of about $1\%$~\cite{THHE09}, and by
\begin{equation}
\left(
    \frac{J}{U}
\right)_{\rm c}
=
\left(
    \frac{J}{U}
\right)_{\rm c}^{\rm MF}
\left(
    1
    +
    \frac{0.35}{d}
    +
    \frac{0.39}{d^2}
    +
    \frac{0.84}{d^3}
\right)
\end{equation}
with the accuracy of $0.15\%$~\cite{TH09}.
It was pointed out~\cite{DP11} that the fit~(\ref{Jc1Dfit}) with the coefficients $(a,b,c)=(5.80,2.66,2.19)$
and $(a,b,c)=(6.70,3.08,2.18)$ in two and three dimensions, respectively, gives also a good approximation for the transition points.

\subsection{Generic transition}

The transition governed by the variation of the filling factor (or chemical potential) belongs to the mean-field universality class (Gaussian model).
The behavior of the spatial correlation length near the critical point is described by Eq.~(\ref{xi_delta})
and the critical exponents are $z=2$ and $\nu=1/2$ in all dimensions~\cite{FWGF89}.
This transition has the upper critical dimension $d=2$.
In one dimension, the Tomonaga-Luttinger parameter takes its universal value ${\cal K}_{\rm TL}={\cal K}_{\rm TL}^*=1$
at the transition points~\cite{Giamarchi04,CCGOR11}.

\begin{figure}[t]
\centering

\stepcounter{nfig}
\includegraphics[page=\value{nfig}]{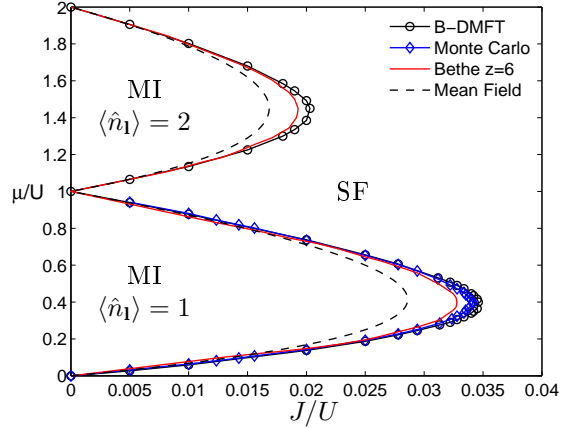}

\caption
{
(Color online).
Ground-state phase diagram of the Bose-Hubbard model in three dimensions showing MI lobes with
$\langle\hat n_{\bf l}\rangle=1,2$
surrounded by the SF.
(Adapted with permission from Ref.~\cite{AGPTW11}).
}
\label{pd3D}
\end{figure}

\begin{figure}[t]

\centering

\stepcounter{nfig}
\includegraphics[page=\value{nfig}]{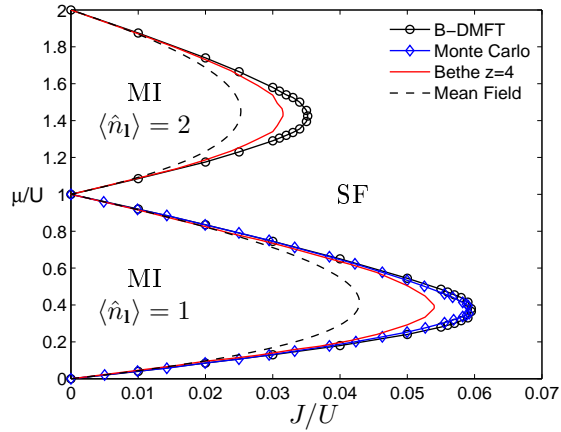}

\caption
{
(Color online)
Ground-state phase diagram of the Bose-Hubbard model in two dimensions showing MI lobes with
$\langle\hat n_{\bf l}\rangle=1,2$
surrounded by the SF.
(Adapted with permission from Ref.~\cite{AGPTW11}).
}
\label{pd2D}
\end{figure}

\begin{figure}[t]
\centering

\stepcounter{nfig}
\includegraphics[page=\value{nfig}]{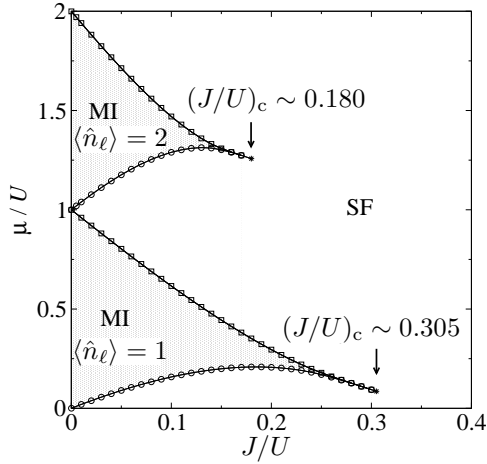}

\caption
{
Ground-state phase diagram of the Bose-Hubbard model in one dimension showing MI lobes with
$\langle\hat n_\ell\rangle=1,2$
calculated by the DMRG method for $L\le 128$.
(Adapted with permission from Ref.~\cite{EFG11}).
}
\label{pd1D}
\end{figure}

The critical points of the generic transition are described by two continuous lines $\mu_\pm(J)$
parametrized by the integer filling factors $\langle\hat n_{\bf l}\rangle$ which
cross at the critical point of the commensurate transition and form the boundaries
of the insulating regions in the $(\mu,J)$ plane (MI lobes).
Within the third-order SCE, the boundaries $\mu_\pm(J)$ are given by Eqs.~(\ref{mu-pm}),~(\ref{mupd}),~(\ref{muhd})
and the corresponding phase diagrams are shown in Fig.~\ref{pd-sc3}.
In Figs.~\ref{pd3D},~\ref{pd2D},~\ref{pd1D}, we show the exact phase diagrams worked out by numerical and semi-analytical methods.
The comparison with Fig.~\ref{pd-sc3} shows that
few lowest orders of strong-coupling expansion appear to be sufficient in order to reproduce the topology of the phase diagram
in three and two dimensions (Figs.~\ref{pd3D},~\ref{pd2D}) except the tips of the lobes.
Higher-order calculations~\cite{EM99,EM99p} and extrapolation to the infinite order~\cite{FM96,EM99,EM99p,FKKKT2009}
allow to reach perfect agreement with exact numerical results.
QMC data for the filling factor $\langle\hat n_{\bf l}\rangle=1$
in two~\cite{CSPS2008} and three~\cite{CPS2007} dimensions were also reproduced with the high accuracy
using the MEP~\cite{SP09,THHE09R,THHE09},
the B-DMFT~\cite{AGPTW10,AGPTW11},
and the NPRG approach~\cite{RD11}.
The POA gives also a high accuracy in three dimensions~\cite{TS11,DTS12}.

In one dimension, the situation is more complicated due to the fact that the critical behavior of the system is of
the Berezinskii-Kosterlitz-Thouless type. The shape of the boundaries $\mu_\pm$ separating the MI from the SF is qualitatively
different (see Fig.~\ref{pd1D}). Starting from certain values of $J/U$, the lowest boundary $\mu_-$ bends down which implies that
in some interval of $\mu$ the MI phase is reentrant, i.e., increasing the tunneling parameter $J$ one returns to the MI.
This feature becomes visible in the SCE calculations up to the 12th order~\cite{EM99,KM98}.

\subsection{Finite temperature}

\begin{figure}[ht]

\centering

\stepcounter{nfig}
\includegraphics[page=\value{nfig}]{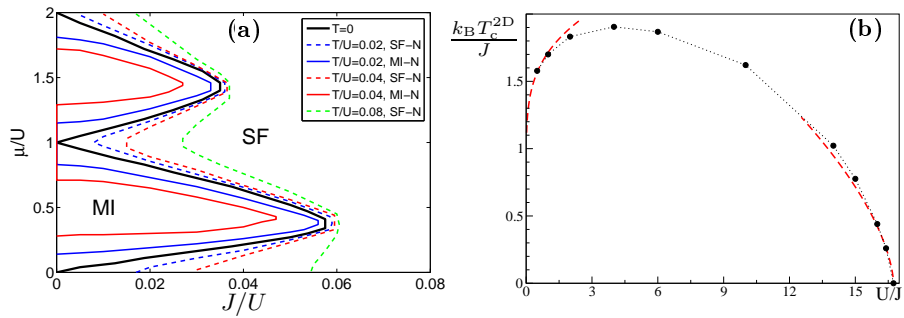}

\caption
{
(color online).
(a)
Finite-temperature phase diagram for the homogeneous Bose-Hubbard
model in two dimensions in the ($\mu/U, t/U$) plane for $k_{\rm B}T/U=0.02$, $0.04$ and $0.08$.
The lines that demarcate the SF and N is a phase boundary, and the lines that demarcate MI and N is a crossover.
At finite temperature, normal phase regions appear between MI and SF.
These normal regions are bigger for higher temperature.
(Adapted with permission from Ref.~\cite{MDKKST11}. \copyright 2011 American Physical Society.)
(b)
Critical temperature in two dimensions at filling $\langle\hat n_{\bf l}\rangle=1$.
Circles are simulation results and the dotted line is to guide the eye.
Dashed lines are analytical results for the weakly interacting gas~(\ref{Tc2D_small_U})
and for the strongly interacting gas near the quantum critical point~(\ref{TcQCP}).
(Adapted with permission from Ref.~\cite{CSPS2008}. \copyright 2008 American Physical Society.)
}
\label{phd2D-T}
\end{figure}

At finite temperature, thermal fluctuations give rise to the normal phase which appears on the phase diagram
in addition to the SF and MI phases. Fig.~\ref{phd2D-T} shows the finite-temperature phase diagram of a two-dimensional system
worked out by QMC simulations~\cite{MDKKST11}, and in three dimensions the topology should remain the same.
Compared to $T=0$, the boundary separating SF from the insulating phases is shifted towards larger values of $J/U$.
Due to the fact that the compressibility never vanishes at finite temperature, there is no drastic difference
between the MI and normal gas. On the other hand, if the compressibility $\kappa$ is small enough, the system can be still
considered as a MI. In the example for two-dimensional system shown in Fig.~\ref{phd2D-T}, the crossover line
between the MI and normal gas was determined from the requirement $\kappa U < 0.04$~\cite{MDKKST11}.

\begin{figure}[ht]
\centering

\stepcounter{nfig}
\includegraphics[page=\value{nfig}]{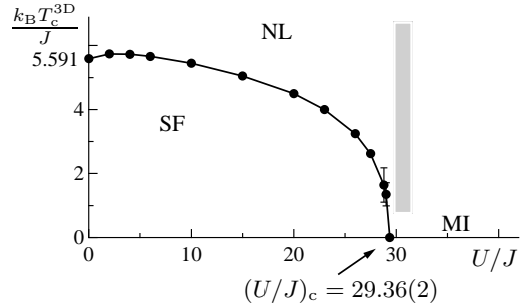}

\caption
{
Critical temperature in three dimensions at filling $\langle\hat n_{\bf l}\rangle=1$.
Circles are QMC simulation results and the line is to guide the eye.
$k_{\rm B}T_{\rm c}^{3D}/J=5.591$ is the critical temperature of the ideal Bose gas with
the tight binding dispersion relation.
(Adapted with permission from Ref.~\cite{CPS2007}. \copyright 2007 American Physical Society.)
}
\label{Tc3D}
\end{figure}

On the other hand,, the compressibility in the insulating phase has a ``thermally activated" form, i.e.,
$\kappa(T)\sim\exp(-\Delta/k_{\rm B}T)$ with a finite energy gap $\Delta$ (see, e.g., Ref.~\cite{CL07}).
This suggests an alternative specification of the crossover line between the normal gas and the MI from the condition
$\Delta=k_{\rm B}T$~\cite{SKPR93,CL07,DODS03,Wang09}.

Figs.~\ref{phd2D-T}(b),~\ref{Tc3D} show the dependences of the critical temperature $T_{\rm c}$ of the superfluid-normal transition
in two and three dimensions for an integer filling $\langle\hat n_{\bf l}\rangle=1$~\cite{CSPS2008,CPS2007}.
Near the quantum critical point of the SF-MI transition $(U/J)_{\rm c}$ at zero temperature, the critical temperature
is related to the superfluid stiffness as~\cite{FWGF89,Sachdev,KC05}
\begin{equation}
T_{\rm c}
=
A
\left(
    f_{L^2}^{\rm s}
\right)^y
\;,\quad
y=\frac{z}{d+z-2}
\;.
\end{equation}
Taking into account Eqs.~(\ref{fs-xi}),~(\ref{xi_delta}), we obtain
\begin{equation}
\label{TcQCP}
\frac{k_{\rm B}T_{\rm c}}{J}
=
A
\left[
    \left(
        \frac{U}{J}
    \right)_{\rm c}
    -
    \frac{U}{J}
\right]^{z\nu}
\;,
\end{equation}
where $z=1$, and $\nu$ is the critical exponent of the correlation length of the $(d+1)$-dimensional XY-model.
In two dimensions, QMC calculations gave $A=0.49\pm0.02$~\cite{CSPS2008}.

In the weakly interacting regime, the dependence of $T_{\rm c}$ in two and three dimensions is qualitatively different.
In two dimensions, the transition is of the Berezinskii-Kosterlitz-Thouless type and for small $U/J$ the critical temperature is given by~\cite{PRS01,CSPS2008}
\begin{equation}
\label{Tc2D_small_U}
\frac{k_{\rm B} T_{\rm c}^{\rm 2D}}{J}
=
\frac
{
 4\pi
 \langle
     \hat n_{\bf l}
 \rangle
}
{
 \ln
 \left(
     2 \xi J/U
 \right)
}
\;,\quad
\xi=380\pm3
\;.
\end{equation}
This equation shows that in the limit of vanishing interaction, $T_{\rm c}^{\rm 2D}$ vanishes.
In three dimensions, the critical temperature in the limit of vanishing interaction tends to a finite value (see section~\ref{sec:TcIBG}).
The critical temperature of the superfluid-insulator transition experimentally measured
in three-dimensional optical lattice~\cite{TPGSBPST10}
shows satisfactory agreement with the QMC data presented in Fig.~\ref{Tc3D}.

Quantum critical behavior was observed in experiments with Cs atoms in a 2D optical lattice at finite temperature near
the normal-to-superfluid transition~\cite{ZHTC12}. In the zero-temperature limit, it connects to the vacuum-to-superfluid transition,
where vacuum can be viewed as a MI with zero occupation number. On the basis of in situ density measurements, the equation of state
$\rho(\mu,T)$ of the sample was determined, which gave the values of the critical exponents $z=2.2 {+1.0 \atop -0.5}$ and $\nu=0.52 {+0.09 \atop -0.10}$
in agreement with theoretical predictions.

\subsection{Criticality in confined systems}

\begin{figure}[ht]

\centering

\stepcounter{nfig}
\includegraphics[page=\value{nfig}]{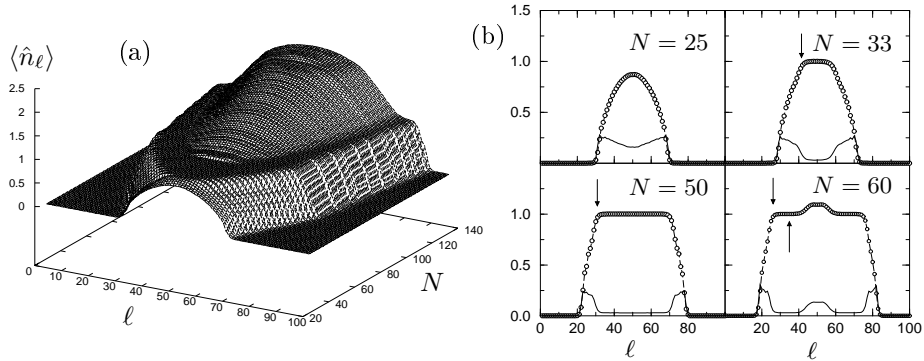}

\caption
{
(a)
Spatial distribution of the mean occupation numbers $\langle\hat n_\ell\rangle$ in a one-dimensional lattice
in the presence of harmonic confinement for increasing total number of bosons $N$ calculated by QMC.
The parameters are $V_{\rm T}/J=0.008$, $U/J=8$, $L=100$.
At low $N$, the system is in a SF phase. MI plateaus appear as $N$ is increased.
(b)
Profiles of the local compressibility $\kappa_\ell$ (solid lines) associated with corresponding mean particle-number distributions (circles)
for the same parameters as in (a) and for fixed $N$.
$\kappa_\ell$ is very small when $\langle\hat n_\ell\rangle=1$.
(Adapted with permission from Ref.~\cite{BRSRMDT02}. \copyright 2002 American Physical Society.)
}
\label{nl_trap_1D}
\end{figure}

In order to create a MI in a homogeneous lattice, it is necessary to have and absolute control over the total number of particles $N$
which has to be a multiple integer of the total number of lattice sites $L^d$. This is hardly achievable with ultracold atoms.
However, real experiments are performed in harmonic traps which allow to create MI in restricted spatial regions.
The presence of harmonic confinement leads to inhomogeneous spatial distribution of $\langle\hat n_{\bf l}\rangle$
in the ground state~\cite{BRSRMDT02,WATB04,BHR04,BKMRS08,RBRS09}, see Fig.~\ref{nl_trap_1D}.
For sufficiently small total number of particles, the density profile is smooth having the form of the inverted confining potential.
With the increase of the particle number a plateau with a local filling of one boson per site develops in the central part of the trap,
provided that the number of particles exceeds some critical value which depends on the system parameters,
similarly to the case of hard-core bosons considered in section~\ref{HCB1Dht}.
In the example shown in Fig.~\ref{nl_trap_1D} this critical number is about $30$.
Within the plateau the local compressibility is small which is considered as an indicator of the MI region.
Further increase of the particle number leads first to the broadening of the plateau but then to the formation of a compressible region
with local fillings larger than one in the center of the trap.
Increasing the particle number further, one can produce a second MI region leading to the so-called wedding-cake (shell) structure
of the density distribution, which can be interpreted as a coexistence of the SF and MI phases.

In the presence of confining potentials it becomes more reasonable to consider state diagrams instead of phase diagrams.
It turns out that the state diagrams are determined by two parameters,
the characteristic density $\tilde\rho$ and $J/U$, and almost independent on the values of $V_{\rm T}/J$~\cite{BKMRS08,RBRS09}.
In the case of harmonic trap, $\tilde\rho=Na^{d}(V_{\rm T}/J)^{d/2}$~\cite{RM04a,BKMRS08}.
It was found that the largest values of $J/U$ that support the local insulator with $\langle\hat n_\ell\rangle=1$
are $0.18$ in one dimension and $0.0575$ in two dimensions~\cite{RBRS09}.
The latter is very close to the critical value of the SF-MI transition in a homogeneous 2D system, but the former is substantially lower
than the corresponding results in 1D (see Tables~\ref{Jc1D},~\ref{Jc2D}).

Since diverging length scales cannot appear in confined systems, it was debated whether quantum criticality can be observed
in experiments~\cite{BHR04,WATB04,NCK06,BKMRS08}.
In earlier papers, the experimental observation of the MI phase was interpreted as a crossover~\cite{WATB04}.
Later it was demonstrated that the critical behavior of trapped systems can be cast in the form of trap-size scaling
in analogy to the finite-size scaling theory for homogeneous systems, which shows that at criticality the spatial scale
depends on the trap size $R$ as $\xi_x\sim R^\theta$ with an additional trap critical exponent $\theta$~\cite{CV09,CV10}.
The value of $\theta$ can be obtained using the renormalization-group analysis.
In the case of power-law potentials with the exponent $p$ ($p=2$ corresponds to the harmonic trap) and for Bose-Hubbard model,
\begin{equation}
\theta
=
\frac{p}{p+1/\nu}
\;,
\end{equation}
where $\nu$ is the critical exponent of the correlation length of the homogeneous system~\cite{CV10}.
It was shown that violation of the local-density approximation is a good indication of the critical behavior,
which can be observed from the measurements of in situ density profiles and quasi-momentum distribution~\cite{PPS10}.
Experiments in two-dimensional lattices confirm this prediction and the critical values of $J/U$ extracted from the measurements
of the fraction of particles with zero momentum are consistent with QMC calculations for trapped systems~\cite{JCLPPS10}.

\section{\label{sec-ENR}Exact numerical results}

In this section we present exact results for the Bose-Hubbard model in different dimensions $d=1,2,3$
for finite systems across the quantum critical point.
They are obtained by different numerical methods such as exact diagonalization,
density-matrix renormalization group~(DMRG) and quantum Monte Carlo~(QMC).

Exact diagonalization is performed in the complete Hilbert space of the Bose-Hubbard Hamiltonian~(\ref{HBH})
and allows to calculate in principle all the eigenvalues and all the eigenstates.
However, due to the rapid growth of the Hilbert space with the number of particles and the number of lattice sites
[see Eq.~(\ref{Fock})] the calculations are limited to the systems of small sizes.
In homogeneous lattices under periodic boundary conditions the dimension of the Hamiltonian matrices can be reduced approximately by a factor $L^d$
using the basis of the eigenstates of the momentum operator that are given by Eq.~(\ref{basis}).
However, this does not allow to increase considerably the size of the systems.
On the other hand, typical eigenstates of quantum many-body systems occupy only a small part of the Hilbert space.
This idea is used in the DMRG method which is very efficient in one dimension and allows to treat large systems.
In higher dimensions, QMC methods based on the stochastic sampling of the complete Hilbert space appear to be superior.

All these methods as well as some others are implemented in the ALPS package
which is freely available~\cite{ALPS11}. It is not difficult to install it and it was already used
in many publications (see a list of references on the homepage of the ALPS project).
The results of the DMRG and QMC calculations reported in this section are obtained with the aid of the package.
However, comparison of our own implementation of the exact diagonalization with that of ALPS showed that the latter is
inefficient and requires too much computer memory, therefore, we used our own exact diagonalization code.
Before we turn to the discussion of results we would like to make some notes about exact diagonalization.

\subsection{Remarks on exact diagonalization}

In simplest situations like in the case of two atoms considered in section~\ref{S_BS2atoms},
one can write down explicitly all the basis states and all the matrix elements of the Hamiltonian.
However, for arbitrary number of atoms and lattice sites this is not possible and we need efficient computer algorithms to deal with this.
In order to generate the Hamiltonian matrix, we have to generate sequentially all the basis states,
act by the Hamiltonian on each of the states and determine the numbers of the resulting states.
This leads us in the field of combinatorics because the basis states~(\ref{Fock})
are nothing but compositions of an integer $N$ into $L$ non-negative parts.
There are efficient algorithms that fulfill the following tasks for given $N$ and $L$:
(1) generate sequentially all compositions;
(2) determine the number of a given composition;
(3) generate a composition with a given number.
The tasks (2) and (3) are called ranking and unranking, respectively.
All these tasks are implemented in the combinatorial package SELECT~\cite{NW78,SELECT}
and can be use for matrix generation as well as for calculations of the observables after the diagonalization.
However, the package does not take into account the spatial symmetries like translational invariance
or discrete reflections and some additional programming is required in order to use those.
Alternatively, one can use hashing techniques~\cite{Liang95,ZD10}.

If we want to make a profit from the translational invariance, we have to use the basis states~(\ref{basis_L}).
The interaction part of the Bose-Hubbard Hamiltonian creates only diagonal matrix elements which is simple.
The hopping part of the Hamiltonian is non-diagonal and in order to understand how to create corresponding matrix
it is sufficient to consider the action of an operator
$\hat H_{\gets}=\sum_{\ell=1}^L\hat a_\ell^\dagger \hat a_{\ell+1}^{\phantom{\dagger}}$,
which describes hopping ``from right to the left", on a basis state~(\ref{basis_L}).
We note that
\begin{eqnarray}
\hat a_\ell^\dagger \hat a_{\ell+1}^{\phantom{\dagger}}
|{\bf n}_\Gamma\rangle
=
\sqrt
{
 \left(
     n_{\Gamma,\ell}+1
 \right)
 n_{\Gamma,\ell+1}
}
\,
\hat{\cal T}^{\,r_{\Gamma\Gamma_\ell}}
|{\bf n}_{\Gamma_\ell}\rangle
\;,
\end{eqnarray}
where $r_{\Gamma\Gamma_\ell}$ is an integer which is uniquely determined by $\Gamma$ and $\ell$.
Then for fixed $K$ it is easy to show that
\begin{eqnarray}
\label{H_left}
\hat H_{\gets}
|{\bf n}_{K\Gamma}\rangle
=
\sum_{\ell=1}^L
\sqrt
{
 \left(
     n_{\Gamma,\ell}+1
 \right)
 n_{\Gamma,\ell+1}
}
\sqrt{\frac{\nu_\Gamma}{\nu_{\Gamma_\ell}}}
{\tau}_K^{r_{\Gamma\Gamma_\ell}}
|{\bf n}_{K\Gamma_\ell}\rangle
\;.
\end{eqnarray}
One has to keep in mind that the states $|{\bf n}_{K\Gamma_\ell}\rangle$ in Eq.~(\ref{H_left}) are not necessarily distinct
and may coincide with $|{\bf n}_{K\Gamma}\rangle$. Calculating all the matrix elements of the operator $\hat H_{\gets}$
one can construct the full Hamiltonian matrix.

\subsection{$(\mu,J)$ diagram}
\label{energyspectrum}

\begin{figure}[t]
\centering

\stepcounter{nfig}
\includegraphics[page=\value{nfig}]{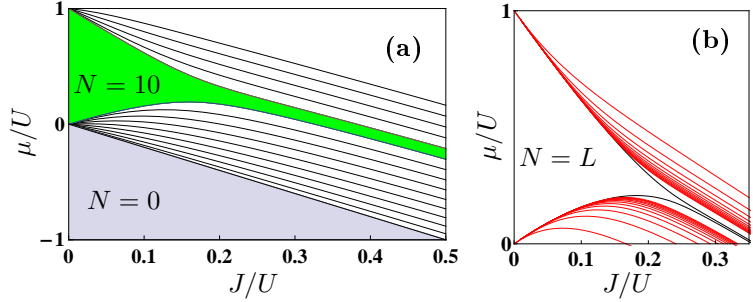}

\caption
{
(a)
Regions with the occupation numbers $N=0,1,\dots,15$ (from bottom to the top) in a one-dimensional lattice of $L=10$ sites.
(b)
The red lines are the boundaries of the region with $N=L$ calculated for $L=2,3,\dots,13$ by exact diagonalization.
The black lines are the boundaries from the DMRG calculations for $128$ sites.
See also Refs.~\cite{PPC04,BV07,PPC13}.
(For interpretation of the references to colour in this figure legend, the reader is referred to the web version of this article.)
}
\label{pd-ed-L10}
\end{figure}

We consider first the boundaries $\mu_\pm(N)$ of the regions in the $(\mu,J)$ plane corresponding to different total particle numbers $N$.
They are determined by Eq.~(\ref{mu-pm}).
The results of numerical calculations for one-dimensional lattices obtained by exact diagonalization are shown in Fig.~\ref{pd-ed-L10},
and the qualitative behavior remains the same in higher dimensions.
Due to the finite system size, the regions of different occupation numbers are always finite
but their boundaries come closer to each other with the increase of the number of lattice sites.
In the thermodynamic limit, they should densely cover the whole $(\mu,J)$ plane except some finite regions
corresponding to the MI phase which exist only for commensurate fillings, provided that the ratio $J/U$ is small enough.
Due to the fact that the number of particles does not depend on $\mu$ within these finite regions, the compressibility~(\ref{kappa}) vanishes,
and we should obtain exactly the same phase diagram as in Fig.~\ref{pd1D}.

\subsection{Superfluid stiffness}

\begin{figure}[t]

\centering

\stepcounter{nfig}
\includegraphics[page=\value{nfig}]{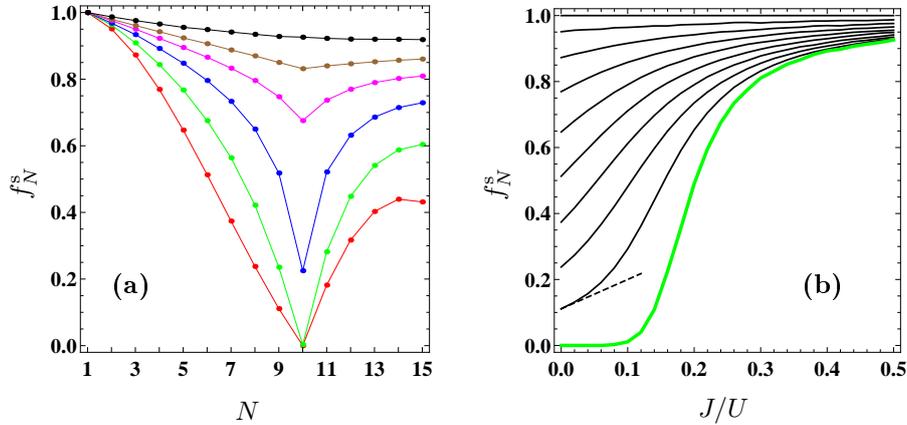}

\caption
{
Superfluid stiffness obtained by numerical diagonalization of the Bose-Hubbard Hamiltonian for $L=10$
in the presence of Peierls factors with $\phi=0.01$.
{\bf(a)}
$f^{\rm s}_N$ as a function of $N$ for
$J/U=0$~(red), $0.08$~(green), $0.16$~(blue), $0.24$~(magenta), $0.32$~(brown), $0.5$~(black).
{\bf(b)}
$f^{\rm s}_N$ as a function of $J/U$ for
$N$ ranging from $1$~($f^{\rm s}_N\equiv 1$) to $10$ from top to the bottom.
Dashed line is the result of the second-order perturbation theory in $J/U$~[Eq.~(\ref{fshole})].
(For interpretation of the references to colour in this figure legend, the reader is referred to the web version of this article.)
}
\label{sf-L10}
\end{figure}

\begin{figure}[th]

\centering

\stepcounter{nfig}
\includegraphics[page=\value{nfig}]{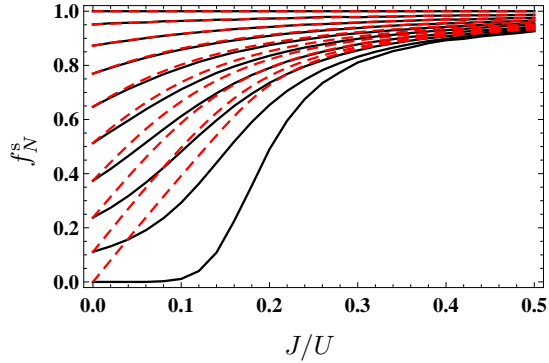}

\caption
{
(color online)
Comparison of the exact results for the superfluid stiffness (black solid lines, the same as in Fig.~\ref{sf-L10}b)
with those obtained from Eq.~(\ref{fNs_hydro}) within the framework of the hydrodynamic approach (red dashed lines).
}
\label{sf_hydro}
\end{figure}

The superfluid stiffness $f_N^{\rm s}$
calculated according to the definition~(\ref{fs-def}) in a finite one-dimensional lattice for different particle numbers $N$
and different values of $J/U$ is shown in Fig.~\ref{sf-L10}. If the filling is not commensurate, the superfluid stiffness
remains finite for all values of $J/U$. However, for commensurate fillings, finite intervals of $J/U$
close to zero appear, where $f^{\rm s}_N$ with a good numerical accuracy can be considered as vanishing, i.e., the system becomes an insulator.
Numerical values of $f^{\rm s}_N$ in the limit $J/U\to0$ are perfectly described by Eqs.~(\ref{fshchom}),~(\ref{fsghchom})
derived for hard-core bosons in one dimension.

In the hydrodynamic regime, the superfluid stiffness is determined by Eq.~(\ref{fs_kappa_cs}).
Together with Eq.~(\ref{csound}) for the sound velocity at $T=0$ derived from the sum rules for the dynamic structure factor, it gives
\begin{equation}
\label{fNs_hydro}
f_N^{\rm s}
=
\frac
{\langle\hat a_{\bf l}^\dagger \hat a_{{\bf l}+{\bf e}_\nu}^{\phantom{\dagger}}\rangle}
{\langle\hat n_{\bf l}\rangle}
\;.
\end{equation}
For the Tonks-Girardeau lattice gas, this expression leads to Eq.~(\ref{fshchom}).
However, in general the validity of Eq.~(\ref{fNs_hydro}) is limited.
This is demonstrated in Fig.~\ref{sf_hydro}, where one can see that Eq.~(\ref{fNs_hydro}) works well only for low fillings.
For fillings close to one, there are strong deviations from the exact results at small $J/U$.

\subsection{Energy spectrum}
\label{energyspectrum}

As it was discussed in Sec.~\ref{T-ivariance},
the eigenstates of the homogeneous system under periodic boundary conditions
are characterized by two indices, ${\bf K}$ and $\Omega$, where the former denotes the total momentum
of $N$ interacting particles in a lattice.
In the case of commensurate filling, the number of eigenstates for ${\bf K}=0$ is always larger
than for other values of ${\bf K}$. For the discussion of the energy spectrum,
it appears to be convenient to start the labeling for $K=0$ with $\Omega=0$ and for other $K$'s with $\Omega=1$.

\begin{figure}[t]
\centering

\stepcounter{nfig}
\includegraphics[page=\value{nfig}]{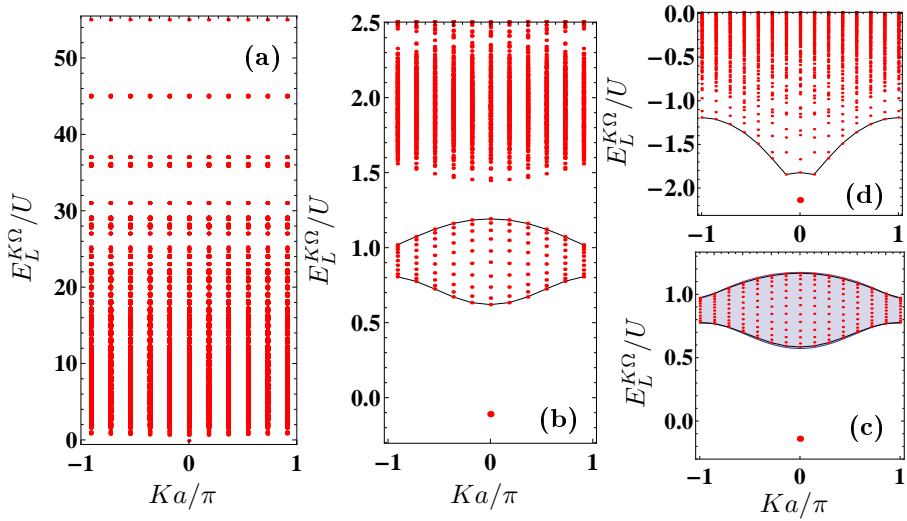}

\caption
{
(color online)
(a)
Full energy spectrum and
(b)
its lowest part for $N=L=11$, $J/U=0.05$.
(c)
Dots: lowest part of the energy spectrum for $N=L=14$, $J/U=0.05$.
Shaded region is the lowest excitation band obtained in the thermodynamic limit in the second order of $J/U$ [Eq.~(\ref{eband-sce})].
Solid lines show the boundaries of the lowest excitation band which is well resolved for this value of $J/U$.
(d)
Lowest part of the spectrum for $N=L=14$, $J/U=0.2$.
Solid line connects the energies of the lowest excited states.
}
\label{sp_01}
\end{figure}

The full energy spectrum calculated by exact diagonalization for $N=L=11$ and $J/U=0.05$,
which corresponds in the thermodynamic limit to the Mott-insulator state, is shown in Fig.~\ref{sp_01}(a).
The lowest dot at $K=0$ is the ground-state energy $E^{00}_{L}$, see also  Fig.~\ref{sp_01}(b).
The energies $E^{K\Omega}_L$, $\Omega=1,\dots,L-1$, form the lowest excitation band shown by the solid lines in Figs.~\ref{sp_01}(b,c).
For such a small value of $J/U$, it does not overlap with the higher excitation bands and is well described by Eq.~(\ref{eband-sce}).
An increase of the system size leads to more dense distribution of the points,
however the structure of the lower part of the spectrum remains the same
[compare Figs.~\ref{sp_01}(b) and~\ref{sp_01}(c)].
At small momenta $K$, the lowest excitation branch can be approximated by a pseudo-relativistic form
\begin{equation}
\label{rel}
\left(
    E^{K1}_L-E^{00}_L
\right)^2
=
\left(
    \Delta{\cal E}
\right)^2
+
v_\mathrm{eff}^2 K^2
\end{equation}
with the energy gap $\Delta{\cal E}$ and the effective velocity $v_\mathrm{eff}$.

If we increase the ratio $J/U$, the excitation bands start to overlap and the energy spectrum becomes qualitatively different
(see Fig.~\ref{sp_01}(d) and Ref.~\cite{Roux10}).
The lowest excited state is not at $K_0=0$ anymore but at $K_{\pm 1}=\pm 2\pi/(La)$ and, therefore, it is degenerate.
In the thermodynamic limit, this should give a sound mode with the linear dispersion relation for small $K$.
In the limit of infinite interaction and for commensurate fillings, the corresponding sound velocity vanishes
according to Eq.~(\ref{cshc1D}) but remains finite for finite values of $J/U$ above the critical point~\cite{KSDZ05}.
In higher-dimensional lattices, the structure of the energy spectrum is expected to be the same,
although no exact results for sufficiently large systems have been reported so far.

Energy-spectrum statistics of the one-dimensional Bose-Hubbard model was studied in Refs.~\cite{BK03,KB04,HKG09,KRBL10}.
While in the special cases $J=0$, $U=0$, as well as in the hard-core limit, the model is obviously integrable,
at intermediate values of the parameters the integrability is lost.
The distribution of spacings between the adjacent energy levels follow the universal Wigner-Dyson law
characteristic for quantum chaotic systems that belong to the Gaussian orthogonal ensemble.
In particular, the probabilities of small level spacings are strongly suppressed which is an indication of level repulsion
due to the avoided crossings.

\begin{figure}[t]

\centering

\stepcounter{nfig}
\includegraphics[page=\value{nfig}]{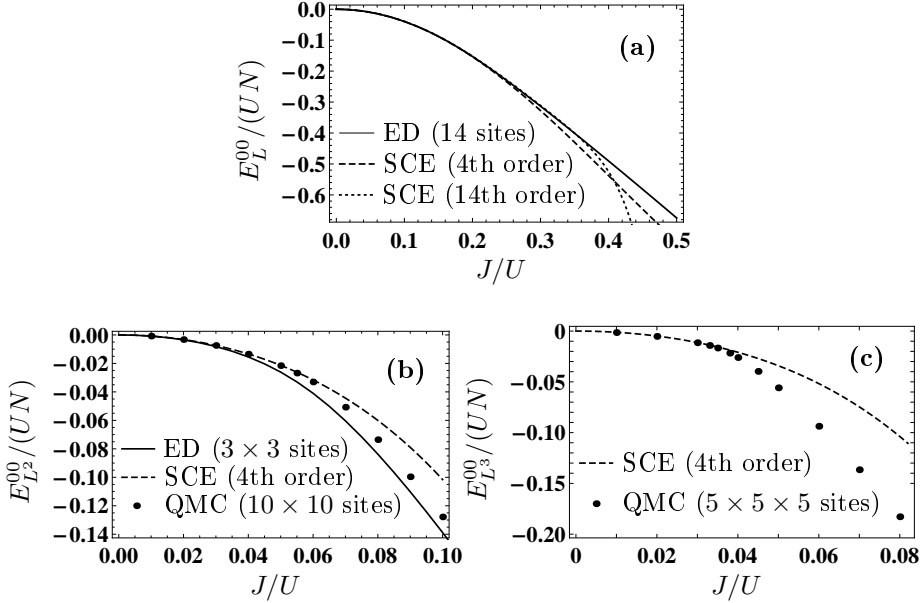}

\caption
{
Ground-state energy for unit filling in one (a), two (b), and three (c) dimensions.
Solid lines are the results of exact diagonalization for $14$ sites in 1D~(a) and for $3\times 3$ sites in 2D~(b).
Dots are QMC data for $10\times 10$ sites in 2D~(b) and $5\times 5\times 5$ sites in 3D~(c).
Dashed lines in all panels are the results of the strong-coupling expansion up to the fourth order [Eq.~(\ref{energy4order})].
Dotted line in (a) is obtained by the strong-coupling expansion up to the 14th order [Eq.~(\ref{energy14order})].
}
\label{onsiteenergy1D}
\end{figure}

In Figs.~\ref{onsiteenergy1D}, we compare exact numerical data for the ground-state energy in different dimensions
with the corresponding results of the strong-coupling perturbation theory~(\ref{energy4order}),~(\ref{energy14order}).
We observe that the $4$th order SCE gives already a satisfactory description up to the critical point of the SF-MI
phase transition in all dimensions.
We have also compared the results of exact diagonalization for $14$ sites in 1D with the DMRG-calculations
for larger systems and did not find any noticeable discrepancy which shows that finite-size effects (at least for local quantities)
are negligible for systems of this size.

The comparison of the ground-state energy in two dimensions calculated by exact diagonalization for lattices of $3\times3$ sites
with QMC data for larger lattices reveals a contribution of the finite-size effects
Note that the numerical values for a lattice of $10\times10$ sites are
better described by the SCE in the thermodynamic limit than those obtained for $3\times3$ sites.
We also compared the results of exact diagonalization and QMC for $3\times3$ sites and found perfect agreement
not only for the ground-state energy but also for other quantities discussed below.

In three dimensions, exact diagonalization does not make much sense due to strong finite-size effects and we have to rely on QMC.
The numerical data obtained for the lattice of $5\times5\times5$ sites are in a good agreement
with the SCE near and below the critical point.

\begin{figure}[t]

\centering

\stepcounter{nfig}
\includegraphics[page=\value{nfig}]{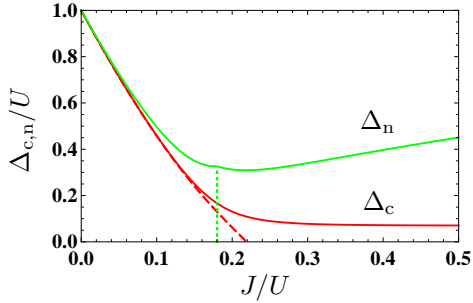}

\caption
{
Energies of neutral (red lines) and charged (green line) excitations in a one dimensional lattice for unit filling.
Solid lines - exact diagonalization for $L=13$, red dashed line - strong-coupling expansion of the 6th order
[Eqs.~(\ref{epk6}),~(\ref{ehk6}) with $K=0$].
Vertical dotted line marks the point, where $E_N^{01}=E_N^{11}$.
(For interpretation of the references to colour in this figure legend, the reader is referred to the web version of this article.)
}
\label{Eexc_cn_1D}
\end{figure}

The energies of neutral and charge excitations in a one-dimensional lattice in the case of unit filling are plotted in Fig.~\ref{Eexc_cn_1D}.
As it was discussed in section~\ref{sec-ph-excitations}, $\Delta_{\rm n}$ and $\Delta_{\rm c}$ coincide at least in the second-order
of the strong-coupling expansion. Numerical results presented in Fig.~\ref{Eexc_cn_1D} confirm this prediction at small values of $J/U$
but show that $\Delta_{\rm n}$ and $\Delta_{\rm c}$ become different for larger $J/U$.
In the thermodynamic limit, both quantities are expected to vanish above the critical point $(J/U)_{\rm c}$.
However, in a finite system, $\Delta_{\rm c}$ remains almost constant for increasing $J/U$ and $\Delta_{\rm n}$ even grows.
In addition, the latter has a point of nonanalyticity marked by a vertical dotted line in Fig.~\ref{Eexc_cn_1D}.
For $J/U$ below the point of nonanaliticity the lowest excited state of the system is at $K_0=0$ and above this point it is at $K=K_{\pm1}$,
i.e., we jump from one excitation branch to the other [see Fig.~\ref{sp_01}(b,d)].
The growth of $\Delta_{\rm n}$ with $J$ is a finite-size effect which can be understood looking at the limit $L\to\infty$
in Eq.~(\ref{Emin}) for the ideal Bose gas.

\subsection{Particle-number distribution}

\begin{figure}[t]

\centering

\stepcounter{nfig}
\includegraphics[page=\value{nfig}]{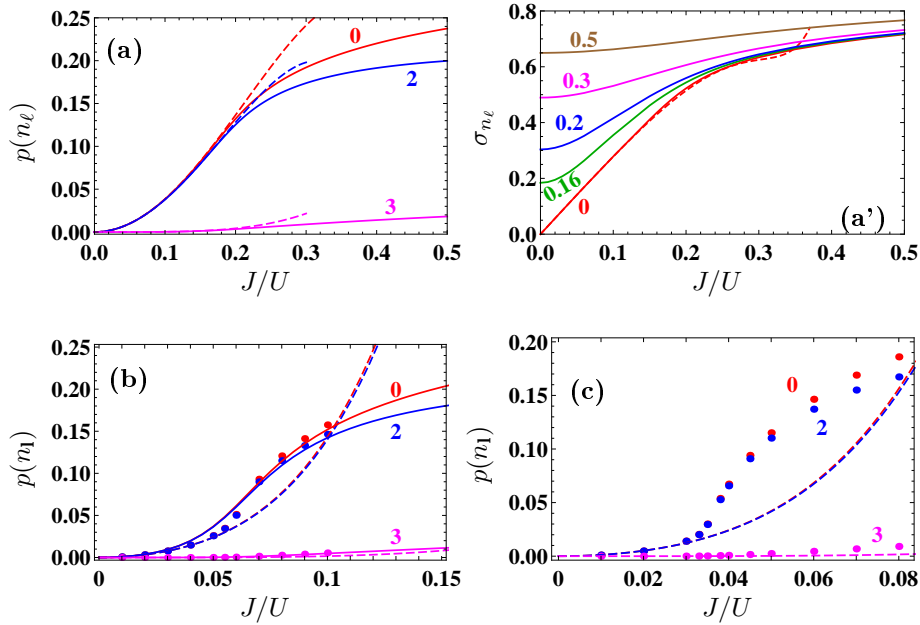}

\caption
{
(color online)
Probabilities of the occupation numbers $n_{\bf l}=0$ (red), $2$ (blue), $3$ (magenta)
of bosons at a lattice site in the case of unit filling in one~(a), two~(b), and three~(c) dimensions at zero temperature.
Solid lines are the results of exact diagonalization for $14$ sites in 1D~(a) and for $3\times 3$ sites in 2D~(b).
Dots are QMC data for $10\times 10$ sites in 2D~(b) and $5\times 5\times 5$ sites in 3D~(c),
see also Fig.~1 in Ref.~\cite{CKPS2007} and Fig.~3 in Ref.~\cite{YMO11}.
Dashed lines in panels (a,b,c) are the results of the strong-coupling expansion up to the fourth order [Eq.~(\ref{energy4order})].
(a')~Particle-number fluctuations in a one-dimensional lattice of $11$ sites at different temperatures:
$k_{\rm B}T/U=0$~(red), $0.16$~(green), $0.2$~(blue), $0.3$~(magenta), $0.5$~(brown).
Red dashed line is the result of SCE at $T=0$ up to $(J/U)^{13}$~\cite{DZ06}.
}
\label{pn}
\end{figure}

In the present section, we discuss the probabilities $p(n_{\bf l}=n)$
to have $n$ atoms at a lattice site
in the case of unit filling, $N=L^d$.
Due to translational invariance, $p(n_{\bf l}=n)$ does not depend on the site index.
At zero temperature and in the limit $J=0$, we have $p(n_{\bf l}=n)=\delta_{n,1}$,
and in the opposite limit, $U=0$, the probabilities are given by the binomial distribution~(\ref{binomial}).
For finite $J$ and $U$, exact numerical results are shown in Fig.~\ref{pn}.
One can clearly see that the probabilities to have three particles or more at one lattice site are very small
for any $J/U$ in the case of unit filling.
For small values of $J/U$, $p(n_{\bf l}=0)$ and $p(n_{\bf l}=2)$ are almost equal to each other,
no matter what the system dimensionality is, which is a manifestation of the particle-hole symmetry.
The exact numerical data are in good agreement with the SCE.
Below $(J/U)_{\rm c}$,
the probabilities of the occupation numbers different from one become smaller if the dimensionality is increased (see Fig.~\ref{pn}).
In Fig.~\ref{pn}(b) one can see that the finite-size effects lead to the larger values of $p(n_{\bf l}=0)$ and $p(n_{\bf l}=2)$
below the critical point and to the lower values of those above $(J/U)_{\rm c}$
compared to the thermodynamic limit.

The particle-number fluctuations
$
\sigma_{n_\ell}
=
\sqrt
{
 \langle
     \hat n_\ell^2
 \rangle
 -
 \langle
     \hat n_\ell
 \rangle^2
}
$
(standard deviation)
at finite temperature calculated by exact diagonalization in one dimension are shown in Fig.~\ref{pn}(a').
Comparison with the zero-temperature result for the same system size
indicates that temperature has stronger influence at smaller values of $J/U$.

\subsection{\label{sec-obdm}One-body density matrix}

Now we turn to the discussion of the two-point correlation functions.
In a homogeneous lattice under periodic boundary conditions they depend on ${\bf l}_{2} - {\bf l}_{1}$.
First, we consider the one-body density matrix with the entries
$
F_a
\left(
    {\bf l}_1, {\bf l}_2
\right)
=
\langle
     \hat a_{{\bf l}_1}^\dagger
     \hat a_{{\bf l}_2}^{\phantom{\dagger}}
\rangle
$.
As one can see in Fig.~\ref{obdm}, it monotonically decreases with $s$ and increases with $J/U$
approaching the asymptotic limit of the ideal gas~(\ref{F_a-ibg}) for $J/U\to\infty$,
and the results of numerical calculations for small $J/U$ are in good quantitative agreement with the SCE.
Higher-order SCE in one dimension~\cite{DZ06} allows to extend the region of validity of the perturbative calculations.

In the MI phase, the presence of the energy gap in the excitation spectrum leads to the exponential asymptotics
of the two-point correlation functions in the ground state at large distances~\cite{HK06}.
This general statement is valid in any finite dimension and the corresponding length scale is inversely
proportional to the energy gap. This type of behavior is demonstrated in Fig.~\ref{obdm}a' for one-dimensional lattices at small values of $J/U$.
In higher dimensions, no exact results have been reported in the literature for the MI at $T=0$.
Nevertheless, high-order SCE~\cite{THHE09} as well as
the quantum rotor approach supplemented by the Bogoliubov theory~\cite{ZK11}
reveal exponential decay of $F_a$ for small $J/U$ in two and three dimensions.
In Refs.~\cite{THHE09,ZK11} it was also shown that the correlations along the lattice diagonals
are weaker than those parallel to the lattice axes in agreement with Eqs.~(\ref{Fa1})-(\ref{Fa3})
and numerical data in Fig.~\ref{obdm}b.

In the SF phase, the two-point correlations decay according to a power law.
In a one-dimensional lattice, the one-body density matrix does not show off-diagonal long-range order and its long-distance
asymptotics~\cite{FWGF89,Giamarchi04}
\begin{eqnarray}
F_a(s)
\approx
A
s^{-{\cal K}_{\rm TL}/2}
\;,
\end{eqnarray}
is determined by the Tomonaga-Luttinger parameter~(\ref{KTL}).
The prefactor $A$ in the case of hard-core bosons coincides with the coefficient $C$ in Eq.~(\ref{ada-asympt})
and for the ideal Bose gas $A=\langle\hat n_\ell\rangle$.
This sort of behavior demonstrated in Fig.~\ref{obdm}(a') was studied in details in Refs.~\cite{ZD08} using the TEBD algorithm.

In higher dimensions, $F_a$ takes constant values at large distances in the SF phase,
which is a manifestation of the Bose-Einstein condensation.
This was confirmed by QMC calculations for two-dimensional systems in discrete~\cite{KKKPS00,YMO11}
and continuum~\cite{AK11} models.

The finite-size effects can be well controlled by comparison with the SCE. This is demonstrated in Fig.~\ref{obdm}(b)
for two-dimensional systems, where one can see that exact diagonalization for the lattice of $3\times 3$ sites overestimates the strength of correlations,
while QMC data obtained for $10\times 10$ sites show a very good agreement with the SCE.
In three dimensions, QMC data for $F_a(1)$ obtained for the lattice of $5^3$ sites are not affected by the finite-size effects.

\begin{figure}

\centering

\stepcounter{nfig}
\includegraphics[page=\value{nfig}]{figures.pdf}

\caption
{
(color online)
Correlation function $F_a(s)$ at zero temperature
in one~(a,a'), two~(b), and three~{(c)} dimensions for unit filling.
{(a)}
Solid lines are obtained by the DMRG calculations in 1D for the lattice of $128$ sites
and $s=1$~(red), $2$~(green), $3$~(blue), $4$~(magenta).
Dotted lines are the results of the strong-coupling expansion up to the $13$th order~\cite{DZ06}.
{(b)}
Solid lines are the results of exact diagonalization in 2D for the lattice of $3\times3$ sites
and $s=1$~(red), $\sqrt{2}$~(green).
Dots are QMC data for the lattice of $10\times10$ sites and $s=1$.
{(c)}
Dots are the results of QMC calculations for the lattice of $5\times5\times5$ sites and $s=1$.
Dashed lines in panels (a,b,c) show corresponding results of the strong-coupling expansion up to the third order 
in $J/U$, see Eqs.~(\ref{Fa1})-(\ref{Fa3}).
{(a')}:
Dependence of $F_a$ on $s$ in 1D for $J/U=0.2,0.25,0.3,0.4,0.5$ from bottom to the top
obtained by the DMRG calculations, see also Refs.~\cite{KSDZ04,ZD08,EFGMKAL12}.
}
\label{obdm}
\end{figure}

\begin{figure}

\centering

\stepcounter{nfig}
\includegraphics[page=\value{nfig}]{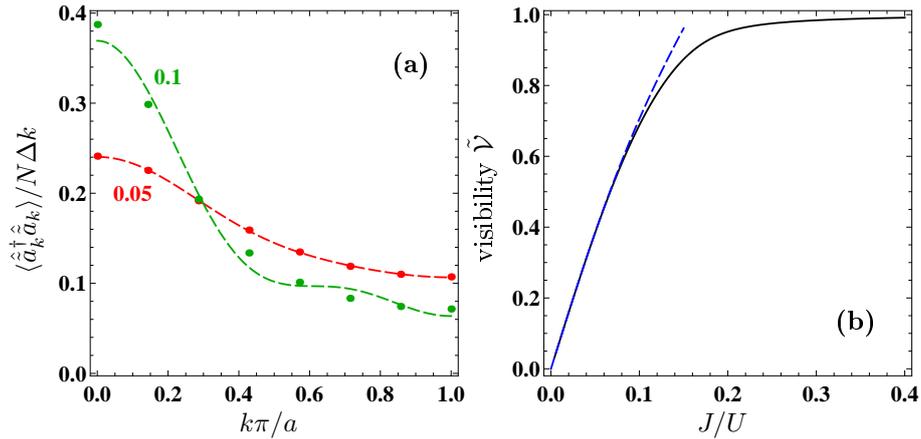}

\caption
{
(color online)
{(a)}
Quasi-momentum distribution at zero temperature in a one-dimensional chain of $L=14$ sites with periodic boundary conditions
in the case of unit filling for $J/U=0.05$~(red), $0.1$~(green).
Dots - exact diagonalization, dashed lines - results of the strong-coupling expansion up to the third order in $J/U$,
see Eqs.~(\ref{md-sce}),~(\ref{md-sce-1D}).
{(b)} Visibility of the interference pattern for the same system obtained by exact diagonalization (solid line)
and strong-coupling expansion (dashed line).
}
\label{md1D}
\end{figure}

The quasi-momentum distribution $\tilde P(k)$ in the ground state calculated for a one-dimensional lattice in the case of unit filling
by exact diagonalization is shown in Fig.~\ref{md1D}. In general, it is periodic and even function of $k$ which monotonically decreases
in the interval $k\in[0,\pi/a]$. For small $J/U$, the distribution is rather flat and this situation is well described by the perturbative expansion
in $J/U$. With the increase of $J/U$, the maximum at $k=0$ grows and becomes very sharp for large $J/U$.
Quantitatively, the form of the quasi-momentum distribution is characterized by visibility defined in analogy to Eq.~(\ref{visibility}) as
\begin{equation}
\tilde{\cal V}
=
\frac
{
  \tilde P(0) - \tilde P(\pi/a)
}
{
  \tilde P(0) + \tilde P(\pi/a)
}
\;.
\end{equation}
It is a monotonic function of $J/U$ which shows a linear dependence for small $J/U$ and becomes close to unity near the quantum critical point.
Quantum Monte Carlo calculations show that this feature remains preserved in higher dimensions and the measurement of visibility can be
used in order to detect not only the quantum critical point but also the critical temperature of the transition from the superfluid into
the normal phase~\cite{KZKT08}.

\subsection{\label{sec-ddcorr}Higher-order correlation functions}

\begin{figure}

\centering

\stepcounter{nfig}
\includegraphics[page=\value{nfig}]{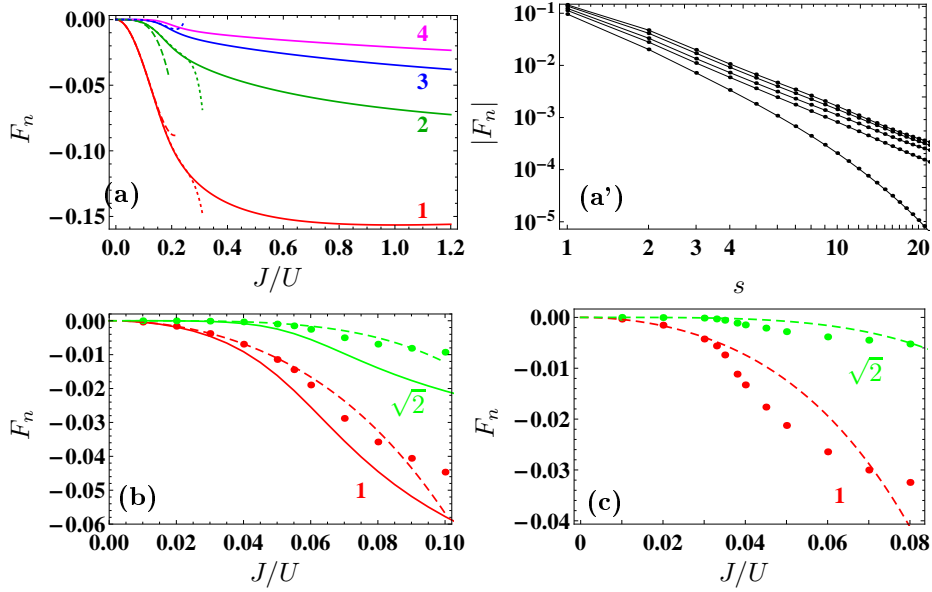}

\caption
{
(color online)
Particle-number correlation function $F_{n}(s)$ at zero temperature
in one {(a,a')}, two {(b)}, and three {(c)} dimensions for unit filling.
{(a)}
Solid lines are obtained by exact diagonalization in 1D for the lattice of $14$ sites
and $s=1$~(red), $2$~(green), $3$~(blue), $4$~(magenta).
Dotted lines are the results of the strong-coupling expansion up to the $14$th order~\cite{DZ06}.
{(b)}
Solid lines are the results of exact diagonalization in 2D for the lattice of $3\times3$ sites
and $s=1$~(red), $\sqrt{2}$~(green).
Dots are QMC data for the lattice of $10\times10$ sites and $s=1$~(red), $\sqrt{2}$~(green).
{(c)}
Dots are the results of QMC calculations for the lattice of $5\times5\times5$ sites and $s=1$~(red), $\sqrt{2}$~(green).
Dashed lines in panels (a,b,c) show corresponding results of the strong-coupling expansion up to the fourth order in $J/U$,
see Eqs.~(\ref{dd1-sce})-(\ref{ddsqrt2-sce}).
{(a')}:
Dependence of $|F_n|$ on $s$ in 1D for $J/U=0.2,0.25,0.3,0.4,0.5$ from bottom to the top obtained by DMRG calculations for $128$ sites.
}
\label{ddcorr}
\end{figure}

\begin{figure}
\centering

\stepcounter{nfig}
\includegraphics[page=\value{nfig}]{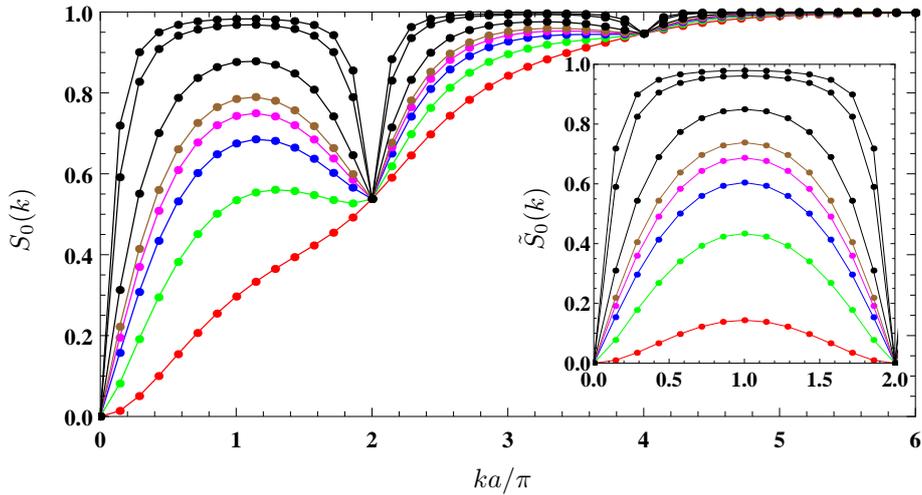}

\caption
{
(color online)
Static structure factor at $T=0$ in a one-dimensional homogeneous lattice of $L=14$ sites with periodic boundary conditions
calculated by exact diagonalization for unit filling and for $J/U=0.1,0.2,0.3,0.4,0.5,1,5,10$ (from bottom to the top).
$\tilde S_0(k_q)$ is defined by Eq.~(\ref{S0discrete}).
$S_0(k_q)$ is obtained from $\tilde S_0(k_q)$ according to Eq.~(\ref{S0lBB}) using $G_0(k)$ for $V_0=10\,E_{\rm R}$.
Lines are guides to the eye.
}
\label{ssf1D}
\end{figure}

\begin{figure}
\centering

\stepcounter{nfig}
\includegraphics[page=\value{nfig}]{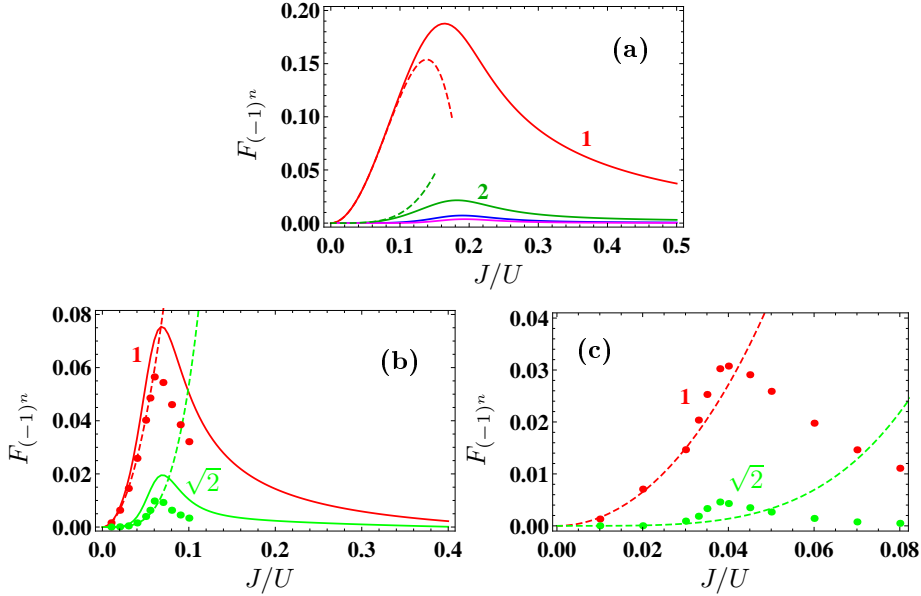}

\caption
{
Parity correlation function $F_{(-1)^n}(s)$ at zero temperature
in one {(a)}, two {(b)}, and three {(c)} dimensions for unit filling.
{(a)}
Solid lines are the results of exact diagonalization in 1D for the lattice of $14$ sites
and $s=1$~(red), $2$~(green), $3$~(blue), $4$~(magenta).
{(b)}
Solid lines are the results of exact diagonalization in 2D for the lattice of $3\times3$ sites
and $s=1$~(red), $\sqrt{2}$~(green).
Dots are QMC data for the lattice of $10\times10$ sites and $s=1$~(red), $\sqrt{2}$~(green).
{(c)}
Dots are the results of QMC calculations for the lattice of $5\times5\times5$ sites and $s=1$~(red), $\sqrt{2}$~(green).
Dashed lines in all panels show corresponding results of the strong-coupling expansion up to the fourth order 
in $J/U$, see Eqs.~(\ref{par1-sce})-(\ref{parsqrt2-sce}).
(For interpretation of the references to colour in this figure legend, the reader is referred to the web version of this article.)
}
\label{parcorr}
\end{figure}

In Figs.~\ref{ddcorr},~\ref{parcorr}, we present exact numerical results for the particle-number and parity correlation functions
$F_n$ and $F_{(-1)^n}$ defined by Eqs.~(\ref{nncorr}),~(\ref{parcorr-def}).
The correlations grow with the increase of $J/U$ up to the maximal value and then decrease approaching the limit of the ideal Bose gas.
Similarly to the one-body density matrix, the second-order correlation functions decrease exponentially with the distance in the MI phase
which is demonstrated in Fig.~\ref{ddcorr}(a') for $F_n$ in one dimension.
This is consistent with the results of the SCE (see Eqs.~(\ref{dd1-sce})-(\ref{parsqrt2-sce}) and Refs.~\cite{DZ06,THHE09})
and remains valid in higher dimensions.
The second-order correlation functions decay also faster along the lattice diagonals than along the axes.

The particle-number correlation function $F_n$ is in general negative because due to the conservation law
the increase of the particle number on one lattice site should be compensated by the corresponding decrease on the other site.
In one dimension, the large-distance behavior of the particle-number correlation function in the SF phase is given by~\cite{Giamarchi04}
\begin{equation}
\label{FnLL}
F_n(s)
\approx
-
\frac{1}{2{\cal K}_{\rm TL} \pi^2 s^2}
+
\frac
{
 A
 \langle
      \hat n_\ell
 \rangle^2
 \cos
 \left(
     2\pi
     \langle
          \hat n_\ell
     \rangle
     s
 \right)
}
{
 \left(
     \langle
          \hat n_\ell
     \rangle
     s
 \right)^{2/{\cal K}_{\rm TL}}
}
\;,
\end{equation}
where the second term does not vanish only for incommensurate fillings $\langle\hat n_\ell\rangle$.
From Eqs.~(\ref{Fn-HCB}),~(\ref{OBDM-Fermi}) it follows that the particle-number correlation function of hard-core bosons
in the thermodynamic limit and for $\ell'\ne\ell$ has exactly the form~(\ref{FnLL}) with ${\cal K}_{\rm TL}=1$ and $A=1/(2\pi^2)$.

Static structure factor at $T=0$ in a one-dimensional lattice in the case of unit filling is shown in Fig.~\ref{ssf1D}.
If $J/U$ is sufficiently small, $S_0(k)$ as well as $\tilde S_0(k)$ are quadratic functions of $k$ for small $k$.
This can be clearly seen in spite of rather coarse discretization in the momentum space.
For larger $J/U$, $S_0(k)$ and $\tilde S_0(k)$ become linear functions of $k$ for small $k$ in agreement with Eq.~(\ref{S0small-k}).
This indicates an advent of the sound mode making a dominant contribution into the structure factor.
In this regime, with the aid of Eqs.~(\ref{rhos_cs}),~(\ref{KTL}) we can rewrite Eq.~(\ref{S0small-k}) in the form
\begin{equation}
\lim_{k\to0}
\tilde S_0(k)
=
\frac
{k}
{
 2 k_{\rm F}
 {\cal K}_{\rm TL}
}
\;,
\end{equation}
where $k_{\rm F}$ is defined by Eq.~(\ref{kF}),
which shows that the behavior of the structure factor in the limit of small $k$ gives an access to the Tomonaga-Luttinger parameter.
Using the condition ${\cal K}_{\rm TL}=1/2$ (see section~\ref{SecTCF}), one can determine the critical value of $J/U$
for the transition from the superfluid into the Mott insulator.
Calculations of $\tilde S_0(k)$ by exact diagonalization for chains up to $L=14$ sites give $(J/U)_{\rm c}\approx 0.28$~\cite{KALM15}
in perfect agreement with other calculations based on the exact diagonalization listed in Table~\ref{Jc1D}.

The parity correlations are of the same order of magnitude as the particle-number correlations.
However, the former possess a more narrow maximum near the critical point.
In one dimension, this maximum is below $(J/U)_{\rm c}$
and captured by the SCE of the 4th order for the nearest-neighboring sites.
In two and three dimensions, the maximum is above $(J/U)_{\rm c}$ and, therefore,
out of reach of the perturbation theory in $J/U$.
Exact results for one and two dimensions presented in Fig.~\ref{parcorr}(a,b) are very similar to that of Ref.~\cite{ECFWSGMBPBK11},
where one can also find a comparison with the experimental data obtained at finite temperatures
in the presence of the harmonic confinement.
Positive values of $F_s$ are due to the fact that a change of the particle number on site ${\bf l}_1$ compensated
by an opposite change on site ${\bf l}_2$ leads to the same change of the parity on both sites.

\section{\label{sec-MFT}Mean-field theory}

Mean-field theory of lattice bosons relies on the concept of spontaneous breaking of U(1) symmetry.
According to the Mermin-Wagner-Hohenberg theorem~\cite{MW66,H67} it is valid in two-dimensional systems at zero temperature
and in higher dimensions at arbitrary temperature.
It gives an exact solution in the limit $d\to\infty$~\cite{RK91,Zwerger03,Zwerger04}, implying that $J\to0$ such that $dJ$ is finite,
as well as in the case $J=0$ and arbitrary $d$.

\subsection{Decoupling approximation}

The mean-field theory is based on the assumption that the second-order fluctuations of the bosonic creation and annihilation operators
are negligible, i.e.,
$
\left(
    \hat a_{{\bf l}_1}^\dagger
    -
    \langle
       \hat a_{{\bf l}_1}^\dagger
    \rangle
\right)
\left(
    \hat a_{{\bf l}_2}^{\phantom \dagger}
    -
    \langle
       \hat a_{{\bf l}_2}^{\phantom \dagger}
    \rangle
\right)
\approx 0
$
for ${\bf l}_1\ne{\bf l}_2$.
This leads to the decoupling approximation of the hopping terms~\cite{FWGF89,SKPR93,Sachdev,OSS01}
\begin{equation}
\label{dec_approx}
\hat a_{{\bf l}_1}^\dagger
\hat a_{{\bf l}_2}^{\phantom \dagger}
\approx
\langle
\hat a_{{\bf l}_1}^\dagger
\rangle
\hat a_{{\bf l}_2}^{\phantom \dagger}
+
\hat a_{{\bf l}_1}^\dagger
\langle
\hat a_{{\bf l}_2}^{\phantom \dagger}
\rangle
-
\langle
\hat a_{{\bf l}_1}^\dagger
\rangle
\langle
\hat a_{{\bf l}_2}^{\phantom \dagger}
\rangle
\;.
\end{equation}
Then the Bose-Hubbard Hamiltonian~(\ref{HBH}) becomes a sum of local operators
\begin{eqnarray}
\label{HBHMF}
\hat H_{\rm BH}
\approx
\hat H_{\rm BH}^{\rm MF}
=
-J
\sum_{\bf l}
\sum_{\nu=1}^d
\left[
    \hat a_{\bf l}^{\phantom{\dagger}}
    \left(
        \psi_{{\bf l}-{\bf e}_\nu}^*
        +
        \psi_{{\bf l}+{\bf e}_\nu}^*
    \right)
    -
    \psi_{\bf l}^{\phantom{*}}
    \psi_{{\bf l}+{\bf e}_\nu}^*
    +
    {\rm h.c.}
\right]
+
\frac{U}{2}
\sum_{\bf l}
\hat a^\dagger_{\bf l}
\hat a^\dagger_{\bf l}
\hat a_{\bf l}^{\phantom{\dagger}}
\hat a_{\bf l}^{\phantom{\dagger}}
\;,
\end{eqnarray}
where $\psi_{\bf l}=\langle \hat a_{\bf l}\rangle$ are c-numbers.
The neighboring lattice sites in the Hamiltonian~(\ref{HBHMF}) are coupled only via the expectation values of the creation and annihilation operators.
Thus, the mean-field theory neglects quantum correlations between different sites and
the states of the Hamiltonian~(\ref{HBHMF}) are tensor products of the local states
\begin{equation}
\label{state}
|\Phi\rangle
=
\bigotimes_{\bf l}
|s_{\bf l}\rangle
\;,\quad
|s_{\bf l}\rangle
=
\sum_{n=0}^\infty
c_{{\bf l}n}
|n\rangle_{\bf l}
\;,
\end{equation}
which is equivalent to the Gutzwiller ansatz~\cite{G63,RK91,KCB92}.
Here $|n\rangle_{\bf l}$ is the Fock state with $n$ atoms at site~${\bf l}$.
Normalization of the $|s_{\bf l}\rangle$ imposes
\begin{equation}
\sum_{n=0}^\infty
\left|
    c_{{\bf l}n}
\right|^2
=
1
\;.
\nonumber
\end{equation}
As it follows from the form of the state (\ref{state}), the Gutzwiller approximation
neglects quantum correlations between different lattice sites but takes into account on-site quantum fluctuations,
provided that $|s_{\bf l}\rangle$ is not a single Fock state.

The mean number of condensed atoms on a lattice site ${\bf l}$ in this approximation is given by $|\psi_{\bf l}|^2$, where
\begin{equation}
\label{psi}
\psi_{\bf l}
=
\langle \hat a_{\bf l}\rangle
=
\sum_{n=1}^\infty
c_{{\bf l},n-1}^* c_{{\bf l}n}
\sqrt{n}
\end{equation}
is the condensate order parameter.
One can easily show that $|\psi_{\bf l}|^2$ cannot be larger than the mean occupation number
\begin{equation}
\label{Nav}
\langle\hat n_{\bf l}\rangle
=
\sum_{n=1}^\infty
n
\left|
    c_{{\bf l}n}
\right|^2
\;.
\end{equation}

Minimization of the functional
\begin{displaymath}
i\hbar
\sum_{n=0}^\infty
(c_{{\bf l}n}^*
\partial_t c_{{\bf l}n}
-
c_{{\bf l}n}
\partial_t c^*_{{\bf l}n})
-
\langle \hat H_{\rm BH}^{\rm MF} \rangle
+
\mu N
\end{displaymath}
leads to the system of Gutzwiller equations (GE)~\cite{Z05,BPVB07}:
\begin{eqnarray}
\label{GEd}
i\hbar
\frac{d c_{{\bf l}n}}{dt}
&=&
\sum_{n'=0}^\infty
H_{\bf l}^{n n'}
c_{{\bf l}n'}
\;,
\\
H_{\bf l}^{n n'}
&=&
\left[
    \frac{U}{2}n(n-1)
    -
    \mu n
\right]
\delta_{n',n}
-
J
\sqrt{n'}
\delta_{n',n+1}
\sum_{\nu=1}^d
\left(
    \psi^*_{{\bf l}+{\bf e}_\nu}
    +
    \psi^*_{{\bf l}-{\bf e}_\nu}
\right)
-
J
\sqrt{n}
\delta_{n,n'+1}
\sum_{\nu=1}^d
\left(
    \psi_{{\bf l}+{\bf e}_\nu}
    +
    \psi_{{\bf l}-{\bf e}_\nu}
\right)
\;.
\nonumber
\end{eqnarray}
Note that these equations are invariant under transformation
$c_{{\bf l}n}\to (-1)^n c_{{\bf l}n}$.

Although the mean-field Hamiltonian~(\ref{HBHMF}) does not satisfy all fundamental commutation relations of the
original Bose-Hubbard Hamiltonian~(\ref{HBH}), the Gutzwiller approximation can be considered as {\it conserving}
because the expectation values of the total number of particles, total energy, and the quasi-momentum remain constant in time.
Equations~(\ref{GEd}) allow to study the properties of the ground state as well as the dynamics of excitations.

In the mean-field approximation, the one-body density matrix takes the form
\begin{equation}
\langle
   \hat a_{{\bf l}_1}^\dagger
   \hat a_{{\bf l}_2}^{\phantom{\dagger}}
\rangle
=
\psi_{{\bf l}_1}^*
\psi_{{\bf l}_2}^{\phantom{*}}
+
\delta_{{\bf l}_1{\bf l}_2}
\left(
    \langle\hat n_{{\bf l}_1}\rangle
    -
    \left|
        \psi_{{\bf l}_1}
    \right|^2
\right)
\;.
\end{equation}
In a homogeneous lattice, $\psi_{{\bf l}_1}\equiv\psi^{(0)}$, and the largest eigenvalue of the one-body density matrix in the thermodynamic limit
is $N_0=L^d\left|\psi^{(0)}\right|^2$.
Therefore, the condensate fraction $f_{\rm c}=\left|\psi^{(0)}\right|^2/\langle\hat n_{{\bf l}}\rangle$
which coincides with the superfluid stiffness $f_{\rm s}$.
This implies that Eq.~(\ref{fNs_hydro}) is valid in this case.
In an inhomogeneous lattice, one cannot write in general an explicit expression for $N_0$ and it has to be determined numerically.
This can be efficiently done by the iteration procedure~\cite{BBWH11}
\begin{equation}
N_0^{(i+1)}
=
N_0^{(i)}
\sum_{\bf l}
\frac
{
  \left|\psi_{\bf l}\right|^2
}
{
  N_0^{(i)}
  -
  \langle\hat n_{\bf l}\rangle
  +
  \left|\psi_{\bf l}\right|^2
}
\end{equation}
obtained from a rearrangement of the eigenvalue equation~(\ref{OBDM-evp}).

As it was shown in Ref.~\cite{KN2011}, in the weakly interacting regime, $U/J\ll 1$,
Eq.~(\ref{GEd}) can be transformed into the discrete Gross-Pitaevskii equation (DGPE)
\begin{eqnarray}
\label{DGPE}
i\hbar
\frac{d \psi_{\bf l}}{dt}
=
-
J
\sum_{\nu=1}^d
\left(
    \psi_{{\bf l}+{\bf e}_\alpha}
    +
    \psi_{{\bf l}-{\bf e}_\alpha}
\right)
-
\mu
\psi_{\bf l}
+
U
\left|
\psi_{\bf l}
\right|^2
\psi_{\bf l}
\end{eqnarray}
assuming that $|s_{\bf l}\rangle$ in Eq.~(\ref{state}) are Glauber coherent states (see also Ref.~\cite{BP08}).
Eq.~(\ref{DGPE}) describes a pure Bose-Einstein condensate in a discrete lattice model,
which implies $\langle\hat n_{\bf l}\rangle\approx\left|\psi_{\bf l}\right|^2$.

\subsection{\label{GS}Ground state}

In the homogeneous lattice, the ground state of the Hamiltonian~(\ref{HBHMF}) is described by the coefficients $c_{{\bf l}n}$
that do not depend on the site index ${\bf l}$. This corresponds to the stationary solution of Eqs.~(\ref{GEd}) of the form
\begin{equation}
c_{{\bf l}n}(t)
\equiv
c_n^{(0)}
\exp
\left(
    -i \omega_0 t
\right)
\;,
\end{equation}
\begin{equation}
\hbar\omega_0
=
- 4dJ
\left|
    \psi^{(0)}
\right|^2
+
\sum_{n=0}^\infty
\left[
    \frac{U}{2}n(n-1) - \mu n
\right]
\left|
c_n^{(0)}
\right|^2
\;.
\end{equation}
The explicit form of $c_n^{(0)}$ depends on the filling factor $\langle \hat n_{\bf l}\rangle$ and the ratio $J/U$.
If the latter is less than the critical value determined as~\cite{Sachdev}
\begin{equation}
\label{crit}
2d(J/U)_{\rm c}
=
\frac
{(n_0-\mu/U)(\mu/U-n_0+1)}
{1+\mu/U}
\end{equation}
for $n_0-1 < \mu/U < n_0$, the solution is given by
\begin{equation}
\label{gsmi}
c_n^{(0)}=\delta_{n,n_0}
\;,
\end{equation}
where $n_0$ is the smallest integer greater than $\mu/U$.
In this case the superfluid order parameter $\psi^{(0)}=\psi_{\bf l}$ defined by Eq.~(\ref{psi}) vanishes
and we have the MI phase with exactly $n_0$ particles at each lattice site.

Eq.~(\ref{crit}) determines the dependence of the critical ratio of $J/U$ on the chemical potential.
It can be also inverted to determine the dependence of the critical chemical potential on $J/U$,
which gives two solutions
\begin{equation}
\label{mu_pm}
\frac
{\mu_\pm(n_0)}
{U}
=
n_0
-
\frac{1}{2}
\left(
    1
    +
    \frac{2dJ}{U}
\right)
\pm
\frac{1}{2}
\sqrt
{
 1
 -
 \left(
     4 n_0 + 2
 \right)
 \frac{2dJ}{U}
 +
 \left(
     \frac{2dJ}{U}
 \right)^2
}
\;,
\end{equation}
that are real, provided that
\begin{equation}
\label{Jcmax}
2d(J/U)
<
2d(J/U)_{\rm c}^{\rm max}
=
\left(
    \sqrt{n_0+1}
    -
    \sqrt{n_0}
\right)^2
\;.
\end{equation}
For $J/U = (J/U)_{\rm c}^{\rm max}$, the two solutions merge into one
\begin{equation}
\frac
{\mu_\pm(n_0)}
{U}
=
\left(
    \frac{\mu}{U}
\right)_{\rm c}
=
\sqrt{n_0(n_0+1)}-1
\end{equation}
which describes the tips of the MI-lobes on the phase diagram.

If we expand Eq.~(\ref{mu_pm}) up to the third order in $2dJ/U$, we immediately observe that it coincides with the results of
the strong-coupling expansion given by Eqs.~(\ref{mupd}),~(\ref{muhd}) in the limit $Z=2d\to\infty$.
This confirms that the mean-field theory becomes exact in infinite dimensions.

\begin{figure}[t]

\centering

\stepcounter{nfig}
\includegraphics[page=\value{nfig}]{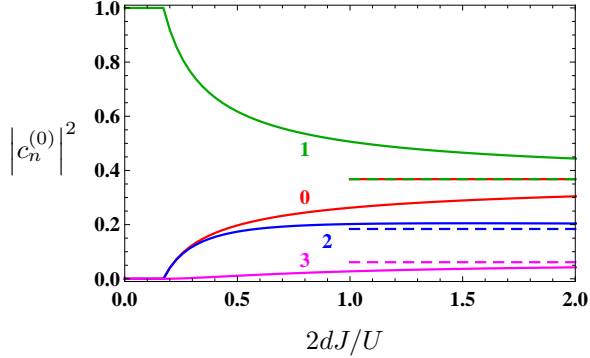}

\caption
{
(color online)
Probabilities of the occupation numbers $n=0$~(red), $1$~(green), $2$~(blue), $3$~(magenta)
in the ground state for the filling factor $\langle \hat n_{\bf l} \rangle=1$.
Solid lines are exact mean-field results and dashed lines show the asymptotic values $\left|c_n^{\rm coh}\right|^2$
in the limit $U\to0$ according to Eq.~(\ref{cs}).
}
\label{Pn-MF}
\end{figure}

If $J/U$ exceeds the critical value~(\ref{crit}), $c_n^{(0)}$ has a broad distribution leading to the fact that
the order parameter $\psi^{(0)}$ does not vanish which corresponds to the SF phase.
In this regime, exact analytical solution for $c_n^{(0)}$ can be obtained only in the limit $U=0$ and has the form
\begin{equation}
\label{cs}
c_n^{(0)}
\equiv
c^{\rm coh}_n
=
\exp
\left(
    -\left|\psi^{(0)}\right|^2/2
\right)
\frac
{\psi^{(0)n}}
{\sqrt{n!}}
\;,\quad
\left|\psi^{(0)}\right|^2
=
\langle \hat n_{\bf l} \rangle
\;,\quad
\mu = - 2dJ
\;.
\end{equation}
Eq.~(\ref{cs}) describes the coherent state and the corresponding particle-number distribution
$\left|c^{\rm coh}_n\right|^2$ coincides with the exact result~(\ref{Poisson}) for the ideal gas.

For finite $J$ and $U$ above the critical value~(\ref{crit}) exact coefficients $c_n^{(0)}$ can be obtained
by numerical diagonalization~\cite{SKPR93,OSS01} and the results for $\langle \hat n_{\bf l} \rangle=1$ are shown in Figs.~\ref{Pn-MF}.
As small values of $J/U$, the probabilities of having $n=0,2$ particles nearly coincide and the probabilities
of the occupation numbers larger than $3$ are negligible. The comparison with the coherent-state distribution~(\ref{cs})
shows that rather large values of $J/U$ are needed in order to approach this asymptotic limit with a good accuracy.

The quasi-momentum distribution~(\ref{md-def}) of the atoms in the ground state is given by (see, e.g., Ref.~\cite{EM03})
\begin{eqnarray}
\label{md-mf}
P_\infty({\bf k})
=
\frac
{
\left|
    \tilde W({\bf k})
\right|^2
}
{\langle\hat n_{\bf l}\rangle}
\left[
    \left(
        \frac{a}{2\pi}
    \right)^d
    \!
    \left(
    \!
    \langle\hat n_{\bf l}\rangle
    -
    \left|\psi^{(0)}\right|^2
    \right)
    +
    \left|\psi^{(0)}\right|^2
    \sum_{\bf n}
    \delta
    \!
    \left(
        {\bf k} - \frac{2\pi}{a}{\bf n}
    \right)
\right]
\;,
\end{eqnarray}
where the lattice is assumed to be infinite.
The whole quasi-momentum distribution is confined within the momentum distribution $\left|\tilde W({\bf k})\right|^2$ of a single site.
The first term in the brackets, $\langle\hat n_{\bf l}\rangle - \left|\psi^{(0)}\right|^2$, describes incoherent part of the system,
and the $\delta$-peaks at the vectors of the reciprocal lattice ${\bf k}=\frac{2\pi}{a}{\bf n}$ are clear signature of the Bose-Einstein condensation.
Similar structure of the quasi-momentum distribution, although with somewhat smeared maxima instead of sharp peaks,
was observed in the experiments~\cite{PPSFBCMMI01,GMEHB02,GWFMGB05,GTFSTWBPTCPS08,SPP07,SPP08,FSSTT09,TPGSBPST10}.

\begin{figure}[t]

\centering

\stepcounter{nfig}
\includegraphics[page=\value{nfig}]{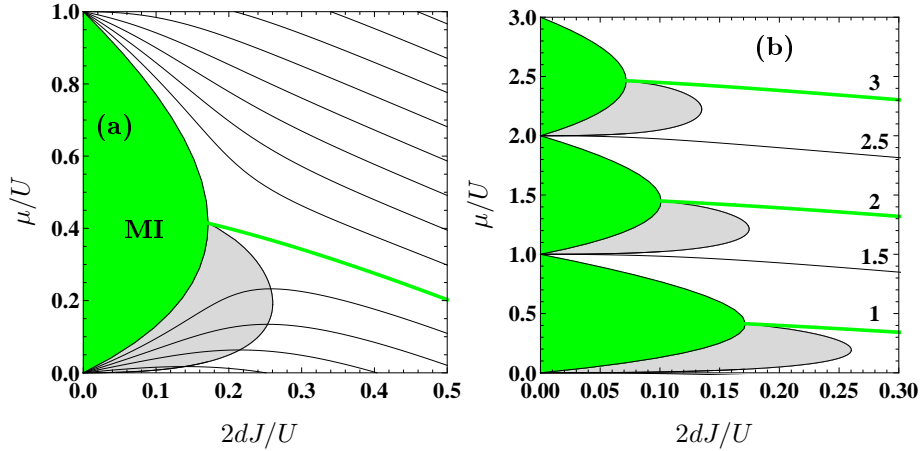}

\caption
{
Mean-field phase diagram.
{(a)}
Green area is the MI region with $n_0=1$ described by Eqs.~(\ref{crit}),~(\ref{mu_pm}).
Thick green line shows the dependence $\mu(J)$ in the superfluid phase for $\langle\hat{n}\rangle=1$.
Other lines show also $\mu(J)$ for non-integer fillings which are multiples of $0.1$ (see also Fig.~13 in Ref.~\cite{KWM00}).
In the gray areas, where $\mu'(J)>0$ at constant $\langle\hat{n}\rangle$ the superfluidity has a hole character.
In the rest part of the diagram, we have a particle SF.
{(b)}
Green areas show the first three MI zones ($n_0=1,2,3$).
The lines of constant $\langle\hat n_{\bf l}\rangle$ are labeled by the corresponding filling factors.
Gray areas are the regions of the hole superfluidity.
In the rest part of the diagram, we have a particle SF.
(For interpretation of the references to colour in this figure legend, the reader is referred to the web version of this article.)
}
\label{hs}
\end{figure}

As discussed in Refs.~\cite{WCHZ04,KLL10}, near the phase boundary one has to distinguish
between particle and hole superfluidity. For the hole SF, the function
$\mu(J)$ at constant filling factor $\langle\hat n\rangle$ has a positive derivative
$\mu'(J)$. This is only possible for fillings $n_0-0.5 < \langle\hat n\rangle < n_0$ as is demonstrated in Fig.~\ref{hs} showing the corresponding
hole SF regions. For the particle SF, on the other hand, $\mu'(J)<0$.
Consequently, far away from the phase boundary, superfluidity has always a particle character.
With the increase of the filling factor, the size of the regions of the hole superfluidity decreases together with the size of the MI-lobes.
As we will see in section~\ref{sec-QS}, the difference between the particle and hole
superfluidity plays an essential role for the character of the topological modes.

\subsection{\label{E}Excitations}

We consider small perturbation of the ground state
\begin{equation}
c_{{\bf l}n}(t)
=
\left[
    c_{n}^{(0)}
    +
    c_{{\bf l}n}^{(1)}(t)
    +
    \dots
\right]
\exp
\left(
    -i \omega_0 t
\right)
\;,
\end{equation}
where
\begin{equation}
\label{sol}
c_{{\bf l}n}^{(1)}(t)
=
    u_{{\bf k}n}
    e^{
        i
        \left(
            {\bf k}\cdot{\bf x}_{\bf l}
            -
            \omega_{\bf k} t
        \right)
      }
    +
    v_{{\bf k}n}^*
    e^{
        -i
        \left(
            {\bf k}\cdot{\bf x}_{\bf l}
            -
            \omega_{\bf k} t
        \right)
      }
\;.
\end{equation}
Substituting this expression into GE and keeping only linear terms
with respect to $u_{{\bf k}n}$ and $v_{{\bf k}n}$, we obtain the system of
linear equations~\cite{TML08},
\begin{equation}
\label{evpexc}
\hbar\omega_{\bf k}
\left(
    \begin{array}{c}
       \vec{u}_{\bf k}\\
       \vec{v}_{\bf k}
    \end{array}
\right)
=
\left(
    \begin{array}{cc}
        A_{\bf k} & B_{\bf k}\\
       -B_{\bf k} & -A_{\bf k}
    \end{array}
\right)
\left(
    \begin{array}{c}
       \vec{u}_{\bf k}\\
       \vec{v}_{\bf k}
    \end{array}
\right)
\;,
\end{equation}
where $\vec{u}_{\bf k}$ and $\vec{v}_{\bf k}$ are infinite-dimensional vectors
with the components $u_{{\bf k}n}$ and $v_{{\bf k}n}$ ($n=0,1,\dots$), respectively.
Matrix elements of $A_{\bf k}$ and $B_{\bf k}$ have the form
\begin{eqnarray}
A_{\bf k}^{nn'}
&=&
-J_{\bf 0}
\psi^{(0)}
\left(
    \sqrt{n'}\,
    \delta_{n',n+1}
    +
    \sqrt{n}\,
    \delta_{n,n'+1}
\right)
+
\left[
    \frac{U}{2}\,n(n-1)
    -
    \mu n
    -
    \hbar\omega_0
\right]
\delta_{n',n}
\nonumber\\
&-&
J_{\bf k}
\left[
    \sqrt{n+1}\,
    \sqrt{n'+1}\,
    c_{n+1}^{(0)}\,
    c_{n'+1}^{(0)}
    +
    \sqrt{n}\,
    \sqrt{n'}\,
    c_{n-1}^{(0)}\,
    c_{n'-1}^{(0)}
\right]
\;,
\nonumber\\
B_{\bf k}^{nn'}
&=&
-
J_{\bf k}
\left[
    \sqrt{n+1}\,
    \sqrt{n'}\,
    c_{n+1}^{(0)}\,
    c_{n'-1}^{(0)}
    +
    \sqrt{n}\,
    \sqrt{n'+1}\,
    c_{n-1}^{(0)}\,
    c_{n'+1}^{(0)}
\right]
\;,
\nonumber
\end{eqnarray}
where $J_{\bf k}=2dJ-\epsilon_{\bf k}$ with $\epsilon_{\bf k}$ being the energy of a free particle~(\ref{e1p-lat}).
This system is valid for both phases and generalizes the Bogoliubov--de~Gennes (BdG) equations
previously derived for coherent states~\cite{JK99,JA2000}.
The dependence on the vector ${\bf k}$ is determined by the variable
\begin{equation}
\label{x}
x
=
\left(
\frac{1}{d}\sum_{\nu=1}^d \sin^2 \frac{k_\nu a}{2}
\right)^{1/2}
\;,
\end{equation}
which varies from $0$ to $1$.
For small $|{\bf k}|$, $x\approx |{\bf k}|a/(2\sqrt{d})$.

The energy increase due to the perturbation is given by~\cite{PS2003}
\begin{equation}
\label{dE}
\Delta E
=
\hbar \omega_{\bf k}
\left(
    |\vec{u}_{\bf k}|^2-|\vec{v}_{\bf k}|^2
\right)
\;.
\end{equation}
Formally, Eqs.~(\ref{evpexc}) have solutions with positive and negative energies
$\pm\hbar\omega_{\bf k}$, which are equivalent because
Eqs.~(\ref{sol}),~(\ref{dE}) are invariant under the transformation
$\omega_{\bf k}\to -\omega_{\bf k}$,
${\bf k}\to -{\bf k}$,
$\vec{u}_{\bf k}\to\vec{v}^*_{\bf k}$,
$\vec{v}^*_{\bf k}\to\vec{u}_{\bf k}$,
so that only solutions with positive energies will be considered in the following.
The eigenvectors are chosen to follow the orthonormality relations
\begin{equation}
\vec{u}^*_{{\bf k},\lambda'}
\cdot
\vec{u}_{{\bf k},\lambda}
-
\vec{v}^*_{{\bf k},\lambda'}
\cdot
\vec{v}_{{\bf k},\lambda}
=
\delta_{\lambda,\lambda'}
\;.
\nonumber
\end{equation}

Perturbation~(\ref{sol}) creates plane waves of the order parameter
$\psi_{\bf l}(t)=\psi^{(0)}+\psi_{\bf l}^{(1)}(t)$, where
\begin{eqnarray}
\label{psiw}
\psi_{\bf l}^{(1)}(t)
&=&
{\cal U}_{\bf k}
e^{
    i
    \left(
        {\bf k}\cdot{\bf x}_{\bf l}
        -
        \omega_{\bf k} t
    \right)
  }
+{\cal V}^*_{\bf k}
e^{
    -i
    \left(
        {\bf k}\cdot{\bf x}_{\bf l}
        -
        \omega_{\bf k} t
    \right)
  }
\;,
\\
{\cal U}_{\bf k}
&=&
\sum_{n=0}^\infty
c_{n}^{(0)}
\left(
    u_{{\bf k},n+1}
    \sqrt{n+1}
    +
    v_{{\bf k},n-1}
    \sqrt{n}
\right)
\;,
\nonumber\\
{\cal V}_{\bf k}
&=&
\sum_{n=0}^\infty
    c_{n}^{(0)}
\left(
    u_{{\bf k},n-1}
    \sqrt{n}
    +
    v_{{\bf k},n+1}
    \sqrt{n+1}
\right)
\;.
\nonumber
\end{eqnarray}

The perturbations for the total density and the condensate density are given by
\begin{eqnarray}
\label{dw}
\langle
    \hat n_{\bf l}
\rangle(t)
&=&
\langle
    \hat n_{\bf l}
\rangle^{(0)}
+
\left[
{\cal A}_{\bf k}
e^{
    i
    \left(
        {\bf k}\cdot{\bf x}_{\bf l}
        -
        \omega_{\bf k} t
    \right)
  }
+
{\rm c.c.}
\right]
\;,
\\
{\cal A}_{\bf k}
&=&
\sum_{n=0}^\infty
c_{n}^{(0)}
\left(
    u_{{\bf k}n}
    +
    v_{{\bf k}n}
\right)
n
\;,
\nonumber
\end{eqnarray}
and
\begin{eqnarray}
\label{cw}
\left|
    \psi_{\bf l}(t)
\right|^2
&=&
\left|
{\psi^{(0)}}
\right|^2
+
\left[
{\cal B}_{\bf k}
e^{
    i
    \left(
        {\bf k}\cdot{\bf x}_{\bf l}
        -
        \omega_{\bf k} t
    \right)
  }
+
{\rm c.c.}
\right]
\;,
\\
{\cal B}_{\bf k}
&=&
\psi^{(0)}
\left(
    {\cal U}_{\bf k}
    +
    {\cal V}_{\bf k}
\right)
\;.
\nonumber
\end{eqnarray}
In what follows we consider the properties of the excitations in the MI and SF phases.

\subsubsection{Mott insulator}

For the MI phase, the coefficients $c_n^{(0)}$
have a simple analytical form~(\ref{gsmi}).
The eigenvalue problem for the infinite-dimensional matrices~(\ref{evpexc})
reduces to the diagonalization of two $2\times 2$-matrices that couple
$u_{{\bf k},n_0-1}$ to $v_{{\bf k},n_0+1}$ and
$u_{{\bf k},n_0+1}$ to $v_{{\bf k},n_0-1}$, respectively.
The lowest-energy excitation spectrum consists of two branches
\begin{equation}
\label{om}
\hbar\omega_{{\bf k}\pm}
=
\frac{1}{2}
\sqrt{
       U^2
       -
       J_{\bf k}
       U
       \left(
           4 n_0 + 2
       \right)
       +
       J_{\bf k}^2
     }
\pm
\left[U
\left(
    n_0-\frac{1}{2}
\right)
    -\frac{J_{\bf k}}{2}
    -\mu
\right]
\;.
\end{equation}
The same result was obtained using the Hubbard-Stratonovich transformation~\cite{OSS01}
and within the Schwinger-boson approach~\cite{HABB07}.

\begin{figure}[t]

\centering

\stepcounter{nfig}
\includegraphics[page=\value{nfig}]{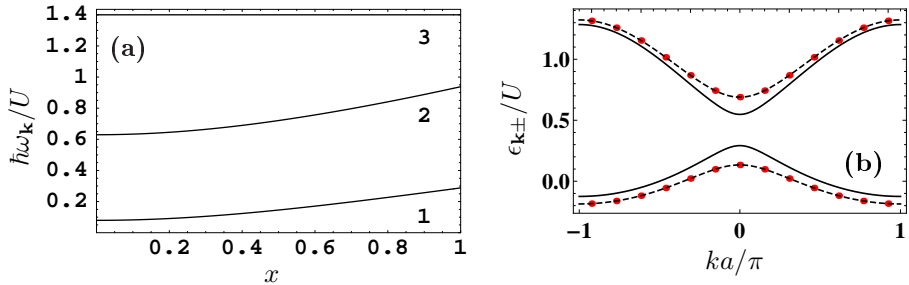}

\caption
{
(color online)
Excitation spectrum of the Mott insulator.
{(a)}
First three branches
$\hbar\omega_{{\bf k}-}$~(1),
$\hbar\omega_{{\bf k}+}$~(2) and
$\hbar\omega_{\lambda=0}$~(3)
of the excitations spectrum of the MI for $\mu/U=1.2$ and $2dJ/U=0.05$, which corresponds to $n_0=2$
[Eqs.~(\ref{om}),~(\ref{om_lambda})].
(Reproduced from Ref.~\cite{KN2011}, \copyright 2011 American Physical Society.)
{(b)}
Energies of particle and hole excitations~(\ref{epm}) calculated for $J/U=0.08$ in one-dimensional lattice in the case of unit filling.
Solid lines -- mean-field theory~[Eq.~(\ref{om})],
dashed lines -- strong-coupling expansion~[Eqs.~(\ref{epk6}),~(\ref{ehk6})],
dots -- exact diagonalization for $N=L=13$.
}
\label{excmi}
\end{figure}

These two branches are shown in Fig.~\ref{excmi} and display a gap.
The sign in front of $\mu$ in Eq.~(\ref{om}) is different for the two modes.
Therefore, the solutions labeled by '$+$' and '$-$' correspond to the situations, when one particle
is added into the system and removed, respectively. Hence, the eigenmodes described by Eq.~(\ref{om})
are called particle and hole excitations~\cite{EM99}.
If the total number of particles is conserved, the two kinds of excitations can be created only in pairs
and the corresponding energies are added.

Nonvanishing coefficients $u_{{\bf k}n}^{(\pm)}$, $v_{{\bf k}n}^{(\pm)}$ for the two modes are given by
\begin{eqnarray}
\label{uv-mi}
\left[
    v_{{\bf k},n_0-1}^{(+)}
\right]^2
=
\left[
    u_{{\bf k},n_0+1}^{(+)}
\right]^2
-1
&=&
\left[
    v_{{\bf k},n_0+1}^{(-)}
\right]^2
=
\left[
    u_{{\bf k},n_0-1}^{(-)}
\right]^2
-1
\\
&&
=
\frac{1}{2}
\left[
    \frac
    {U-2J_{\bf k}\left(n_0+\frac{1}{2}\right)}
    {
      \sqrt{U^2-4J_{\bf k}U\left(n_0+\frac{1}{2}\right)+J_{\bf k}^2}
    }
-1
\right]
\nonumber
\;.
\end{eqnarray}
According to Eqs.~(\ref{cw}),~(\ref{dw}), no density wave is created in the two modes.
However, the order parameter~(\ref{psiw}) does not vanish and takes the form
\begin{eqnarray}
\psi_{{\bf l}+}^{(1)}(t)
&=&
{\cal U}_{{\bf k}+}
e^{
    i
    \left(
        {\bf k}\cdot{\bf x}_{\bf l}
        -
        \omega_{{\bf k}+} t
    \right)
  }
\nonumber\\
\psi_{{\bf l}-}^{(1)}(t)
&=&
{\cal V}^*_{{\bf k}-}
e^{
    -i
    \left(
        {\bf k}\cdot{\bf x}_{\bf l}
        -
        \omega_{{\bf k}-} t
    \right)
  }
\;.
\end{eqnarray}
In the complex plane $({\rm Re}(\psi),{\rm Im}(\psi))$, this corresponds to the motion on the circles with radii
$\left|{\cal U}_{{\bf k}+}\right|$ and $\left|{\cal V}^*_{{\bf k}-}\right|$ around $\psi=0$.

Other solutions of Eq.~(\ref{evpexc}) are independent of ${\bf k}$ with the energies
\begin{equation}
\label{om_lambda}
\hbar\omega_{\lambda}
=
\frac{U}{2}
\left[
    \lambda(\lambda-1)
    -
    n_0(n_0-1)
\right]
    -
    \mu
    (\lambda-n_0)
\;,
\end{equation}
They are denoted by $\lambda$ which are non-negative integers different from $n_0,n_0\pm 1$.
Since $n_0$ is greater than $\mu/U$, the excitation energies are always positive.
The eigenvectors of these modes have the form
$u_{{\bf k}n\lambda}=\delta_{n,\lambda}$, $v_{{\bf k}n\lambda}=0$,
and the amplitudes of all the waves defined by Eqs.~(\ref{psiw}),~(\ref{dw}),~(\ref{cw}) vanish.

In Fig.~\ref{excmi}(b) we compare the energies of the particle and hole excitations
\begin{eqnarray}
\label{epm}
\epsilon_{{\bf k}\pm}
=
\pm
\left(
    \hbar\omega_{{\bf k}\pm}
    \pm
    \mu
\right)
\end{eqnarray}
determined by Eq.~(\ref{om}) with the analogous quantities
\begin{equation}
\epsilon_{{\bf k}\pm}
=
\pm
\left(
    E_{n_0 L^d\pm1}^{{\bf k}0}
    -
    E_{n_0 L^d}^{{\bf 0}0}
\right)
\end{equation}
calculated by exact diagonalization and strong-coupling expansion for $n_0=1$ and $d=1$.
We see that the strong-coupling expansion of the 6th order is in perfect agreement with the exact numerical results.
The mean-field theory based on the Gutzwiller ansatz gives qualitatively correct predictions but underestimates
the gap for the particle-hole excitations.

The boundary between the SF and MI phases is determined from the disappearance of the gap in the excitation spectrum, i.e.,
when $\omega_{{\bf 0}-}=0$.
Under this condition, we recover the critical ratio~(\ref{crit}).
For $J/U>(J/U)_{\rm c}$, the lowest frequency $\omega_{{\bf 0}-}$ in Eq.~(\ref{om}) 
becomes negative leading to a negative expression for Eq.~(\ref{dE}),
so that the Mott-phase solution (\ref{gsmi}) does not correspond to the ground state anymore.

The excitation spectrum has peculiar features on the boundary between the MI and SF.
For $(J/U)_{\rm c}=(J/U)_{\rm c}^{\rm max}$, the excitation energies~(\ref{om}) can be rewritten as
\begin{equation}
\hbar\omega_{{\bf k}\pm}
=
\left[
    \sqrt{n_0(n_0+1)}
    \,
    U
    \epsilon_{\bf k}
    +
    \frac{\epsilon^2_{\bf k}}{4}
\right]^{1/2}
\pm
\frac{\epsilon_{\bf k}}{2}
\;.
\end{equation}
For small $|{\bf k}|$, the two branches are degenerate and have linear dependence
$\omega_{{\bf k}\pm}=c_{\rm s}^{\rm tip}|{\bf k}|$ with the sound velocity
\begin{equation}
\label{cs0}
c_{\rm s}^{\rm tip}
=
\frac{U}{\hbar\sqrt{d}}
\,
\frac
{\sqrt{n_0+1}-\sqrt{n_0}}
{\sqrt{2}}
\left[
    n_0(n_0+1)
\right]^{1/4}
\end{equation}
expressed in units of the number of sites per second.
For other points on the boundary, i.e., $(J/U)_{\rm c} < (J/U)_{\rm c}^{\rm max}$,
no degeneracy appears and the sound velocity vanishes leading to
the quadratic dispersion $\omega_{{\bf k}\pm}\sim {\bf k}^2$ for small $|{\bf k}|$.

From the expression for the excitation energies~(\ref{om}) one can obtain the values of the mean-field
critical exponents for the MI-SF phase transition. According to the scaling theory~\cite{FWGF89},
the energy gap should be proportional to $|t-t_{\rm c}|^{z\nu}$, where $t$ is a control parameter which
approaches its critical value $t_{\rm c}$, $z$ is the dynamical critical exponent and $\nu$ is the critical exponent
of the correlation length.

First we consider the transition at fixed $J/U$ under variation of the particle number.
In this case, the role of the control parameter is played by $\mu$.
Since the energies $\hbar\omega_{{\bf k}\pm}$ in Eq.~(\ref{om}) are linear functions of $\mu$,
we get $z\nu=1$ which is consistent with the fact that $z=2$ and $\nu=1/2$.

Now we consider the transition at fixed particle number controlled by the ratio $J/U$.
Using Eqs.~(\ref{om}),~(\ref{Jcmax}), we can write the expression for the total energy required to create
one particle-hole excitation at ${\bf k}={\bf 0}$ in the form
\begin{equation}
\hbar\omega_{{\bf 0}-}
+
\hbar\omega_{{\bf 0}+}
=
U
\left[
    4
    \sqrt{n_0(n_0+1)}
    +
    {\tilde J}_{\rm c} - {\tilde J}
\right]^{1/2}
\left(
    {\tilde J}_{\rm c} - {\tilde J}
\right)^{1/2}
\;,
\end{equation}
where ${\tilde J}=2dJ/U$. When ${\tilde J}$ is close to ${\tilde J}_{\rm c}$, the excitation energy is well approximated by
the lowest-order term
\begin{equation}
\hbar\omega_{{\bf 0}-}
+
\hbar\omega_{{\bf 0}+}
\approx
2U
\left[
    n_0(n_0+1)
\right]^{1/4}
\left(
    {\tilde J}_{\rm c} - {\tilde J}
\right)^{1/2}
\;,
\end{equation}
which gives $z\nu=1/2$. This is consistent with the result $z=1$, $\nu=1/2$ for the classical XY-model
in the dimensions larger than three.

\subsubsection{Superfluid}

\begin{figure}[t]

\centering

\stepcounter{nfig}
\includegraphics[page=\value{nfig}]{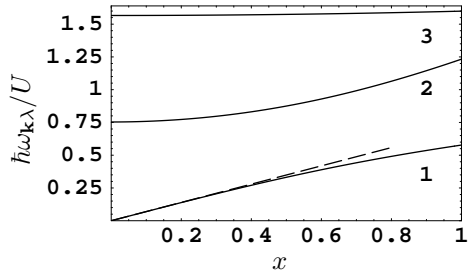}

\caption
{
First three branches $\hbar\omega_{{\bf k},\lambda}$ ($\lambda=1,2,3$) of
the excitations spectrum of the SF for $\mu/U=1.2$ and $2dJ/U=0.15$.
The straight dashed line represents the linear approximation with the sound velocity~(\ref{c_s}).
(Reproduced from Ref.~\cite{KN2011}, \copyright 2011 American Physical Society.)
}
\label{excsf}
\end{figure}

In the SF phase, the eigenvalue problem~(\ref{evpexc}) is solved using
the numerical values of $c_n^{(0)}$ for each $J/U$ and $\mu/U$.
The energies of the lowest-energy excitations are shown in Fig.~\ref{excsf}.
The excitation spectrum consists of several branches which form a band structure.
We note that only the first two lowest-energy branches have a strong dependence on ${\bf k}$.
In the complex plane $({\rm Re}(\psi),{\rm Im}(\psi))$ different modes correspond to the motion around $\psi^{(0)}\ne0$
on the ellipses with the axes
\begin{equation}
b'_{{\bf k}\lambda}
=
\left|
    {\cal U}_{{\bf k}\lambda}
    +
    {\cal V}^*_{{\bf k}\lambda}
\right|
\;,\quad
b''_{{\bf k}\lambda}
=
\left|
    {\cal U}_{{\bf k}\lambda}
    -
    {\cal V}^*_{{\bf k}\lambda}
\right|
\;.
\end{equation}
In order to distinguish between different modes, it is convenient to introduce a flatness parameter~\cite{EFPCSGDKB12}
\begin{equation}
\label{flat-param}
f_{{\bf k}\lambda}
=
(b'_{{\bf k}\lambda}-b''_{{\bf k}\lambda})/(b'_{{\bf k}\lambda}+b''_{{\bf k}\lambda})
\;,\quad
\left|
    f_{{\bf k}\lambda}
\right|
\le1
\;.
\end{equation}

In contrast to the MI, the lowest branch has no gap.
It is a Goldstone mode that appears due to the spontaneous breaking of the U(1) symmetry.
The flatness parameter~(\ref{flat-param}) for this mode is negative which is interpreted as
phase-like oscillations~\cite{HTAB08,EFPCSGDKB12}.
As is shown in Fig.~\ref{AB},
the amplitude of the total-density wave is larger than the amplitude of the condensate-density
wave for this mode. A value greater than unity for the ratio
${\cal A}_{{\bf k}1}/{\cal B}_{{\bf k}1}$ means that the condensed part and the normal part
oscillate in phase.

\begin{figure}[t]

\centering

\stepcounter{nfig}
\includegraphics[page=\value{nfig}]{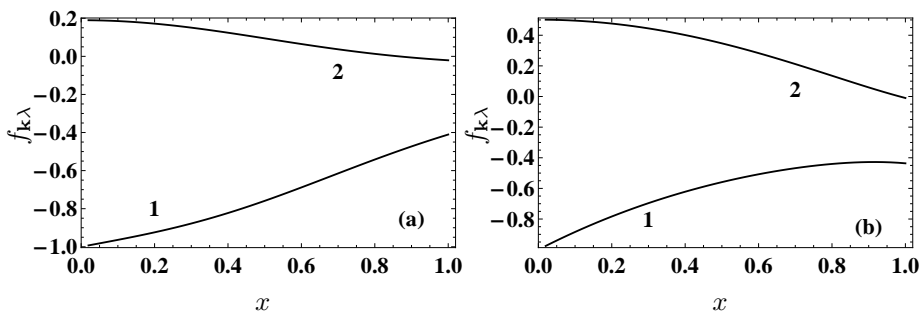}

\caption
{
Flatness parameter~(\ref{flat-param}) for
$\mu/U=1.2$ and $2dJ/U=0.15$~(a), $1$~(b).
}
\label{fig-flat-param}
\end{figure}

\begin{figure}[t]

\centering

\stepcounter{nfig}
\includegraphics[page=\value{nfig}]{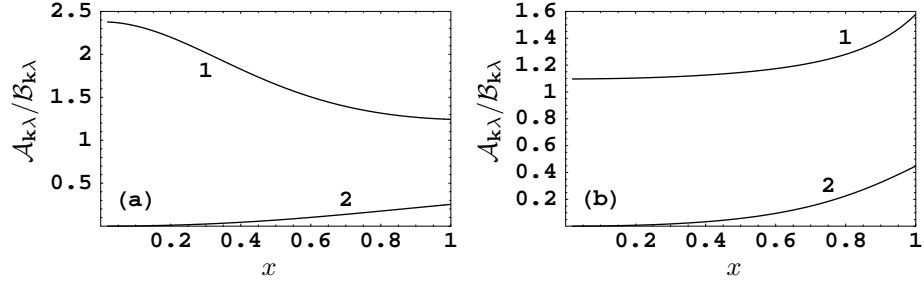}

\caption
{
${\cal A}_{\bf k}/{\cal B}_{\bf k}$ for
$\mu/U=1.2$ and $2dJ/U=0.15$~(a), $1$~(b).
(Reproduced from Ref.~\cite{KN2011}, \copyright 2011 American Physical Society.)
}
\label{AB}
\end{figure}

The lowest-energy branch has a linear form
$\omega_{{\bf k},1}=c_{\rm s}^0|{\bf k}|$
for small ${\bf k}$ with the sound velocity given by~\cite{KN2011}
\begin{equation}
\label{c_s}
c_{\rm s}^0
=
\sqrt{\frac{2J}{\kappa}}
\left|
    \psi^{(0)}
\right|
/{\hbar}
\;,
\end{equation}
where $\kappa$ is the compressibility~(\ref{kappa}).
This result proves that the Gutzwiller approximation is {\it gapless}.

\begin{figure}[t]

\centering

\stepcounter{nfig}
\includegraphics[page=\value{nfig}]{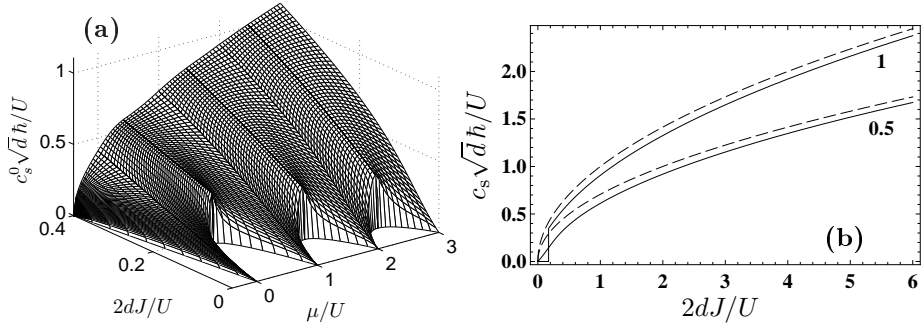}

\caption
{
{(a)}
Sound velocity calculated numerically from Eq.~(\ref{c_s}).
Note the discontinuities at the points
$[(J/U)_{\rm c}^{\rm max},(\mu/U)_{\rm c}]$
described by Eq.~(\ref{cs0}).
(Reproduced from Ref.~\cite{KN2011}, \copyright 2011 American Physical Society.)
{(b)}
Comparison of the sound velocity calculated numerically
from Eq.~(\ref{c_s}) (solid lines) with the analytical expression
(\ref{csB}) (dashed lines) for
$\langle \hat n_{\bf l}\rangle=0.5,1$.
}
\label{sv}
\end{figure}

Fig.~\ref{sv} shows the dependence of the sound velocity on $\mu$ and $J$
calculated numerically using Eq.~(\ref{c_s}). If we approach the boundary of the MI,
the sound velocity goes to zero everywhere except the tips of the lobes,
where it is perfectly described by Eq.~(\ref{cs0}).
This behavior can be understood considering the properties of $\psi^{(0)}$
and $\kappa$. If we approach the SF-MI transition from the SF part of the phase diagram,
the order parameter $\psi^{(0)}$ tends always continuously to zero.
The compressibility $\kappa$ reaches a finite value at every point of the boundary
except the tips of the MI-lobes where it tends continuously to zero such that
the ratio $\psi^{(0)}/\sqrt{\kappa}$ is finite. Therefore, the sound velocity
vanishes at any point of the phase boundary except the tips of the lobes~\cite{FWGF89,MT08}.

For a weakly interacting gas ($U \ll J$),
$|\psi^{(0)}|^2\approx \langle\hat n_{\bf l}\rangle$ and $\kappa\approx 1/U$.
In this limit, we recover the Bogoliubov dispersion relation
\begin{equation}
\label{Bogdisp}
\hbar\omega_{\bf k}^{\rm B}
=
\sqrt
{
 \epsilon_{\bf k}
 \left(
     \epsilon_{\bf k}
     +
     2 U
     \langle\hat n_{\bf l}\rangle
 \right)
}
\end{equation}
and the expression for the sound velocity~\cite{MKSPS04,TZ03},
\begin{equation}
\label{csB}
c_{\rm s}^{\rm B}
=
\sqrt{2JU\langle\hat n_{\bf l}\rangle}/\hbar
\;.
\end{equation}
A comparison with the exact numerical values calculated according to Eq.~(\ref{c_s}) shows that the approximation (\ref{csB})
overestimates the sound velocity and predicts completely different behavior
at small tunneling rates and integer fillings [see Fig.~\ref{sv}(b)].

In the opposite limit ($J \ll U$), the sound velocity is given by~\cite{KN2011}
\begin{eqnarray}
\label{csu}
c_{\rm s}^0
=
2J(n_0+1)
\sqrt{2d(\langle \hat n_{\bf l} \rangle-n_0)(n_0+1-\langle \hat n_{\bf l} \rangle)}
/\hbar
\;.
\end{eqnarray}
It vanishes at $\langle\hat n_{\bf l}\rangle=n_0,n_0+1$
and takes maximal values at $\langle\hat n_{\bf l}\rangle=n_0+1/2$.
This qualitative behavior is the same as in the case of hard-core bosons in 1D, where the sound velocity is given by Eq.~(\ref{cshc1D}).

Experimentally the speed of sound can be measured with the aid of an external potential
which creates a density perturbation of the gas~\cite{AKMDTIK97,MKS09}.
Corresponding numerical simulations of the sound waves propagation for the lattice Bose gas were performed
within the framework of the mean-field theory on the basis of the DGPE~\cite{MKSPS04}
as well as GE~\cite{KN2011} and show perfect agreement with Eqs.~(\ref{csB}) and~(\ref{c_s}), respectively.
Exact numerical simulations for soft-core bosons in 1D were also done making use of the DMRG method~\cite{KSDZ05}.

Another possibility is to employ the Bragg spectroscopy which gives an access to the momentum-resolved excitation spectrum.
The experimental data for the lowest excitation branch of the superfluid obtained in Ref.~\cite{EGKPLPS2010}
agree qualitatively very well with Eq.~(\ref{Bogdisp}).
However, the energies for small momenta appear to be slightly underestimated and for high momenta overestimated
which might be an indication that the calculations beyond the Bogoliubov theory are needed in this regime.

Higher modes ($\lambda \geq 2$) have gaps that grow with the increase of $J$.
For the second mode ($\lambda=2$), the flatness parameter~(\ref{flat-param}) is positive (see Fig.~\ref{fig-flat-param}).
This type of excitations corresponds to the amplitude-like oscillations of the order parameter
which is called `Higgs' amplitude mode~\cite{HABB07,HTAB08,BLGHKWBSH11}.
The amplitude of the total-density wave is much less than that
of the condensate-density wave (see Fig.~\ref{AB})
meaning that the oscillations of the condensate and normal components are out-of-phase.
The properties of this mode have been studied in details in the context of the Bose-Hubbard model
within the framework of the mean-field theory~\cite{AA02,HCG04,SD05,CHG06,OKM06,HABB07,HTAB08,MT08,BLGHKWBSH11,GSP11,KN2011}
and in quantum Monte Carlo calculations~\cite{PEH09,PP12,LCDEPP15}.
It was experimentally observed with the aid of Bragg spectroscopy in the non-linear regime~\cite{BLGHKWBSH11}
and with lattice modulation in the linear-response regime~\cite{EFPCSGDKB12}.
However, the DMRG calculations of the dynamic structure factor in one dimension did not reveal any distinct gapped modes
in the superfluid phase~\cite{EFG12,EFGMKAL12}.

\subsection{Bragg scattering}

Within the Gutzwiller approximation, the susceptibility in the lattice version of Eq.~(\ref{lin-resp}) can be written in the form~\cite{KN2011}
\begin{eqnarray}
\label{chi2}
\chi({\bf k},\omega)
=
\frac{2}{\hbar}
\sum_\lambda
\frac
{\chi_{{\bf k}\lambda} \omega_{{\bf k}\lambda}}
{
 \left(
     \omega + i0
 \right)^2
 -
 \omega_{{\bf k}\lambda}^2
}
\;,
\end{eqnarray}
where $\lambda$ denotes various excitation branches discussed in the previous section associated to the eigenvalues $\omega_{{\bf k}\lambda}$
and the corresponding amplitude of the Bragg scattering is determined by the amplitude of the density wave~(\ref{dw}) as
\begin{equation}
\label{chik}
\chi_{{\bf k}\lambda}
=
\left|
    {\cal A}_{{\bf k}\lambda}
\right|^2
\;.
\end{equation}

\begin{figure}[t]

\centering

\stepcounter{nfig}
\includegraphics[page=\value{nfig}]{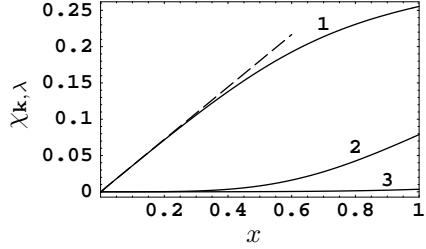}

\caption
{
Transition amplitudes $\chi_{{\bf k},\lambda}$ associated with the 
transition frequency $\omega_{{\bf k},\lambda}$ for the lowest excitation
branches ($\lambda=1,2,3$) and
for $\mu/U=1.2$ and $2dJ/U=0.15$.
The dashed line corresponds to the approximation~(\ref{chiapp}).
(Reproduced from Ref.~\cite{KN2011}, \copyright 2011 American Physical Society.)
}
\label{chisf}
\end{figure}

\begin{figure}[t]

\centering

\stepcounter{nfig}
\includegraphics[page=\value{nfig}]{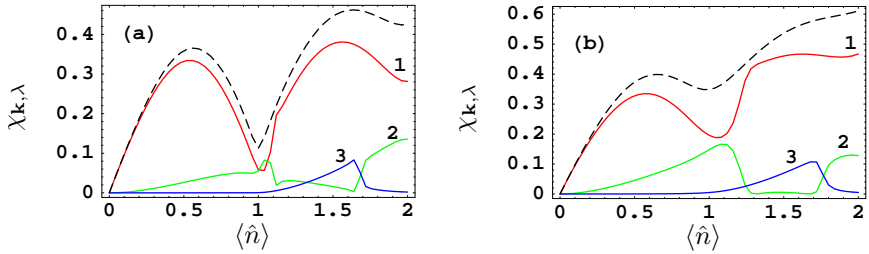}

\caption{
(color online)
Transition amplitudes $\chi_{{\bf k},\lambda}$ versus the density for the first excitation
branches ($\lambda=1,2,3$) and
for $x=1$ and $2dJ/U=0.2$~(a), $0.3$~(b).
Dashed lines show the static structure factor $\tilde S_0({\bf k})$.
(Reproduced from Ref.~\cite{KN2011}, \copyright 2011 American Physical Society.)
}
\label{chisf1}
\end{figure}

The dependences of $\chi_{{\bf k},\lambda}$ on the variable $x$ defined by Eq.~(\ref{x})
for the excitation branches with $\lambda=1,2,3$ in the SF phase are shown in Fig.~\ref{chisf}.
For the chosen values of parameters, only the two lowest branches display noticeable amplitudes.
In the long-wavelength limit, only the lowest mode has a nonvanishing amplitude
\begin{eqnarray}
\label{chiapp}
\chi_{{\bf k},1}
\stackrel
{
 {\bf k} \rightarrow {\bf 0}
}
{=}
\frac{\kappa}{2}
c_{\rm s}^0
|{\bf k}|
\;.
\end{eqnarray}
Similar results have been also obtained in Ref.~\cite{HABB07}.
However, the calculations in Ref.~\cite{HABB07} are valid only close to the boundaries
of the MI-SF phase transition
because the occupation numbers $n$ in Eq.~(\ref{state}) were restricted to $n=n_0,n_0\pm 1$.
In Fig.~\ref{chisf1} instead, we see that the amplitude for the third excitation branch
as well as for the second one can become significant at certain densities.

As in the case of a Bose gas in continuum, the sum rules~(\ref{S0-lat}),~(\ref{fsum-hom}),~(\ref{comp-sum-rule-lat})
allow us to deduce the static structure factor. We find indeed
\begin{eqnarray}
\tilde S_0({\bf k})
=
\sum_\lambda \chi_{{\bf k},\lambda}
\stackrel{{\bf k} \rightarrow {\bf 0}}
{=}
\frac{\kappa}{2} c_{\rm s}^0 |{\bf k}|
\;.
\end{eqnarray}
This result shows an interesting feature of the sum-rule approach.
Starting from the lowest-order Gutzwiller approximation that does not contain any
correlation, the two-point correlation function is determined as a next-order
contribution. Similarly, starting from the time-dependent DGPE,
an analogous procedure has been successfully used to recover the static structure factor
predicted from the Bogoliubov theory~\cite{PS2003,PN89}.

In the MI phase, the Gutzwiller approximation does not allow us to observe any
branches since $\chi_{{\bf k},\lambda}\equiv 0$. No Bragg response is possible,
although the excitations exist in the mean-field approach.
In order to allow a nonvanishing response, correlations between
different sites should be included, which goes beyond the Gutzwiller approximation~\cite{NS10,HABB07}.
In such a description, excitations in the Bragg process are created
as particle-hole pairs~\cite{HABB07,ODFSS05,NS10}.
However, the latter appears to be of the second order in the inverse of coordination number $Z=2d$~\cite{SUF08,NS10,QNS12,QKNS14}
and, therefore, is not taken into account by the standard Gutzwiller ansatz.

\subsection{One-particle Green's function}

In the context of the Gutzwiller approximation, the one-particle Green's function can be determined
as a linear response to the perturbation~\cite{KN2011}
\begin{eqnarray}
\label{Hint2}
\hat H'(t)
&=&
\sum_{\bf l} \eta_{{\bf k},\omega}
e^{i({\bf k}\cdot{{\bf x}_{\bf l}}-\omega t)}
\hat a^\dagger_{\bf l}
+
{\rm h.c.}
\;,
\end{eqnarray}
which explicitly breaks the U(1) symmetry.
This induces fluctuations of the order parameter,
\begin{eqnarray}
\left(
\begin{array}{c}
\psi_{\bf l}- \psi^{(0)}
\\
\psi^*_{\bf l}- (\psi^{(0)})^*
\end{array}
\right)
=
\underline{\underline {G}}({\bf k},\omega).
\left(
\begin{array}{c}
\eta_{{\bf k},\omega}
e^{i({\bf k}\cdot{\bf x}_{\bf l}-\omega t)}
\\
\eta^*_{{\bf k},\omega}
e^{-i({\bf k}\cdot{\bf x}_{\bf l}-\omega t)}
\end{array}
\right)
\quad.
\end{eqnarray}
The proportionality term is the one-particle $2\times2$ matrix Green's function with the general expression
\begin{eqnarray}
\label{G}
\underline{\underline{G}}({\bf k},\omega)
=
\sum_\lambda
\frac{\underline{\underline{g}}_{{\bf k},\lambda}}
{\omega +i0 -  \omega_{{\bf k},\lambda}}
\;,
\end{eqnarray}
where the matrix transition amplitude is defined as
\begin{equation}
\underline{\underline{g}}_{{\bf k},\lambda}
=
\left(
\begin{tabular}{cc}
$
\left|
    {\cal U}_{{\bf k},\lambda}
\right|^2
$
&
$
{\cal U}_{{\bf k},\lambda}
{\cal V}_{{\bf k},\lambda}
$
\\
$
{\cal V}^*_{{\bf k},\lambda}
{\cal U}^*_{{\bf k},\lambda}
$
&
$
\left|
    {\cal V}_{{\bf k},\lambda}
\right|^2
$
\end{tabular}
\right)
\end{equation}
and the functions ${\cal U}_{{\bf k},\lambda}$, ${\cal V}_{{\bf k},\lambda}$
are determined by Eq.~(\ref{psiw}).

In the MI phase, matrix $\underline{\underline{g}}_{{\bf k},\lambda}$ is diagonal and according to
Eqs.~(\ref{psiw}),~(\ref{uv-mi}) its nonvanishing entries are given by
\begin{eqnarray}
\left|
    {\cal V}_{{\bf k}-}
\right|^2
=
\frac{1}{2}
\left[
\frac
{(2n_0+1)U- J_{\bf k}}
{
 \sqrt{U^2-4J_{\bf k}U\left(n_0+\frac{1}{2}\right)+J_{\bf k}^2}
}
-1
\right]
\;,\quad
\left|
    {\cal U}_{{\bf k}+}
\right|^2
=
\left|
    {\cal V}_{{\bf k}-}
\right|^2
+1
\;.
\end{eqnarray}
Thus, the time-dependent Gutzwiller approach allows to recover the results previously
established in the context of quantum field theory~\cite{ODFSS05,SD05}.
The spectral weight $\left|{\cal V}_{{\bf k}-}\right|^2$ of the hole branch yields
the quasi-momentum distribution~\cite{SD05,GWFMGB05a}.
For $J=0$, $\left|{\cal V}_{{\bf k}-}\right|^2=n_0$ in agreement with Eq.~(\ref{md-mf}),
but for any finite $J<J_{\rm c}$ it has maxima at the vectors of the reciprocal lattice ${\bf k}=\frac{2\pi}{a}{\bf q}$
which grow as we approach the critical point.

\subsection{Finite-temperature phase diagram}

At finite temperature, the expectation values of the operators are calculated for the thermal state
and the definition of the order parameter $\psi_{\bf l}$ should be generalized as
\begin{equation}
\overline{\psi}_{\bf l}
=
{\cal Z}_{\bf l}^{-1}(\mu)
\sum_{s_{\bf l}}
\exp
\left(
    - \frac{E_{s_{\bf l}}}{k_{\rm B}T}
\right)
\psi_{s_{\bf l}}
\;,
\end{equation}
where $\psi_{s_{\bf l}}$'s are defined by Eq.~(\ref{psi}) for a particular state $|s_{\bf l}\rangle$
with the eigenvalue of the local mean-field Hamiltonian~(\ref{HBHMF}) in the grand-canonical ensemble
and ${\cal Z}_{\bf l}(\mu)$ is the corresponding partition function.
In the homogeneous case, all the $\overline{\psi}_{\bf l}$ are equal to each other.
The region in the $(\mu,J)$ plane with vanishing $\overline{\psi}$ is described
by the equation~\cite{BV04,KPG06}
\begin{equation}
\label{Jmu-T0}
\frac{2dJ}{U}
{\cal Z}_0^{-1}(\mu)
\sum_{n=0}^\infty
\frac
{
 \left(
     1 + \mu/U
 \right)
 \exp
 \left(
     - \frac{E_n-\mu n}{k_{\rm B}T}
 \right)
}
{
 \left(
     n - \mu/U
 \right)
 \left(
     \mu/U - n + 1 
 \right)
}
< 1
\;,
\end{equation}
where the partition function ${\cal Z}_0(\mu)$ and the energies $E_n$ are exactly the same as in the absence of hopping [see Eq.~(\ref{Z0})].
In the limit of vanishing temperature, Eq.~(\ref{Jmu-T0}) reduces to~(\ref{crit}).

\begin{figure}[t]

\centering

\stepcounter{nfig}
\includegraphics[page=\value{nfig}]{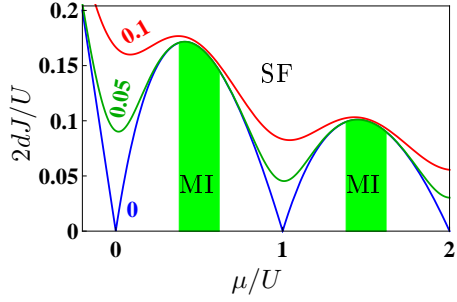}

\caption
{
(color online)
Finite-temperature mean-field phase diagram.
The boundaries of the superfluid-insulator transition~(\ref{Jmu-T0}) are shown by solid lines for
$k_{\rm B}T/U=0$~(blue), $0.05$~(green), $0.1$~(red).
Green stripes are the MI regions for $k_{\rm B}T/U=0.05$ obtained from the requirement $\kappa U < 0.01$.
Outside of the stripes within the insulating phase we have a normal gas.
}
\label{Fig:muJ-T-mf}
\end{figure}

The boundaries of the superfluid-insulator transition described by Eq.~(\ref{Jmu-T0}) are shown in Fig.~\ref{Fig:muJ-T-mf}
for different temperatures. The size of the insulating regions grows with temperature and the topology is the same as
in QMC simulations [see Fig.~\ref{phd2D-T}(a)]. The insulating region is divided into two parts, MI and normal gas,
separated by the crossover lines determined from the condition that the compressibility is fixed by a small arbitrary number.
In the mean-field approximation, the latter is given by
\begin{equation}
\kappa
=
\left(
    \langle\hat n_{\bf l}^2\rangle
    -
    \langle\hat n_{\bf l}\rangle^2
\right)
/(k_{\rm B}T)
\end{equation}
which follows from Eq.~(\ref{kappa-fluct}) if we neglect correlations of the particle numbers at different sites.
Since in the mean-field theory the properties of the insulator do not depend on $J$, the crossover lines are parallel
to the $J/U$ axis in contrast to the QMC calculations.

\begin{figure}[t]

\centering

\stepcounter{nfig}
\includegraphics[page=\value{nfig}]{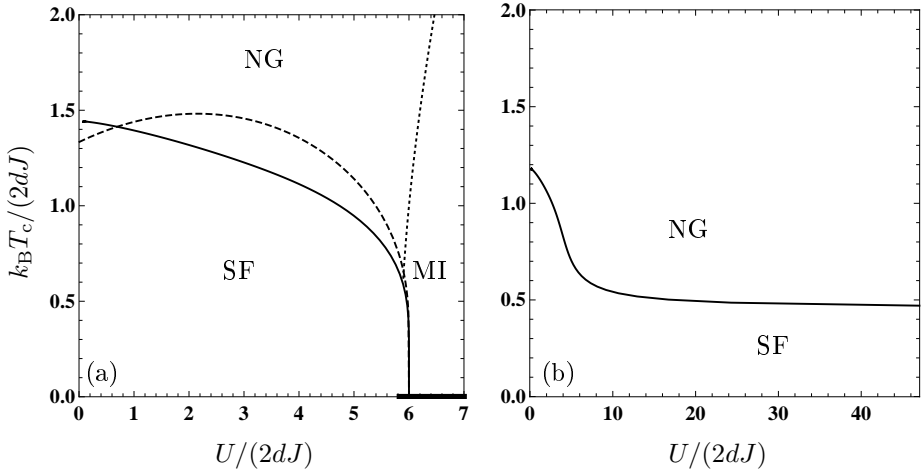}

\caption
{
Critical temperature of the superfluid-insulator transition
for the filling factor $\langle\hat n_{\bf l}\rangle=1$~{(a)}, $0.75$~{(b)}.
Solid lines are numerical solutions of Eqs.~(\ref{Jmu-T0}),~(\ref{pnTJ0}),~(\ref{mean-pn}).
Dashed line in panel (a) is approximate analytical solution~(\ref{Tc-slave}) obtained within the slave-boson approach~\cite{DODS03}.
Dotted line in panel (a) is the crossover line between the MI and normal gas determined from the condition
$\Delta=\hbar\omega_{{\bf 0}+}+\hbar\omega_{{\bf 0}-}=k_{\rm B}T_{\rm c}$,
where $\hbar\omega_{{\bf 0}\pm}$ are the energies of particle and hole excitations~(\ref{om}),
see also Refs.~\cite{SKPR93,Wang09}.
}
\label{Fig:Tc-mf}
\end{figure}

Eq.~(\ref{Jmu-T0}) allows also to determine the critical temperature $T_{\rm c}$ of the superfluid-insulator transition.
For a given filling $\langle\hat n_{\bf l}\rangle$, the chemical potential can be eliminated with the aid of
Eqs.~(\ref{pnTJ0}),~(\ref{mean-pn}) and the result for $T_{\rm c}$ is shown in Fig.~\ref{Fig:Tc-mf}.
In the case of unit filling, the dependence of $T_{\rm c}$ on $U$ is qualitatively similar to that obtained in QMC
calculations in three dimensions (see Fig.~\ref{Tc3D}).
However, the mean-field theory predicts larger values of $T_{\rm c}$
compared to QMC and cannot reproduce correctly the ideal-gas limit ($k_{\rm B}T_{\rm c}/J=5.591$).

Other versions of the mean-field theory yield somewhat different results for the critical temperature.
For instance, if the occupation numbers are first restricted to $n_0$, $n_0\pm1$ and then the mean-field approximation is applied,
the critical temperature becomes lower, although the behavior of $T_{\rm c}(U)$ remains qualitatively the same~\cite{Wang09}.
Different form of $T_{\rm c}(U)$ was obtained in the slave-boson approach, where $T_{\rm c}(U)$ has a maximum~\cite{DODS03}.
If the occupation numbers are again restricted to $n_0$, $n_0\pm1$, the latter approach gives an analytical expression~\cite{DODS03}
\begin{equation}
\label{Tc-slave}
\frac{k_{\rm B}T_{\rm c}}{2dJ}
=
\frac{\tilde U}{2}
\ln^{-1}
\frac
{
 \left[
     \tilde U
     -
     8
     \left(
         2 n_0 + 1
     \right)
 \right]
 \left[
     \tilde U
     +
     2 n_0 + 1
 \right]
}
{
 \left[
     \tilde U
     -
     2
     \left(
         2 n_0 + 1
     \right)
 \right]
 \left[
     \tilde U
     +
     4
     \left(
         2 n_0 + 1
     \right)
 \right]
}
\;,\quad
\tilde U
=
\frac{U}{2dJ}
\;.
\end{equation}
This result is also shown in Fig.~\ref{Fig:Tc-mf}(a) for comparison.

\subsection{Quantum solitons}
\label{sec-QS}

Superfluid phase of the ultracold bosons in optical lattices far from the transition into the MI is well described
by the DGPE~(\ref{DGPE}) if the lattice is deep or by its continuum counterpart~\cite{PS2003} in the case of shallow lattices.
One of the remarkable features of these equations is that they allow soliton solutions in analogy to nonlinear optics~\cite{KA2003}.
This has triggered theoretical interest in discrete (lattice) solitons in the context of ultracold atoms~\cite{TS01,AS05},
and has led to the seminal observations of gap solitons, i.e., lattice solitons with repulsive
interactions, but with an appropriate dispersion management~\cite{EAATTKO04}.

While most of the studies of solitons were concentrated on their classical aspects, more recently, considerable interest has been
devoted to the effect of thermal noise~\cite{MSESL02,AS05}, quantum properties of solitons, and the role of quantum fluctuations~\cite{MR10}.
The latter may cause filling up of the dark soliton core in the quantum detection process, as was shown using the Bogoliubov-de~Gennes
equations~\cite{D2004}. The same method was also employed to study the stability of solitons~\cite{YS03,JK99,KCTFM03}, excitations caused by
the trap opening~\cite{C2009}, and entanglement generation in collisions of two bright solitons~\cite{LM09}. A noisy version of
the standing bright solitons was studied using the exact diagonalization and quantum Monte Carlo method~\cite{JS08}. Bright
solitons in 1D were considered in Ref.~\cite{C2009}, where exact Lieb-Liniger solutions were used to calculate the internal
correlation function of the particles positions. Making use of the DGPE, and the
time-evolving block decimation algorithm~\cite{V2004} it was demonstrated that quantum effects cause the soliton to fill in, and
that soliton collisions become inelastic~\cite{MC09,MC2009,MDCC09}.
In the next section, we consider the properties of the discrete dark solitons near the SF-MI transition
within the framework of the Gutzwiller mean-field theory.

\subsubsection{Standing modes}

In the present section, we consider low-energy excited states, where the coefficients $c_{{\bf l}n}$ as well as
the order parameters $\psi_{\bf l}$ depend only on one spatial direction $\nu$.
Without loss of generality, we can assume that this is $\nu=1$. Then
\begin{equation}
\psi_{{\bf l}\pm{\bf e}_\nu}
=
\left\{
    \begin{tabular}{ll}
    $\psi_{l\pm 1}$ &, if $\nu=1$, \\
    $\psi_{l}$      &, if $\nu>1$.
    \end{tabular}
\right.
\end{equation}

We are interested in the stationary solutions of Eqs.~(\ref{GEd})
\begin{eqnarray}
c_{l n}(t)
&=&
c_{l n}^{(0)}
\exp
\left(
    - i \omega_l t
\right)
\;,
\nonumber\\
\label{omegal}
\hbar\omega_l
&=&
-2J
\left[
    \psi_{l-1}^{(0)}
    +
    \psi_{l+1}^{(0)}
    +
    2(d-1)
    \psi_{l}^{(0)}
\right]
\psi_{l}^{(0)}
+
\sum_{n=0}^\infty
\left[
    \frac{U}{2}n(n-1) - \mu n
\right]
\left|
    c_{l n}^{(0)}
\right|^2
\;,
\end{eqnarray}
where $\psi_l^{(0)}$ is determined by the coefficients $c_{l n}^{(0)}$ according to Eq.~(\ref{psi}).
We require that the condensate order parameter $\psi_l^{(0)}$
is an antisymmetric function with respect to the middle point of the lattice $l_0$.
These are the kink states which can be treated as standing dark solitons.
In contrast to the ground state discussed in Sec.~\ref{GS}, all the quantities
which describe the solitons are labeled by the site index $l$.

In general, one has to distinguish between the two cases:
when the middle point $l_0$ is on the lattice site
(on-site modes) and in the middle of two neighboring sites (off-site modes).
The two modes have different energies, and the difference
defines the Peierls-Nabarro barrier~\cite{KKC1994,ASPSL2004}, which may affect the mobility of solitons.
In addition, the stability of the on-site and off-site modes can in general be different,
as it will be shown in the next section.

\begin{figure}[t]

\centering

\stepcounter{nfig}
\includegraphics[page=\value{nfig},angle=90]{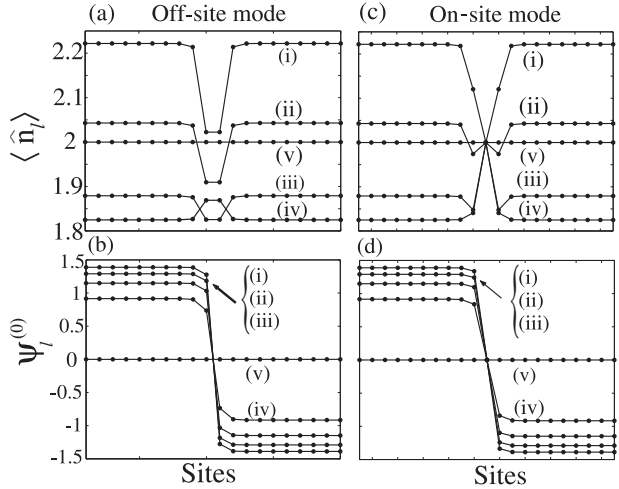}

\caption
{
Mean number of atoms $\langle\hat{n}_l\rangle$ (a) and (c),
and mean-field order parameter $\psi_l^{(0)}$ (b) and (d).
The scaled chemical potential $\mu/U=1.2$ and the tunneling rates
$2dJ/U$: 0.7~(i), 0.5~(ii), 0.3~(iii), 0.15~(iv), and 0.05~(v).
(Reproduced from Ref.~\cite{KLL10}, \copyright 2010 American Physical Society.)
}
\label{smodes}
\end{figure}

We consider first the SF phase.
Typical behavior of the kink modes with the lowest energy is displayed in Fig.~\ref{smodes}~\cite{KLL10}.
The mean occupation numbers
$\langle\hat{n}_l\rangle$ calculated according to Eq.~(\ref{Nav})
are shown in (a) and (c), while (b) and (d) give the associated $\psi_l^{(0)}$
defined by Eq.~(\ref{psi}).
The individual curves correspond to different tunneling rates $J$.
Far from the middle point of the lattice,
$\langle\hat{n}_l\rangle$ as well as $\psi_l^{(0)}$ tend to the same values as
in the ground state. Near the middle point, on the other hand, they have nontrivial position dependence.

For the considered chemical potential, $\mu/U=1.2$, the MI-SF transition occurs according to Eq.~(\ref{crit}) at
$2d(J/U)_{\rm c}\approx0.0727$. Much above this value,
$\langle\hat{n}_l\rangle$ has only one extremum which is a global minimum.
It is doubly degenerate in the case of the off-site modes
[Fig.~\ref{smodes}(a), curves (i)-(iii); Fig.~\ref{smodes}(c), curve (i)].
Expectedly, these solutions reproduce the well-known standing soliton of the
DGPE~\cite{KKC1994,ASPSL2004}. For smaller values of $J$, when we come closer
to the phase boundary, the global minimum turns into a maximum
[Fig.~\ref{smodes}(a), curve (iv); Fig.~\ref{smodes}(c), curves (ii)-(iv)].
For the off-site modes, this maximum is always a global extremum.
In the case of the on-site modes, the maximum of $\langle\hat{n}_l\rangle$
is either a global extremum [Fig.~\ref{smodes}(c), curve (iv)]
or a local one which is accompanied by side minima
[Fig.~\ref{smodes}(c), curves (ii),~(iii)].
Contrary to the results deep in the SF region, these types of the atomic distributions cannot be described by the DGPE.
Similar features were also found for vortices~\cite{WCHZ04,Lundh08,GM09} and the underlying physical mechanism is essentially the same.

The existence of the modes with maxima and minima of $\langle\hat{n}_l\rangle$ can be easier understood in the case of on-site modes.
Let us assume that $\langle\hat{n}_l\rangle\in\left(n_0-0.5,n_0+0.5\right)$, where $n_0$ is a natural number including zero.
Since $\psi_{l=l_0}^{(0)}\equiv0$, the number of bosons at site $l_0$ is fixed by some integer (local Mott insulator).
From the minimization of the interaction energy, it follows that this integer coincides with $n_0$ which is either larger or smaller
than $\langle\hat{n}_l\rangle$. This leads to the two types of solutions.
For the off-site modes the situation is more involved but qualitative picture remains the same.

\begin{figure}[t]

\centering

\stepcounter{nfig}
\includegraphics[page=\value{nfig}]{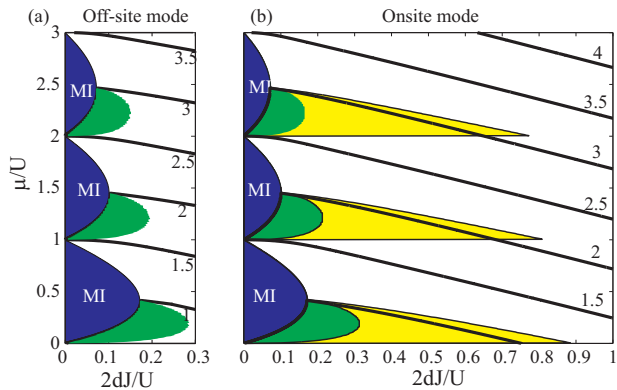}

\caption
{
{(a)}
{\it Off-site modes}.
Dark (blue) areas show the first three MI zones ($n_0=1,2,3$). Gray (green) areas indicate the regions where the
off-site modes have a global maximum of $\langle\hat n_l\rangle$ at the middle lattice sites around $l_0$, see curve (iv) of Fig.~\ref{smodes}~(a).
In the rest part of the diagram, $\langle\hat n_l\rangle$ has only one extremum and takes the minimal value at the middle sites,
see curves (i),~(ii),~(iii) of Fig.~\ref{smodes}~(a). The lines of constant $\langle\hat{n}\rangle$ corresponding to the ground-state
densities are shown as well and labeled by the numerical values.
{(b)}
{\it On-site modes}.
Gray (green) areas depict the regions where the on-site modes have a global maximum of $\langle\hat n_l\rangle$ at the middle lattice site $l_0$,
(in these regions $\langle\hat n_l\rangle$ have only one extremum which is a global maximum), see curve (iv) of Fig.~\ref{smodes}~(c).
In the light-gray (yellow) areas, the on-site modes have side minima near the maximum of $\langle\hat n_l\rangle$,
see curves (ii) and (iii) of Fig.~\ref{smodes}~(c). In the rest part of the diagram, $\langle\hat n_l\rangle$ has only one extremum
and takes the minimal value at the middle site, see curve (i) of Fig.~\ref{smodes}~(c).
(Reproduced from Ref.~\cite{KLL10}, \copyright 2010 American Physical Society.)
(For interpretation of the references to colour in this figure legend, the reader is referred to the web version of this article.)
}

\label{fig2}
\end{figure}

In order to have a better understanding of the modes with the maxima of $\langle\hat{n}_l\rangle$, we depict in Fig.~\ref{fig2}
a $(\mu,J)$-diagram identifying the various types of solutions. The anomalous regions where $\langle\hat{n}_l\rangle$ attains a global
maximum are almost the same for the off-site and on-site modes. They are, to a very good approximation, located in the ``hole" areas of the
$(\mu,J)$-plane as displayed in Fig.~\ref{hs}, and hence, the corresponding modes can be interpreted as dark solitons of holes.
The anomalous regions of the on-site modes which have minima of $\langle\hat{n}_l\rangle$ near the middle lattice point are located
in the intermediate regions between particle and hole-areas and can thereby be interpreted as a mixture of dark solitons of holes and particles.
With the increase of the filling factor the size of the MI lobes as well as of the anomalous regions decrease.

In the MI phase, there is only trivial solution $\psi_l^{(0)}\equiv 0$, i.e., soliton modes do not exist.
This follows from the fact that in the Gutzwiller ansatz, the excited states of the MI are products of
local Fock states, where the occupation numbers $n_l$ can be locally different from the homogeneous filling $n_0$.
As a consequence, all $\psi_l^{(0)}$ must identically vanish and no soliton solutions are therefore possible within these parameter regimes.

Experimentally, dark (or gray) solitons are typically created via a phase-imprinting method~\cite{BBDESAS99,DSFCCCDHHRRSP00}.
Initially ($t=0$) the system of atoms is assumed to be in its ground state. During a short time $t_{\rm imp}$ one applies a spatially dependent
potential on top of the lattice. In the Bose-Hubbard Hamiltonian, it is described by the term
$\sum_l \epsilon_l \hat a_l^\dagger \hat a_l^{\phantom{\dagger}}$.
If the time $t_{\rm imp}$ is much shorter than other characteristic time scales, from Eqs.~(\ref{GEd}) we get that the additional term induces a shift in the
phase of the atomic states
\begin{eqnarray}
c_{ln}(t_{\rm imp})
=
c_n^{(0)}
\exp
\left(
    - i \phi_l n
\right)
\;,\quad
\psi_{l}(t_{\rm imp})
=
\psi^{(0)}
\exp
\left(
    - i \phi_l
\right)
\;.
\end{eqnarray}
For the creation of dark solitons it is appropriate to choose a hyperbolic tangent imprinting potential, such that
\begin{equation}
\phi_l
=
\frac{\epsilon_l t_{\rm imp}}{\hbar}
=
\frac{\Delta\phi}{2}
\left[
    1+
    \tanh
    \left(
        \frac{l-l_0}{0.45\, l_{\rm imp}}
    \right)
\right]
\;,
\end{equation}
where $l_0$ is the middle point of the lattice. Here, $l_{\rm imp}$ is the width of the interval around $l=l_0$ where
$\phi_l/\Delta\phi$ grows from $0.1$ to $0.9$, and $\Delta\phi$ is the amplitude of the imprinted phase~\cite{BCOSS02}.
Apart from the moving gray soliton, the phase imprinting also induces a density wave propagating in the opposite direction to the
soliton, which appears due to the impulse imparted by the imprinting potential~\cite{BBDESAS99,DSFCCCDHHRRSP00,BCOSS02}.
Numerical simulations performed in Ref.~\cite{KLL10} according to this procedure show that the form of the propagating
dark solitons becomes qualitatively different from those predicted by the DGPE if the system parameters are in the green (gray)
regions of Fig.~\ref{fig2}, where the standing solitons have global maxima of $\langle\hat n_l\rangle$.

\subsubsection{Stability of standing solitons}

We consider small perturbation of the soliton state determined by the coefficients
$c_{ln}^{(0)}$ as follows:
$
c_{ln}(t)
=
\left[
    c_{ln}^{(0)}
    +
    c_{ln}^{(1)}(t)
\right]
\exp
\left(
    -i \omega_l t
\right)
$,
where $\omega_l$ is given by Eq.~(\ref{omegal}) and
\begin{equation}
\label{sol}
c_{ln}^{(1)}(t)
=
u_{ln}
e^{-i \omega t}
+
v_{ln}^*
e^{i \omega t}
\;.
\end{equation}
Substituting this expression into the Gutzwiller equations and keeping only linear terms
with respect to $u_{ln}$ and $v_{ln}$, we obtain the system of linear equations:
\begin{eqnarray}
\label{evpexc-inh}
\hbar\omega
u_{ln}
&=&
\sum_{n',l'}
\left(
    A_{nl}^{n'l'}
    u_{l'n'}
    +
    B_{nl}^{n'l'}
    v_{l'n'}
\right)
\;,
\nonumber\\
-
\hbar\omega
v_{ln}
&=&
\sum_{n',l'}
\left(
    B_{nl}^{n'l'}
    u_{l'n'}
    +
    A_{nl}^{n'l'}
    v_{l'n'}
\right)
\;,
\end{eqnarray}
where
\begin{eqnarray}
A_{nl}^{n'l'}
&=&
\left[
    \frac{U}{2}\,n(n-1)
    -
    \mu n
    -
    \hbar\omega_{l}
\right]
\delta_{l',l}
\delta_{n',n}
-
J
\left[
    \psi_{l-1}^{(0)}
    +
    \psi_{l+1}^{(0)}
    +
    2(d-1)
    \psi_{l}^{(0)}
\right]
\delta_{l',l}
\left(
    \sqrt{n'}\,
    \delta_{n',n+1}
    +
    \sqrt{n}\,
    \delta_{n,n'+1}
\right)
\nonumber\\
&-&
J
\left[
    \sqrt{n+1}\,
    \sqrt{n'+1}\,
    c_{l,n+1}^{(0)}\,
    c_{l',n'+1}^{(0)}
+
\sqrt{n}\,
\sqrt{n'}\,
c_{l,n-1}^{(0)}\,
c_{l',n'-1}^{(0)}
\right]
\left[
    \delta_{l,l'+1}
    +
    \delta_{l',l+1}
    +
    2(d-1)
    \delta_{l',l}
\right]
\;,
\nonumber\\
B_{nl}^{n'l'}
&=&
-J
\left[
    \sqrt{n+1}\,
    \sqrt{n'}\,
    c_{l,n+1}^{(0)}\,
    c_{l',n'-1}^{(0)}
    +
    \sqrt{n}\,
    \sqrt{n'+1}\,
    c_{l,n-1}^{(0)}\,
    c_{l',n'+1}^{(0)}
\right]
\left[
    \delta_{l,l'+1}
    +
    \delta_{l',l+1}
    +
    2(d-1)
    \delta_{l',l}
\right]
\;.
\nonumber
\end{eqnarray}
Eqs.~(\ref{evpexc-inh}) generalize Eqs.~(\ref{evpexc}) to the inhomogeneous case and are analogous
to the Bogoliubov-de~Gennes equations which were employed
for the stability analysis of the dark solitons governed by the DGPE~\cite{JK99,KCTFM03}.

The stationary modes are linearly stable, if all the eigenvalues $\hbar\omega$ are real.
Numerical solutions of the eigenvalue problem (\ref{evpexc-inh}) show that most of the eigenvalues are real
but there are always few ones, which contain nonvanishing imaginary part $\omega_i$.
The magnitude of $\omega_i$ determines the inverse lifetime of the solitons,
which can be almost equal or drastically different for the off-site and on-site modes and
there is no any principal difference in this respect between the normal and anomalous modes.

\begin{figure}[t]

\centering

\stepcounter{nfig}
\includegraphics[page=\value{nfig}]{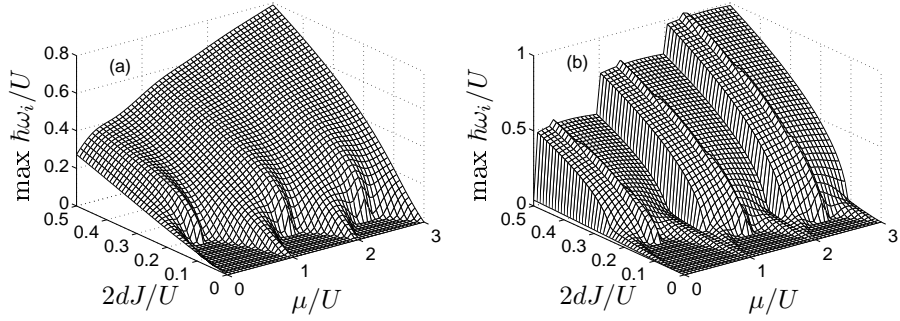}

\caption
{
Maximal imaginary part of the complex eigenvalues $\omega$ for the
(a) off-site and (b) on-site modes.
(Reproduced from Ref.~\cite{KLL10}, \copyright 2010 American Physical Society.)
}
\label{maxim}
\end{figure}

Figure~\ref{maxim} shows the maximal imaginary part of the complex eigenvalues $\omega$
which vanishes in the MI regions, where solitons do not exist, but does not vanish in the SF region.
With the increase of $\mu$ and $J$, maximal $\omega_i$ increases for both types of soliton modes meaning that the instability grows.
There is, however, one important qualitative difference between the off-site and on-site modes. For the on-site modes,
there are rather small regions between the MI lobes, where $\omega_i$ is close to zero and
much smaller than that for the off-site modes, i.e., the on-site solitons are much more stable.
This feature has some similarity to the stability of the standing dark solitons governed by the DGPE,
where it was found~\cite{JK99} that on-site modes are stable if the tunneling $J$
does not exceed a certain critical value, while off-site modes are unstable for all tunnelings.

\section{\label{sec-SBNFR}Spinless bosons near Feshbach resonance}

Magnetic Feshbach resonances provide and important tool to tune the interaction strength in ultracold atomic gases.
They found numerous applications and are often used in experiments for the observation of different phenomena,
some of them were already mentioned in this review.
Feshbach resonances occur when the energy of two scattering atoms is close to the energy of their molecular bound state.
In essence, it is a quantum interference effect of these two states.
The energy difference between the states can be controlled by an external magnetic field when the corresponding magnetic moments
are different which gives a possibility of tuning.
Excellent reviews on Feshbach resonances in the context of ultracold atoms were given in Refs.~\cite{TTHK99,DS04,HS06,KGJ2006,CGJT10,Kokkelmans14}.
However, the lattice problems were not addressed there and the aim of the present section is to fill this gap.

\subsection{Hamiltonian}

We consider a system of identical atoms, each having the mass $M$, in an optical lattice created by a far-detuned standing laser wave.
In addition, the atoms are subject to the external magnetic field $B$,
with $B=B_0$ corresponding to the Feshbach resonance of the width $\Delta B$.
This allows to convert pairs of atoms into molecule of mass $2M$
and there is a reverse process when a molecule dissociates into two atoms.
In order to describe Feshbach resonance we adopt two-channel model with
one open channel that describes asymptotically free atoms and one closed channel
corresponding to the two-atom molecule~\cite{KGJ2006,NSJ2006}.
The model is adequate to describe an isolated (narrow) resonance.

The far-detuned optical laser field creates an effective potential not only for the atoms
but also for the molecules that has the same form as for the atoms but the amplitude is doubled.
Using the basis of the atomic and molecular Wannier functions $W_{\rm a}({\bf x})$,
$W_{\rm m}({\bf x})$ one can derive the discrete Hamiltonian of the system.
In the tight-binding approximation, the microscopic Hamiltonian is given by~\cite{DKOS05,DH06,DOS06}
\begin{eqnarray}
\label{H_am}
\hat H_{\rm am}
&=&
-
J_{\rm a}
\sum_{\bf l}
\sum_{\nu=1}^d
\left(
    \hat a^{\dagger}_{\bf l}
    \hat a^{\phantom \dagger}_{{\bf l}+{\bf e}_\nu}
    +
    {\rm h.c.}
\right)
-
J_{\rm m}
\sum_{\bf l}
\sum_{\nu=1}^d
\left(
    \hat b^{\dagger}_{\bf l}
    \hat b^{\phantom \dagger}_{{\bf l}+{\bf e}_\nu}
    +
    {\rm h.c.}
\right)
+
\delta
\sum_{\bf l}
\hat b^{\dagger}_{\bf l}
\hat b^{\phantom \dagger}_{\bf l}
\nonumber\\
&+&
\frac{U}{2}
\sum_{\bf l}
\hat a^{\dagger}_{\bf l}
\hat a^{\dagger}_{\bf l}
\hat a^{\phantom \dagger}_{\bf l}
\hat a^{\phantom \dagger}_{\bf l}
+
\tilde g
\sum_{\bf l}
\left(
\hat b^{\dagger}_{\bf l}
\hat a^{\phantom \dagger}_{\bf l}
\hat a^{\phantom \dagger}_{\bf l}
+
\hat a^{\dagger}_{\bf l}
\hat a^{\dagger}_{\bf l}
\hat b^{\phantom \dagger}_{\bf l}
\right)
+
\frac{U_{\rm m}}{2}
\sum_{\bf l}
\hat b^{\dagger}_{\bf l}
\hat b^{\dagger}_{\bf l}
\hat b^{\phantom \dagger}_{\bf l}
\hat b^{\phantom \dagger}_{\bf l}
+
U_{\rm am}
\sum_{\bf l}
\hat a^{\dagger}_{\bf l}
\hat a^{\phantom \dagger}_{\bf l}
\hat b^{\dagger}_{\bf l}
\hat b^{\phantom \dagger}_{\bf l}
\;,
\end{eqnarray}
where
$\hat a^{\dagger}_{\bf l}$~($\hat b^{\dagger}_{\bf l}$)
and
$\hat a^{\phantom \dagger}_{\bf l}$~($\hat b^{\phantom \dagger}_{\bf l}$)
are creation and annihilation operators of a single atom (molecule) at a lattice site ${\bf l}$,
$\delta=\Delta\mu(B-B_0)$ is a detuning from the Feshbach resonance.
Here, $\Delta\mu$ is the difference in magnetic moments of the two atoms and a molecule.
The atom-molecule conversion is determined by~\cite{DS04}
\begin{equation}
\tilde g
=
\hbar
\sqrt
{
 \frac{2\pi a_{\rm s} \Delta B \Delta\mu}{M}
}
\int
W_{\rm a}^2({\bf x})
W_{\rm m}({\bf x})
d{\bf x}
\;,
\end{equation}
where $W_{\rm a}({\bf x})$ and $W_{\rm m}({\bf x})$ are the Wannier functions for the atoms and molecules, respectively.
In the three-dimensional lattice and in the Gaussian approximation [see Eq.~(\ref{WGauss})],
it has the form~\cite{DKOS05,SBLDVDR07}
\begin{equation}
\tilde g
=
\hbar
\sqrt
{
 \frac
 {2\pi a_{\rm s} \Delta B \Delta\mu}
 {M(2 \pi a_{\rm ho}^2)^{3/4}}
}
\;.
\end{equation}
The on-site molecule-molecule and the atom-molecule interaction parameters $U_{\rm m}$, $U_{\rm am}$
are defined by the expressions similar to that for the atomic parameter $U$.
Due to the differences in the physical properties of the atoms and molecules discussed above,
the molecular tunneling parameter $J_{\rm m}$ is much smaller than the atomic one.
The Hamiltonian~(\ref{H_am}) does not conserve the number of atoms and molecules separately
but the total number of the atomic constituents
\begin{equation}
\hat N_{\rm t}
=
\sum_{\bf l}
\hat n_{\bf l}
=
\sum_{\bf l}
\left(
\hat a^{\dagger}_{\bf l}
\hat a^{\phantom \dagger}_{\bf l}
+
2
\hat b^{\dagger}_{\bf l}
\hat b^{\phantom \dagger}_{\bf l}
\right)
\end{equation}
is preserved.

In the regime of small tunneling and in the case of two atomic constituents, the low-energy properties of the system
can be described restricting the local Hilbert space of the Hamiltonian~(\ref{H_am}) by the states with two atoms or one molecule.
This leads to the mapping on the spin-$1/2$ quantum Ising model~\cite{BEHSEFS11}:
\begin{equation}
\label{H_Ising}
\hat H_{\rm am}
\approx
\hat H_{\rm Ising}
=
J_z
\sum_{\bf l}
\sum_{\nu=1}^d
\hat S^z_{{\bf l}}
\hat S^z_{{\bf l}+{\bf e}_\nu}
+
\sum_{\bf l}
\left(
    h_\parallel
    \hat S^z_{{\bf l}}
    +
    h_\perp
    \hat S^x_{{\bf l}}
\right)
+
{\rm const}
\;.
\end{equation}
The spin operators are given by
\begin{equation}
\hat S^+
=
\frac
{\hat b^\dagger \hat a \hat a}
{\sqrt{2}}
\;,\quad
\hat S^-
=
\frac
{\hat a^\dagger \hat a^\dagger \hat b}
{\sqrt{2}}
\;,\quad
\hat S^x
=
\frac{\hat S^+ + \hat S^-}{2}
\;,\quad
\hat S^z
=
\frac
{
\hat b^\dagger \hat b
-
\hat a^\dagger \hat a/2
}
{2}
\;,
\end{equation}
where we omitted the site index ${\bf l}$.
The first term in the Hamiltonian~(\ref{H_Ising}) stems from the virtual hoppings of atoms and molecules between the neighboring sites
and corresponds to an effective magnetic exchange interaction with the coupling constant
\begin{equation}
J_z
=
\frac{4 J_{\rm a}^2}{U_{\rm am}-U}
+
\frac{J_{\rm m}^2}{U_{\rm am}}
\;.
\end{equation}
Other parameters
\begin{equation}
h_\parallel
=
\delta - U
\;,\quad
h_\perp
=
2\tilde g \sqrt{2}
\end{equation}
correspond to the longitudinal and transverse magnetic fields.
This mapping facilitates understanding of the phase diagram of the system which will be discussed in Section~\ref{T0_am_diagram}.

\subsection{Two atomic constituents on one lattice site}

In the limit of vanishing tunneling, the Hamiltonian~(\ref{H_am}) becomes a sum of local terms.
Then the on-site problem for two atomic constituents can be easily solved
analytically. In this case there are two eigenmodes which are superpositions of the two-atom and
molecular states with the energies
\begin{equation}
\label{Eonsite}
E_\pm
=
\frac
{\delta+U}
{2}
\pm
\sqrt
{
  \left(
      \frac
      {\delta-U}
      {2}
  \right)^2
  +
  2
  \tilde g^2
}
\;,
\end{equation}
and the probability to find a molecule is given by
\begin{equation}
\label{pm}
p_{{\rm m}\pm}
=
\frac{1}{2}
\left[
    1
    \pm
    \frac
    {\delta-U}
    {
      \sqrt
      {
       \left(
           \delta-U
       \right)^2
       +
       8
       \tilde g^2
      }
    }
\right]
\;.
\end{equation}

\begin{figure}[tb]
\centering

\stepcounter{nfig}
\includegraphics[page=\value{nfig}]{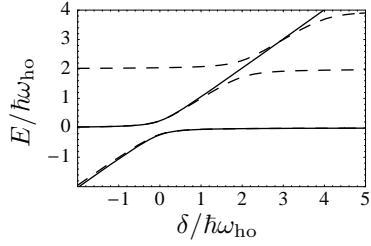}

\caption
{
Eigenenergies in the case of two atoms on the same lattice site.
Solid lines show the results given by Eq.~(\ref{Eonsite}) which correspond
to the lowest-band approximation. The results for the infinite number of bands
[Eq.~(\ref{emo})] are shown by dashed lines.
$U=0$, $2 \sqrt{\pi} \tilde g^2 / \left(\hbar\omega_{\rm ho}\right)^2=0.1$.
}
\label{ee}
\end{figure}

If the system is initially prepared in the state with two atoms at each site, the probability to find a molecule
will oscillate in time according to
\begin{equation}
p_{{\rm m}\pm}
=
\frac
{4\tilde g^2}
{\omega^2}
\left(
    1 - \cos \omega t
\right)
\;,\quad
\omega
=
\frac{E_+-E_-}{\hbar}
=
\sqrt
{
\left(
    \delta-U
\right)^2
+
8
\tilde g^2
}
\;.
\end{equation}
This kind of Rabi oscillations was observed in the experiments with $^{87}$Rb in deep optical lattices~\cite{SBLDVDR07}.

It is interesting to compare this solution with that obtained in Ref.~\cite{DKOS05} for two atomic constituents
on one site of a deep lattice for the infinite number of bands neglecting the atom-atom interaction.
In this case, each site can be described by a harmonic potential with the frequency $\hbar\omega_{\rm ho}$ and
the eigenenergies $E$ are determined by the equation
\begin{equation}
\label{emo}
E - \delta
=
\frac{2\sqrt{\pi}\tilde g^2}{\hbar \omega_{\rm ho}}
\frac
{\Gamma (- E / 2 \hbar \omega_{\rm ho})}
{\Gamma (- E / 2 \hbar \omega_{\rm ho} - 1/2)}
\;,
\end{equation}
where $\Gamma(x)$ is the Gamma-function.
The eigenenergies given by Eq.~(\ref{Eonsite}) for $U=0$ and the solutions of Eq.~(\ref{emo})
are plotted in Fig.~\ref{ee}. As we see, the lower branch $E_-$ in Eq.~(\ref{Eonsite})
is in excellent agreement with the lowest branch of Eq.~(\ref{emo}) for arbitrary $\delta$.
The upper branch $E_+$ fails to reproduce the second branch of Eq.~(\ref{emo})
if $\delta$ is far above the Feshbach resonance where the contribution of the second band
becomes significant, remaining however in a very good agreement near the resonance
and below it. This implies
that the lowest-band approximation is valid if the detuning
$\delta$ is less than
the gap between the two lowest Bloch bands, which is the quantity of the order of $\hbar\omega_{\rm ho}$,
and/or if we are interested in the eigenmodes of the Hamiltonian (\ref{H_am})
with the energies less than the energy of the second Bloch band.
In addition, the parameters $U$ and $\tilde g$ must be much smaller than the bands separation.

\subsection{\label{S_Bound-am}Two-body eigenmodes and bound states}

As in section~\ref{S_BS2atoms}, we consider a one-dimensional model with $L$ lattice sites assuming that $L$ is odd.
Under periodic boundary conditions the eigenstates of the Hamiltonian~(\ref{H_am}) for two atomic constituents have the form
\begin{equation}
\label{psi_k}
|K\Omega\rangle_{\rm am}
=
|K\Omega\rangle
+
c^{\rm m}_{K\Omega}
\sum_{j=0}^{L-1}
\left(
    \frac{\hat {\cal T}}{\tau_K}
\right)^j
|1_{\rm m} \underbrace{0 \dots 0}_{L-1} \rangle
\;,
\end{equation}
where $|K\Omega\rangle$ is the atomic part defined by Eqs.~(\ref{apsi_k}),~(\ref{nk2}) and
$|1_{\rm m} 0 \dots 0\rangle$ is a state with one molecule
on the first lattice site and all the other sites being unoccupied.
The eigenvalue problem for the Hamiltonian~(\ref{H_am}) can be written down as follows:
\begin{eqnarray}
\label{evp-am}
&&
\delta_K c_{K\Omega}^{\rm m}
+
\sqrt{2} \tilde g c_{K\Omega 0}
=
E^{K\Omega}_2
c_{K\Omega}^{\rm m}
\;,
\\
&&
\sqrt{2} \tilde g c_{K\Omega}^{\rm m}
+
H_K^{00} c_{K\Omega 0}
+
H_K^{01} c_{K\Omega 1}
=
E^{K\Omega}_2
c_{K\Omega 0}
\;,
\nonumber\\
&&
\sum_{\Gamma'=0}^{(L-1)/2}
H_K^{\Gamma,\Gamma'}
c_{K\Omega\Gamma'}
=
E^{K\Omega}_2
c_{K\Omega\Gamma}
\;,\quad
\Gamma=1,\dots,\frac{L-1}{2}
\;,
\nonumber
\end{eqnarray}
where
$
 \delta_K
 =
 \delta
 -
 2 J_{\rm m}
 \cos
 \left(
     K a
 \right)
$
and the nonvanishing matrix elements $H_K^{\Gamma,\Gamma'}$ are determined by Eq.~(\ref{HK-2a}).
The normalization condition takes the form
\begin{eqnarray}
\label{norma}
\left|
    c_{K\Omega}^{\rm m}
\right|^2
+
\sum_{\Gamma=0}^{(L-1)/2}
\left|
    c_{K\Omega\Gamma}
\right|^2
&=&
1
\;.
\end{eqnarray}
The aim of the present section is to investigate the influence of the molecular mode on
the bound state~\cite{KS2006,NPM2008,NPM08,SOJM2011,SGRR2011}.

As in the case of two atoms considered in section~\ref{S_BS2atoms},
the spectrum consists of two types of modes.
The scattering modes have the same energies as in Eq.~(\ref{EKk}).
The bound states have the form~(\ref{cKG}).
Substituting this ansatz into Eq.~(\ref{evp-am}), in the limit $L\to\infty$
we obtain the equation for the eigenenergy
\begin{equation}
\label{EK-am}
{\cal E}_K^2
=
U_K^2
+
q_K^2
\;,\quad
U_K
=
U
+
\frac{2\tilde g^2}{{\cal E}_K-\delta_K}
\;,
\end{equation}
where $q_K$ is defined in Eq.~(\ref{EKk}).
Formal comparison with Eq.~(\ref{E_k}) shows that
$U_K$ plays a role of the effective atomic interaction.
The corresponding values of $c_{K0}$ and $b_K$ are given by
\begin{equation}
\label{cK0-am}
c_{K0}
=
\sqrt
{
 \frac
 {
   \left(
       {\cal E}_K
       -
       \delta_K
   \right)^2
   \left(
       1
       -
       b_K^2
   \right)
 }
 {
   \left(
       {\cal E}_K
       -
       \delta_K
   \right)^2
   \left(
       1
       +
       b_K^2
   \right)
   +
   2 \tilde g^2
   \left(
       1
       -
       b_K^2
   \right)
 }
}
\;,\quad
b_K
=
\frac
{
 U_K
 -
 {\cal E}_K
}
{q_K}
\;,
\end{equation}
and the amplitude of the molecular state takes the form
\begin{equation}
c^m_{K}
=
\frac
{\sqrt{2} \tilde g c_{K0}}
{{\cal E}_K-\delta_K}
\;.
\end{equation}
Eqs.~(\ref{EK-am}),~(\ref{cK0-am}) remain unchanged under transformation
$U\to -U$, $\delta_K\to -\delta_K$, ${\cal E}_K\to -{\cal E}_K$, $b_K\to -b_K$.
Therefore, in order to get the complete solution it is enough to study the case of
attractive interaction in the whole range of $\delta_K$ and $\tilde g$.
The latter can be considered as positive because Eq.~(\ref{EK-am}) contains $\tilde g^2$ only.

Eq.~(\ref{EK-am}) can be multiplied by
$
 \left(
     {\cal E}_K
     -
     \delta_{K}
 \right)^2
$
and treated as quartic equation for ${\cal E}_K$ which contains always four roots.
However, depending on the values of the parameters only one or two roots are real
and provide normalized eigenstates with
$
  \left|
      b_K
  \right|
  < 1
$
implying that the others are unphysical and should be rejected.
In the special case $\tilde g=0$, $c_K^m$ vanishes and Eqs.~(\ref{EK-am}),~(\ref{cK0-am})
reduce to the solution~(\ref{E_k}),~(\ref{cKG}) in the absence of the Feshbach resonance.

Although analytical solutions of the quartic equation are well known,
simple expressions for ${\cal E}_K$ can be obtained only in some special cases.
For instance, one can easily show that in the special case $J_{\rm a}=J_{\rm m}=0$ the physical solutions
of Eq.~(\ref{EK-am}) are given by (\ref{Eonsite}).

In the limit of large detuning, $\left|\delta_K\right|\gg\left|{\cal E}_K\right|$,
the effective interaction parameter takes the form
\begin{equation}
\label{cond}
U_K
=
U - 2\tilde g^2/\delta_K
\end{equation}
which is equivalent to the expression for the effective scattering length
\begin{equation}
a_{\rm s}^{\rm eff}
=
a_{\rm s}
(1-\Delta B \Delta\mu/\delta)
\end{equation}
that appears in the mean-field theory as a result of the adiabatic elimination
of the molecular field~\cite{MVA95}.
In this limit, the solution for ${\cal E}_K$ is given by Eq.~(\ref{E_k})
with $U$ being replaced by $U_K$.
In the special case $U=\delta_K=0$, we obtain
\begin{equation}
{\cal E}_{K\pm}
=
\pm
\sqrt
{
 \sqrt
 {
  \frac{q_K^4}{4}
  +
  4 \tilde g^4
 }
 +
 \frac{q_K^2}{2}
}
\;.
\end{equation}

\begin{figure}[tb]
\centering

\stepcounter{nfig}
\includegraphics[page=\value{nfig}]{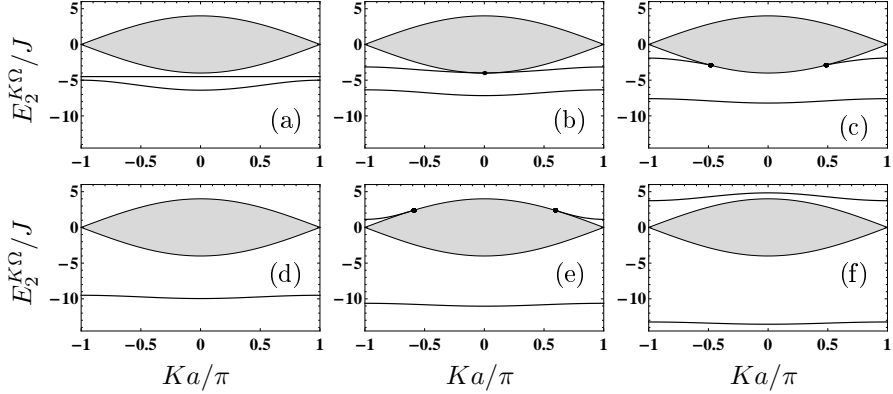}

\caption
{
Energy spectrum of two atoms in the presence of Feshbach resonance.
$U/J=-5$, $\delta/J=-4.5$;
$\tilde g/J=0$~(a), $\sqrt{\frac{U\delta+4UJ}{2J^2}}\approx 1.118$~(b), $2$~(c),
$\sqrt{\frac{U\delta}{2J^2}}\approx 3.35$~(d), $4.15$~(e), $6$~(f).
}
\label{E-Bound-am}
\end{figure}

In general, Eq.~(\ref{EK-am}) allows at least one bound state which corresponds
to that for the two atoms without Feshbach resonance in the limit of vanishing $\tilde g$.
With the increase of $\tilde g$, the absolute value of the energy of this state grows but
the sign remains unchanged.
The second bound state exists, provided that the absolute value of the quantity
$
Q_K
=
\left(
    U\delta_K - 2\tilde g^2
\right)
/
\left(
    U q_K
\right)
$
is larger than one.
$Q_K>1$ leads to positive $U_K$ and $Q_K<-1$ gives negative $U_K$.
As a consequence, the second mode exists for any value of $K$, if $\left|Q_0\right|>1$.
Otherwise, the second mode exists only for
$
|K|
>
K_{\rm c}
=
\frac{2}{a}
\arccos |Q_0|
$
within the first Brillouin zone.
In the expression for $K_{\rm c}$ we have neglected the molecular tunneling $J_{\rm m}$.
For $Q_0=0$, $K_{\rm c}=\pi/a$, i.e., the second bound state does not exist.
Different types of solutions are shown in Fig.~\ref{E-Bound-am}.

The properties of the bound states are determined by the values of the effective interaction
parameter. The state is attractively bound for $U_K<0$, even if the interaction parameter $U$ is
positive or vanishes. If $U_K>0$, the state is repulsively bound, no matter what the parameter
$U$ is. Moreover, attractively and repulsively bound states can coexist which is not possible
in the absence of Feshbach resonance.

\subsection{\label{T0_am_diagram}Zero-temperature phase diagram}

\begin{figure}[tb]
\centering

\stepcounter{nfig}
\includegraphics[page=\value{nfig}]{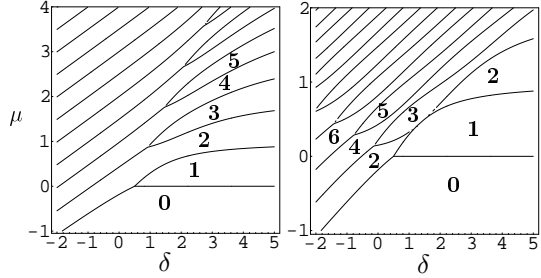}

\caption
{
Phase diagram in the local limit ($J_a=J_m=0$)  {\it vs} chemical potential $\mu$ 
and detuning $\delta$ from the Feshbach resonance.
$\tilde g=-0.5$, $U=U_{\rm m}=1$. $U_{\rm am}=1$ (left), $U_{\rm am}=0.25$ (right).
Each phase is labeled by the number $n_0$ of atomic constituents per site.
(Adapted with permission from Ref.~\cite{SD2005epl}).
}
\label{mu_delta}
\end{figure}

In the absence of tunnelings, $J_{\rm a}=J_{\rm m}=0$, and for an arbitrary integer number
$\langle\hat n_{\bf l}\rangle=n_0$
of the atomic constituents at each lattice site, the local eigenstates of the Hamiltonian are superpositions of the form
\begin{equation}
|\psi^{}\rangle
=
\sum_{n_m=0}^{[n_0/2]}
C_{n_m}
|n_{\rm a},n_{\rm m}\rangle
\;,
\end{equation}
where $n_{\rm m}$ is the number of molecules and the number of atoms $n_{\rm a}=n_0-2n_{\rm m}$.
The $(\mu,\delta)$ - diagram worked out in Ref.~\cite{SD2005epl} by numerical diagonalization of the Hamiltonian matrix
for different values of $n_0$ is shown in Fig.~\ref{mu_delta}. It consists of MI regions labeled by the values of $n_0$.
If the detuning is large ($|\delta/\tilde g|\gg1$) and $n_0$ is even, the MI state exists on both sides of the resonance.
For positive $\delta$ it is dominated by $|n_{\rm a}=n_0,n_{\rm m}=0\rangle$ and for negative $\delta$ by $|n_{\rm a}=0,n_{\rm m}=n_0/2\rangle$.
For odd $n_0$ and still for large $|\delta/\tilde g|$, the state exists only for positive $\delta$.
For large negative $\delta$, the ground state for odd $n_0$ is a superposition of Fock states with $n_m=(n_0\pm1)/2$
and is expected to be unstable against superfluidity for any finite tunneling.
The phase diagram for small detuning strongly depends on $U_{\rm am}$:
For small $U_{\rm am}$, the MI phase with odd $n_0$ can appear also for negative $\delta$.

In order to understand the phase diagram for nonvanishing tunneling parameters, it is useful to pay attention
to the fact that the Hamiltonian~(\ref{H_am}) is invariant under U(1)$\times Z_2$ transformation~\cite{LL06,EBHEFS11,BEEFHS12}:
\begin{equation}
\hat b_{\bf l}
\to
\hat b_{\bf l}
e^{i2\theta}
\;,\quad
\hat a_{\bf l}
\to
\pm
\hat a_{\bf l}
e^{i\theta}
\;,
\end{equation}
where $\theta$ is real. In the dimensions larger than one, the continuous U(1) symmetry may be broken without
breaking the discrete $Z_2$ symmetry $\hat a_{\bf l}\to\pm\hat a_{\bf l}$.
This leads to the molecular condensate phase (MC) with the order parameters $\langle\hat b_{\bf l}\rangle\ne0$,
$\langle\hat a_{\bf l}\rangle=0$ and corresponds to the Ising degree of freedom in the disordered phase coexisting with the molecular superfluidity.
On the other hand, the U(1)$\times Z_2$ symmetry can be completely broken which gives rise to the atomic-molecular condensate (AC+MC)
with nonvanishing $\langle\hat a_{\bf l}\rangle$, $\langle\hat b_{\bf l}\rangle$ and corresponds to a $Z_2$ ordered Ising degree of freedom
coexisting with the atomic and molecular superfluidity. If the U(1) symmetry is not broken, the system will be an insulator as
in the case of one-component bosons.
Mean-field studies reported in Refs.~\cite{SD2005epl,DKOS05,FRR15} for commensurate and incommensurate fillings are in agreement
with these general considerations.

In one dimension, breaking of the U(1) symmetry is prohibited and the formation of nonvanishing expectation values
$\langle\hat a_{\bf l}\rangle$ and $\langle\hat b_{\bf l}\rangle$ is excluded.
Low-energy effective theory based on the ``bosonization" approach~\cite{Giamarchi04} shows that in the atomic-molecular superfluid
phase which is analogous to AC+MC in higher dimensions the atomic and molecular one-body correlation functions
have a power-law decay at large distances~\cite{LL06,EBHEFS11,BEEFHS12}:
\begin{equation}
\langle
    \hat a^{\dagger}_{\ell_1}
    \hat a^{\phantom{\dagger}}_{\ell_2}
\rangle
\sim
|\ell_1-\ell_2|^{-\alpha_{\rm a}}
\;,\quad
\langle
    \hat b^{\dagger}_{\ell_1}
    \hat b^{\phantom{\dagger}}_{\ell_2}
\rangle
\sim
|\ell_1-\ell_2|^{-\alpha_{\rm m}}
\;.
\end{equation}
Due to the phase locking of the atomic and molecular components arising from the Feshbach term in the Hamiltonian~(\ref{H_am}),
$\alpha_{\rm m}=4\alpha_{\rm a}$.
In the other superfluid phase corresponding to MC, the molecular correlation function has the same power-law decay
but the atomic function decays exponentially~\cite{EBHEFS11,BEEFHS12}
\begin{equation}
\langle
    \hat a^{\dagger}_{\ell_1}
    \hat a^{\phantom{\dagger}}_{\ell_2}
\rangle
\sim
|\ell_1-\ell_2|^{-\alpha_{\rm a}-1/2}
e^{-|\ell_1-\ell_2|/\xi}
\;,
\end{equation}
where $\xi$ is the Ising correlation length. However, the correlation function of atomic pairs exhibits
a power-law behavior~\cite{EBHEFS11,BEEFHS12}:
\begin{equation}
\langle
    \hat a^{\dagger}_{\ell_1}
    \hat a^{\dagger}_{\ell_1}
    \hat a^{\phantom{\dagger}}_{\ell_2}
    \hat a^{\phantom{\dagger}}_{\ell_2}
\rangle
\sim
|\ell_1-\ell_2|^{-\alpha_{\rm m}}
\;.
\end{equation}
In both superfluid phases, the atomic-molecular correlation function
$\langle\hat b^{\dagger}_{\ell_1}\hat a^{\phantom{\dagger}}_{\ell_2}\hat a^{\phantom{\dagger}}_{\ell_2}\rangle$
decays as a power law with the exponent $\alpha_{\rm m}$ and all particle-number correlations have the same asymptotics
\begin{equation}
\langle
    \hat n_{\alpha\ell_1}
    \hat n_{\beta\ell_2}
\rangle
\approx
\langle
    \hat n_{\alpha\ell_1}
\rangle
\langle
    \hat n_{\beta\ell_2}
\rangle
+
\frac
{C_{\alpha\beta}}
{|\ell_1-\ell_2|^2}
\;,
\end{equation}
where $\alpha,\beta={\rm a},{\rm m}$ and $C_{\alpha\beta}$ are nonuniversal constants.

These field-theory predictions were successfully tested by DMRG calculations in one-dimensional chains up to $L=512$ sites
with two atomic constituents per site~\cite{EBHEFS11,BEEFHS12}. It was found that MC (AC+MC) undergoes MI transition, when the molecular (atomic)
correlation exponent $\alpha_{\rm m}$ ($\alpha_{\rm a}$) reaches the value $1/4$ which is exactly the critical value
for the Berezinskii-Kosterlitz-Thouless transition.
At the transition point from the AC+MC phase into the MI, the molecular exponent takes the value $\alpha_{\rm m}=1$.
This is an indication of the molecular superfluidity which signals the absence of a single-component atomic superfluidity close
to the MI boundary in contrast to Ref.~\cite{RD09} which was claiming the opposite based on QMC calculations.
The latter was attributed to finite-size effects~\cite{BEEFHS12}.

The transition between the two SF phases is expected to be in the universality class of the $(d+1)$-dimensional Ising model.
The values of the critical exponents of the correlation length and of the order parameter for the two-dimensional Ising model
($\nu=1$, $\beta=1/8$) are in perfect agreement with the DMRG calculations in one dimension~\cite{EBHEFS11,BEEFHS12}.

\begin{figure}[tb]
\centering

\stepcounter{nfig}
\includegraphics[page=\value{nfig}]{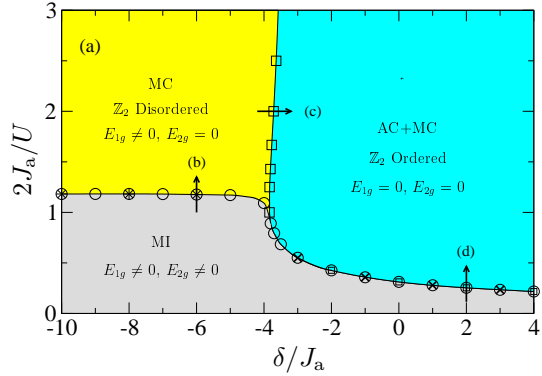}

\caption
{
(color online)
Phase diagram of the 1D Hamiltonian (\ref{H_am}) with two atomic constituents per site,
showing a Mott insulator (MI), a molecular condensate (MC), and a coupled atomic-molecular condensate (AC+MC).
The parameters are $J_{\rm m}=J_{\rm a}/2$, $\tilde g=U/2$, $U_{\rm m}=U$, $U_{\rm am}=U/4$.
The squares and circles indicate the vanishing of the one-particle and two-particle gaps, $E_{1g}$ and $E_{2g}$,
respectively, as $L\rightarrow\infty$. The stars and crosses indicate where the
molecular and atomic correlation exponents, $\alpha_{\rm m}$ and $\alpha_{\rm a}$
reach $1/4$ in the MC and AC+MC phases respectively. These values
correspond to a molecular and an atomic Berezinskii-Kosterlitz-Thouless transition, respectively.
(Adapted with permission from Ref.~\cite{BEEFHS12}. \copyright 2012, American Physical Society.)
}
\label{phd_am}
\end{figure}

The phase diagram of the one-dimensional chain with two atomic constituents per site obtained in Ref.~\cite{EBHEFS11} by DMRG method
(see also Ref.~\cite{BEEFHS12})
is shown in Fig.~\ref{phd_am}. The phase boundaries correspond to the vanishing of one-particle and two-particle excitation gaps ($n=1,2$)
$E_{ng}=\mu_{n+}(N_{\rm t})-\mu_{n-}(N_{\rm t})$ extrapolated to the thermodynamic limit, where
\begin{equation}
\mu_{n\pm}(N_{\rm t})
=
\pm
\left[
    E_0(N_{\rm t}\pm n) - E_0(N_{\rm t})
\right]/n
\;,
\end{equation}
and $E_0(N_{\rm t})$ is the ground-state energy for a system of fixed size $L$ with the total number $N_{\rm t}$ of the atomic constituents.
In the second-order of strong-coupling expansion, the MI phase reveals a phase transition from $Z_2$ disordered phase with
vanishing staggered magnetization
$\sum_\ell (-1)^\ell \langle \hat S_\ell^z\rangle/L$, where $\hat S^z=(\hat n_{\rm m}-\hat n_{\rm a}/2)/2$,
to the ordered phase with a finite staggered magnetization and long-range antiferromagnetic correlations~\cite{BEHSEFS11}.

The phase diagram similar to that shown in Fig.~\ref{phd_am} was also obtained in Ref.~\cite{RD09} by means of QMC simulations.
However, the MI phase was named ``super Mott" (see also~\cite{RD08}) based on the observation that the superfluid stiffnesses
of the atomic and molecular components calculated from the fluctuations of the corresponding winding numbers do not vanish,
although the superfluid stiffness of the whole system does. This interpretation was criticized as having no sense due to violation
of the particle-number conservation in the atomic and molecular subsystems~\cite{ER10}.

Zero-temperature phase diagram of the Hamiltonian~(\ref{H_am}) was recently studied in the case of two atomic constituents per site
for two-dimensional and three-dimensional lattices within the framework of the decoupling mean-field approximation [see Eq.~(\ref{dec_approx})]
complemented by QMC calculations for two-dimensional lattices~\cite{FRR15}. These studies reveal the same superfluid phases as
in one dimension accompanied by true Bose-Einstein condensation in agreement with the previous works~\cite{DKOS05,SD2005epl}.
It was pointed out that one has to distinguish between three types of insulators: molecular, atomic-molecular and Feshbach insulator.
While the molecular insulator is a conventional one with the same properties as in the case of one-component spinless bosons,
the atomic-molecular and Feshbach insulators are basically products of the superpositions of local states with two atoms and one molecule
on each site. The main difference between the atomic-molecular and Feshbach insulators is that the two-particle excitation gap $E_{2g}$
decreases with the increase of $\tilde g$ for the former but increases for the latter starting from $E_{2g}=0$ at $\tilde g=0$
and the transition between the two insulators is rather a crossover.
The transition from the superfluid into the Feshbach insulator as well as the transitions from the atomic-molecular superfluid to
all the insulating phases were found to be strongly first order.
It was also established that the mean-field theory captures correctly the succession of phases in the system.

In the case of two-component bosons corresponding to different atomic species near the Feshbach resonance which supports
the creation of heteronuclear molecules, the zero-temperature phase diagram has a reacher structure~\cite{BSHS09,HSBS10}.
It contains MI regions surrounded by the regions of single-component atomic and molecular superfluids
(in contrast to the homonuclear case discussed above) as well as the regions
with all three superfluids. However, there are no phases with just two superfluid components.
In the regime of small tunneling with two atomic constituents per site, the underlying Hamiltonian can be mapped
to the quantum Ising model with longitudinal and transverse fields as in the homonuclear case and, therefore,
reveals again the Ising phase transition.

Up to now the complete phase diagram of ultracold atoms in lattices near Feshbach resonance was not explored experimentally.
So far only the creation of the homonuclear molecular Mott insulator has been reported~\cite{VSBEDR06,DVSBHR06}.

\section{\label{sec-S1B}Spin-1 bosons}

If the atoms a trapped by purely optical means, the spin degree of freedom is not frozen.
We consider a dilute gas of bosonic atoms with hyperfine spin $F=1$ possessing three Zeeman-degenerate
internal ground states with magnetic quantum numbers $\alpha\equiv m_F=0,\pm 1$ in the field of an optical laser
described by a $3\times3$ matrix $V^{\rm las}({\bf x})$.
The system is governed by the following Hamiltonian~\cite{Ho,OM}:
\begin{eqnarray}
\label{Hs1}
&&
\hat H_{F=1}
=
\int
\left[
\hat\Psi_\alpha^\dagger({\bf x})
\left(
    -
    \frac{\hbar^2}{2M} \nabla^2
\right)
\hat\Psi_\alpha({\bf x})
+
\hat\Psi_\alpha^\dagger({\bf x})
V^{\rm las}_{\alpha\beta}({\bf x})
\hat\Psi_\beta({\bf x})
\right.
\\
&&
+
\left.
\frac{g_{\rm s}}{2}
\hat\Psi_\alpha^\dagger({\bf x})
\hat\Psi_\beta^\dagger({\bf x})
\hat\Psi_\beta({\bf x})
\hat\Psi_\alpha({\bf x})
+
\frac{g_{\rm a}}{2}
\hat\Psi_\alpha^\dagger({\bf x})
\hat\Psi_{\alpha'}^\dagger({\bf x})
{\bf F}_{\alpha\beta}
\cdot
{\bf F}_{\alpha'\beta'}
\hat\Psi_{\beta'}({\bf x})
\hat\Psi_\beta({\bf x})
\right]
d{\bf x}
\;,
\nonumber
\end{eqnarray}
where $\hat\Psi_\alpha({\bf x})$ is the bosonic field annihilation
operator for the atom in the hyperfine ground state $|F=1,\alpha\rangle$.
${\bf F}$ is a vector of traceless spin-1 matrices:
\begin{eqnarray}
F_1
=
\frac{1}{\sqrt{2}}
\left(
    \begin{tabular}{ccc}
    $0$ & $1$ & $0$\\
    $1$ & $0$ & $1$\\
    $0$ & $1$ & $0$
    \end{tabular}
\right)
\;,\quad
F_2
=
\frac{i}{\sqrt{2}}
\left(
    \begin{tabular}{ccc}
    $0$ & $-1$ & $0$\\
    $1$ & $0$  & $-1$\\
    $0$ & $1$  & $0$
    \end{tabular}
\right)
\;,\quad
F_3
=
\left(
    \begin{tabular}{ccc}
    $1$ & $0$ & $0$\\
    $0$ & $0$ & $0$\\
    $0$ & $0$ & $-1$
    \end{tabular}
\right)
\;.
\end{eqnarray}
In Eq.~(\ref{Hs1}), summation over repeated spin indices $\alpha,\beta$ is implied.

The parameters $g_{\rm s,a}$ describe the strength of the repulsive interactions of the atoms and the spin-changing collisions.
In three dimensions, they are related to the scattering lengths $a_0$ and $a_2$ of two colliding bosons of mass $M$
with total angular momenta $0$ and $2$ (singlet and quintuplet channels) as
\begin{equation}
g_{\rm s}
=
\frac{4\pi\hbar^2}{3M}
\left(
    a_0 + 2 a_2
\right)
\;,\quad
g_{\rm a}
=
\frac{4\pi\hbar^2}{3M}
\left(
    a_2 - a_0
\right)
\;.
\end{equation}
The values of $a_0$ and $a_2$ for atoms usually used in the experiments have been reviewed in Refs.~\cite{SU13,KU12}.
In the case of $^{7}$Li, $^{41}$K and $^{87}$Rb, $g_{\rm a}$ turns out to be negative, while for $^{23}$Na it is positive.

As long as all the atoms are in the internal state with $\alpha=-1$ or $\alpha=+1$,
the spin degrees of freedom do not play any role and all the physical properties
will be the same as for spinless atoms.
The spin degrees of freedom come into play when the internal states with different values
of $\alpha$ are populated by many atoms.
This property can be used to probe the particle-number statistics across the MI-SF transition~\cite{GFWMB06}.
If the atoms are initially prepared in the internal state $|F=1,m_F=-1\rangle$ and then transferred to the state with $m_F=0$,
a reversible exchange between the populations in the $m_F=0$ and $m_F=\pm1$ Zeeman sublevels is observed
unless the initial state is a product of local Fock states with the occupation numbers equal to one.

\subsection{Bose-Hubbard model}

If the detuning of all the lasers creating an optical lattice is much larger than the fine splitting of the electronic energy levels,
matrix $V^{\rm las}$ in Eq.~(\ref{Hs1}) becomes a scalar, i.e., all spin-components $\alpha=0,\pm1$
experience the same lattice potential as in the case of spinless atoms.
Expanding the field operators in the Wannier basis and using the tight-binding approximation,
we obtain~\cite{DZ02,SC03,ILD03,LSA12,SU13}
\begin{eqnarray}
\label{HBHs1}
\hat H_{F=1}^{\rm BH}
=
-
J
\sum_{\nu=1}^d
\sum_{\bf l}
\left(
    \hat a_{{\bf l}\alpha}^\dagger
    \hat a_{{\bf l}+{\bf e}_\nu,\alpha}^{\phantom\dagger}
    +
    {\rm h.c.}
\right)
+
\sum_{\bf l}
\left[
    \frac{U_{\rm s}}{2}
    \hat n_{\bf l}
    \left(
        \hat n_{\bf l} - 1
    \right)
    +
    \frac{U_{\rm a}}{2}
    \left(
        \hat{\bf L}_{\bf l}^2 - 2\hat n_{\bf l}
    \right)
\right]
\;,
\end{eqnarray}
where
\begin{equation}
\hat n_{\bf l}
=
\hat a_{{\bf l}\alpha}^\dagger
\hat a_{{\bf l}\alpha}^{\phantom\dagger}
\end{equation}
is an operator of the total number of atoms on site ${\bf l}$ and
\begin{equation}
\hat{\bf L}_{\bf l}
=
\hat a_{{\bf l}\alpha}^\dagger
{\bf F}_{\alpha\beta}
\hat a_{{\bf l}\beta}^{\phantom\dagger}
\end{equation}
is the spin operator on site ${\bf l}$. Its components obey the standard commutation relations for the angular momentum
\begin{equation}
\left[
    \hat L_{{\bf l}a_1} , \hat L_{{\bf l}a_2}
\right]
=
i
\epsilon_{a_1 a_2 a_3}
\hat L_{{\bf l}a_3}
\;,
\end{equation}
where $\epsilon$ is the completely antisymmetric Levi-Civita tensor.
The operator $\hat n_{\bf l}$ commutes with $\hat{\bf L}_{\bf l}$,
and the total spin operator $\hat{\bf L}=\sum_{\bf l}\hat{\bf L}_{\bf l}$ commutes with the Hamiltonian~(\ref{HBHs1}).

The interaction parameters $U_{\rm s,a}$ are given by Eq.~(\ref{U_d}) with $g$ replaced by $g_{\rm s,a}$ and their ratio is limited by
\begin{equation}
\label{Urestriction}
-1
<
\frac{U_{\rm a}}{U_{\rm s}}
<
\frac{1}{2}
\;,
\end{equation}
provided that both $a_0$ and $a_2$ are positive.
The tunneling matrix element for the nearest-neighboring sites $J$ is exactly the same as in the case of spinless atoms.
The term of the Hamiltonian~(\ref{HBHs1}) with the prefactor $U_{\rm a}$ has an explicit form
(the site index is omitted)
\begin{eqnarray}
\hat{\bf L}^2 - 2\hat n
&=&
\hat a_1^\dagger
\hat a_1^\dagger
\hat a_1^{\phantom\dagger}
\hat a_1^{\phantom\dagger}
+
\hat a_{-1}^\dagger
\hat a_{-1}^\dagger
\hat a_{-1}^{\phantom\dagger}
\hat a_{-1}^{\phantom\dagger}
-2
\hat a_{1}^\dagger
\hat a_{-1}^\dagger
\hat a_{1}^{\phantom\dagger}
\hat a_{-1}^{\phantom\dagger}
+2
\hat a_{1}^\dagger
\hat a_{0}^\dagger
\hat a_{1}^{\phantom\dagger}
\hat a_{0}^{\phantom\dagger}
\nonumber\\
&+&
2
\hat a_{-1}^\dagger
\hat a_{0}^\dagger
\hat a_{-1}^{\phantom\dagger}
\hat a_{0}^{\phantom\dagger}
+2
\hat a_{0}^\dagger
\hat a_{0}^\dagger
\hat a_{1}^{\phantom\dagger}
\hat a_{-1}^{\phantom\dagger}
+2
\hat a_{-1}^\dagger
\hat a_{1}^\dagger
\hat a_{0}^{\phantom\dagger}
\hat a_{0}^{\phantom\dagger}
\;,
\end{eqnarray}
where the last two summands describe the spin-changing collisions.
The latters were observed in experiments with $^{87}$Rb atoms in deep optical lattices~\cite{WGFGMB05,WGFGMB06}
and the measured differences of the scattering lengths $a_0-a_2$ agree with the theoretical predictions~\cite{KKHV02}
within $20\%$.

Sometimes it is convenient to work with the operators $\hat b_{a}$, $a=1,2,3$,~\cite{ILD03,PS07}
\begin{equation}
\label{b-op}
\hat b_1
=
\frac
{\hat a_{-1} - \hat a_{1}}
{\sqrt{2}}
\;,\quad
\hat b_2
=
-i
\frac
{\hat a_{-1} + \hat a_{1}}
{\sqrt{2}}
\;,\quad
\hat b_3
=
\hat a_0
\;,
\end{equation}
which satisfy the standard bosonic commutation relations and transform as vectors under spin rotations.
In therms of the operators $\hat b_a$, the particle-number operator can be expressed in a usual way,
$\hat n_{{\bf l}}=\hat b^\dagger_{{\bf l}a}\hat b^{\phantom{\dagger}}_{{\bf l}a}$,
and the spin operator is given by
\begin{eqnarray}
\hat L_{{\bf l}a_1}
&=&
-i\epsilon_{a_1 a_2 a_3}
\hat b_{{\bf l}a_2}^\dagger
\hat b_{{\bf l}a_3}^{\phantom\dagger}
\;,
\nonumber\\
\hat {\bf L}_{{\bf l}}^2
&=&
\hat n_{{\bf l}}
\left(
     \hat n_{{\bf l}} + 1
\right)
-
\hat b_{{\bf l}a_1}^\dagger
\hat b_{{\bf l}a_1}^\dagger
\hat b_{{\bf l}a_2}^{\phantom\dagger}
\hat b_{{\bf l}a_2}^{\phantom\dagger}
\;.
\label{L2b}
\end{eqnarray}
The transformation properties of $\hat b_{a}$, $a=1,2,3$, can be verified with the aid of the commutation relations
\begin{equation}
\left[
    \hat L_{a_1} , \hat b_{a_2}
\right]
=
i
\epsilon_{a_1 a_2 a_3}
\hat b_{a_3}
\;,\quad
\left[
    \hat L_{a_1} , \hat b_{a_2}^\dagger
\right]
=
i
\epsilon_{a_1 a_2 a_3}
\hat b_{a_3}^\dagger
\;.
\end{equation}
Using~(\ref{b-op}) and~(\ref{L2b}) the Hamiltonian~(\ref{HBHs1}) can be rewritten in the form
\begin{eqnarray}
H_{F=1}^{\rm BH}
=
-
J
\sum_{\nu=1}^d
\sum_{\bf l}
\left(
    \hat b_{{\bf l}a}^\dagger
    \hat b_{{\bf l}+{\bf e}_\nu,a}^{\phantom\dagger}
    +
    {\rm h.c.}
\right)
+
\sum_{\bf l}
\left[
    \frac{U_{\rm s}+U_{\rm a}}{2}
    \hat n_{\bf l}
    \left(
        \hat n_{\bf l} - 1
    \right)
    -
    \frac{U_{\rm a}}{2}
    \hat b_{{\bf l}a_1}^\dagger
    \hat b_{{\bf l}a_1}^\dagger
    \hat b_{{\bf l}a_2}^{\phantom{\dagger}}
    \hat b_{{\bf l}a_2}^{\phantom{\dagger}}
\right]
\end{eqnarray}
which is invariant under global spin rotations.

As in the case of spinless bosons, the eigenstates of the spin-1 lattice system can be studied
using the basis of local Fock states $|n_{{\bf l}1},n_{{\bf l}0},n_{{\bf l}-1}\rangle$.
On the other hand, magnetic properties are better described in the basis of spin states
$|n_{\bf l},L_{\bf l},L_{{\bf l}3}\rangle$ where the quantum numbers label
the eigenstates of the operators $\hat n_{\bf l}$ and $\hat{\bf L}_{\bf l}$.
The spin states can be uniquely expressed in terms of the Fock states~\cite{Wu96,TKK04}
and the Hamiltonian~(\ref{HBHs1}) imposes two constraints on the quantum number $L_{\bf l}$.
First, the total spin cannot be larger than the total number of particles, i.e., $L_{\bf l}\le n_{\bf l}$.
Second, bosonic symmetry under permutation of any two particles leads to the requirement that
$n_{\bf l}$ and $L_{\bf l}$ should have the same parity, i.e., $L_{\bf l}$ must be even (odd), if $n_{\bf l}$ is even (odd).
A rigorous proof of this statement was given in Ref.~\cite{Wu96}.

\subsection{Single-particle states}

Single-particle eigenstates in a homogeneous lattice with periodic boundary conditions can be written as
$|{\bf k}\alpha\rangle = \hat{\tilde a}_{{\bf k}\alpha}^\dagger|0\rangle$, $\alpha=0,\pm1$.
Although being a trivial generalization of the spinless case, they show already different magnetic properties.
$|{\bf k}0\rangle$ is a simplest example of a nematic state that has vanishing expectation values of all
spin-components, i.e., $\langle\hat L_{a}\rangle=0$ for $a=1,2,3$, but breaks the spin symmetry because
$\langle\hat L_1^2\rangle=\langle\hat L_2^2\rangle=1$ and $\langle\hat L_3^2\rangle=0$.

\subsection{Eigenstates of two atoms}

We consider first the eigenstates of two atoms on a single lattice site.
In this case there are six states and four of them are not influenced by the spin-changing collisions.
In terms of the Fock states the latters are given by
\begin{eqnarray}
|\psi_1\rangle = |2,0,0\rangle
\;,\quad
|\psi_2\rangle = |0,0,2\rangle
\;,\quad
|\psi_3\rangle = |1,1,0\rangle
\;,\quad
|\psi_4\rangle = |0,1,1\rangle
\;.
\end{eqnarray}
Two other eigenstates have the form
\begin{eqnarray}
|\psi_5\rangle
&=&
\sqrt{\frac{2}{3}}|0,2,0\rangle
+
\sqrt{\frac{1}{3}}|1,0,1\rangle
\label{psi_5}
\\
|\psi_6\rangle
&=&
\sqrt{\frac{1}{3}}|0,2,0\rangle
-
\sqrt{\frac{2}{3}}|1,0,1\rangle
\;.
\label{singlet-state}
\end{eqnarray}
In terms of the spin states, the six eigenstates are given by
\begin{eqnarray}
&&
|\psi_1\rangle
=
|2,2,2\rangle
\;,\quad
|\psi_2\rangle
=
|2,2,-2\rangle
\;,\quad
|\psi_3\rangle
=
|2,2,1\rangle
\;,\quad
\nonumber\\
&&
|\psi_4\rangle
=
|2,2,-1\rangle
\;,\quad
|\psi_5\rangle
=
|2,2,0\rangle
\;,\quad
|\psi_6\rangle
=
|2,0,0\rangle
\;.
\label{psi123456}
\end{eqnarray}
The state $|\psi_6\rangle$ is unique and has the energy $E=U_{\rm s}-2U_{\rm a}$,
while all the others are degenerate and their energy is $E=U_{\rm s}+U_{\rm a}$.
For positive $U_{\rm a}$, $|\psi_6\rangle$ is the ground state and for $U_{\rm a}<0$ it becomes an excited state.

Note that $|\psi_6\rangle$ is the only state among the others which has equal populations of all spin components $\alpha=0,\pm1$, i.e.,
$\langle\hat n_1\rangle=\langle\hat n_0\rangle=\langle\hat n_{-1}\rangle=2/3$.
It is a spin singlet and Eq.~(\ref{singlet-state}) defines the creation operator of a singlet pair~\cite{ILD03,SZ04}
\begin{eqnarray}
\hat A_{\rm sg}^\dagger
=
\frac{1}{\sqrt{6}}
\left(
    \hat a_0^\dagger \hat a_0^\dagger
    -
    2 \hat a_1^\dagger \hat a_{-1}^\dagger
\right)
\equiv
\frac{1}{\sqrt{6}}
\hat b_a^\dagger
\hat b_a^\dagger
\;.
\end{eqnarray}
Since $\hat A_{\rm sg}^\dagger$ commutes with the spin operator $\hat {\bf L}$, it changes neither the total spin nor the spin components $a=1,2,3$.
Moreover, from Eq.~(\ref{L2b}) one can see that the eigenstates of the total spin operator must be always eigenstates
of the ``singlet counting operator" $\hat A_{\rm sg}^\dagger\hat A_{\rm sg}^{\phantom{\dagger}}$.

The eigenstates of two atoms in the case of nonvanishing tunneling can be readily constructed from the solutions
for distinguishable and indistinguishable atoms discussed in section~\ref{S_BS2atoms}.
In the present case there are six bound states with the effective interaction parameters $U=U_{\rm s}+U_{\rm a}$
and $U=U_{\rm s}-2U_{\rm a}$. Due to the restriction~(\ref{Urestriction}), $U$ is always positive and
the energies of all bound states appear to be above the scattering (quasi-)continuum.

\subsection{Ground-state phase diagram}

In this section we shall discuss the ground-state phase diagram of the lattice spin-1 system in the absence of an external magnetic field.
As in the case of spinless bosons, it consists of the MI and SF phases. However, the spin degree of freedom
leads to new interesting aspects related to magnetic properties which appear to be completely different in the case of
positive and negative $U_{\rm a}$.

The difference between positive and negative $U_{\rm a}$ can be already expected looking at the ground state
of the Hamiltonian~(\ref{HBHs1}) in the limit of small tunneling.
In the case of negative $U_{\rm a}$ the ground state should prefer the largest possible values of $L_{\bf l}$,
while in the case of positive $U_{\rm a}$ the smallest possible values of $L_{\bf l}$ are favorable.
Moreover, we can expect differences between even and odd fillings in the case of positive $U_{\rm a}$
because the smallest value of $L_{\bf l}$ is zero in the former case and one in the latter.

\subsubsection{$U_{\rm a}=0$}

In the special case $U_{\rm a}=0$, the spin-dependent interaction is absent and the numbers of bosons in all components
$\alpha=0,\pm1$ are conserved separately. The ground states are highly degenerate and exhibit ``SU(3)-ferromagnetism"~\cite{KT13}.
In the case of integer fillings and for small $J/U_{\rm s}$, the ground state is predicted to be a nematic insulator~\cite{DZ02}.
This state has vanishing expectation values of all spin-components: $\langle\hat L_a\rangle=0$, $a=1,2,3$,
but the spin-rotational symmetry SO(3) is broken, while the time reversal symmetry is preserved.
The order parameter for the nematic state is a traceless symmetric tensor $Q$ with the entries~\cite{ILD03,DZ02,SZ04}
\begin{equation}
Q_{ab}
=
\langle
    \hat L_a
    \hat L_b
\rangle
-
\frac{\delta_{ab}}{3}
\langle\hat {\bf L}^2 \rangle
\;,\quad
a,b=1,2,3.
\end{equation}
In the case of integer fillings and large $J/U_{\rm s}$ as well as in the case of fractional fillings and arbitrary $J/U_{\rm s}$,
the ground state is a polar superfluid. In two and three dimensions, it breaks both U(1) and SO(3) symmetries~\cite{DZ02}.

Mean-field theory predicts the same zero-temperature phase diagram as in the spinless case and, in particular,
continuous transition from the MI into SF~\cite{PSP08} which agrees with QMC calculations in two dimension~\cite{PHRB13}.
However, at small but finite temperatures, the transition becomes first order and if the temperature is increased further
it is again continuous~\cite{PSP08}.

\subsubsection{$U_{\rm a}<0$}

According to the theorem proven in Ref.~\cite{KT13}, the ground state in the case of negative $U_{\rm a}$ exhibits
saturated ferromagnetism ($L_{\bf l}$ takes the largest possible value) in any dimension and both in the MI and SF phases.
Mean-field theory predicts in this case continuous transitions form the MI into SF with the phase boundary described by Eq.~(\ref{crit}),
where $U=U_{\rm s}+U_{\rm a}$~\cite{LPASZ11}. Continuous character of the phase transition was confirmed by QMC calculations
in one~\cite{BRS09} and two dimensions~\cite{PHRB13}. It was also shown that the phase diagram in one dimension can be
described very accurately using the third-order strong-coupling expansion for spinless bosons with $U=U_{\rm s}+U_{\rm a}$~\cite{BRS09}.
In two dimensions, the MI lobe for $\langle\hat n_{\bf l}\rangle=2$ obtained by the QMC method is well reproduced by the mean-field result,
whereas the MI lobe for $\langle\hat n_{\bf l}\rangle=1$ in QMC calculations is significantly larger than in the mean-field theory~\cite{PHRB13}.
QMC calculations also demonstrated that the populations of the spin-components in the MI and SF phases are
$\langle\hat n_{{\bf l}0}\rangle=2\langle\hat n_{{\bf l}1}\rangle=2\langle\hat n_{{\bf l}-1}\rangle=\langle\hat n_{\bf l}\rangle/2$~\cite{PHRB13}
in agreement with the mean-field theory~\cite{PSP08}.
Note that in the case of two atoms on one lattice site the state $|\psi_5\rangle$ determined by Eqs.~(\ref{psi_5}),~(\ref{psi123456}),
which has the largest possible spin $L=2$ and vanishing $L_3$, shows the same ratio of populations.

\subsubsection{$U_{\rm a}>0$}

In the case of positive $U_{\rm a}$ the system has a rich variety of phases with different magnetic ordering and the dimensionality
plays an important role. We consider first the case of even integer filling $\langle\hat n_{\bf l}\rangle$.
In the limit of vanishing tunneling, the ground state for each isolated site is a spin singlet (which is unique) and the excitations
are spin $L_{\bf l}=2,4,\dots,\langle\hat n_{\bf l}\rangle$ which are gapped by the energies of the order of $U_{\rm a}$.
Therefore, in all dimensions the ground state in this limit is a spin singlet Mott insulator~\cite{Z03}.
QMC calculations in one and two dimensions for $\langle\hat n_{\bf l}\rangle=2$ show that the populations of the components
$\alpha=0,\pm1$ are equal to each other within this phase~\cite{BRS09,PHRB13} which is a general property of the singlet states~\cite{KU12}.

If the tunneling grows, the spin singlet state transforms into the insulating nematic state~\cite{ILD03,DZ02,SZ04,BRS09,PHRB13},
provided that the ratio $U_{\rm a}/U_{\rm s}$ is small enough~\cite{ILD03,PHRB13}. Mean-field theory predicts in this case
a first-order phase transition~\cite{ILD03,SZ04} and provides an estimate of the transition point $J_{\rm c}$ which depends on the filling.
For two particles per site, $J_{\rm c}^2=U_{\rm s}U_{\rm a}/(4d)$, and for large even fillings
$J_{\rm c}^2=9U_{\rm s}U_{\rm a}/(2d\langle\hat n_{\bf l}\rangle^2)$~\cite{ILD03}.
The insulating nematic state exists only for small enough values of $U_{\rm a}/U_{\rm s}$, which are less than $0.05/d$ in the case of two
particles per site. These predictions of the mean-field theory were confirmed by QMC calculations in two dimensions~\cite{PHRB13}.
However, in one dimension QMC calculations reveal that the advent of the insulating nematic state is a crossover~\cite{BRS09}.

If the tunneling is increased further, the system undergoes a transition into the SF phase. However, the character of the transition
depends on $U_{\rm a}/U_{\rm s}$: if it is small, the transition is first order but for larger values it becomes second order.
This result was obtained first within the framework of the mean-field theory~\cite{KTK05,PSP08} and then confirmed by QMC calculations
in one and two dimensions~\cite{BRS09,PHRB13}.
Earlier mean-field studies provided an estimate of the largest value of $U_{\rm a}/U_{\rm s}$ which still allows the first-order
transition in the case of $\langle\hat n_{\bf l}\rangle=2$: $U_{\rm a}/U_{\rm s}\sim0.32$~\cite{KTK05},
which was later corrected to $U_{\rm a}/U_{\rm s}\sim0.2$~\cite{LPASZ11}.
The latter is in a better agreement with QMC calculations in two dimensions: $U_{\rm a}/U_{\rm s}\sim 0.15$~\cite{PHRB13}.

If $\langle\hat n_{\bf l}\rangle$ is odd, the situation becomes different. In the absence of tunneling, the ground state
for each individual site is a spin $L_{\bf l}=1$ state with a three-fold degeneracy. For two decoupled sites,
the ground state has a nine-fold degeneracy corresponding to the states with the total spin $L_{\rm tot}=0,1,2$.
Finite tunneling lifts this degeneracy but the form of the resulting state depends on the dimensionality
and other details~\cite{DZ02,Z03,ILD03}.

In one dimension, the MI phase in the case of odd fillings is always dimerized~\cite{Y03,Z03,ILD03,RRCMF05,RRCMF06,AS06,AS06b,LST06}.
The dimer state breaks translational symmetry and favors singlets on every second bond. Its distinguishing property
is the doubling of the unit cell of the lattice, while the spin long-range order is absent.
The simplest dimerized state can be written as~\cite{ILD03,RRCMF06}
\begin{equation}
|D\rangle
=
\bigotimes_{\ell\ {\rm odd}}
|L_\ell=1,L_{\ell+1}=1,L_\ell+L_{\ell+1}=0\rangle
\;,
\end{equation}
where the product could be also over even $\ell$, i.e., the state is doubly degenerate.
The dimerization can be described by looking at the expectation values of a pair Hamiltonian $\hat H_{\ell_1\ell_2}$ on adjacent bonds
($\hat H=\sum_{\langle\ell_1\ell_2\rangle}\hat H_{\ell_1\ell_2}$).
The corresponding order parameter reads $|\langle\hat H_{\ell-1,\ell}-\hat H_{\ell,\ell+1}\rangle|$~\cite{RRCMF05,RRCMF06}.

If the parameter $U_{\rm a}/U_{\rm s}$ tends to zero but remains finite, the amplitude of the dimer state becomes very small~\cite{LST06}
and the spectrum of excitations shows qualitative changes~\cite{PVC06,RES07}.
This might be an indication that the MI state in a one-dimensional lattice becomes nematic.
However, the limitations of different analytical and numerical methods applied in this regime do not allow to make
a certain statement (see, e.g., discussion in Ref.~\cite{LST06}).
In higher dimensions, insulating phases with an odd number of particles per site are always nematic~\cite{ILD03,SZ04,PHRB13,TTYIY12}.

With the increase of the tunneling parameter, the system enters again into the SF regime.
For $\langle\hat n_{\bf l}\rangle=1$, the transition is always continuous~\cite{BRS09,PHRB13,KTK05,PSP08,LPASZ11}.
For larger odd fillings and small values of $U_{\rm a}/U_{\rm s}$, it can also become first order,
although in this case the effect is not so pronounced as for even fillings~\cite{KTK05}.

The formation of the singlet pairs in the case of even fillings stabilizes the MI phase against the transition into the SF.
This leads to the asymmetry of the phase diagram: the size of the MI lobes is larger than for odd fillings.
This feature was demonstrated by DMRG~\cite{RRCMF05,RRCMF06,AS06,AS06b} and QMC calculations~\cite{PHRB13}
and also captured by the mean-field theory~\cite{TKK04,KTK05,LPASZ11}.

\subsection{Effective spin-$1/2$ Bose-Hubbard model}

Lin-$\theta$-lin laser configuration discussed in section~\ref{lin-t-lin} leads to two sets
of orthogonal Bloch eigenmodes denoted by the indices $0$ and $\Lambda$. This allows to derive effective
spin-$1/2$ Bose-Hubbard model from the Hamiltonian~(\ref{Hs1}).
In the tight-binding regime, the atoms stay always
in the lowest Bloch bands with the dispersion relations $E_0^{(0)}({\bf k})$ and $E_0^{(\Lambda)}({\bf k})$.
Then the spinor-field operator $\hat{\bf\Psi}({\bf x})$ can be decomposed as
\begin{equation}
\label{Psi}
\hat{\bf\Psi}({\bf x})=
\sum_{\bf l}
\sum_{\sigma=0,\Lambda}
{\bf W}_{\bf l}^{(\sigma)}({\bf x})
\hat a_{\sigma {\bf l}}
\,,
\end{equation}
where $\hat a_{\sigma{\bf l}}$ is the Bose annihilation operator for the $\sigma$-mode attached to the ${\bf l}$th lattice site.
${\bf W}_{\bf l}^{(\sigma)}({\bf x})\equiv{\bf W}^{(\sigma)}({\bf x}-{\bf x}_{\bf l})$
are three-component Wannier spinors for the lowest energy bands localized at the minima of the lattice potential labeled by ${\bf l}$,
which have the form similar to Eq.~(\ref{3c-spinors}).
They are obtained by the solution of the eigenvalue problem for the single atom discussed in section~\ref{lin-t-lin} and satisfy the
orthonormality condition
\begin{equation}
\int
{\bf W}^{(\sigma)\dagger}_{{\bf l}_1}({\bf x})
\cdot
{\bf W}^{(\sigma')}_{{\bf l}_2}({\bf x})
\,d{\bf x}
=
\delta_{{\bf l}_1 {\bf l}_2}
\delta_{\sigma\sigma'}
\;.
\end{equation}

Substituting Eq.~(\ref{Psi}) into Eq.~(\ref{Hs1}) and taking into account only the hopping between the nearest lattice sites and the
on-site atomic interactions, we obtain the two-component Bose-Hubbard Hamiltonian~\cite{KG04}
\begin{eqnarray}
\label{BHH}
&&
\hat H_{\rm BH}
=
-\sum_{\sigma}
    J_\sigma
\sum_{\nu=1}^d
\sum_{\bf l}
\left(
    \hat a_{\sigma{\bf l}}^\dagger
    \hat a_{\sigma{\bf l}+{\bf e}_\nu}^{\phantom{\dagger}}
    +
    {\rm h.c.}
\right)
+
\sum_{\sigma}
\frac{U_\sigma}{2}
\sum_{\bf l}
\hat n_{\sigma {\bf l}} (\hat n_{\sigma {\bf l}}-1)
\nonumber\\
&&
+
K
\sum_{\bf l}
\hat n_{0{\bf l}} \hat n_{\Lambda {\bf l}}
+
\frac{U_{\rm a}}{2}
\sum_i
\left(
    \hat a_{0{\bf l}}^\dagger
    \hat a_{0{\bf l}}^\dagger
    \hat a_{\Lambda {\bf l}}
    \hat a_{\Lambda {\bf l}}
    +
    \hat a_{\Lambda {\bf l}}^\dagger
    \hat a_{\Lambda {\bf l}}^\dagger
    \hat a_{0{\bf l}}
    \hat a_{0{\bf l}}
\right)
-
\delta
\sum_{\bf l}
\hat n_{0{\bf l}}
\,.
\end{eqnarray}
The tunneling matrix elements $J_\sigma$ already discussed in section~\ref{lin-t-lin}
as well as the atomic interaction parameters
\begin{eqnarray}
U_\Lambda
&=&
\int
\left[
    \left(
        g_s + g_a
    \right)
    \left(
        \left|
            W_{+{\bf l}}
    \right|^2
    +
        \left|
            W_{-{\bf l}}
    \right|^2
    \right)^2
    -
    4 g_a
    \left|
        W_{+{\bf l}}
    \right|^2
    \left|
        W_{-{\bf l}}
    \right|^2
\right]
\,d{\bf x}
\;,
\nonumber\\
U_0
&\equiv&
U_{\rm s}
=
g_s
\int
\left|
    W_{0{\bf l}}
\right|^4
\,d{\bf x}
\;,
\\
K
&=&
\left(
    g_s + g_a
\right)
\int
\left|
    W_{0{\bf l}}
\right|^2
\left(
    \left|
        W_{+{\bf l}}
    \right|^2
    +
    \left|
        W_{-{\bf l}}
    \right|^2
\right)
\,d{\bf x}
\;,
\nonumber\\
U_{\rm a}
&=&
2 g_a
\int
\left(
    W_{0{\bf l}}^*
\right)^2
W_{+{\bf l}}
W_{-{\bf l}}
\,d{\bf x}
\;,
\nonumber
\end{eqnarray}
and the relative shift of the mean energies of the eigenmodes
\begin{eqnarray}
\delta
=
\frac{1}{L^d}
\sum_{{\bf k}\in 1{\rm BZ}}
\left[
    E_0^{(\Lambda)}({\bf k})
    -
    E_0^{(0)}({\bf k})
\right]
\end{eqnarray}
can be simultaneously changed by varying the laser intensity and/or the angle $\theta$,
but the variations of $J_\sigma$ and $\delta$ are much faster.
The parameter $U_{\rm a}$ can be either positive or negative depending
on the sign of the antisymmetric coupling $g_a$.
We will consider the case of repulsive interactions when $U_0$ and $U_\Lambda$ are positive and of about equal size
but $K$ can be larger or smaller than $U_\sigma$ depending on the sign of the antisymmetric coupling $g_a$.

\subsubsection{"Ferromagnetic" and "antiferromagnetic" superfluid states}

In the case when only $\Lambda$-mode is populated the Hamiltonian~(\ref{BHH}) becomes equivalent to that of spinless bosons.
The only difference is that the tunneling matrix element $J_\Lambda$ can take not only positive but also negative values.
This leads to two different superfluid regimes which can be easily understood in the limit of the ideal gas.
In the usual situation of positive $J_\Lambda$ the eigenstates of non-interacting bosons are given by Eqs.~(\ref{e1p}),~(\ref{1pp})
and the ground state corresponds to ${\bf k}=0$, i.e., the phases of the wavefunction for any lattice site are the same.
In the case of negative $J_\Lambda$, the sign in Eq.~(\ref{e1p}) is reversed but Eq.~(\ref{1pp}) remains unchanged.
Now the ground state corresponds to $k_\nu a=\pi$ which leads to the phase shift of the wavefunction for the neighboring lattice sites.
In analogy to spin ordering in magnetic systems, one can call this ``ferromagnetic" and ``antiferromagnetic" phase ordering~\cite{KG03}.

In the presence of nonvanishing interactions this qualitative picture is preserved and the two different superfluid phases
are readily distinguishable experimentally via the spatial interference pattern generated by the
coherent matter waves which one obtains in the time-of-flight images:
The interference maxima obtained in the ferromagnetic case turn into
minima in the antiferromagnetic case and vice versa.
Deep in the MI phase, the phase coherence is lost and the sign of $J_\Lambda$ does not play any role.

\begin{figure}[t]
\centering

\stepcounter{nfig}
\includegraphics[page=\value{nfig}]{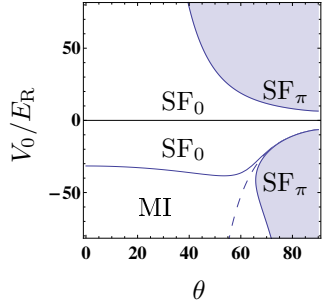}

\caption
{
Phase diagram in the $(\theta,V_0)$ plane for $(\mu/U)_{\rm c}=\sqrt{2}-1$.
$V_0<0$:
The boundaries between the SF and the MI ($\langle\hat n_{\bf l}\rangle=1$) phases are shown by solid lines.
The dashed line corresponds to $J_\Lambda=0$.
It lies always in the MI phase, separating the regions of the ferromagnetic (${\rm SF}_0$) and antiferromagnetic (${\rm SF}_\pi$) superfluid phases.
$V_0>0$:
The line $J_\Lambda=0$ as well as the two boundaries separating ${\rm SF}_0$ and ${\rm SF}_\pi$ superfluid phases
from the Mott phase are indistinguishable on the large-scale plot. The MI phase is located in the extremely narrow region between
the SF phases.
}
\label{pd_lambda}
\end{figure}

The phase diagram of the system is determined by the ratio $J_\Lambda/U_\Lambda$.
In the case of spinless bosons the ratio of the tunneling rate to the interaction parameter is a monotonic function of $|V_0|$.
In the two-component case with the $\Lambda$-coupling we are dealing with, it has quite different properties.
Its dependence on the control parameters $V_0$ and $\theta$ is not monotonic and there is a change of sign.
Therefore, it is reasonable to draw $\mu-V_0-\theta$ diagrams instead of $\mu-J$ diagrams.
At the points $(V_0,\theta)$, where $J_\Lambda/U_\Lambda$ vanishes, we have the MI phase for any values of $\mu/U$.
These points define the boundary between the ferromagnetic and antiferromagnetic states.
The mean-field phase diagram in the plane spanned by $\theta$ and $V_0$ for $(\mu/U)_{\rm c}=\sqrt{2}-1$ corresponding to the tip
of the MI lobe for $\langle\hat n_{\bf l}\rangle=1$ is shown in Fig.~\ref{pd_lambda}.
The MI phase is strongly suppressed in the case $V_0>0$ due to the dominant contribution of the dark component into the ground state.

\subsubsection{First- and second-order phase transitions}

In the present section, we are interested in the situations when both modes are occupied.
Therefore we have to restrict ourselves to negative $V_0$ and small values of $\theta$.
In this case, $J_\Lambda$ as well as $J_0$ are positive quantities.
If $\theta=0$, the $\Lambda$ and $0$ modes are degenerate and $\delta$ vanishes.
In addition, the components of the Wannier spinors satisfy the relation
$W_{+{\bf l}}=W_{-{\bf l}}=W_{0{\bf l}}/\sqrt{2}$ and we have
\begin{equation}
\label{simpl}
J_0 = J_\Lambda \equiv J
\;,\;
U_0 = U_\Lambda \equiv U_{\rm s}
\;,\;
K = U_{\rm s} + U_{\rm a}
\;,
\end{equation}
i.e., the Hamiltonian becomes symmetric with respect to the exchange of the indices $0$ and $\Lambda$ of the bosonic operators.
Very useful representation of the Hamiltonian can be obtained in this case with the aid of the isospin operator
$\hat{\bf S}_{\bf l}$ with the components
\begin{eqnarray}
\hat{S}_{1{\bf l}}
&=&
\frac{1}{2}
\left(
\hat{a}_{\Lambda {\bf l}}^\dagger
\hat{a}_{0 {\bf l}}^{\phantom\dagger}
+
\hat{a}_{0 {\bf l}}^\dagger
\hat{a}_{\Lambda {\bf l}}^{\phantom\dagger}
\right)
\:,
\nonumber\\
\hat{S}_{2{\bf l}}
&=&
\frac{i}{2}
\left(
\hat{a}_{\Lambda {\bf l}}^\dagger
\hat{a}_{0 {\bf l}}^{\phantom\dagger}
-
\hat{a}_{0 {\bf l}}^\dagger
\hat{a}_{\Lambda {\bf l}}^{\phantom\dagger}
\right)
\;,
\\
\hat{S}_{3{\bf l}}
&=&
\frac{1}{2}
\left(
    \hat{a}_{0 {\bf l}}^\dagger
    \hat{a}_{0 {\bf l}}^{\phantom\dagger}
    -
    \hat{a}_{\Lambda {\bf l}}^\dagger
    \hat{a}_{\Lambda {\bf l}}^{\phantom\dagger}
\right)
\;,
\nonumber
\end{eqnarray}
which has the property
\begin{equation}
\hat{\bf S}_{\bf l}^2
=
\frac{\hat n_{\bf l}}{2}
\left(
    \frac{\hat n_{\bf l}}{2} + 1
\right)
\;,
\end{equation}
where
\begin{equation}
\hat n_{\bf l}
=
\hat{a}^\dagger_{0 {\bf l}}
\hat{a}_{0 {\bf l}}^{\phantom\dagger}
+
\hat{a}_{\Lambda {\bf l}}^\dagger
\hat{a}_{\Lambda {\bf l}}^{\phantom\dagger}
\end{equation}
is an operator of the total particle number on site ${\bf l}$.
Note that the components of the operator $\hat{\bf S}_{\bf l}$ generate the SU(2) algebra.
In this notations, the Hamiltonian takes the form
\begin{eqnarray}
\label{BHHsim}
\hat H_{\rm BH}
=
- J
\sum_{\sigma}
\sum_{\nu=1}^d
\sum_{\bf l}
\left(
    \hat a_{\sigma{\bf l}}^\dagger
    \hat a_{\sigma{\bf l}+{\bf e}_\nu}^{\phantom{\dagger}}
    +
    {\rm h.c.}
\right)
+
\frac{U_{\rm s}}{2}
\sum_{\bf l}
\hat n_{{\bf l}} (\hat n_{{\bf l}}-1)
-
\frac{U_{\rm a}}{2}
\sum_{\bf l}
\hat n_{{\bf l}}
+2 U_{\rm a}
\sum_{\bf l}
\hat S_{1 {\bf l}}^2
\,.
\end{eqnarray}

If the tunneling is negligible, the local eigenstates of the Hamiltonian~(\ref{BHHsim})
coincide with the eigenstates $|n/2,{\cal M}\rangle$ of the isospin operator with the corresponding eigenenergies given by
\begin{eqnarray}
\label{e}
E^{(0)}_{n/2,{\cal M}}
=
\frac{U_{\rm s}}{2}
n (n - 1)
+
U_{\rm a}
\left(
    2
    {\cal M}^2
    -
    \frac{n}{2}
\right)
\;,
\end{eqnarray}
where $n$ is the number of atoms on a lattice site and
${\cal M}=-n/2,\dots,n/2$ is the isospin projection on the direction $1$ in the isospin space.
Note that $n/2$ plays the role of the isospin quantum number.
If $n$ is even, the states with ${\cal M}=0$ are unique, while the others with ${\cal M} \ne 0$ are doubly degenerate.
If $n$ is odd, the states with ${\cal M}=0$ do not exist and, therefore, all the eigenstates in the absence of tunneling are doubly degenerate.

If $U_{\rm a}<0$, the ground state is described by ${\cal M}=\pm n/2$.
In the case of $U_{\rm a}>0$, one has to distinguish between odd and even $n$,
where the ground state corresponds to ${\cal M}=0$ and ${\cal M}=\pm 1/2$, respectively.
In the basis of the eigenstates of the isospin operator, nonvanishing matrix elements of the bosonic operators are given by
\begin{eqnarray}
\left\langle
\frac{n-1}{2},{\cal M} \mp \frac{1}{2}
\right|
\hat a_\Lambda
\left|
\frac{n}{2},{\cal M}
\right\rangle
&=&
\pm
\left\langle
\frac{n-1}{2},{\cal M} \mp \frac{1}{2}
\right|
\hat a_0
\left|
\frac{n}{2},{\cal M}
\right\rangle
\nonumber\\
&=&
\frac{\sqrt{n \pm 2{\cal M}}}{2}
\exp
\left(
    \mp i
    \frac{\pi}{4}
\right)
\;.
\end{eqnarray}

Zero-temperature mean-field phase diagram of the symmetric spin-$1/2$ Bose-Hubbard model described by the Hamiltonian~(\ref{BHHsim})
was studied in details in Refs.~\cite{KG04,KTG05,PTHRSB10,PHRSB11}.
The formalism is based on the decoupling approximation similar to Eq.~(\ref{dec_approx})
\begin{equation}
\label{mfa}
\hat a_{\sigma {\bf l}_1}^\dagger
\hat a_{\sigma {\bf l}_2}^{\phantom\dagger}
\approx
\psi_{\sigma {\bf l}_1}^*
\hat a_{\sigma {\bf l}_2}^{\phantom\dagger}
+
\hat a_{\sigma {\bf l}_1}^\dagger
\psi_{\sigma {\bf l}_2}^{\phantom*}
-
\psi_{\sigma {\bf l}_1}^*
\psi_{\sigma {\bf l}_2}^{\phantom*}
\;,
\end{equation}
where
$\psi_{\sigma {\bf l}}=\langle \hat a_{\sigma {\bf l}} \rangle$
is the order parameter for Bose-Einstein condensation
in the component $\sigma=0,\Lambda$, which can be considered as a real and position-independent quantity.
The free energy per lattice site ${\cal F}$ is in general independent of the sign of $\psi_\sigma$,
$
{\cal F}(\psi_\Lambda,\psi_0)={\cal F}(|\psi_\Lambda|,|\psi_0|)
$,
and it is a symmetric function of $\psi_0$ and $\psi_\Lambda$:
${\cal F}(\psi_0,\psi_\Lambda)={\cal F}(\psi_\Lambda,\psi_0)$.

\begin{figure}[t]
\centering

\stepcounter{nfig}
\includegraphics[page=\value{nfig}]{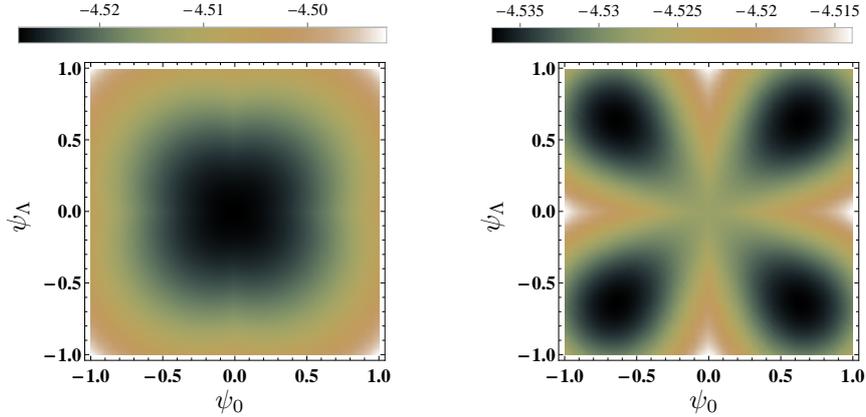}

\caption
{
(color online)
Free energy of the Hamiltonian (\ref{BHHsim}) for $^{87}$Rb
($U_{\rm a}/U_{\rm s} \approx -0.009$~\cite{Ho}), $\mu/U_{\rm s}=2.5$.
$2 d J/U_{\rm s}=0.03$~{(a)}, $0.09$~{(b)}.
}
\label{egn}
\end{figure}

Typical dependences of ${\cal F}$ on the order parameters $\psi_\Lambda$ and $\psi_0$
in the case $U_{\rm a}<0$ are shown in Fig.~\ref{egn}.
As long as the ratio $J/U_{\rm s}$ is small, there is a single minimum at $\psi_0=\psi_\Lambda=0$ corresponding to the MI phase.
For larger values of $J/U_{\rm s}$ the single minimum transforms into four equal minima
located on the lines $\psi_\Lambda=\pm \psi_0$.
This corresponds to the SF phase with equal contributions of both components.
As in the case of spinless bosons, the transition is continuous, i.e., second order,
and the phase boundary is described by Eq.~(\ref{crit}) with $U=U_{\rm s}-|U_{\rm a}|$.

\begin{figure}[t]
\centering

\stepcounter{nfig}
\includegraphics[page=\value{nfig}]{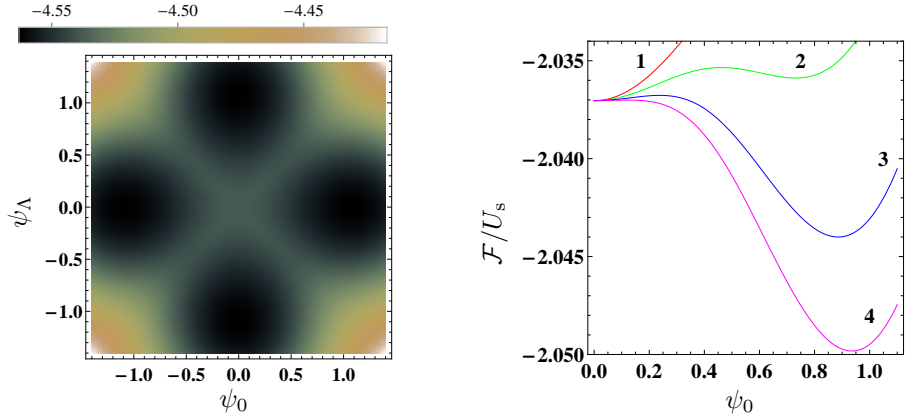}

\caption
{
(color online)
Free energy of the Hamiltonian~(\ref{BHHsim}) for $^{23}$Na ($U_{\rm a}/U_{\rm s} \approx 0.037$~\cite{Ho}).
{(a)}
$\mu/U_{\rm s}=2.5$, $2 d J/U_{\rm s}=0.12$.
{(b)}
$\psi_\Lambda=0$, $\mu/U_{\rm s}=1.5$,
$2 d J/U_{\rm s}=0.125(1),0.148(2),0.157(3),0.167(4)$.
}
\label{egp}
\end{figure}

The case of positive $U_{\rm a}$ is quite different as one can see in Figs.~\ref{egp}
which shows typical dependences of the ground-state energy per lattice site on the order parameters.
At small values of $J/U_{\rm s}$, there is again only one minimum at $\psi_\Lambda=\psi_0=0$ which corresponds to the MI phase.
If the ratio $J/U_{\rm s}$ is increased, four equal minima appear on the lines $\psi_\Lambda=0$ and $\psi_0=0$
which means that the superfluid is polarized. In addition, the minimum at $\psi_\Lambda=\psi_0=0$ does not always disappear
immediately, but only if $J/U_{\rm s}$ is further increased.
This implies that in such cases the phase transition is discontinuous, i.e., of the first order,
and in a certain range of $J/U_{\rm s}$ the two phases coexist.

\begin{figure}[t]
\centering

\stepcounter{nfig}
\includegraphics[page=\value{nfig}]{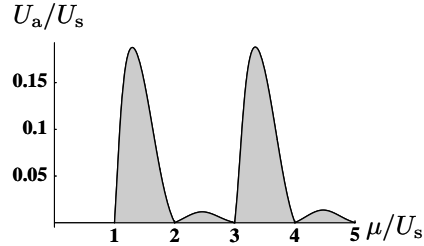}

\caption
{
In the shaded regions of this diagram ${\cal F}(\psi)$ has two minima at certain values of $J$.
In the remaining part ${\cal F}(\psi)$ has only one minimum.
(Reproduced from Ref.~\cite{KTG05}, \copyright 2005 American Physical Society.)
}
\label{pmu}
\end{figure}

\begin{figure}[t]
\centering

\stepcounter{nfig}
\includegraphics[page=\value{nfig}]{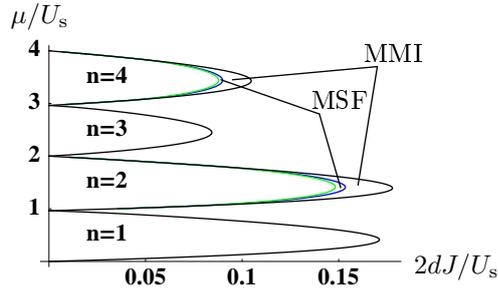}

\caption
{
(color online)
Phase diagram for $^{23}$Na ($U_{\rm a}/U_{\rm s} \approx 0.037$~\cite{Ho}).
The regions of metastable SF phase coexisting with the stable MI phase and that of metastable MI phase
coexisting with the stable SF phase are denoted by MSF and MMI, respectively.
(Adapted from Ref.~\cite{KTG05}, \copyright 2005 American Physical Society.)
}
\label{pdp}
\end{figure}

The values of $\mu$ and $U_{\rm a}$ which allow the first-order transition are shown in Fig.~(\ref{pmu}).
For $n=1$, the transition is always second order.
First-order transition is possible for $n \ge 2$ and $U_{\rm a}/U_{\rm s}$ smaller than some critical value,
which is about $0.188$ for even $n$ and grows from $0.012$ ($n=3$) to $0.015$ ($n\to\infty$) for odd $n$.
In the case of $^{23}$Na shown in Fig.~\ref{pdp}, an interesting regime is achieved,
when the QPT for odd $n$ is second order, but for even $n$ it is first order.

The phase diagram for $^{23}$Na is presented in Fig.~\ref{pdp}.
It consists of a series of (internal) lobes corresponding to the stable Mott phase
and external regions corresponding to the stable superfluid phase. However,
in the case of even $n$, the two regions are separated from one another by intermediate
ones, where the stable and metastable superfluid and Mott phases coexist.
The boundary separating the region of the stable superfluid phase from other ones
can be calculated exactly in the mean-field theory using the second-order perturbation theory.
In the case of odd $n$, the boundary is given by
\begin{eqnarray}
J'
&=&
4
\left[
    \frac{n-1}{\mu'-n+1+2U_{\rm a}'}
    +
    \frac{n+3}{n+U_{\rm a}'-\mu'}
    +
    2
    (n+1)
    \left(
        \frac{1}{n-U_{\rm a}'-\mu'}
	+
        \frac{1}{\mu'-n+1}
    \right)
\right]^{-1}
\;,
\end{eqnarray}
where $J'=2 d J/U_{\rm s}$, $U_{\rm a}'=U_{\rm a}/U_{\rm s}$, and $\mu'=\mu/U_{\rm s}$ with
$U_{\rm s}(n-1)<\mu<U_{\rm s} n - U_{\rm a}$.
For even $n$, it is described by the equation
\begin{equation}
J'
=
\frac
{
 \left(
     \mu'-n+1+U_{\rm a}'
 \right)
 \left(
     n-\mu'
 \right)
}
{
 \mu'-n+1+U_{\rm a}'
 +
 \left(
     1+U_{\rm a}'
 \right)
 n/2
}
\;.
\end{equation}
where $U_{\rm s}(n-1)-U_{\rm a}<\mu<U_{\rm s} n$.

The predictions of the mean-field theory have been tested by QMC calculations with one and two bosons per site
in one and two dimensions~\cite{PTHRSB10,PHRSB11,PHRB12}. It was verified that for ferromagnetic interactions ($U_{\rm a}<0$)
the phases are unpolarized and the transitions are continuous. In the case of antiferromagnetic interactions ($U_{\rm a}>0$)
it was confirmed that the SF phase is always dominated by one component.
The first-order transitions were found in two dimensions with two bosons per lattice site and for $U_{\rm a}/U_{\rm s}\lesssim0.25$
which is in good agreement with the results shown in Fig.~\ref{pmu}.
However, the mean-field theory fails to describe all the features of the MI phase at finite tunneling and of the one-dimensional
systems in general. QMC calculations demonstrated that all the transitions in one dimension are continuous.
In addition, it was shown that the populations of the components in the case of one atoms per site become unbalanced already
in the MI phase at finite ratios $J/U_{\rm s}$ and this polarization persists through the MI-SF transition.
It was also shown that thermal fluctuations immediately destroy the polarization of the MI phase, while the SF phase
is less sensitive to that at least for small temperatures~\cite{PHRB12}.

\section{Concluding remarks}

In this review, we presented studies of equilibrium properties of ultracold bosons with short-range interactions in optical lattices
of the simplest hypercubic geometries. Although this kind of systems displays already quite rich physics, this is just a tiny part of activities
in a huge area of research on cold atoms in optical lattices. Other topics include atoms with long-range dipole-dipole interactions,
disordered systems, mixtures of different bosonic species, fermions as well as Bose-Fermi mixtures, lattices with more complicated
geometric structures, e.g., hexagonal and triangular.
We just mention intriguing experimental observations of Anderson localization and Bose-glass phase
in incommensurate lattices~\cite{FLGFI07,REFFFZMMI08},
observation of effective multi-body interactions up to the six-body case in a three-dimensional lattice~\cite{WBSHLB10},
experimental realization of strong effective magnetic fields in a two-dimensional optical superlattice~\cite{AANTCB11,AANTCB13},
quantum simulations of a one-dimensional chain of interacting Ising spins in the presence of longitudinal and transverse fields~\cite{SBMTPG11},
in situ studies of photon-assisted tunneling in a Mott insulator by amplitude modulation of a tilted optical lattice~\cite{MTPBSG11},
realization of a finite-momentum superfluid in the lowest $P$-band of a square lattice with two different depths of the potential wells
arranged in a checkerboard pattern~\cite{WOH11},
studies of quantum magnetism in high-spin systems~\cite{PSCMHPSGVL13},
observation of spin-exchange interactions with polar molecules~\cite{YMGCHRJY13}.

Another field of research which became very popular in the last years is the study of nonequilibrium phenomena.
This is due to the fact that ultracold atoms are very well isolated from the environment which makes possible experimental investigations
of relaxation and thermalization in closed quantum systems~\cite{KWW06,TCFMSEB12,GKLKRSMSADS12}
as well as the dynamics of nonlocal quantum correlations~\cite{CBPESFGBKK12}.
This is a challenge for theorists because exact calculations of the dynamics of interacting quantum systems remain a difficult problem.
In spite of a great progress achieved for one-dimensional systems with the aid of matrix-product states,
the methods for higher-dimensional systems are not so well developed.
Some progress in this direction is achieved by the dynamical mean-field theory~\cite{EKW09,EKW10,AGPTW10,WTE12}
and Monte Carlo methods~\cite{GA12,CBSSF14}.
An alternative approach was recently suggested in Refs.~\cite{NS10,QNS12,QKNS14,KNQS14} which deals with the dynamical equations for the reduced
density matrices of different number of lattice sites.
However, the full theoretical description of sufficiently large systems
in the whole parameter range and for arbitrarily long times is still not reached.

\section*{Acknowledgment}

I am grateful to
\mbox{G.~Astrakharchik},
\mbox{N.~ten~Brinke},
\mbox{R.~Egger},
\mbox{L.~de~Forges~de~Parny},
\mbox{R.~Graham},
\mbox{D.~A.~W.~Hutchinson},
\mbox{P.~Navez},
\mbox{J.~Oertel},
\mbox{A.~Osterloh},
\mbox{J.~Larson},
\mbox{M.~Lewenstein},
\mbox{A.~Pelster},
\mbox{C.~Schneider},
\mbox{R.~Sch\"utzhold},
\mbox{D.~V.~Skryabin},
\mbox{N.~Szpak},
\mbox{M.~Thorwart}
for useful discussions and fruitful collaborations.
Special thanks to F.~Queisser for critical reading of the manuscript and helpful comments.
This work was supported by the SFB/TR 12 of the German Research Foundation (DFG).

\bibliographystyle{elsarticle-num-modified}
\bibliography{review}

\end{document}